\RequirePackage{etoolbox}
\csdef{input@path}%
{%
 {sty/},
 {img/},
}
\documentclass{elsevierbook}

\usepackage{natbib}
\begin{document}


  
\begin{frontmatter}

\chapter{Digital Reconfigurable Intelligent Surfaces: On the Impact of Realistic Reradiation Models}\label{chap1}
\subchapter{Analysis of sub-wavelength implementations, quantization of the reflection coefficient, interplay between the amplitude and phase of the reflection coefficient, near-field and far-field regions, electromagnetic interference}
\vspace{0.75cm}
\subchapter{Marco Di Renzo$^{(1)}$, Abdelhamed Mohamed$^{(1)}$, Alessio Zappone$^{(2)}$, Vincenzo Galdi$^{(3)}$, Gabriele Gradoni$^{(4)}$, Massimo Moccia$^{(3)}$, and Giuseppe Castaldi$^{(3)}$}
\vspace{0.15cm}
\noindent $^{(1)}$ Universit\'e Paris-Saclay, CNRS, CentraleSup\'elec, Laboratoire des Signaux et Syst\`emes, 3 Rue Joliot-Curie, 91192 Gif-sur-Yvette, France. (marco.di-renzo@universite-paris-saclay.fr). \\
\noindent $^{(2)}$ University of Cassino and Lazio Meridionale, Italy. \\
\noindent $^{(3)}$ University of Sannio, Italy. \\
\noindent $^{(4)}$ University of Nottingham and University of Cambridge, UK. \\


\begin{abstract}
Reconfigurable intelligent surface (RIS) is an emerging technology that is under investigation for different applications in wireless communications. RISs are often analyzed and optimized by considering simplified electromagnetic reradiation models. Also, the existence of possible electromagnetic waves that are reradiated towards directions different from the desired ones are often ignored. Recently, some testbed platforms have been implemented, and experimentally validated reradiation models for RISs have been reported in the literature. In this chapter, we aim to study the impact of realistic reradiation models for RISs as a function of the sub-wavelength inter-distance between nearby elements of the RIS, the quantization levels of the reflection coefficients, the interplay between the amplitude and phase of the reflection coefficients, and the presence of electromagnetic interference. Furthermore, we consider both case studies in which the users may be located in the far-field and near-field regions of an RIS. Our study shows that, due to design constraints, such as the need of using quantized reflection coefficients or the inherent interplay between the phase and the amplitude of the reflection coefficients, an RIS may reradiate power towards unwanted directions that depend on the intended and interfering electromagnetic waves. Therefore, it is in general important to optimize an RIS by taking into account the entire reradiation pattern by design, in order to maximize the reradiated power towards the desired directions of reradiation while keeping the power reradiated towards other unwanted directions at a low level. Among the considered designs for RISs, our study shows that a 2-bit digitally controllable RIS with an almost constant reflection amplitude as a function of the applied phase shift, and whose scattering elements have a size and an inter-distance between $(1/8)$th and $(1/4)$th of the signal wavelength may be a good tradeoff between performance, implementation complexity and cost. However, the presented results are preliminary and pave the way for further research into the performance of RISs based on accurate and realistic electromagnetic reradiation models.
\end{abstract}

\begin{keywords}
\kwd{Wireless communications}
\kwd{reconfigurable intelligent surfaces}
\kwd{dynamic metasurfaces}
\kwd{modeling}
\kwd{performance evaluation}
\kwd{optimization.}
\end{keywords}

\end{frontmatter}

\section{Introduction}\label{sec0:Intro}

\subsection{Reconfigurable Intelligent Surfaces and Holographic Surfaces}

In the last few years, intelligent surfaces have been the subject of extensive research activities in the context of wireless communications and networks \cite{MDR_EURASIP}, \cite{Rui_COMMAG}, \cite{Liaskos_ACM}, \cite{MDR_Relays}, \cite{Cunhua_Magazine}. A recent roadmap can be found in \cite{arxiv.2111.08676}. In essence, an intelligent surface is a dynamic metasurface, which shapes the reradiated electromagnetic waves as desired, thanks to a careful design of elementary scattering elements and to an appropriate optimization of simple electronic circuits \cite{Cui_CodingMeta}. In wireless communications, intelligent surfaces have been researched for several applications, and mainly for two possible uses:
\begin{enumerate}
\item \textbf{Nearly-passive reconfigurable devices} that are capable of shaping the electromagnetic waves that impinge upon them \cite{MDR_JSAC}. Two typical examples are surfaces that reflect or refract, e.g., smart windows, the electromagnetic waves towards non-specular directions. These surfaces are usually referred to as reconfigurable intelligent surfaces (RISs) \cite{ETSI_ISG-RIS}.
\item \textbf{Low-complexity active transceivers} that are capable of realizing extremely massive multiple-input multiple-output communications \cite{Wankai_Transmitter}. These surfaces are usually referred to as dynamic metasurface antennas (DMA) \cite{9324910} or holographic surfaces (HoloS) \cite{MDR_Holos}.
\end{enumerate}

\begin{figure}[!t]
\includegraphics[width=0.94\columnwidth]{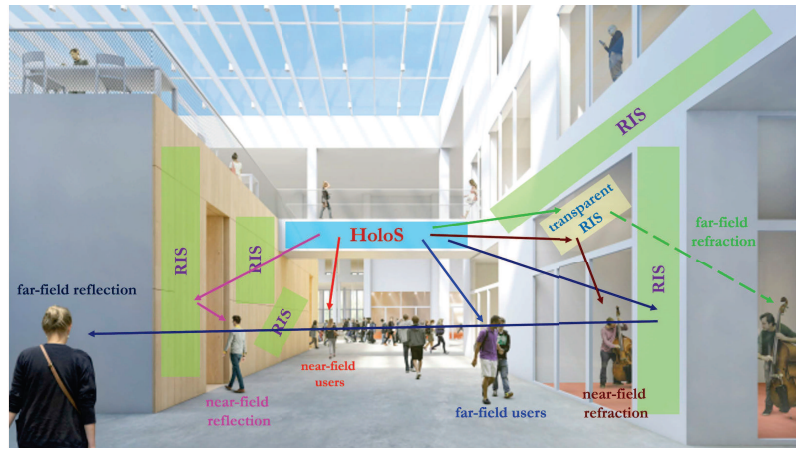}
\caption{Emerging wireless communication scenario with far-field and near-field users.}\label{Fig_WirelessFuture} \vspace{-0.45cm}
\end{figure}

The main advantage of an RIS consists of controlling the propagation environment besides the end points of a transmission link, i.e., transmitters and receivers, without the need of requiring power amplifiers, radio frequency chains, and digital signal processors. An RIS operates in the analog domain directly on the electromagnetic waves. The main advantage of a HoloS consists of being equipped with a very large number of reconfigurable metamaterial elements (like an RIS) but with a limited number of radio frequency chains. This feature is highly desirable, since it reduces the number of radio frequency chains while offering higher beamforming and spatial multiplexing gains \cite{9324910}. Notably, these surfaces, if sufficiently large in size, may provide spatial multiplexing gains, i.e., multiple orthogonal communication modes, even in free-space line-of-sight propagation environments \cite{9139337}, \cite{9475156}, \cite{8289332}. In wireless communication systems and networks, RIS and HoloS are jointly deployed in the environment to boost the communication performance. An illustration of this emerging communication scenario is shown in Fig. \ref{Fig_WirelessFuture}.

Specifically, thanks to their expected large electrical size, RIS and HoloS may bring fundamentally new challenges to the design and optimization of wireless networks. One of them is the need of engineering and optimizing wireless communication networks whose devices may likely operate in the near-field of each other and, therefore, the electromagnetic waves can no longer be assumed to be characterized by a planar wavefront, but a spherical wavefront needs to be accounted for at the design stage \cite{arxiv.2112.05989}. Also, the possibility of packing on an intelligent surface hundreds or thousands of radiating elements at sub-wavelength inter-distances requires new communication models that account, at the optimization stage, for the mutual coupling among the elements \cite{MDR_MutualImpedances}, \cite{MDR_MutualImpedancesOpt}, \cite{MDR_MutualImpedancesMIMO}, \cite{9762020}.

\subsection{Electromagnetically Consistent Modeling of Reconfigurable Intelligent Surfaces}

In this chapter, we focus our attention on RISs. The deployment and optimization of RISs in wireless networks need several challenges to be tackled. Interested readers can consult, e.g., \cite{Cunhua_Magazine}, \cite{MDR_JSAC} for a comprehensive discussion. One of the major and open research challenges in RIS-aided wireless communications lies in developing and utilizing electromagnetically-consistent models that account for the practical implementation of RISs. A comprehensive summary of the  communication models most widely utilized in wireless communications for RISs is reported in \cite{arxiv.2110.00833}. From the overview in \cite{arxiv.2110.00833}, it is apparent that three main communication models are typically utilized:
\begin{enumerate}
\item The locally periodic discrete model;
\item The mutually coupled antenna-elements model;
\item The inhomogeneous sheets of surface impedance model.
\end{enumerate}

Interested readers are referred to \cite{arxiv.2110.00833} for a comprehensive discussion of the main characteristics, strengths, and limitations of these models. In this contribution, we focus our attention on the \textbf{locally periodic discrete model}, since it is the most widely used model in wireless communications and in the field of \textbf{digital metasurfaces} \cite{adom.201700455}. According to this model, an RIS is modeled as an ensemble of reconfigurable elements that can be configured in a finite number of states. From an implementation standpoint, each RIS reconfigurable element is made of one or several engineered scattering elements and some electronic circuits. From a signal and system (or communication) standpoint, each RIS reconfigurable element is associated with a discrete-valued alphabet, sometimes referred to as lookup table or codebook, which determines the finite number of wave manipulations that each RIS reconfigurable element can apply to the incident electromagnetic waves.

According to \cite{arxiv.2110.00833}, each value of the alphabet (or state of the RIS) can be interpreted as the (electric field) reflection coefficient of an infinite and homogeneous surface whose constituent elements are all configured to the same state. This definition and characterization of the alphabet of each RIS reconfigurable element introduces some limitations on the applicability of the locally periodic discrete model: The model may not be accurate if the desired wave transformation is not ``slowly-varying'' at the scale of the wavelength of the electromagnetic waves. In other words, the model can be applied if a ``not-too-small'' number of neighboring RIS reconfigurable elements is configured to the same state, so as to ensure that, locally, each RIS reconfigurable element ``sees'' other RIS reconfigurable elements configured to the same value of the alphabet. This ``slowly-varying'' or ``locally periodic'' condition needs to be carefully evaluated when utilizing this communication model for RISs.

\subsection{Realistic Scattering Models for Reconfigurable Intelligent Surfaces}

The numerical results in \cite{arxiv.2110.00833} show, in addition, that the presence of imperfections (non-ideal effects), with respect to the theoretically optimum, of the RIS configuration that is required to realize the desired wave transformations may result in the existence of higher-order harmonics (or grating lobes in antenna theory) towards undesired directions. Also, any periodic RIS that is illuminated by a plane wave may reradiate several electromagnetic waves according to Floquet's theorem, whose intensity depends on how the RIS surface impedance is engineered \cite{Sergei_MacroscopicModel}, \cite{Vittorio_RayTracing}. In the context of the locally periodic discrete model, inaccuracies in the design of the alphabet of the RIS scattering elements, i.e., the use of a non-ideal alphabet, with respect to the theoretically optimum, may result in the presence of reradiated but undesired electromagnetic waves. Specifically,  three major practical issues can be identified when engineering the RIS reconfigurable elements (which encompass the scattering elements and the electronic circuits):
\begin{enumerate}
\item \textit{The phases of the complex-valued elements of the alphabet of the RIS reconfigurable elements are not exactly the same as those identified at the design stage}. Assume, for example, that one wants to realize a three-bit digitally controllable RIS. In theory, the phase difference between the complex-valued elements of the alphabet should be an integer multiple of 45 degrees. However, it may not be possible to realize eight different phases with this constraint \cite{9632392}. Also, it may not be possible to realize any phase shifts, i.e., the range of admissible phase shifts may be smaller than 360 degrees.
\item \textit{The amplitudes of the complex-valued elements of the alphabet of the RIS reconfigurable elements are not exactly unitary and they are not independent of the corresponding phases of the complex-valued elements of the alphabet}. For example, if one wants to apply a given phase shift to an incident electromagnetic wave, the corresponding amplitude may be much smaller than one. This results in a phase-dependent signal attenuation \cite{9115725}, \cite{Romain_RIS-Prototype}.
\item \textit{Due to the implementation and power constraints associated with any electronic circuit, it may be possible to realize alphabets for the RIS reconfigurable elements with only a finite number of complex-valued elements, i.e., the number of RIS states is finite}. An often convenient implementation, due to the ease of realization, reduced cost, and limited power consumption, is the design of binary surfaces whose elements can only take two possible states and whose nominal phase shifts differ by 180 degrees. Recent results have shown that these extremely quantized surfaces may result in far-field reradiatiated beams towards undesired directions, e.g., towards the direction that is symmetrical with respect to the desired direction of radiation for the case study of anomalous reflectors \cite{9424172}.
\end{enumerate}


\subsection{Modeling Reconfigurable Intelligent Surfaces in Wireless Communications}

In wireless communications, in spite of these considerations, the typical assumptions made on the complex-valued elements of the alphabet of the RIS reconfigurable elements (usually referred to as the RIS reflection coefficients in wireless communications) when considering the locally periodic discrete model for RIS are the following \cite{9326394}:
\begin{enumerate}
\item The amplitudes of the RIS reflection coefficients are assumed to be unitary or constant as a function of the corresponding phases;
\item The phase shifts of the RIS reflection coefficients are either assumed to be continuous-valued variables or discrete-valued variables with equal phase differences;
\item In some cases, the amplitudes and phases of the RIS reflection coefficients are assumed to be optimized independently of one another.
\end{enumerate}

Most probably, however, \textbf{the main assumption made in the context of wireless communications lies in optimizing a utility function at particular locations of one or multiple receivers, while disregarding the reradiation pattern of an RIS towards other directions}. Inherently, therefore, the presence and impact of undesired reradiated beams are not explicitly investigated. The presence of unwanted reradiations may, however, not be negligible if realistic alphabets (reflection coefficients) for the RIS reconfigurable elements are utilized. The existence of these undesired beams has, in particular, two major and negative consequences on a communication system:
\begin{enumerate}
\item Some power is directed towards directions that are different from the nominal ones. This implies that the power efficiency of an RIS towards the target directions is reduced. Since an RIS does not usually amplify the incident signals, this may further limit the transmission coverage;
\item The unwanted beams result in interference that an RIS injects into the network, which may negatively affect the performance of other network users.
\end{enumerate}

\subsection{Chapter Contribution}

Motivated by these considerations, we evince that it is necessary to comprehensively study the reradiation pattern of an RIS when practical reflection models (alphabets) are utilized, according to the assumption of the locally periodic discrete model. The aim of the present chapter is to study this open issue, which, to the best of our understanding, is not sufficiently understood by the wireless community. More precisely, we consider some recently reported alphabets for RISs that operate at different frequencies and study the reradiation pattern of each of them. Special focus is put on comparing currently available RIS alphabets based on existing hardware prototypes against ``ideal'' alphabets whose elements have unit amplitudes and evenly spaced phases, as is often assumed in wireless communications. Our numerical results show that major differences in the reradiation pattern of an RIS usually exist, especially either when binary RIS reconfigurable elements are utilized, or when the amplitudes and phases of the reflection coefficients are not independent of one another and the variations of the amplitudes with the phases are not negligible.

\subsection{Chapter Organization}

The rest of this chapter is organized as follows. In Section \ref{sec:SYS_Model}, the considered system model is described. To focus our attention on the main contribution of the chapter, i.e., the impact of practical alphabets of the RIS reconfigurable elements, the canonical system model with a single transmit and a single receive antenna is considered. In Section \ref{sec:OPT_Algo}, we describe the algorithm that is utilized for optimizing an RIS under the assumption that each RIS reconfigurabale element is characterized by complex-valued elements (alphabet) whose amplitudes and phases are not necessarily independent of one another. In Section \ref{sec:NUM_Results}, several numerical results are illustrated by utilizing experimentally validated alphabets for RISs. In Section \ref{sec:Conclusions}, finally, we provide concluding remarks.

\section{System Model}\label{sec:SYS_Model}

We consider a single-user system in which a single-antenna transmitter and a single-antenna receiver communicate through an RIS. For simplicity, we assume that no direct link between the transmitter and the receiver exists. We denote by $\bf{H}$ and $\bf{G}$ the channel matrices from the transmitter to the RIS and from the RIS to the receiver, respectively.

As far as the RIS is concerned, we model it as a uniform planar array with $NM$ RIS reconfigurable elements that are arranged and are equally spaced on $N$ rows and $M$ columns. The RIS is assumed to be centered at the origin and to lie in the $xy$ plane (i.e., $z=0$). The inter-distance between the RIS reconfigurable elements on each row and column is denoted by $d_x$ and $d_y$, respectively. The inter-distances $d_x$ and $d_y$ are referred to as the geometric periods of the RIS in the context of dynamic metasurfaces. The surface area of each RIS reconfigurable element is $d_x d_y$, and it encompasses one or multiple scattering elements and the associated tuning circuits. For simplicity, we can assume that each RIS reconfigurable element is made of a single radiating element and one or more positive-intrinsic-negative (PIN) diodes \cite{Linglong_Testbed} or varactors \cite{Romain_RIS-Prototype}. Each RIS reconfigurable element can be optimized independently of the others.

As mentioned in Section \ref{sec0:Intro}, we adopt the locally periodic discrete model for an RIS \cite{arxiv.2110.00833}. Accordingly, each RIS reconfigurable element is associated with a set of $L$ complex-valued coefficients (the RIS alphabet) denoted by $\Gamma_1$, $\Gamma_2$, \ldots, $\Gamma_L$. Each element of the alphabet is obtained by appropriately configuring the electronic circuits of the RIS reconfigurable element. For ease of description, we assume that the RIS operates as a reflecting surface. From the physical standpoint, therefore, the complex-valued coefficient $\Gamma_l$ has the meaning of a reflection coefficient, i.e., the ratio between the reflected electric field and the incident electric field, of an infinite RIS whose elements are all configured to the same state. Therefore, the corresponding equivalent structure is a homogeneous surface that realizes specular reflection. According to this definition, each RIS reconfigurable element is characterized by means of locally periodic boundary conditions, and, since an RIS is not endowed with power amplifiers, the reflection coefficients $\Gamma_l$ for $l=1,2,\ldots,L$ have an amplitude that is, by definition, less than one, i.e., $\left|{\Gamma_l}\right| \le 1$ for $l=1,2,\ldots,L$. However, this neither necessarily implies that the amplitude of $\Gamma_l$ is a constant independent of the phase of $\Gamma_l$ nor that the amplitude and the phase of $\Gamma_l$ can be optimized independently of one another.

Examples of alphabets that can be found in the literature and that correspond to available hardware platforms, i.e., are experimentally validated, can be found in Table \ref{Table_RelectionCoefficientDiscrete} and Table \ref{Table_RelectionCoefficientContinuous}, which were originally reported in \cite{arxiv.2110.00833}. As far as the RIS prototype introduced in \cite{Hongliang_OmniSurface} is concerned, we have reported only the reflection coefficients in Table \ref{Table_RelectionCoefficientDiscrete} and have ignored the transmission coefficients, since they are of no interest for our study. The reflection coefficients in Table \ref{Table_RelectionCoefficientDiscrete} and Table \ref{Table_RelectionCoefficientContinuous} are utilized in Section \ref{sec:NUM_Results} to obtain the numerical results. It is worth mentioning that the ``typical'' model utilized in wireless communications for a reflecting-type RIS assumes that (i) either  $\left|{\Gamma_l}\right| =1$ for any possible phase of ${\Gamma_l}$ and that the phase of ${\Gamma_l}$ can be adjusted to any continuous values or (ii) $\left|{\Gamma_l}\right| =1$ for any possible phase of ${\Gamma_l}$ and that the phase of ${\Gamma_l}$ can be adjusted to a finite number of phase shifts that are evenly spaced within the range $0-360$ degrees. Further details are provided in Section \ref{sec:NUM_Results}.
\begin{table*}[!t]
		\centering
		\caption{Examples of reflection coefficients for an RIS with discrete-valued phase shifts (two-state and four-state RIS elements). $f$ and $\lambda$ denote the frequency and the wavelength, respectively.}
		\label{Table_RelectionCoefficientDiscrete}
		\footnotesize
		\newcommand{\tabincell}[2]{\begin{tabular}{@{}#1@{}}#2\end{tabular}}
		\begin{tabular}{c|c||c} \hline
			
			Reference & Size of the unit cell &  Reflection Coefficient \\ \hline \hline
			
			\cite{Wankai_PathLoss-mmWave} ($f= 33$ GHz) & $0.418\lambda \times 0.418\lambda$ & $\begin{array}{l} \left| {{\Gamma _1}} \right| = 0.8,\quad \angle {\Gamma _1} = {150^\circ }\\ \left| {{\Gamma _2}} \right| = 0.8,\quad \angle {\Gamma _2} = {0^\circ }\end{array}$ \\ \hline
			
			\cite{Wankai_PathLoss-mmWave} ($f= 27$ GHz) & $0.126\lambda \times 0.252\lambda$ & $\begin{array}{l} \left| {{\Gamma _1}} \right| = 0.9,\quad \angle {\Gamma _1} = {165^\circ }\\ \left| {{\Gamma _2}} \right| = 0.7,\quad \angle {\Gamma _2} = {0^\circ }\end{array}$ \\ \hline
			
			\cite{Hongliang_OmniSurface} ($f= 3.6$ GHz) &  $0.345\lambda \times 0.170\lambda$ & $\begin{array}{l} \left| {{\Gamma _1}} \right| = 0.46,\quad \angle {\Gamma _1} = {20^\circ }\\ \left| {{\Gamma _2}} \right| = 0.55,\quad \angle {\Gamma _2} = {215^\circ }\end{array}$ \\ \hline
			
			\cite{Linglong_Testbed} ($f= 2.3$ GHz) & $0.286\lambda \times 0.286\lambda$ & $\begin{array}{l} \left| {{\Gamma _1}} \right| = -1.2  \, {\rm{dB}},\quad \angle {\Gamma _1} = {-205.5^\circ }\\ \left| {{\Gamma _2}} \right| = -1.2 \, {\rm{dB}},\quad \angle {\Gamma _2} = {-383.2^\circ } \\ \left| {{\Gamma _3}} \right| = -0.8  \, {\rm{dB}},\quad \angle {\Gamma _3} = {-290.2^\circ } \\ \left| {{\Gamma _4}} \right| = -0.7  \, {\rm{dB}},\quad \angle {\Gamma _4} = {-110.3^\circ }\end{array}$ \\ \hline
			
		\end{tabular}
	\end{table*}
\begin{table}[!t]
		\centering
		\caption{Example of reflection coefficient for an RIS with continuous-valued phase shifts \cite{Romain_RIS-Prototype}. The operating frequency is $5.15-5.75$ GHz and the size of the unit cells is approximately a quarter of the signal wavelength.}
		\label{Table_RelectionCoefficientContinuous}
		\footnotesize
		\newcommand{\tabincell}[2]{\begin{tabular}{@{}#1@{}}#2\end{tabular}}
		\begin{tabular}{c||c|c} \hline
			
			Voltage &  $\begin{array}{c} {\rm{Reflection \; coefficient}}\\ {\rm{amplitude \; (}}\left| \Gamma  \right|{\rm{)}} \end{array}$ & $\begin{array}{c} {\rm{Reflection \; coefficient}}\\ {\rm{phase \; (}}\angle \Gamma  {\rm{)}} \end{array}$ \\ \hline \hline
			
			0 $V$ & -1.517 dB & 32.798$^\circ$ \\ \hline
            0.25 $V$ & -1.807 dB & 40.854$^\circ$ \\  \hline
            0.5 $V$ & -3.156 dB & 46.807$^\circ$ \\   \hline
            0.75 $V$ & -5.59 dB & 53.543$^\circ$ \\   \hline
            1 $V$ & -9.576 dB & 70.32$^\circ$ \\  \hline
            1.25 $V$ & -20.563 dB & -167.158$^\circ$ \\   \hline
            1.5 $V$ & -6.615 dB & -73.171$^\circ$ \\  \hline
            1.75 $V$ & -3.029 dB & -49.627$^\circ$ \\ \hline
            2 $V$ & -1.959 dB & -35.908$^\circ$ \\    \hline
            2.5 $V$ & -0.874 dB & -23.263$^\circ$ \\  \hline
            3 $V$ & -0.749 dB & -16.087$^\circ$ \\    \hline
            3.5 $V$ & -0.469 dB & -12.663$^\circ$ \\  \hline
            4 $V$ & -0.528 dB & -9.925$^\circ$ \\ \hline
            5 $V$ & -0.439 dB & -6.906$^\circ$ \\ \hline
			
		\end{tabular}
	\end{table}

Under the considered modeling assumptions, the achievable rate per unit bandwidth can be formulated as follows: 
\begin{equation}
R = {\log _2}\left( {1 + \frac{p}{{{\sigma ^2}}}\left|\sum\limits_{n = 1}^N {\sum\limits_{m = 1}^M {{g_{nm}}{\gamma _{nm}}{h_{nm}}} } \right|^2 } \right) \label{Rate}
\end{equation}
where $p$ is the transmitted power, $\sigma ^2$ it the noise power at the receiver, $g_{nm}$ is the $(n,m)$th entry of the channel matrix $\bf{G}$, $h_{nm}$ is the $(n,m)$th entry of the channel matrix $\bf{H}$, and $\gamma_{nm}$ is the value of the reflection coefficient of the $(n,m)$th RIS reconfigurable element, with $\gamma_{nm} \in \left\{ {{\Gamma _1},{\Gamma _2}, \ldots ,{\Gamma _L}} \right\}$ for $n=1,2,\ldots,N$ and $m=1,2,\ldots,M$.

Since we are interested in characterizing the reradiation pattern of an RIS as a function of the alphabet of the RIS reconfigurable elements, we focus our attention on free-space propagation. Accordingly, it was recently proved in \cite{Wankai_PathLoss-mmWave},  \cite{Wankai_PathLoss} that the power received at the generic location ${{\bf{r}}}^{({\rm{rx}})} = \left( x_{\rm{rx}}, y_{\rm{rx}}, z_{\rm{rx}} \right)$ can be formulated as follows:
\begin{equation}
{p_{\rm{rx}}}\left( {{\bf{r}}^{({\rm{rx}})}} \right)
= \frac{{{p}}}{{16{\pi ^2}}}{\left| {\sum\limits_{n = 1}^N {\sum\limits_{m = 1}^M { \gamma_{nm} \left[ \begin{array}{l}
\sqrt {{G_{{\rm{tx}}}}\left( {\theta _{nm}^{({\rm{tx}})}} \right)} \left( {\frac{{{z_{\rm{tx}}}}}{{\left| {{\bf{r}}_{nm}^{(\rm{tx})}} \right|}}\frac{{\exp \left( { - jk\left| {{\bf{r}}_{nm}^{(\rm{tx})}} \right|} \right)}}{{\left| {{\bf{r}}_{nm}^{(\rm{tx})}} \right|}}} \right) \\
\sqrt {{G_{{\rm{rx}}}}\left( {\theta _{nm}^{({\rm{rx}})}} \right)}
\left( {\frac{{{z_{\rm{rx}}}}}{{\left| {{\bf{r}}_{nm}^{(\rm{rx})}} \right|}}\frac{{\exp \left( { - jk\left| {{\bf{r}}_{nm}^{(\rm{rx})}} \right|} \right)}}{{\left| {{\bf{r}}_{nm}^{(\rm{rx})}} \right|}}} \right)\\
\left( {{d_x}{\mathop{\rm sinc}\nolimits} \left( {k\left( {\frac{{\left( {{x_{{\rm{rx}}}} - {x_n}} \right)}}{{\left| {{\bf{r}}_{nm}^{({\rm{rx}})}} \right|}} + \frac{{\left( {{x_{{\rm{tx}}}} - {x_n}} \right)}}{{\left| {{\bf{r}}_{nm}^{({\rm{tx}})}} \right|}}} \right)\frac{{{d_x}}}{2}} \right)} \right)\\
\left( {{d_y}{\mathop{\rm sinc}\nolimits} \left( {k\left( {\frac{{\left( {{y_{{\rm{rx}}}} - {y_m}} \right)}}{{\left| {{\bf{r}}_{nm}^{({\rm{rx}})}} \right|}} + \frac{{\left( {{y_{{\rm{tx}}}} - {y_m}} \right)}}{{\left| {{\bf{r}}_{nm}^{({\rm{tx}})}} \right|}}} \right)\frac{{{d_y}}}{2}} \right)} \right)
\end{array} \right]} } } \right|^2}\label{Received_power}
\end{equation}
where the following notation is used:
\begin{itemize}
\item $j=\sqrt{-1}$ is the imaginary unit, ${\rm{sinc}}\left( x \right) = \sin \left( x \right)/x$, and $k=2\pi/\lambda$ where $\lambda$ is the wavelength;
    \item $(x_{\rm{tx}},y_{\rm{tx}},z_{\rm{tx}})$  is the location of the transmitter;
    \item $(x_{n},y_{m})$ is the center point of the $(n,m)$th RIS reconfigurable element;
    \item ${p_{\rm{rx}}}\left( {{\bf{r}}^{({\rm{rx}})}} \right)$ is the received power at ${{\bf{r}}^{({\rm{rx}})}}$;
    \item ${{G_{{\rm{tx}}}}\left( {\theta _{nm}^{({\rm{tx}})}} \right)}$ and ${{{G_{{\rm{rx}}}}\left( {\theta _{nm}^{({\rm{rx}})}} \right)}}$ are the antenna gains of the transmitter and receiver, respectively, where ${\theta _{nm}^{({\rm{tx}})}}$ is the angle between the transmitter and the center point of the $(n,m)$th RIS reconfigurable element and ${\theta _{nm}^{({\rm{rx}})}}$ is the angle between the transmitter and the center point of the $(n,m)$th RIS reconfigurable element;
    \item $\left|{{\bf{r}}_{nm}^{\left( {{\rm{tx}}} \right)}}\right|$ is the distance between the transmitter and the center point of the $(n,m)$th RIS reconfigurable element;
    \item $\left|{{\bf{r}}_{nm}^{\left( {{\rm{rx}}} \right)}}\right|$ is the distance between the center point of the $(m,n)$th RIS reconfigurable element and the receiver.
\end{itemize}

In the next section, based on this signal model, we propose a simple algorithm for optimizing the link rate in \eqref{Rate}. It is worth mentioning that the proposed optimization algorithm can be applied to any channel model, and that the free-space model is considered only for illustration and because we are interested in characterizing the reradiation pattern of an RIS.

\section{Optimization Algorithm}\label{sec:OPT_Algo}
Let $\boldsymbol{\Gamma}$ be the $N \times M$ matrix of complex values $\gamma_{nm}$ and let ${\mathcal{A}} = \left\{ {{\Gamma _1},{\Gamma _2}, \ldots ,{\Gamma _L}} \right\}$ denote the alphabet of each RIS reconfigurable element, i.e., $\mathcal{A}$ is the feasible set of $\gamma_{nm}$ for $n=1,2,\ldots,N$ and $m=1,2,\ldots,M$. The maximization of the rate $R$ in \eqref{Rate} in a free-space channel model is equivalent to the following constrained optimization problem:
\begin{subequations}
\begin{align}
&\mathop {\max }\limits_{\boldsymbol{\Gamma }} {p_{{\rm{rx}}}}\left( {{\bf{r}}^{\left( {{\rm{rx}}} \right)}} \right) = {p\left| {\sum\limits_{n = 1}^N {\sum\limits_{m = 1}^M {{g_{nm}}{\gamma _{nm}}{h_{nm}}} } } \right|^2}\\
& \mathrm{s.t.}~~{\gamma _{nm}} \in {\mathcal{A}},\quad\forall~ n = 1,2,\dots,N, ~ m = 1,2,\dots,M
\end{align}\label{Local_Optimization_Problem}
\end{subequations}
where the following two identities hold:
\begin{equation}
\begin{split}
g_{nm} &= \frac{{{1}}}{{4{\pi}}}\sqrt {{G_{{\rm{tx}}}}\left( {\theta _{nm}^{({\rm{tx}})}} \right)} \left( {\frac{{{z_{\rm{tx}}}}}{{\left| {{\bf{r}}_{nm}^{(\rm{tx})}} \right|}}\frac{{\exp \left( { - jk\left| {{\bf{r}}_{nm}^{(\rm{tx})}} \right|} \right)}}{{\left| {{\bf{r}}_{nm}^{(\rm{tx})}} \right|}}} \right) \\
&\times \left( {{d_x}{\mathop{\rm sinc}\nolimits} \left( {k\left( {\frac{{\left( {{x_{{\rm{rx}}}} - {x_n}} \right)}}{{\left| {{\bf{r}}_{nm}^{({\rm{rx}})}} \right|}} + \frac{{\left( {{x_{{\rm{tx}}}} - {x_n}} \right)}}{{\left| {{\bf{r}}_{nm}^{({\rm{tx}})}} \right|}}} \right)\frac{{{d_x}}}{2}} \right)} \right)
\end{split}
\end{equation}
\begin{equation}
\begin{split}
h_{nm} &= \frac{{{1}}}{{4{\pi}}}
\sqrt {{G_{{\rm{rx}}}}\left( {\theta _{nm}^{({\rm{rx}})}} \right)}
\left( {\frac{{{z_{\rm{rx}}}}}{{\left| {{\bf{r}}_{nm}^{(\rm{rx})}} \right|}}\frac{{\exp \left( { - jk\left| {{\bf{r}}_{nm}^{(\rm{rx})}} \right|} \right)}}{{\left| {{\bf{r}}_{nm}^{(\rm{rx})}} \right|}}} \right)\\
& \times \left( {{d_y}{\mathop{\rm sinc}\nolimits} \left( {k\left( {\frac{{\left( {{y_{{\rm{rx}}}} - {y_m}} \right)}}{{\left| {{\bf{r}}_{nm}^{({\rm{rx}})}} \right|}} + \frac{{\left( {{y_{{\rm{tx}}}} - {y_m}} \right)}}{{\left| {{\bf{r}}_{nm}^{({\rm{tx}})}} \right|}}} \right)\frac{{{d_y}}}{2}} \right)} \right)
\end{split}
\end{equation}

The formulated optimization problem in \eqref{Local_Optimization_Problem} is characterized by the optimization variables $\gamma_{nm}$ that can only take discrete values, which prevents us from using gradient-based optimization methods. Moreover, the discrete values in the set $\mathcal{A}$ do not possess any specific structure that can be exploited for optimization purposes: They are generic complex numbers. Finally, the fact that only discrete-valued phase shifts can be applied by an RIS that is characterized by the feasible set $\mathcal{A}$ makes it impossible to perfectly compensate the phases of the channels $g_{nm}$ and $h_{nm}$ for any $n=1,2,\ldots,N$ and $m=1,2,\ldots,M$. In principle, given the discrete nature of
the feasible set, it is always possible to resort to an
exhaustive search over the set $\mathcal{A}^{NM}$. However, this would require searching over $(NM)^L$ configurations of the matrix $\boldsymbol{\Gamma}$, which is computationally infeasible for practical values of $(N,M)$ and $L$. In fact, typical values for $NM$ are of the order of hundreds or thousands, while $L$ is of the order of units. This motivates us to develop optimization algorithms with a complexity that is linear in $NM$, i.e., with the total number of RIS reconfigurable elements.

The approach that we propose to solve the formulated optimization problem in \eqref{Local_Optimization_Problem} is based on the alternating optimization principle, i.e., the $NM$ reflection coefficients are optimized sequentially one at a time. To elaborate, the optimization of the generic reflection
coefficient ${\gamma_{qp}}$, while all other reflection coefficients are kept fixed, is stated as follows:
\begin{subequations}
\begin{align}
&\mathop {\max }\limits_{{\gamma _{qp}}} {\left| {{g_{qp}}{\gamma _{qp}}{h_{qp}} + \sum\limits_{n = 1 \ne q}^N {\sum\limits_{m = 1 \ne p}^M {{g_{nm}}{\gamma _{nm}}{h_{nm}}} } } \right|^2}\\
& \mathrm{s.t.}~~{\gamma _{qp}} \in {\mathcal{A}}
\end{align} \label{Reformulated_Local_Optimization_Problem}
\end{subequations}

Defining ${\alpha _{qp}} = \sum\nolimits_{n = 1 \ne q}^N {\sum\nolimits_{m = 1 \ne p}^M {{g_{nm}}{\gamma _{nm}}{h_{nm}}} }$, the objective function in (\ref{Reformulated_Local_Optimization_Problem}) can be expanded as follows:
\begin{equation}
{\left| {{g_{qp}}{\gamma _{qp}}{h_{qp}} + {\alpha _{qp}}} \right|^2} = {\left| {{\gamma _{qp}}} \right|^2}{\left| {{g_{qp}}{h_{qp}}} \right|^2} + 2{\mathop{\Re}\nolimits} \left\{ {{\gamma _{qp}}{g_{qp}}{h_{qp}}\alpha _{qp}^*} \right\} + {\left| {{\alpha _{qp}}} \right|^2}\label{expanding_objective_function}
\end{equation}
where $\Re$ denotes the real part of a complex number and $(\cdot)^*$ denote the complex conjugate operator.

\begin{algorithm}[!t]
\caption{Algorithm to solve the optimization problem in \eqref{Local_Optimization_Problem}}\label{alg:cap}
\begin{algorithmic}
\State \textbf{Initialize}  $\boldsymbol{\Gamma}$ to a feasible value $\boldsymbol{\Gamma}_0$ ; $k=0$; $F^{(k)}=F(\boldsymbol{\Gamma}_0)$, $F^{(k-1)}=0$;
\State Set a convergence tolerance $0 \le {\varepsilon} \le F^{(0)}$;
\While{$\left| {{F^{(k)}} - {F^{(k - 1)}}} \right| >{\varepsilon} $ }
\For{$q=1,2,\dots,N$}
\For{$p=1,2,\dots,M$}
    \State  Set $\gamma_{qp}$ as the maximizer of (\ref{expanding_objective_function});
\EndFor
\EndFor
 \State $k=k+1$; $F^{(k-1)}=F^{(k)}$, $F^{(k)}=F{(\boldsymbol{\Gamma})}$;
\EndWhile
\State \Return $\boldsymbol{\Gamma}$.
\end{algorithmic}
\end{algorithm}
Since the third addend in (\ref{expanding_objective_function}) is independent of the optimization variable $\gamma_{qp}$, the objective function is maximized by searching among the $L$ elements of the set $ \mathcal{A}$ and by selecting the element of the set that provides the largest value of $\left| {{\gamma _{qp}}} \right|$ whose phase is the closest to the term ${{g_{qp}}{h_{qp}}\alpha _{qp}^*}$. The complete iterative algorithm for solving the formulated optimization problem in \eqref{Local_Optimization_Problem} is reported in Algorithm \ref{alg:cap}. Specifically, the function $F^{(k)}$ denotes the objective function in \eqref{Reformulated_Local_Optimization_Problem} at the $k$th iteration of the algorithm and the maximizer of \eqref{expanding_objective_function} is obtained through a simple exhaustive search over the $L$ elements of $\mathcal{A}$.

By direct inspection, we see that the complexity of Algorithm \ref{alg:cap} scales linearly with $NM$, since the $NM$ reflection coefficients of the RIS are optimized sequentially. Therefore, Algorithm \ref{alg:cap} can be considered to be a low-complexity optimization solution. Specifically, the overall complexity of Algorithm \ref{alg:cap} is $\mathcal{O}(I_{\varepsilon}NML)$, where $I_{\varepsilon}$ is the number of iterations required to reach the convergence tolerance ${\varepsilon}$. As far as the quality of the solution obtained by Algorithm \ref{alg:cap} is concerned, the discrete nature of the optimization problem in \eqref{Local_Optimization_Problem} prevents us from making any formal claim. It is known, however, that alternating optimization methods converge to a first-order optimal solution when the objective function and the constraint functions are differentiable, and the constraints are decoupled in the optimization variables, which is the case of the problem in \eqref{Local_Optimization_Problem} \cite{Bertsekas/99}. Thus, we can claim that Algorithm \ref{alg:cap} converges to a first-order optimal point of the optimization problem that is obtained by relaxing the optimization variables in (\ref{Local_Optimization_Problem}) to the continuous set
$\left\{ {{\gamma _{nm}} \in \mathbb{C} :\left| {{\gamma _{nm}}} \right| \le 1,\forall n = 1,2, \ldots,N \; \text{and} \; m = 1,2, \ldots,M} \right\}$.

\section{Numerical Results}\label{sec:NUM_Results}
In this section, we illustrate several numerical results in order to show the impact that realistic alphabets of the RIS reconfigurable elements may have in wireless communications. To this end, we consider the following canonical setup unless stated otherwise:
\begin{itemize}
\item The direct link between the transmitter and the receiver is assumed to be sufficiently weak and is ignored to focus on the role of the RIS;
\item The RIS is assumed to be in the far field region of both the transmitter and the receiver, and, therefore, the electromagnetic waves have a planar wavefront across the RIS and at the receiver;
\item The electromagnetic wave emitted by the transmitter impinges normally upon the RIS (normal incidence), which is consistent with the alphabets reported in Table \ref{Table_RelectionCoefficientDiscrete} and Table \ref{Table_RelectionCoefficientContinuous} that are obtained for normal incidence;
\item The RIS is assumed to be a square surface, which is centered at the origin and that lies in the $xy$ plane at $z=0$, whose size is $Nd_x =Md_y = 1$ meter;
\item The inter-distances along the rows and the columns of the RIS are assumed to be identical, i.e., $d_x=d_y=d$. Specifically, we consider different values of $d \le \lambda/2$ in order to analyze the impact of sub-wavelength implementations;
\item Two desired angles of reflection are considered: 45 degrees and 75 degrees. This choice allows us to analyze the impact of moderate and large angles of reflection with respect to the incident electromagnetic wave (whose angle of incidence is equal to zero degrees).
\end{itemize}

As far as the parametric study as a function of $d$ is concerned, the following remarks are in order. The RIS alphabets reported in Table \ref{Table_RelectionCoefficientDiscrete} and Table \ref{Table_RelectionCoefficientContinuous} are obtained for normal incidence (i.e., the angle of incidence is zero degrees) and the corresponding reflection coefficients are obtained for a given size of the RIS reconfigurable elements. Therefore, rigorously, these alphabets should be utilized only for the sizes given in Table \ref{Table_RelectionCoefficientDiscrete} and Table \ref{Table_RelectionCoefficientContinuous}. In this chapter, however, we are interested in analyzing how \textit{qualitatively} the reradiation capabilities of an RIS change as a function of the size of the RIS reconfigurable elements, while assuming that their reflection characteristics are kept unchanged as their size changes (either shrinks or increases). This provides us with an indication of how the spatial sampling affects the reradiation pattern of an RIS while keeping fixed the quantization of the phase shift and the relation between the amplitude and the phase shift of the reflection coefficient. It needs to be emphasized that it may sometimes be difficult to design RIS reconfigurable elements with a very small size (e.g., 16 or 32 times smaller than the wavelength) while ensuring that they can realize a wide range of phase shifts. If the number of quantization bits, i.e., the cardinality ${L}$ of the RIS alphabet, is limited to one or two as in Table \ref{Table_RelectionCoefficientDiscrete}, this may, in general, be easier to be obtained in practice.

In order to evaluate the impact of realistic alphabets for the RIS reconfigurables elements, the following optimization criteria are considered:
\begin{itemize}
\item \textbf{Unit Amplitude and Continuous Phase (UACP)}. Based on this optimization criterion, which is usually considered in wireless communications, the reflection coefficient $\gamma_{nm}$ in \eqref{Local_Optimization_Problem} is set as follows:
\begin{equation}
\begin{array}{l}
\left| {{\gamma _{nm}}} \right| = 1\\
\angle {\gamma _{nm}} =  - \angle {g_{nm}} - \angle {h_{nm}}
\end{array}
\end{equation}
\item \textbf{Unit Amplitude and Discrete Evenly-Spaced Phase (UADP)}. Based on this optimization criterion, which is usually considered in wireless communications, the reflection coefficient $\gamma_{nm}$ in \eqref{Local_Optimization_Problem} is set as follows:
\begin{equation}
\begin{array}{l}
\left| {{\gamma _{nm}}} \right| = 1\\
\angle {\gamma _{nm}} = \mathop {\arg \min }\limits_{\phi  \in {\mathcal{P}}\left( L \right)} \left\{ {\left| {\phi  - \left( {\angle {g_{nm}} + \angle {h_{nm}}} \right)} \right|} \right\}
\end{array}
\end{equation}
where ${\mathcal{P}}\left( L \right)$ is the set of $L$ evenly-spaced phases as follows:
\begin{equation}
{\mathcal{P}}\left( L \right) = \left\{ {\phi :\frac{{2\pi }}{L}n,\quad n = 0,1, \ldots ,L - 1} \right\}
\end{equation}
\item \textbf{Unit Amplitude and Experimental Phase (UAEP)}. Based on this optimization criterion, the reflection coefficients $\gamma_{nm}$ in \eqref{Local_Optimization_Problem} are set as follows:
\begin{equation}
\begin{array}{l}
\left| {{\gamma _{nm}}} \right| = 1\\
\angle {\gamma _{nm}} = \mathop {\arg \min }\limits_{\phi  \in {\mathcal{A}}} \left\{ {\left| {\phi  - \left( {\angle {g_{nm}} + \angle {h_{nm}}} \right)} \right|} \right\}
\end{array}
\end{equation}
where the set $\mathcal{A}$ is the actual alphabet in Table \ref{Table_RelectionCoefficientDiscrete} and Table \ref{Table_RelectionCoefficientContinuous}.
\item \textbf{Experimental Alphabet (Alphabet)}. This corresponds to the case of interest, in which the actual reflection coefficients, in amplitude and phase, in Table \ref{Table_RelectionCoefficientDiscrete} and Table \ref{Table_RelectionCoefficientContinuous} are considered.
\end{itemize}

In order to analyze the reradiation pattern of an RIS, we plot the received power in \eqref{Received_power} as a function of the angle of observation (elevation) ${\theta _{{\rm{obs}}}} \in \left[ { - 90,90} \right]$ for the four case studies ``UACP'', ``UADP'', ``UAEP'', and ``Alphabet''. As far as the UADP case study is concerned, we indicate the number ${L}$ of evenly-spaced phases as well. Due to the large number of configuration setups that can be considered, we are not able to show all the simulation results. As an example, therefore, we report all the considered case studies only for the RIS alphabet in \cite{Linglong_Testbed}, and we then show a subset of figures for the other considered RIS alphabets in order to highlight differences and similarities.

\begin{figure}[!t]
\includegraphics[width=0.67\columnwidth]{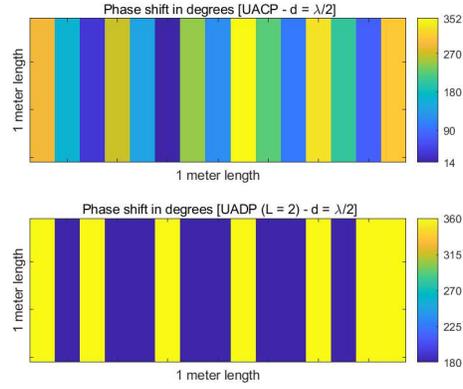}
\vspace{-0.60cm} \caption{Color map representation of $\boldsymbol{\Gamma}$ corresponding to the UACP and UADP (${L}=2$) case studies. The desired angle of reflection is 45 degrees and the inter-distance is $d=\lambda/2$.}\label{fig:Fig1__45deg_05lambda_Q1} \vspace{-0.30cm}
\end{figure}

\begin{figure}[!t]
\includegraphics[width=0.67\columnwidth]{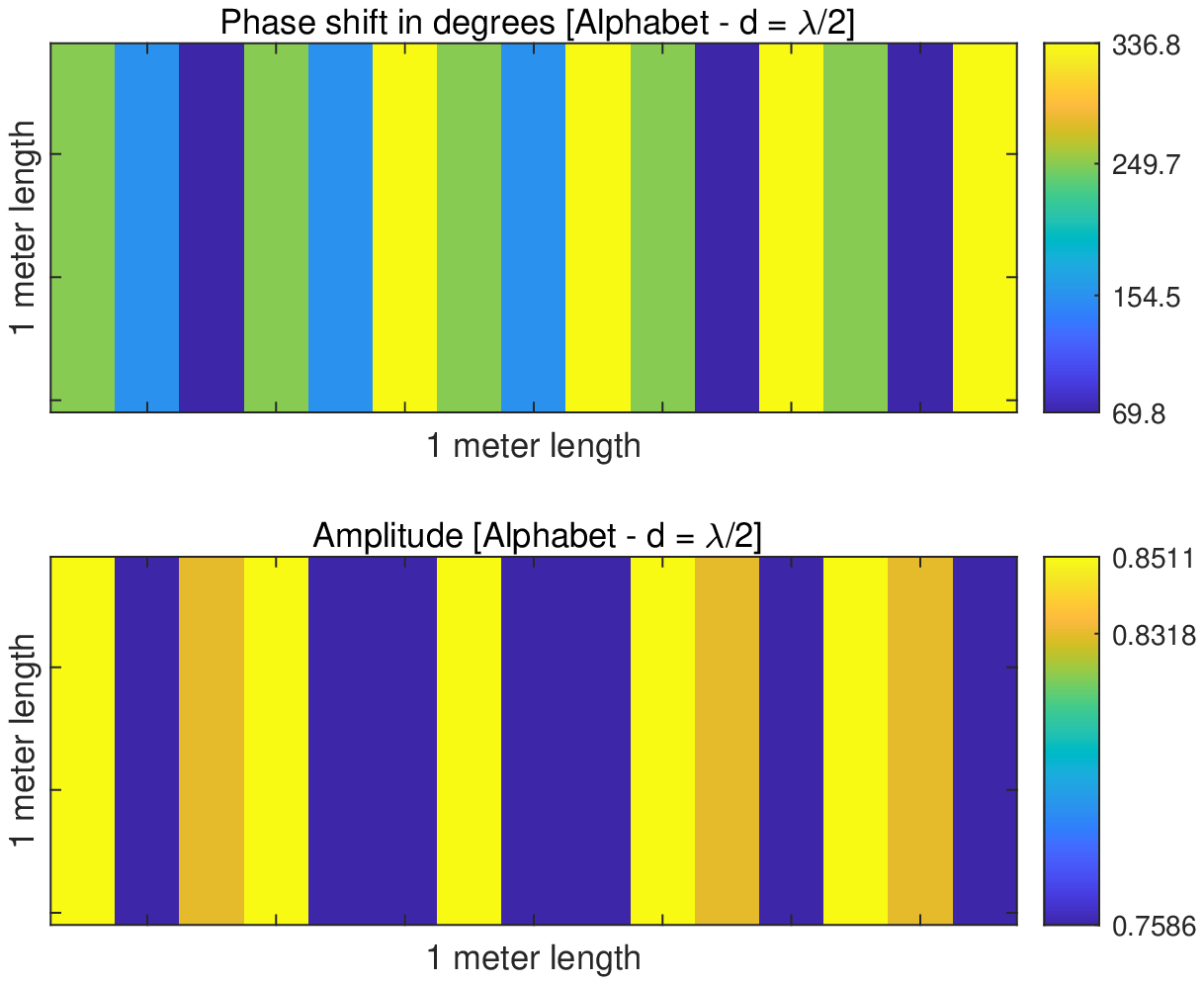}
\vspace{-0.60cm} \caption{Color map representation of $\boldsymbol{\Gamma}$ corresponding to the RIS alphabet in \cite{Linglong_Testbed}. The desired angle of reflection is 45 degrees and the inter-distance is $d=\lambda/2$.}\label{fig:Fig2__45deg_05lambda_Q1} \vspace{-0.30cm}
\end{figure}

\begin{figure}[!t]
\includegraphics[width=0.67\columnwidth]{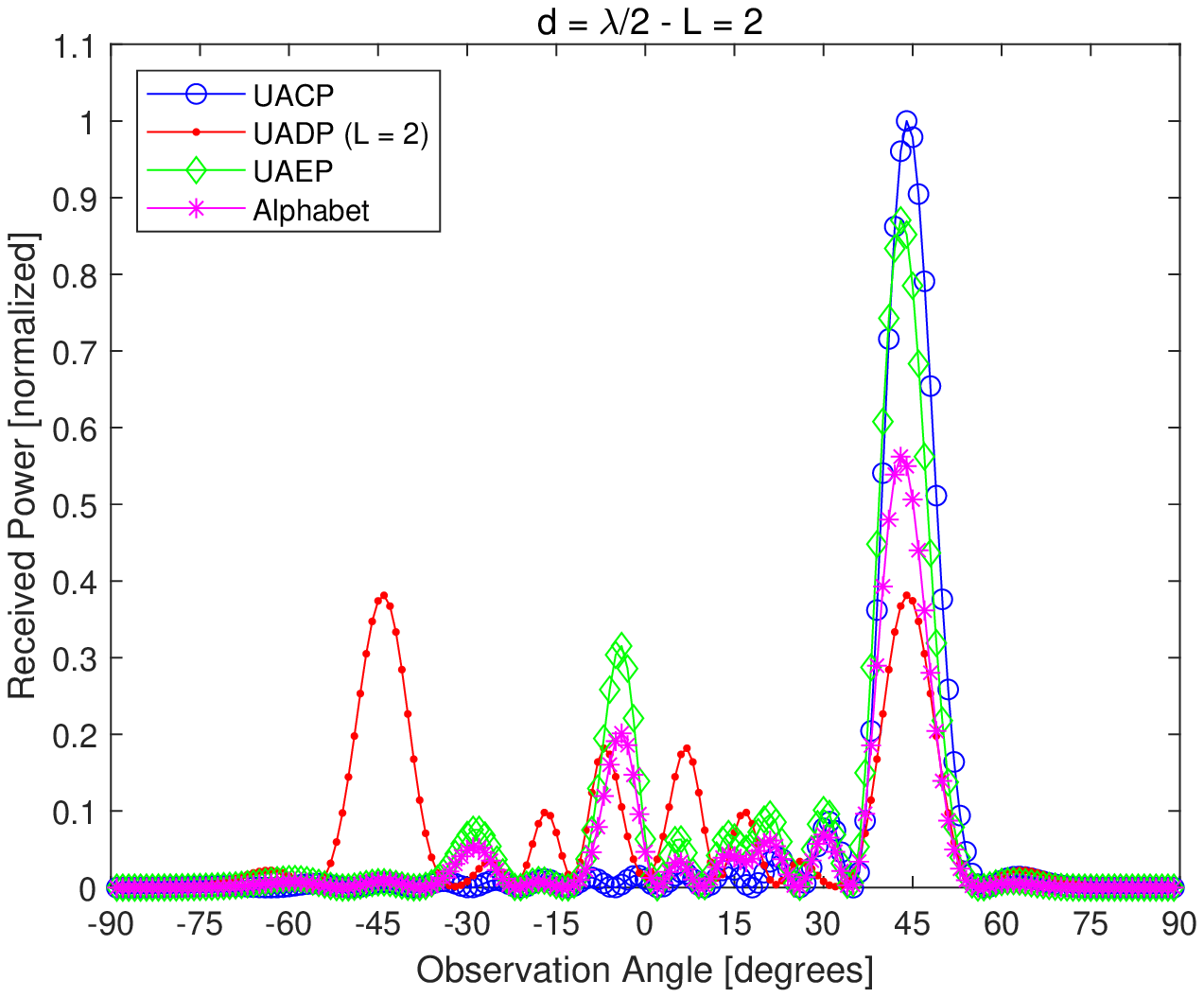}
\vspace{-0.25cm} \caption{Received power as a function of the angle of observation. The RIS alphabet is \cite{Linglong_Testbed}, the desired angle of reflection is 45 degrees, and the inter-distance is $d=\lambda/2$.}\label{fig:Fig3__45deg_05lambda_Q1} \vspace{-0.30cm}
\end{figure}

\begin{figure}[!t]
\includegraphics[width=0.67\columnwidth]{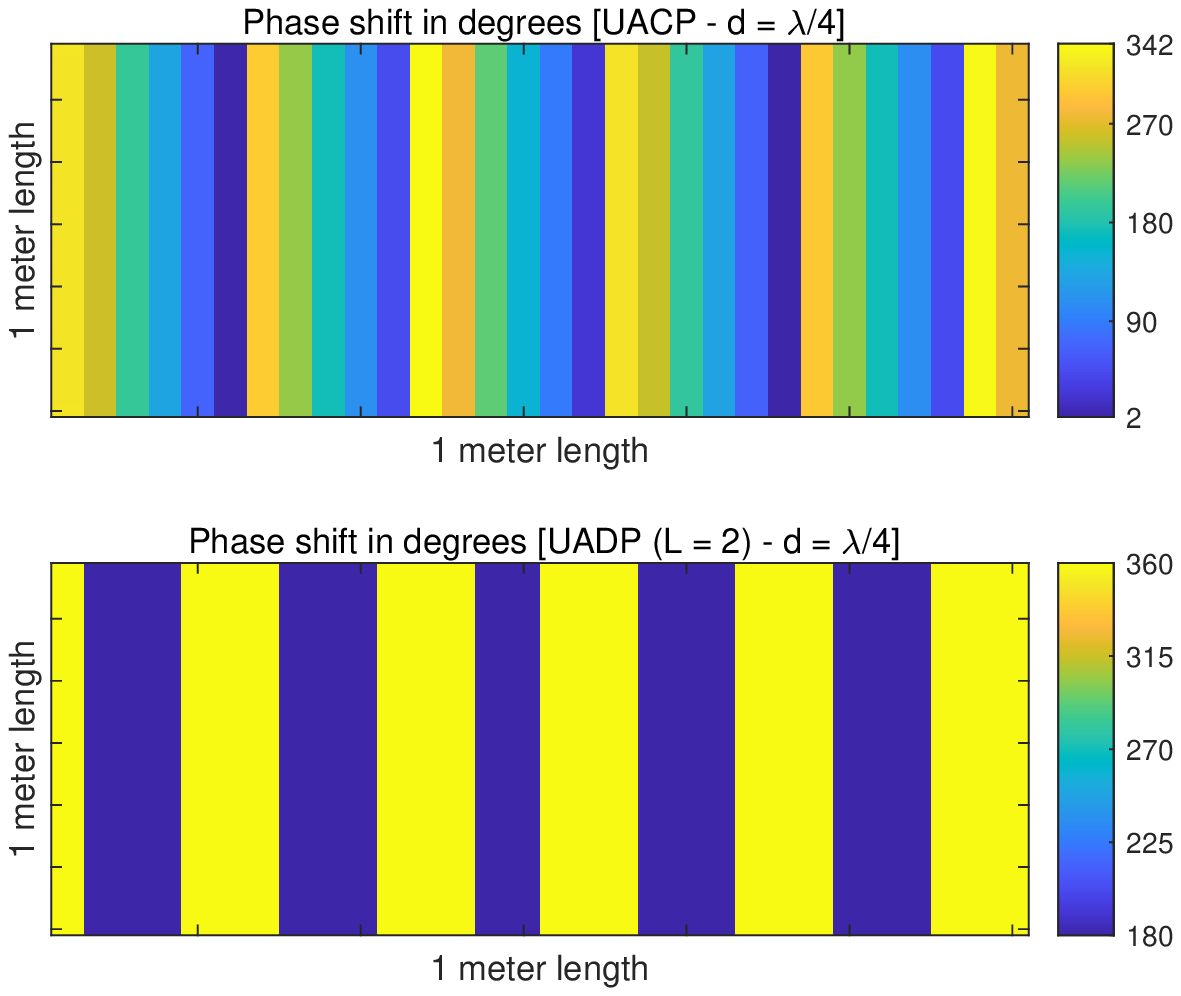}
\vspace{-0.60cm} \caption{Color map representation of $\boldsymbol{\Gamma}$ corresponding to the UACP and UADP (${L}=2$) case studies. The desired angle of reflection is 45 degrees and the inter-distance is $d=\lambda/4$.}\label{fig:Fig1__45deg_025lambda_Q1} \vspace{-0.30cm}
\end{figure}

\begin{figure}[!t]
\includegraphics[width=0.67\columnwidth]{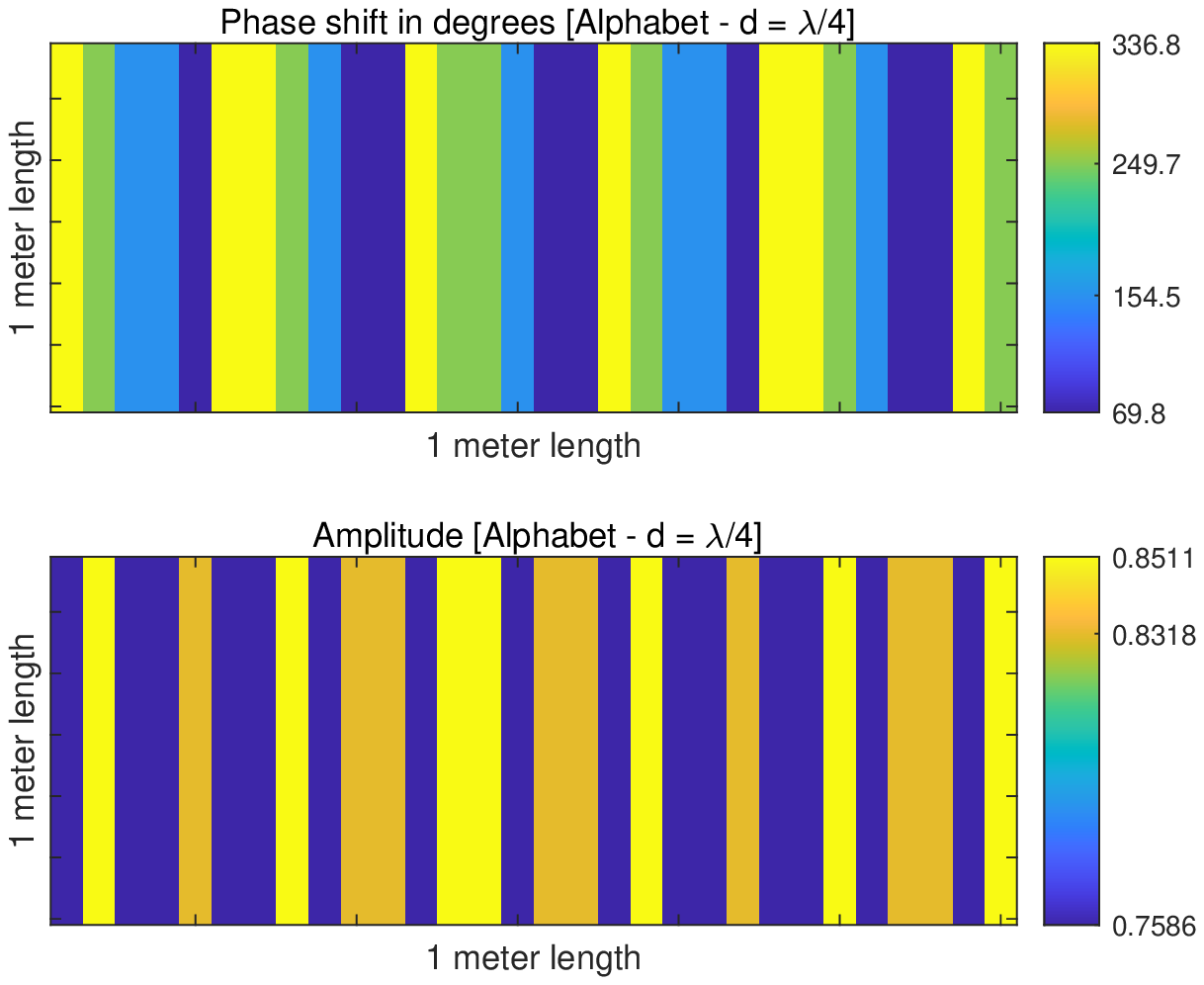}
\vspace{-0.60cm} \caption{Color map representation of $\boldsymbol{\Gamma}$ corresponding to the RIS alphabet in \cite{Linglong_Testbed}. The desired angle of reflection is 45 degrees and the inter-distance is $d=\lambda/4$.}\label{fig:Fig2__45deg_025lambda_Q1} \vspace{-0.30cm}
\end{figure}

\begin{figure}[!t]
\includegraphics[width=0.67\columnwidth]{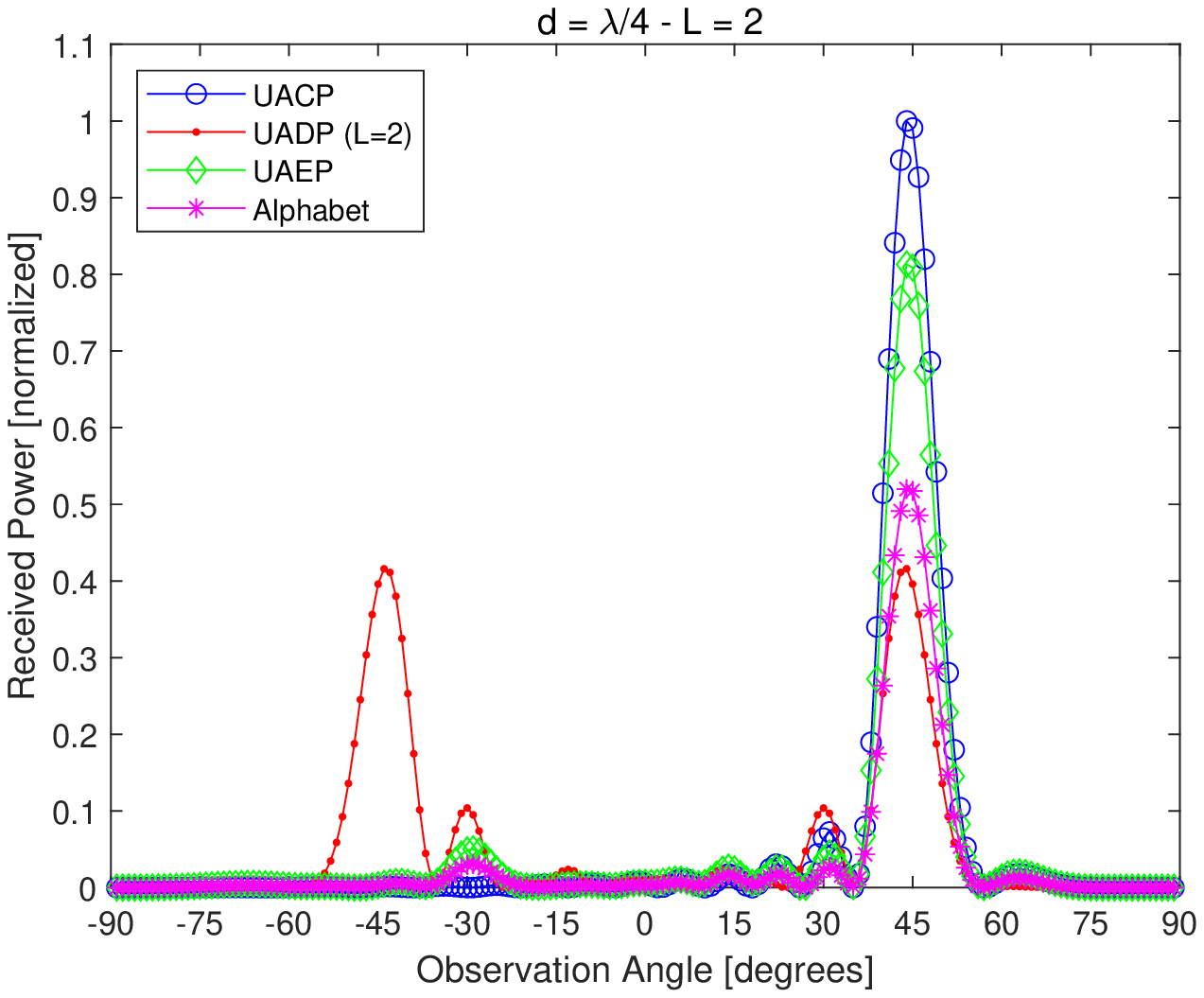}
\vspace{-0.25cm} \caption{Received power as a function of the angle of observation. The RIS alphabet is \cite{Linglong_Testbed}, the desired angle of reflection is 45 degrees, and the inter-distance is $d=\lambda/4$.}\label{fig:Fig3__45deg_025lambda_Q1} \vspace{-0.30cm}
\end{figure}

\begin{figure}[!t]
\includegraphics[width=0.67\columnwidth]{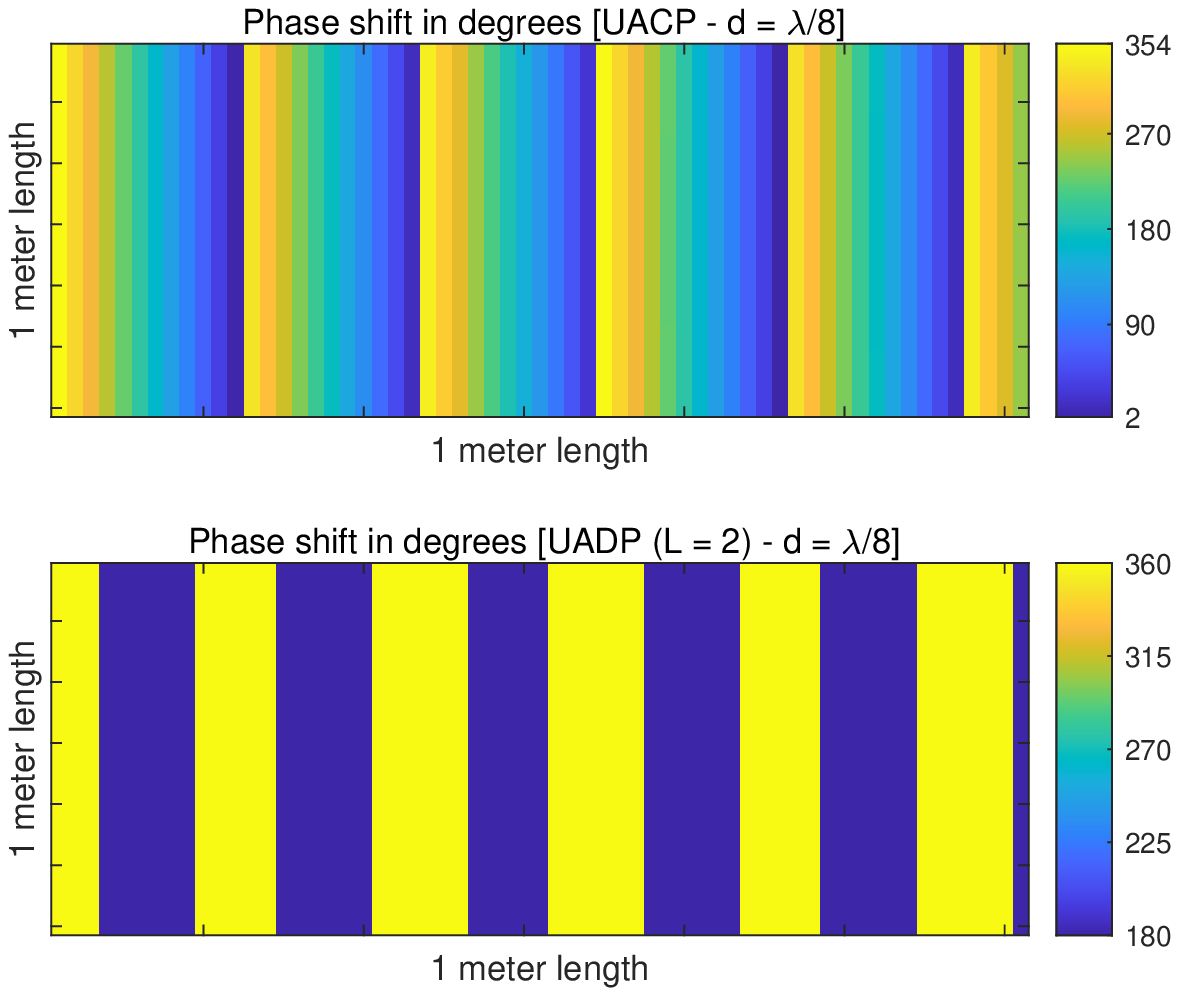}
\vspace{-0.60cm} \caption{Color map representation of $\boldsymbol{\Gamma}$ corresponding to the UACP and UADP (${L}=2$) case studies. The desired angle of reflection is 45 degrees and the inter-distance is $d=\lambda/8$.}\label{fig:Fig1__45deg_0125lambda_Q1} \vspace{-0.30cm}
\end{figure}

\begin{figure}[!t]
\includegraphics[width=0.67\columnwidth]{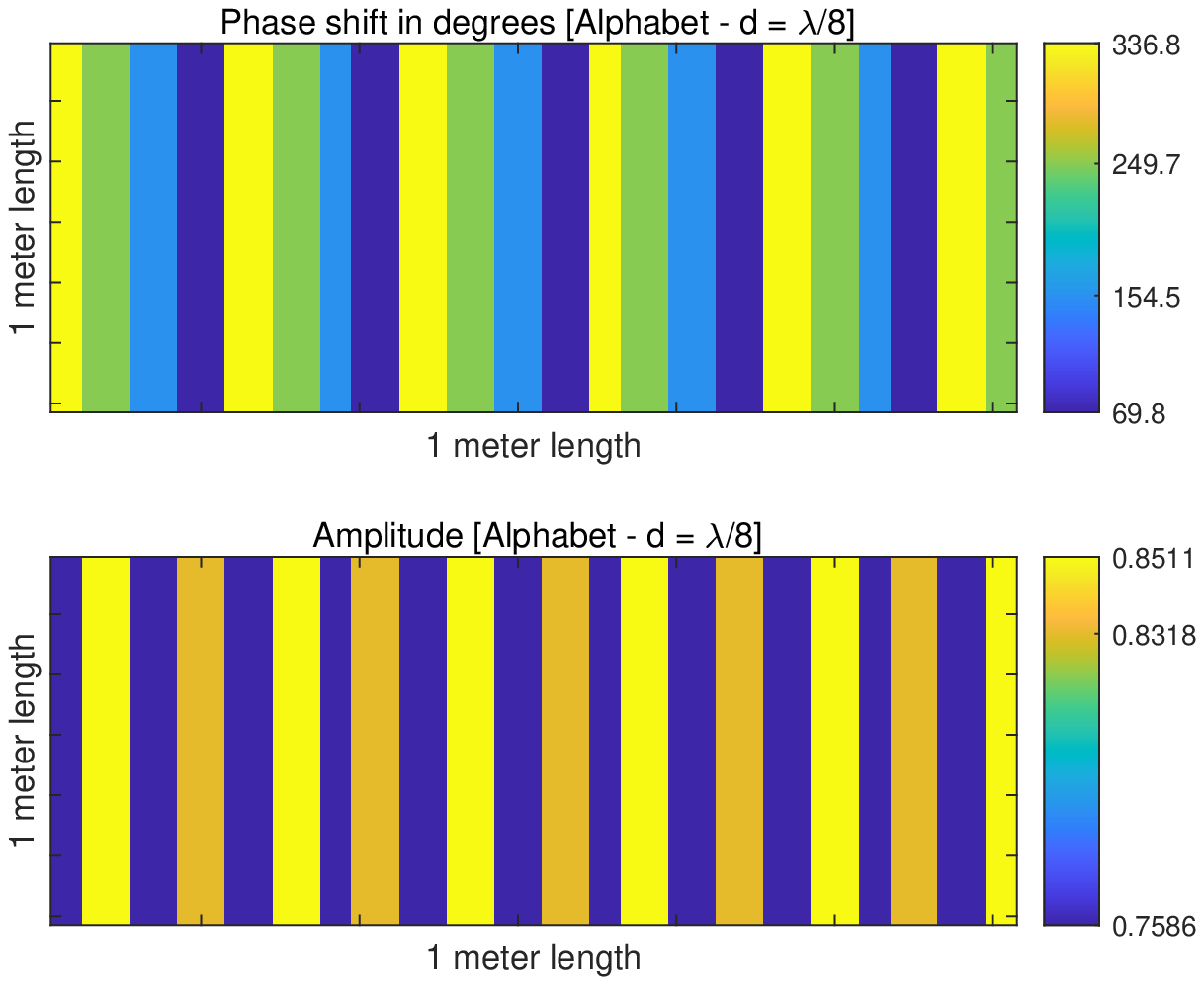}
\vspace{-0.60cm} \caption{Color map representation of $\boldsymbol{\Gamma}$ corresponding to the RIS alphabet in \cite{Linglong_Testbed}. The desired angle of reflection is 45 degrees and the inter-distance is $d=\lambda/8$.}\label{fig:Fig2__45deg_0125lambda_Q1} \vspace{-0.30cm}
\end{figure}

\begin{figure}[!t]
\includegraphics[width=0.67\columnwidth]{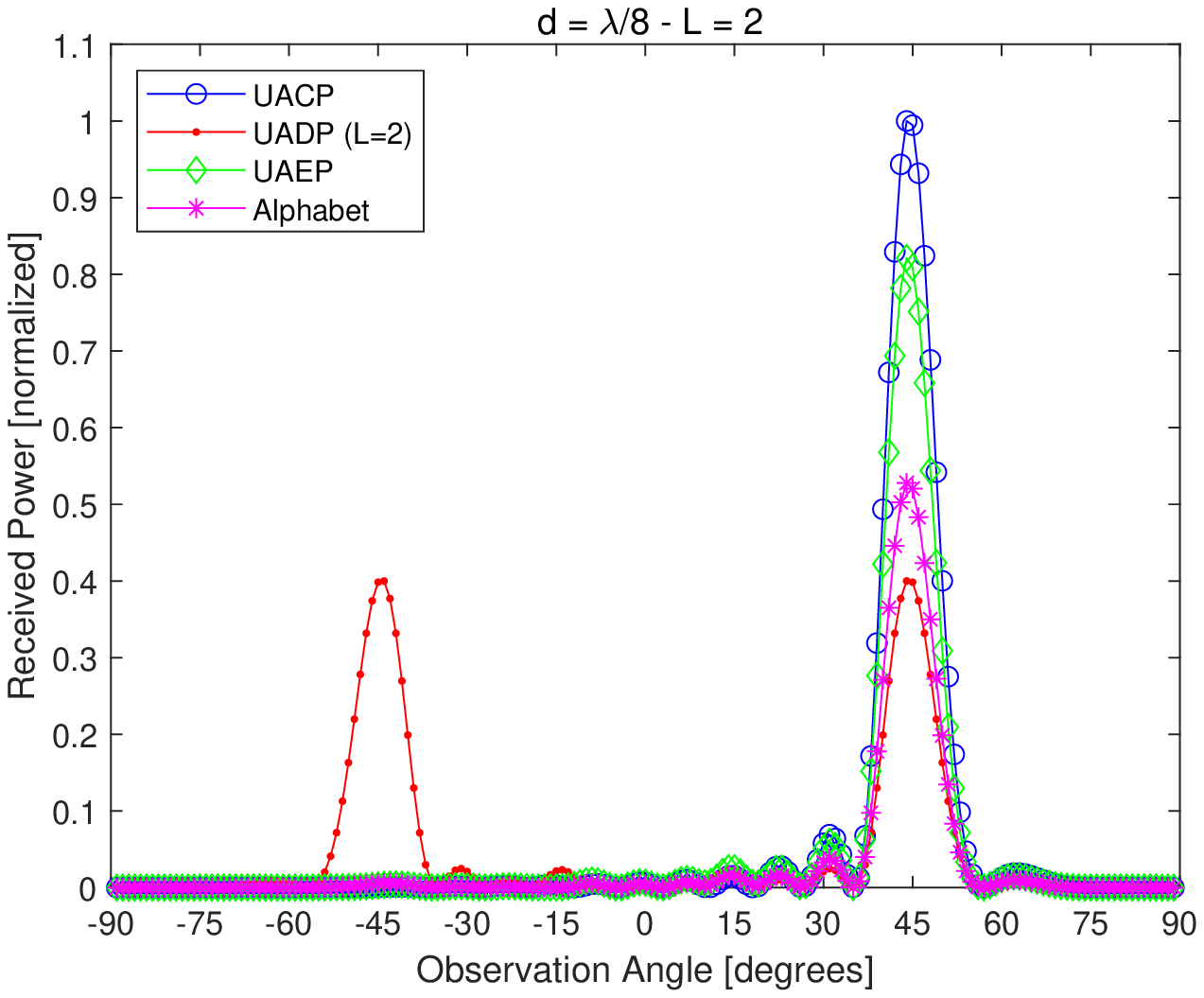}
\vspace{-0.25cm} \caption{Received power as a function of the angle of observation. The RIS alphabet is \cite{Linglong_Testbed}, the desired angle of reflection is 45 degrees, and the inter-distance is $d=\lambda/8$.}\label{fig:Fig3__45deg_0125lambda_Q1} \vspace{-0.30cm}
\end{figure}

\begin{figure}[!t]
\includegraphics[width=0.67\columnwidth]{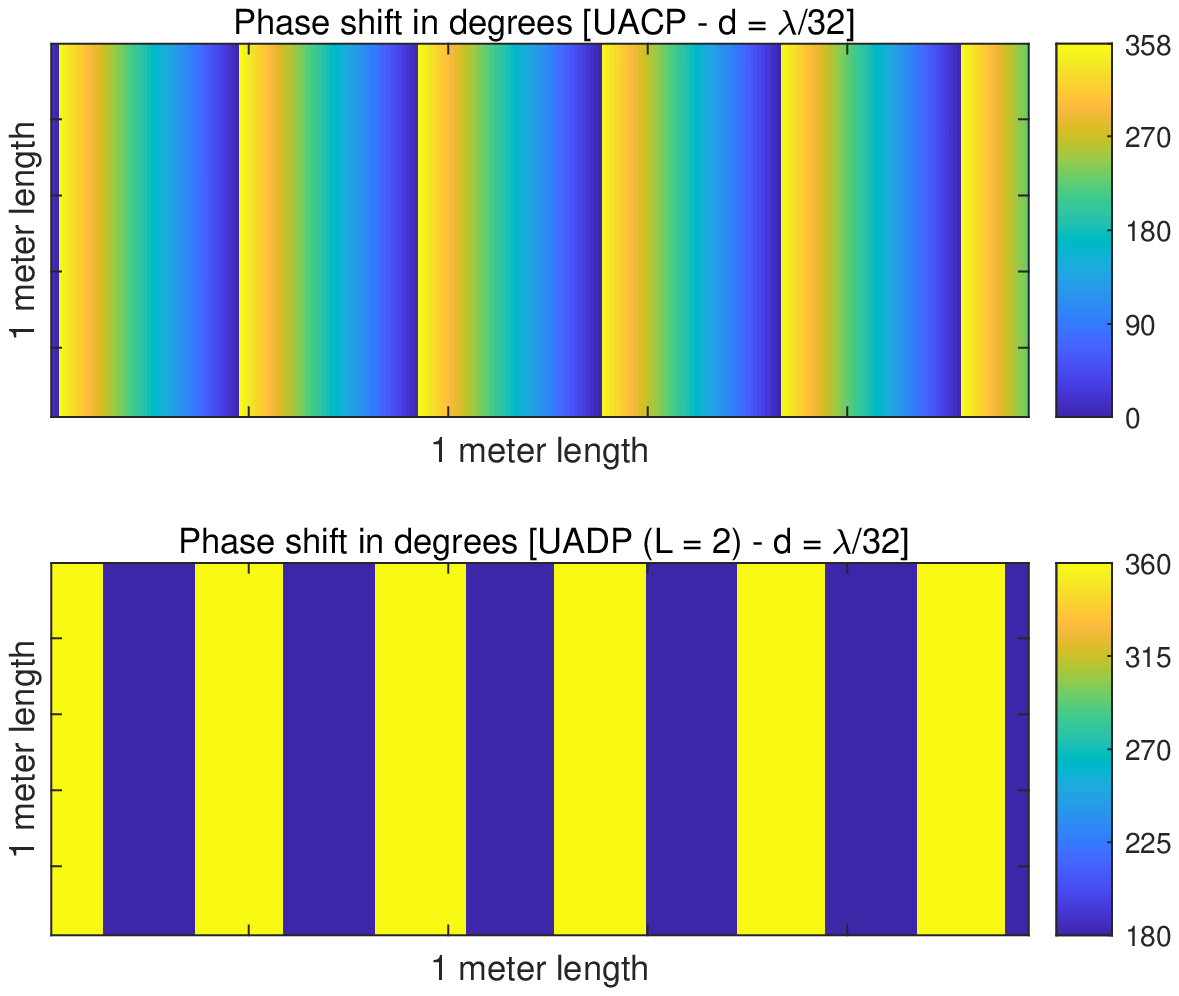}
\vspace{-0.60cm} \caption{Color map representation of $\boldsymbol{\Gamma}$ corresponding to the UACP and UADP (${L}=2$) case studies. The desired angle of reflection is 45 degrees and the inter-distance is $d=\lambda/32$.}\label{fig:Fig1__45deg_lambda32_Q1} \vspace{-0.30cm}
\end{figure}

\begin{figure}[!t]
\includegraphics[width=0.67\columnwidth]{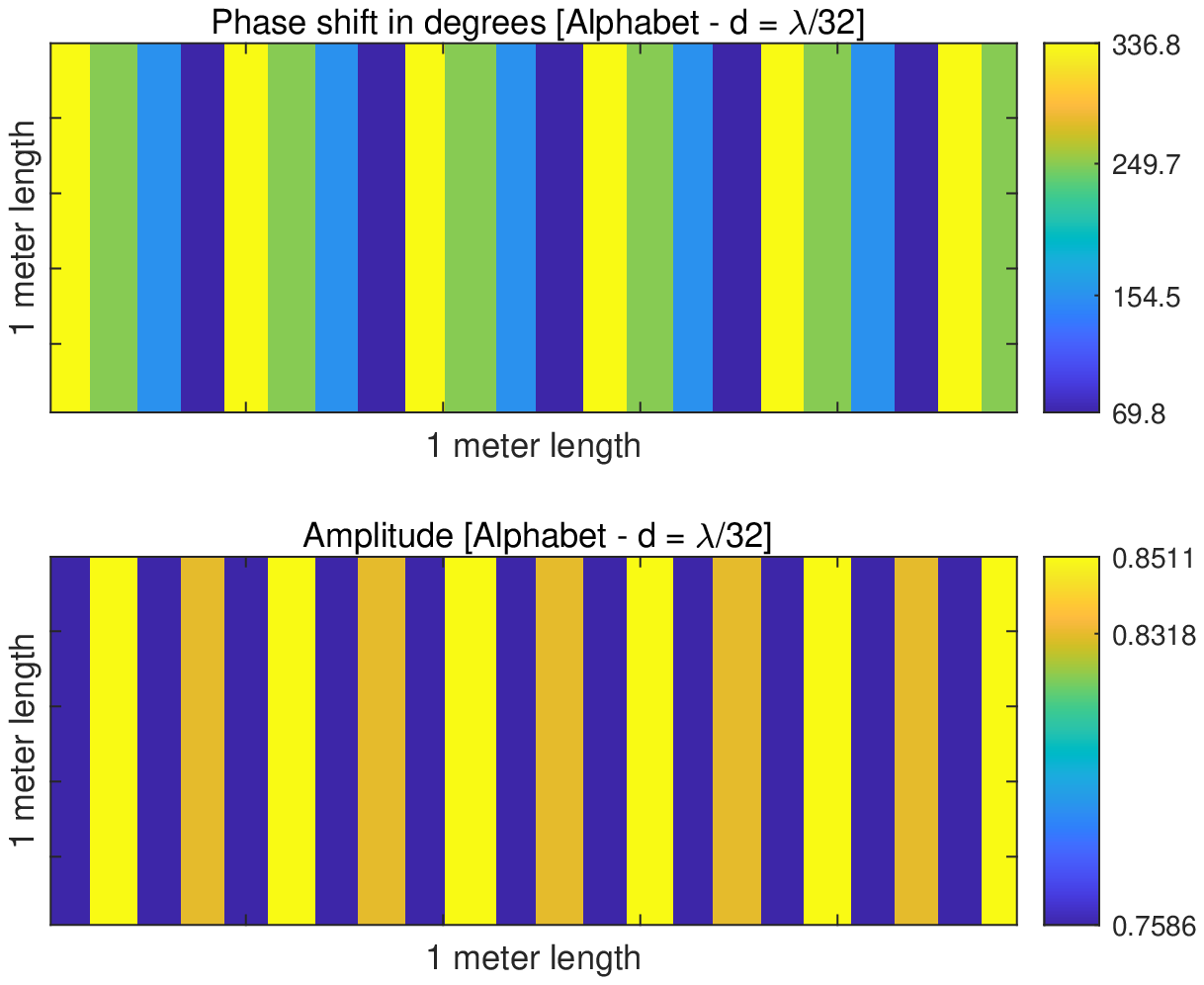}
\vspace{-0.60cm} \caption{Color map representation of $\boldsymbol{\Gamma}$ corresponding to the RIS alphabet in \cite{Linglong_Testbed}. The desired angle of reflection is 45 degrees and the inter-distance is $d=\lambda/32$.}\label{fig:Fig2__45deg_lambda32_Q1} \vspace{-0.30cm}
\end{figure}

\begin{figure}[!t]
\includegraphics[width=0.67\columnwidth]{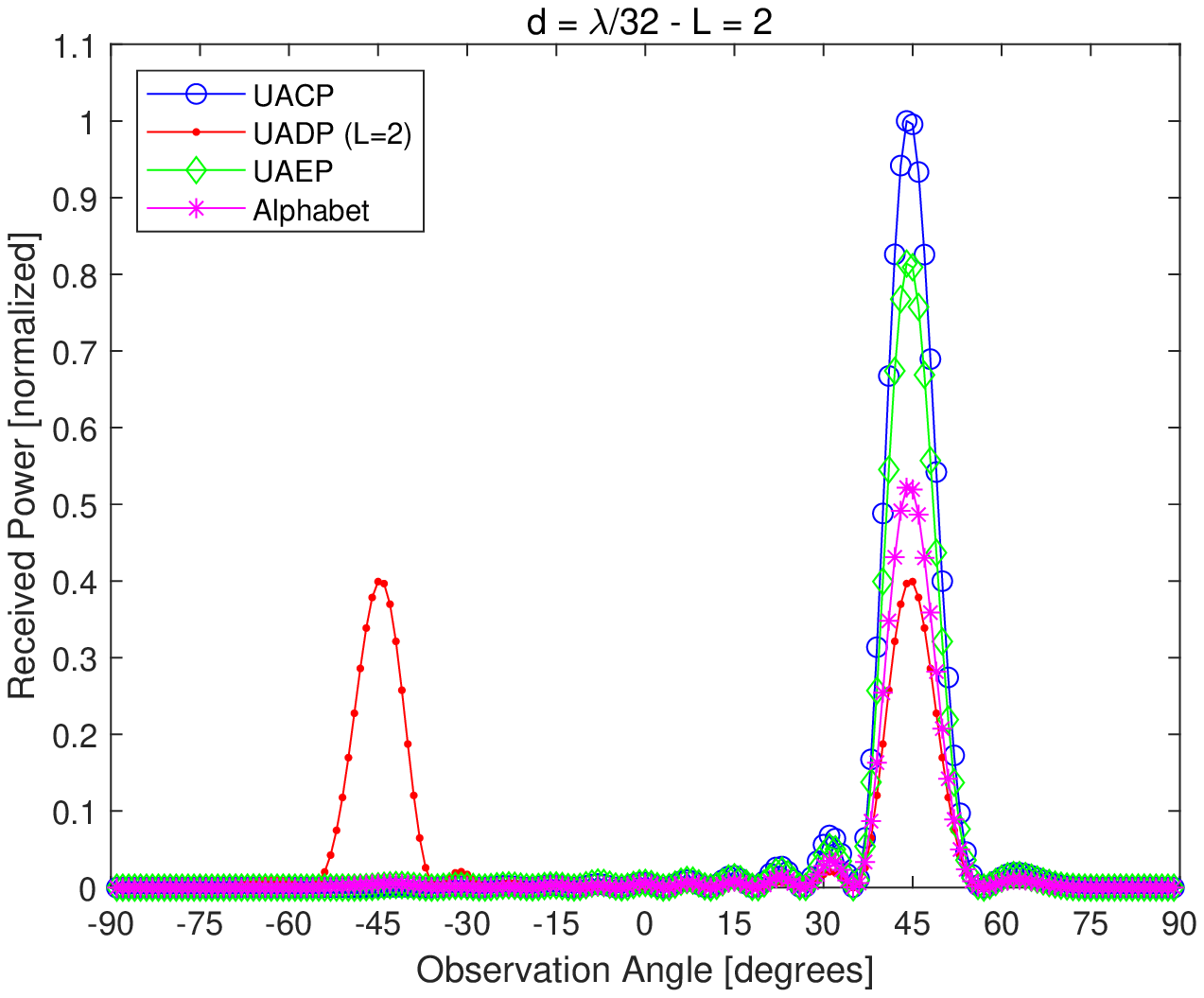}
\vspace{-0.25cm} \caption{Received power as a function of the angle of observation. The RIS alphabet is \cite{Linglong_Testbed}, the desired angle of reflection is 45 degrees, and the inter-distance is $d=\lambda/32$.}\label{fig:Fig3__45deg_lambda32_Q1} \vspace{-0.30cm}
\end{figure}

\begin{figure}[!t]
\includegraphics[width=0.67\columnwidth]{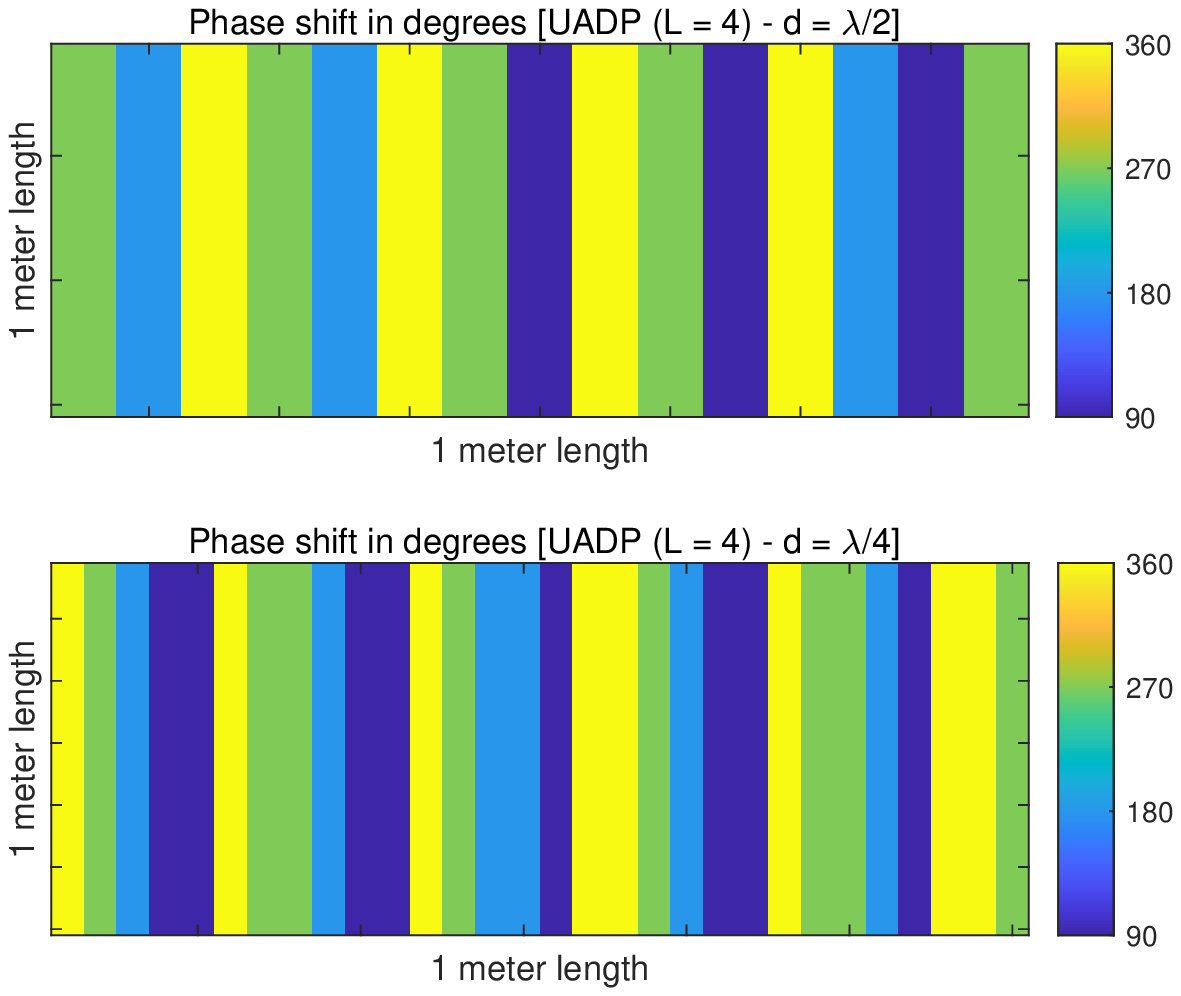}
\vspace{-0.60cm} \caption{Color map representation of $\boldsymbol{\Gamma}$ corresponding to the UADP (${L}=4$) case study. The desired angle of reflection is 45 degrees and the inter-distance is $d=\lambda/2$ and $d=\lambda/4$.}\label{fig:ColorMap__45deg_lambda2and4_Q2} \vspace{-0.30cm}
\end{figure}

\begin{figure}[!t]
\includegraphics[width=0.67\columnwidth]{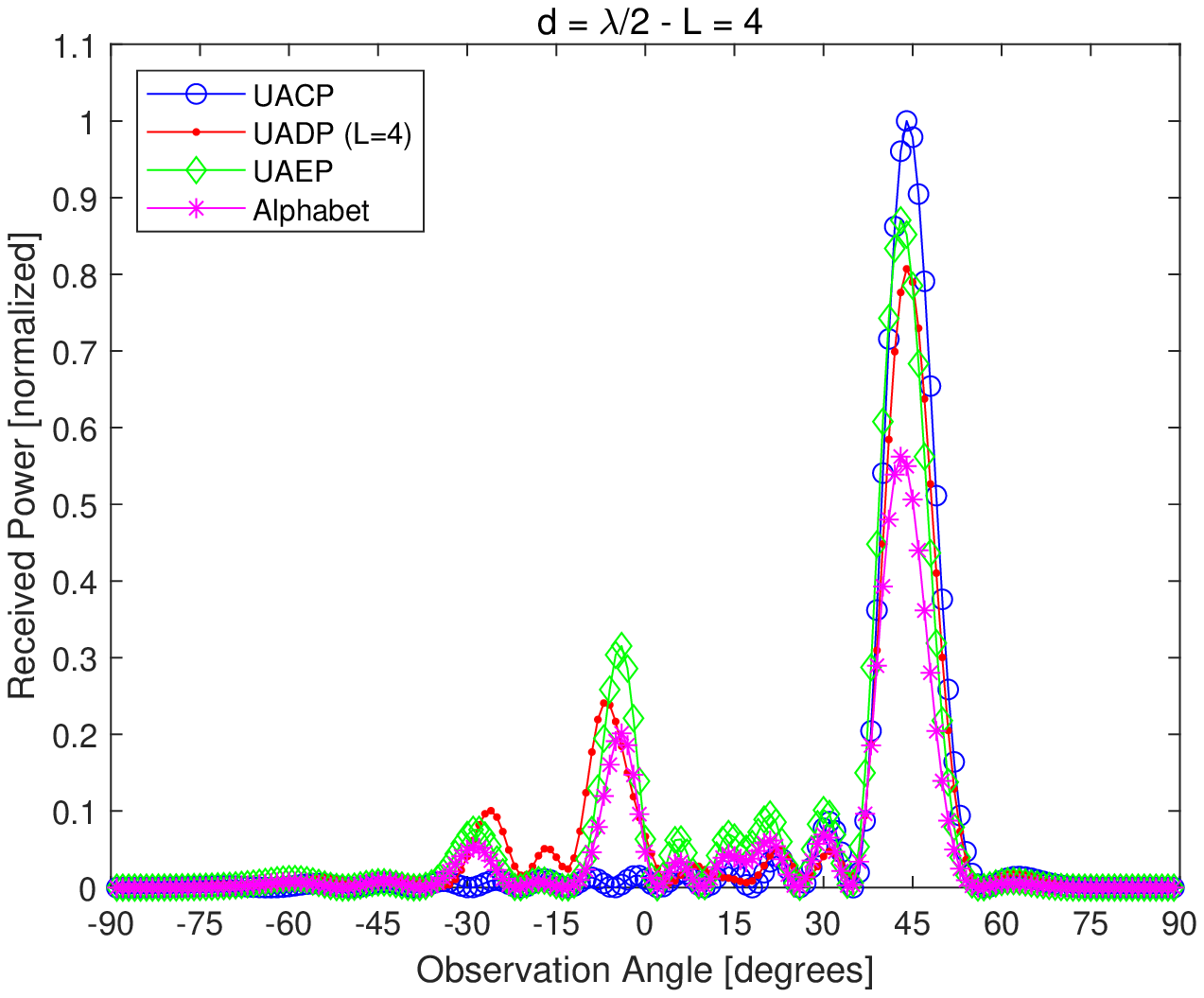}
\vspace{-0.25cm} \caption{Received power as a function of the angle of observation. The RIS alphabet is \cite{Linglong_Testbed}, the desired angle of reflection is 45 degrees, and the inter-distance is $d=\lambda/2$.}\label{fig:Prx__45deg_lambda2_Q2} \vspace{-0.30cm}
\end{figure}

\begin{figure}[!t]
\includegraphics[width=0.67\columnwidth]{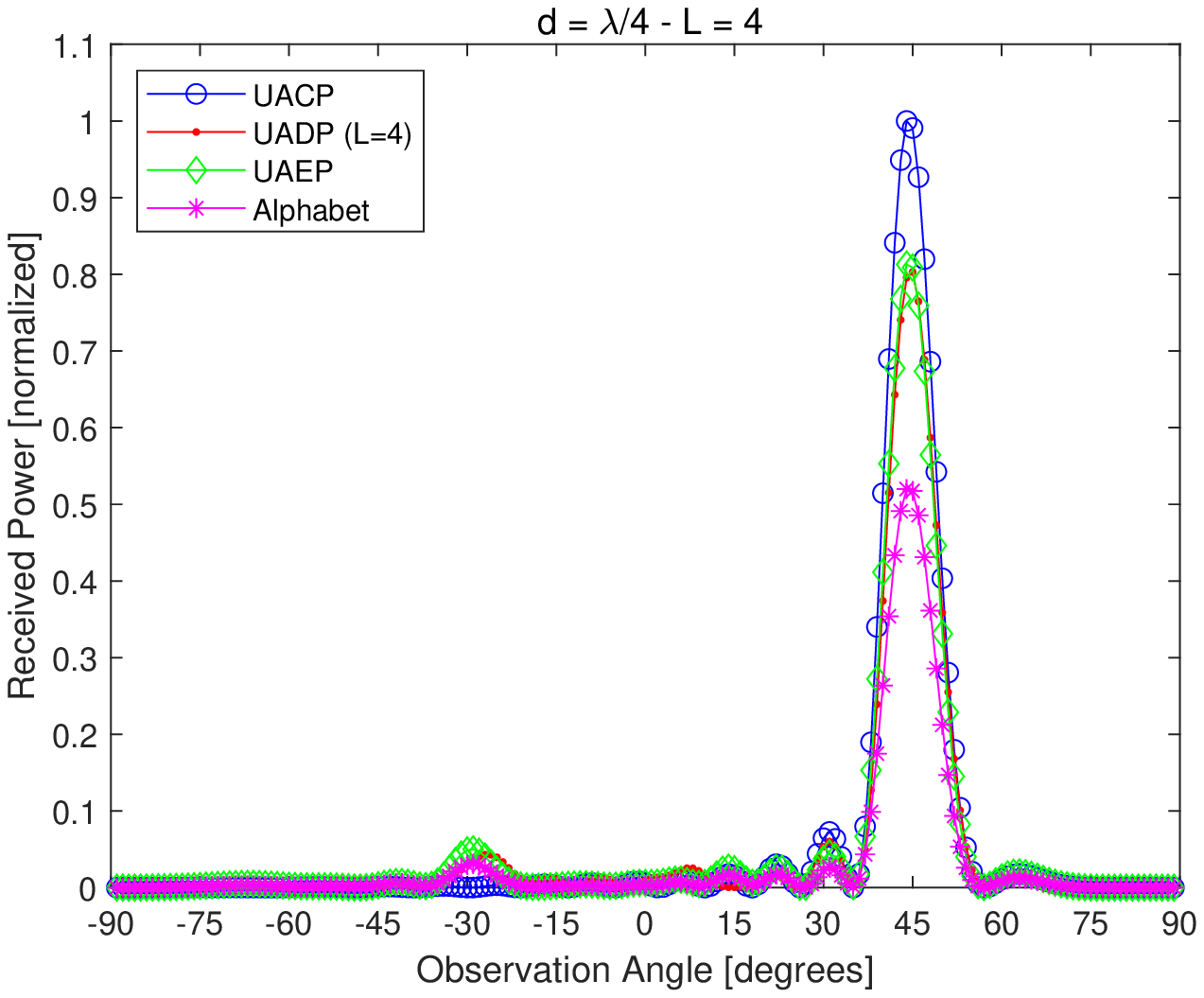}
\vspace{-0.25cm} \caption{Received power as a function of the angle of observation. The RIS alphabet is \cite{Linglong_Testbed}, the desired angle of reflection is 45 degrees, and the inter-distance is $d=\lambda/4$.}\label{fig:Prx__45deg_lambda4_Q2} \vspace{-0.30cm}
\end{figure}

\begin{figure}[!t]
\includegraphics[width=0.67\columnwidth]{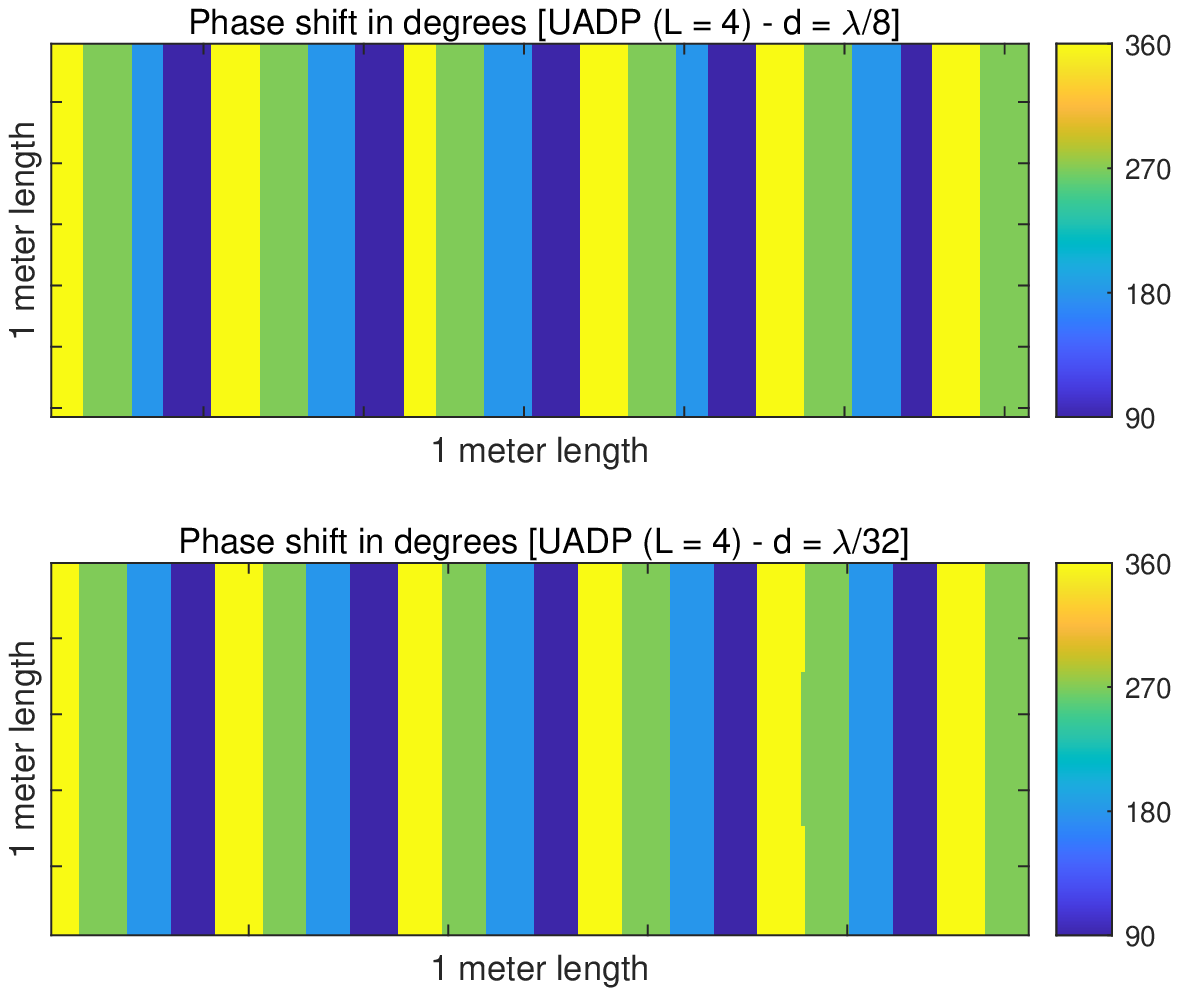}
\vspace{-0.60cm} \caption{Color map representation of $\boldsymbol{\Gamma}$ corresponding to the UADP (${L}=4$) case study. The desired angle of reflection is 45 degrees and the inter-distance is $d=\lambda/8$ and $d=\lambda/32$.}\label{fig:ColorMap__45deg_lambda8and32_Q2} \vspace{-0.30cm}
\end{figure}

\begin{figure}[!t]
\includegraphics[width=0.67\columnwidth]{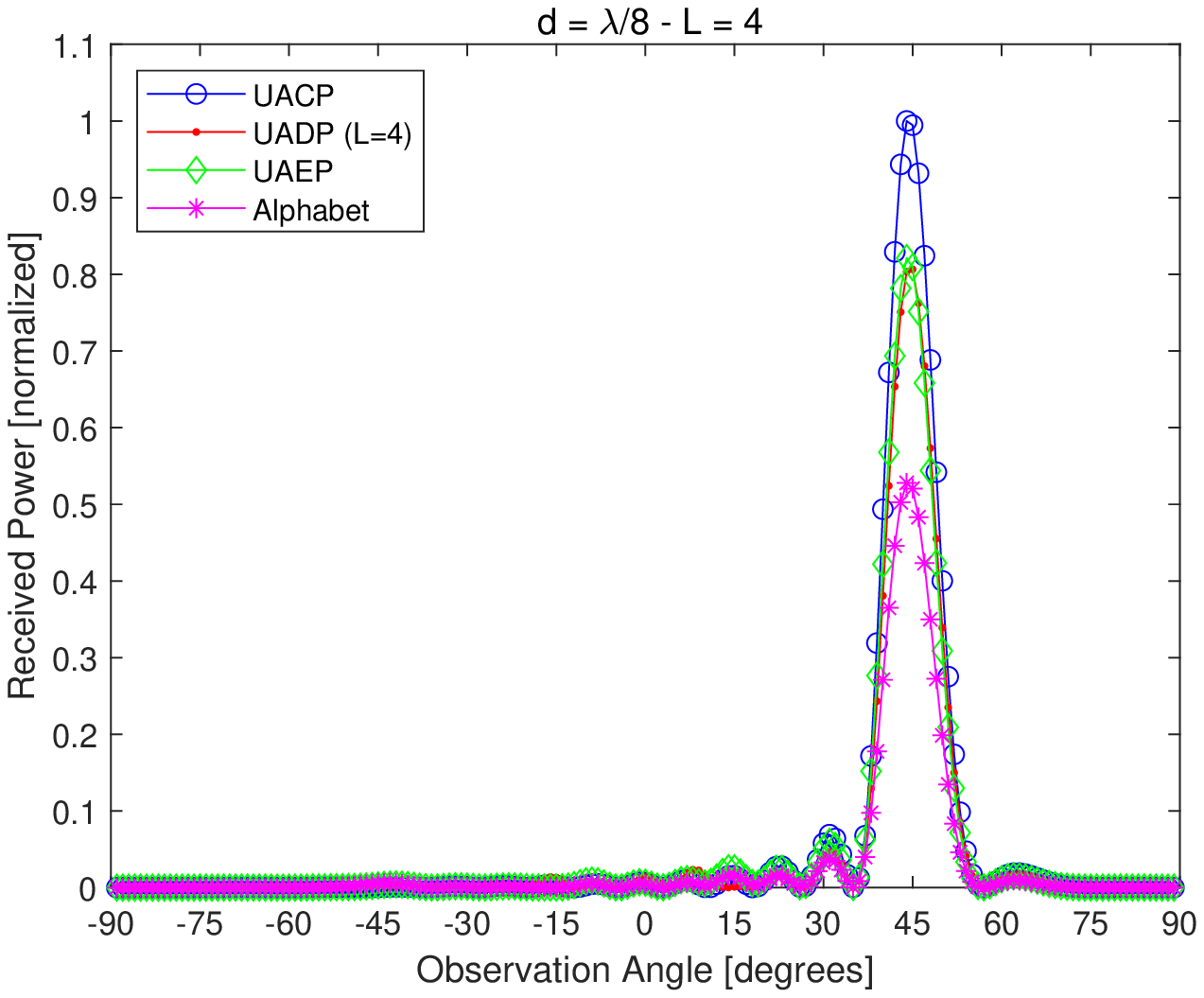}
\vspace{-0.25cm} \caption{Received power as a function of the angle of observation. The RIS alphabet is \cite{Linglong_Testbed}, the desired angle of reflection is 45 degrees, and the inter-distance is $d=\lambda/8$.}\label{fig:Prx__45deg_lambda8_Q2} \vspace{-0.30cm}
\end{figure}

\begin{figure}[!t]
\includegraphics[width=0.67\columnwidth]{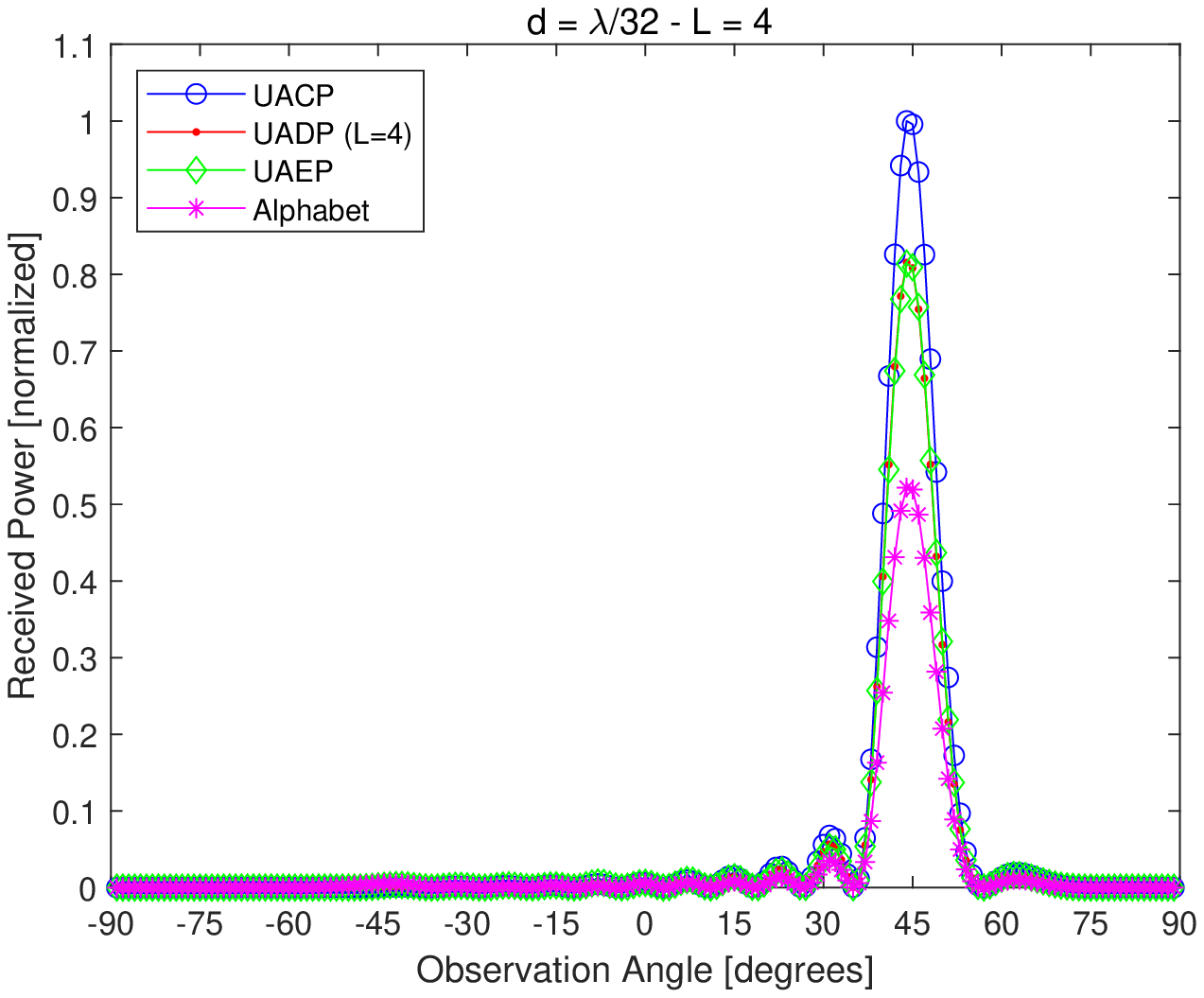}
\vspace{-0.25cm} \caption{Received power as a function of the angle of observation. The RIS alphabet is \cite{Linglong_Testbed}, the desired angle of reflection is 45 degrees, and the inter-distance is $d=\lambda/32$.}\label{fig:Prx__45deg_lambda32_Q2} \vspace{-0.30cm}
\end{figure}

\begin{figure}[!t]
\includegraphics[width=0.67\columnwidth]{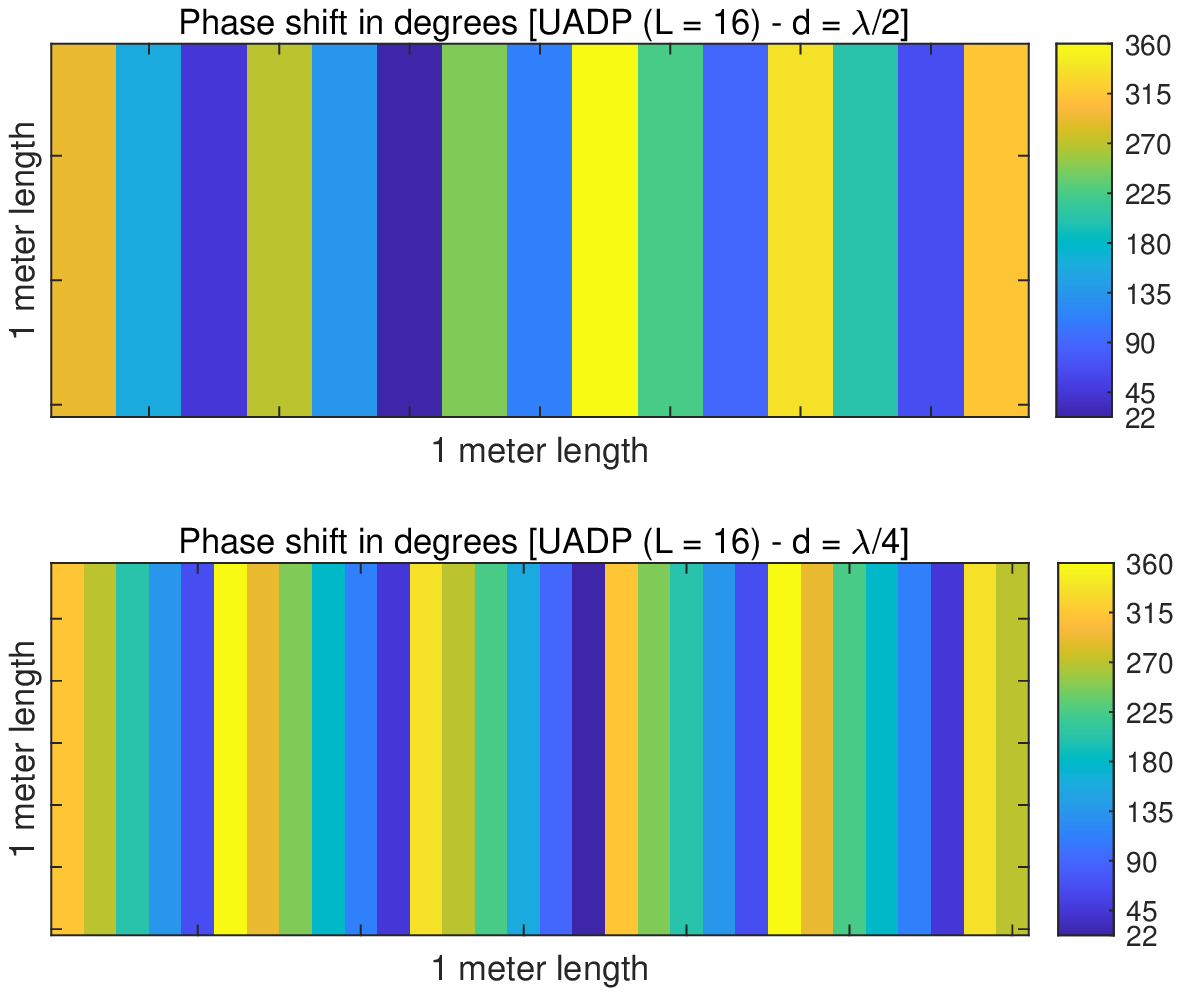}
\vspace{-0.60cm} \caption{Color map representation of $\boldsymbol{\Gamma}$ corresponding to the UADP (${L}=16$) case study. The desired angle of reflection is 45 degrees and the inter-distance is $d=\lambda/2$ and $d=\lambda/4$.}\label{fig:ColorMap__45deg_lambda2and4_Q4} \vspace{-0.30cm}
\end{figure}

\begin{figure}[!t]
\includegraphics[width=0.67\columnwidth]{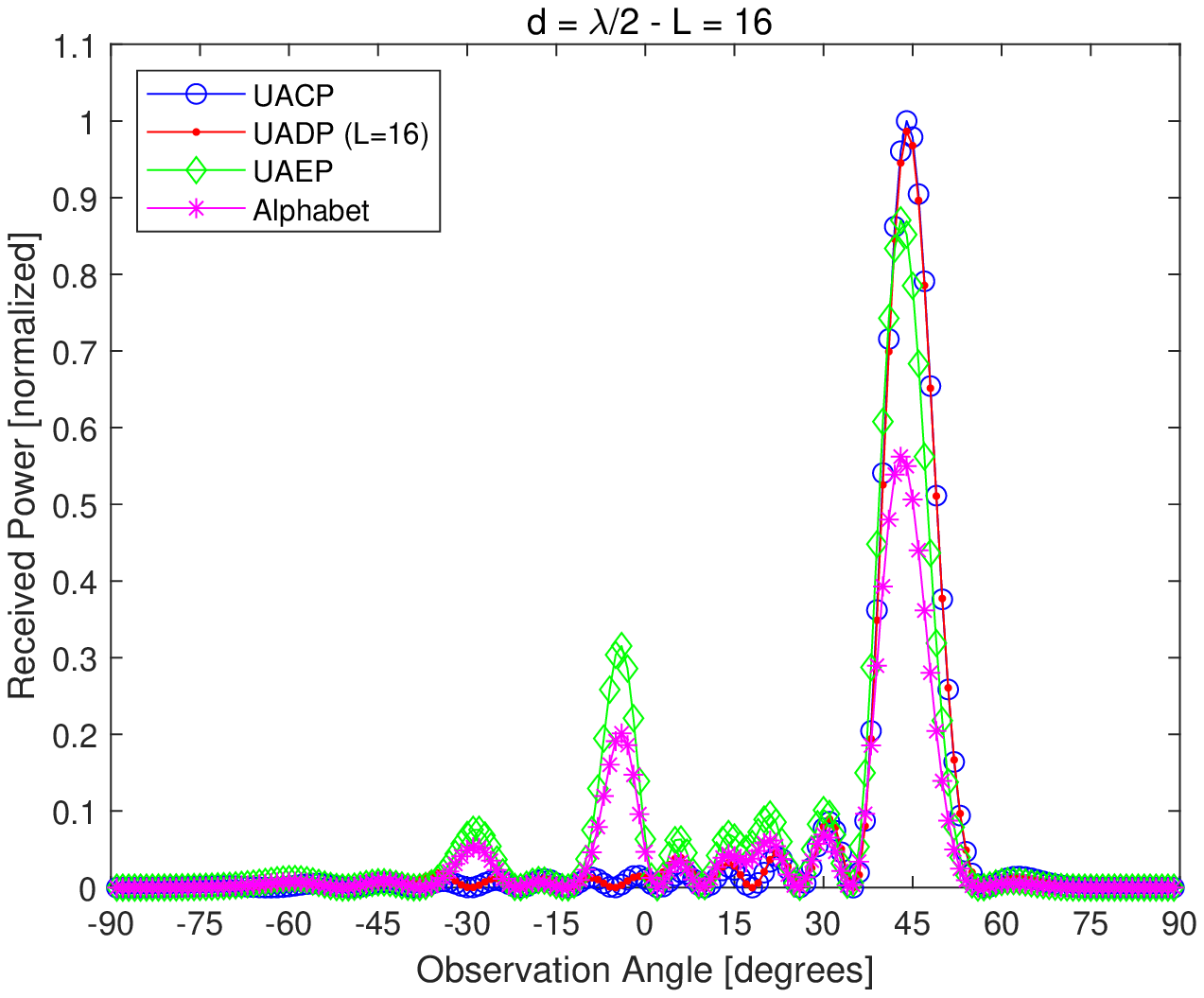}
\vspace{-0.25cm} \caption{Received power as a function of the angle of observation. The RIS alphabet is \cite{Linglong_Testbed}, the desired angle of reflection is 45 degrees, and the inter-distance is $d=\lambda/2$.}\label{fig:Prx__45deg_lambda2_Q4} \vspace{-0.30cm}
\end{figure}

\begin{figure}[!t]
\includegraphics[width=0.67\columnwidth]{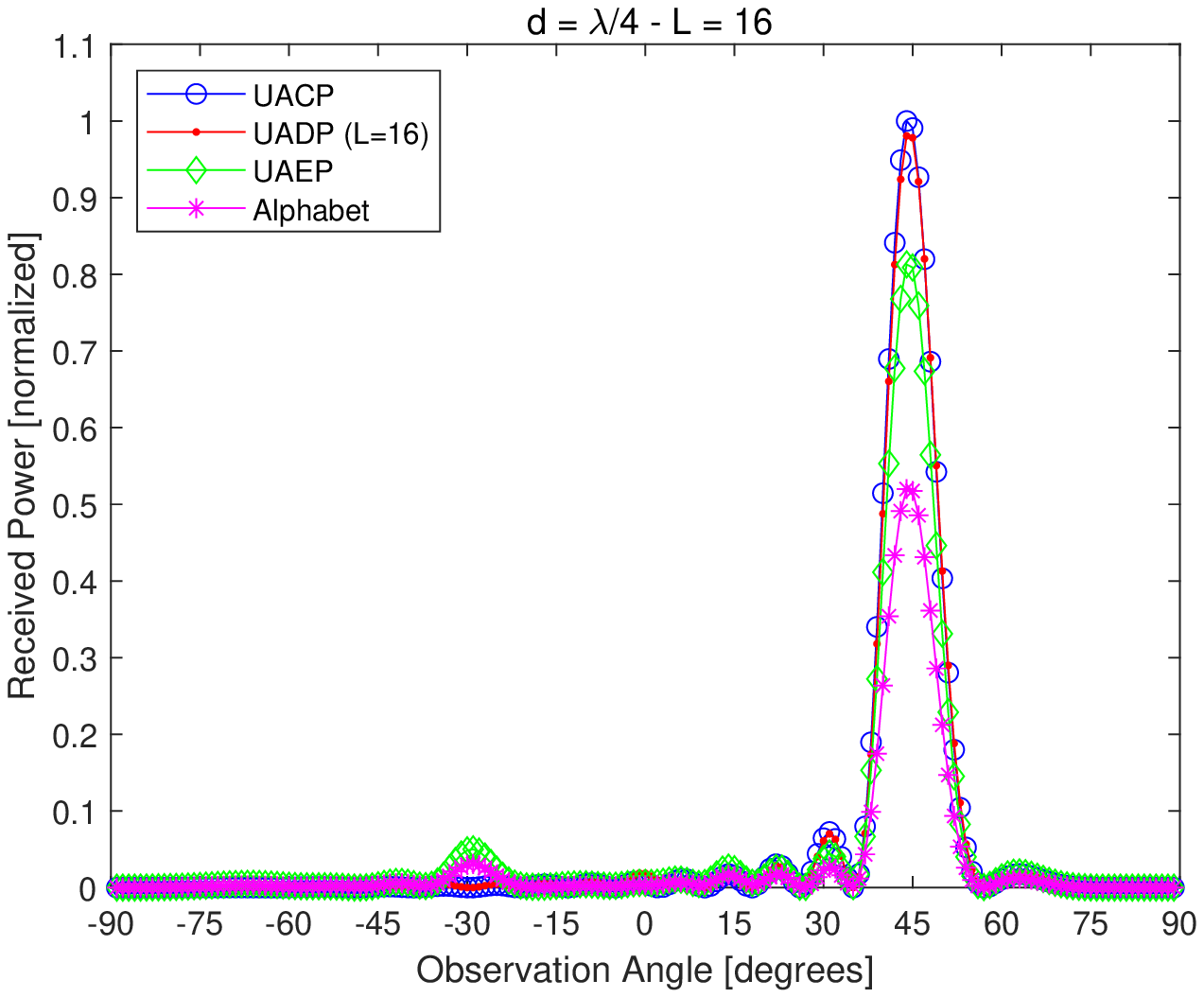}
\vspace{-0.25cm} \caption{Received power as a function of the angle of observation. The RIS alphabet is  \cite{Linglong_Testbed}, the desired angle of reflection is 45 degrees, and the inter-distance is $d=\lambda/4$.}\label{fig:Prx__45deg_lambda4_Q4} \vspace{-0.30cm}
\end{figure}

\begin{figure}[!t]
\includegraphics[width=0.67\columnwidth]{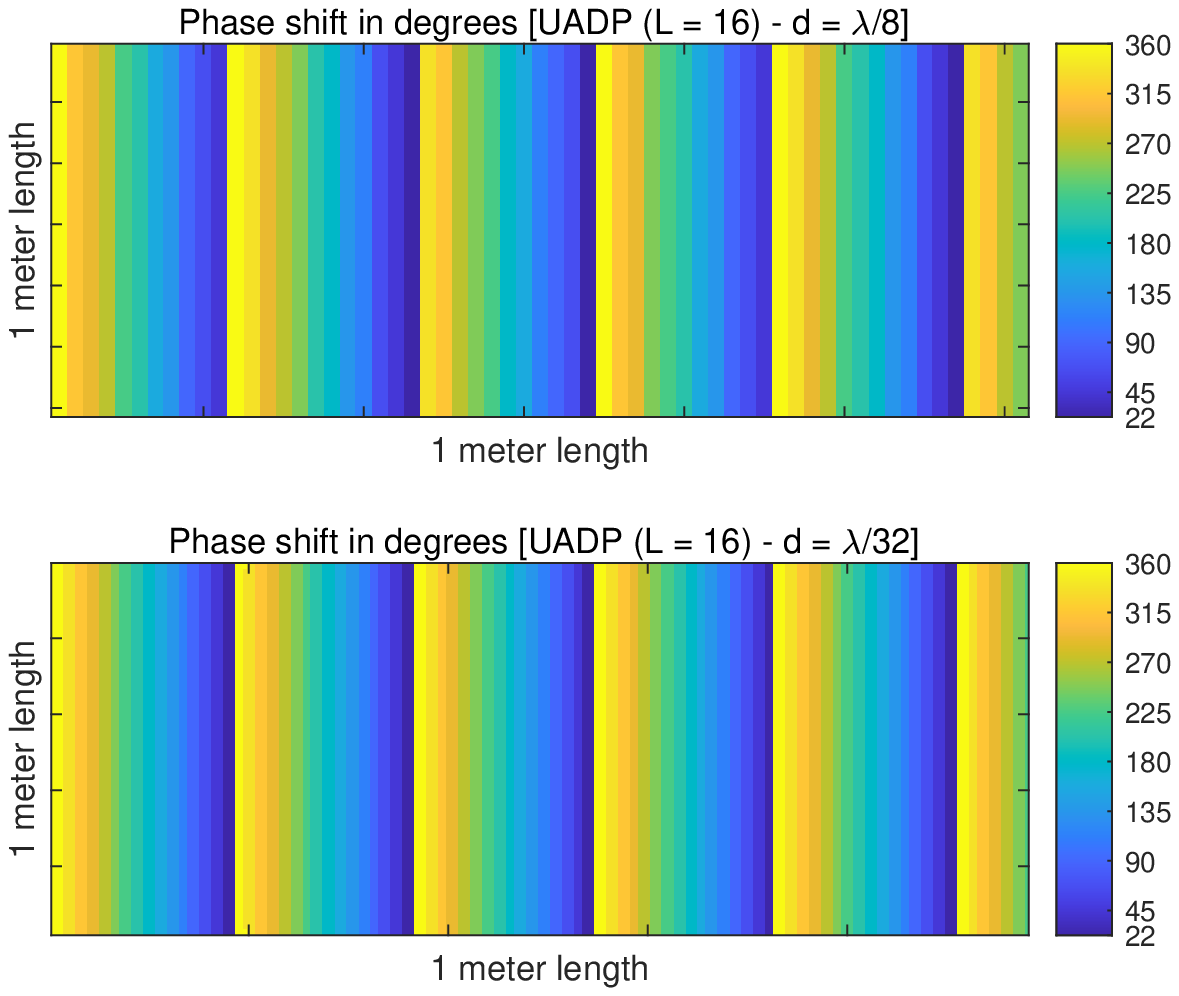}
\vspace{-0.60cm} \caption{Color map representation of $\boldsymbol{\Gamma}$ corresponding to the UADP (${L}=16$) case study. The desired angle of reflection is 45 degrees and the inter-distance is $d=\lambda/8$ and $d=\lambda/32$.}\label{fig:ColorMap__45deg_lambda8and32_Q4} \vspace{-0.30cm}
\end{figure}

\begin{figure}[!t]
\includegraphics[width=0.67\columnwidth]{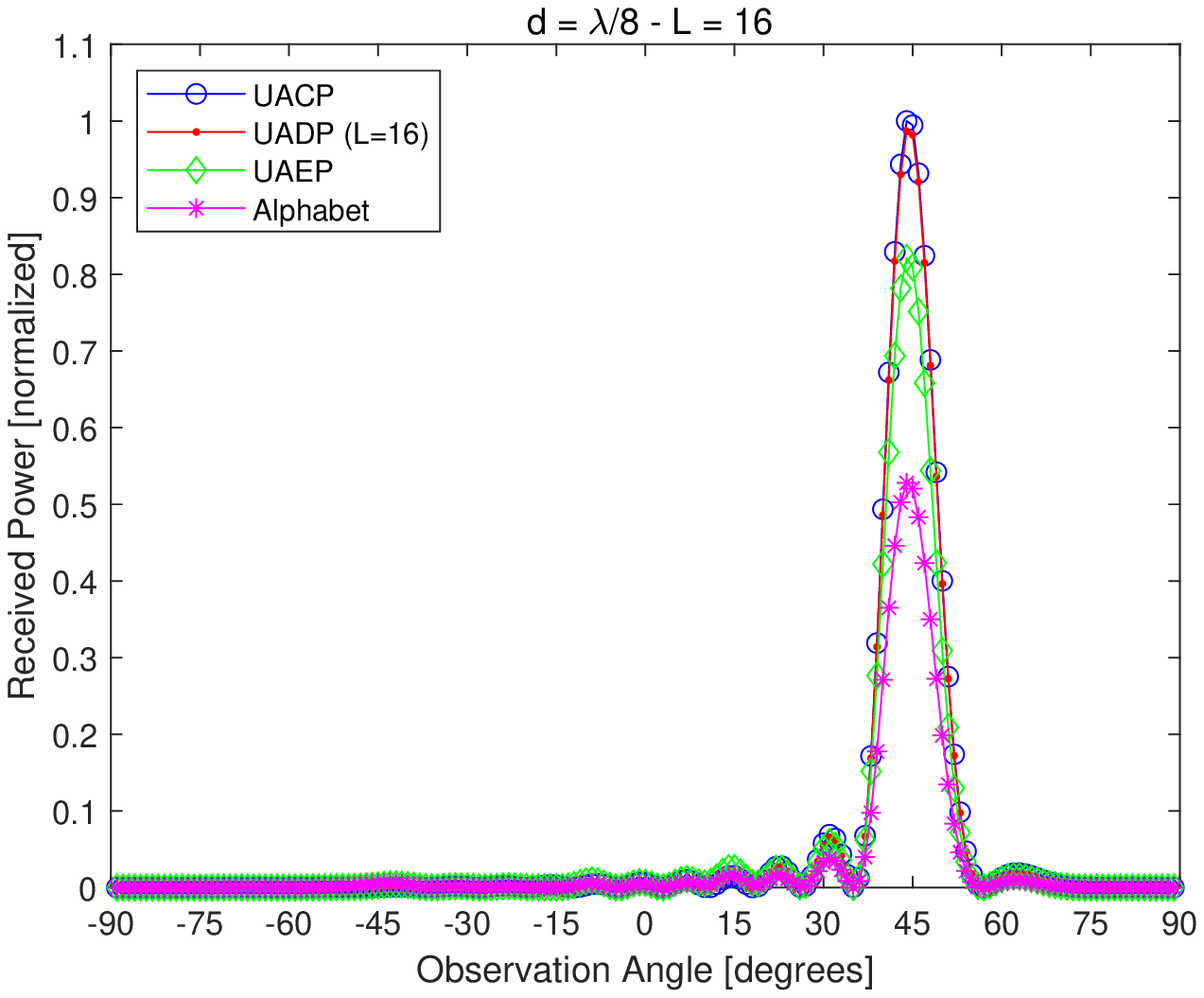}
\vspace{-0.25cm} \caption{Received power as a function of the angle of observation. The RIS alphabet is \cite{Linglong_Testbed}, the desired angle of reflection is 45 degrees, and the inter-distance is $d=\lambda/8$.}\label{fig:Prx__45deg_lambda8_Q4} \vspace{-0.30cm}
\end{figure}

\begin{figure}[!t]
\includegraphics[width=0.67\columnwidth]{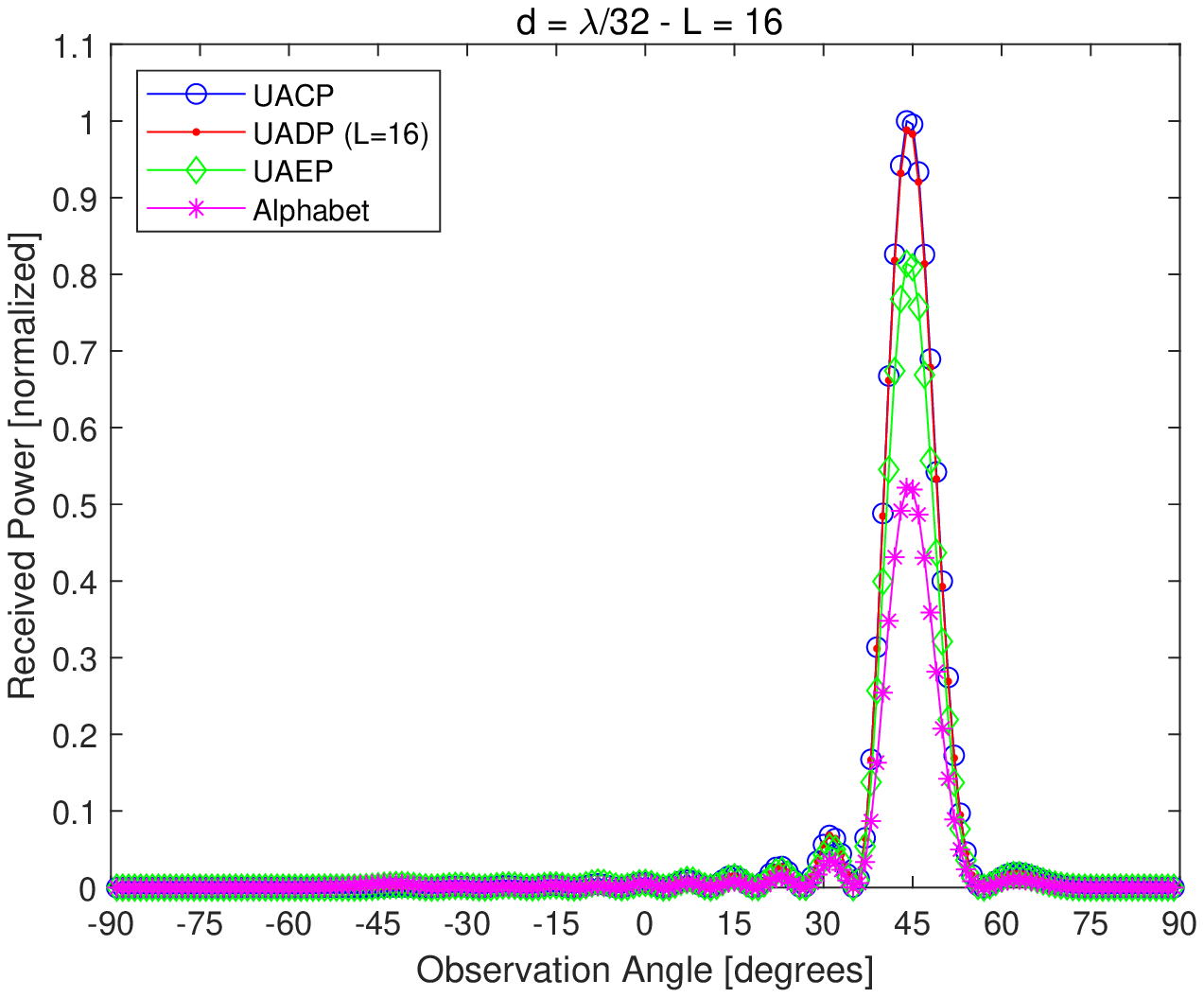}
\vspace{-0.25cm} \caption{Received power as a function of the angle of observation. The RIS alphabet is \cite{Linglong_Testbed}, the desired angle of reflection is 45 degrees, and the inter-distance is $d=\lambda/32$.}\label{fig:Prx__45deg_lambda32_Q4} \vspace{-0.30cm}
\end{figure}

\begin{figure}[!t]
\includegraphics[width=0.67\columnwidth]{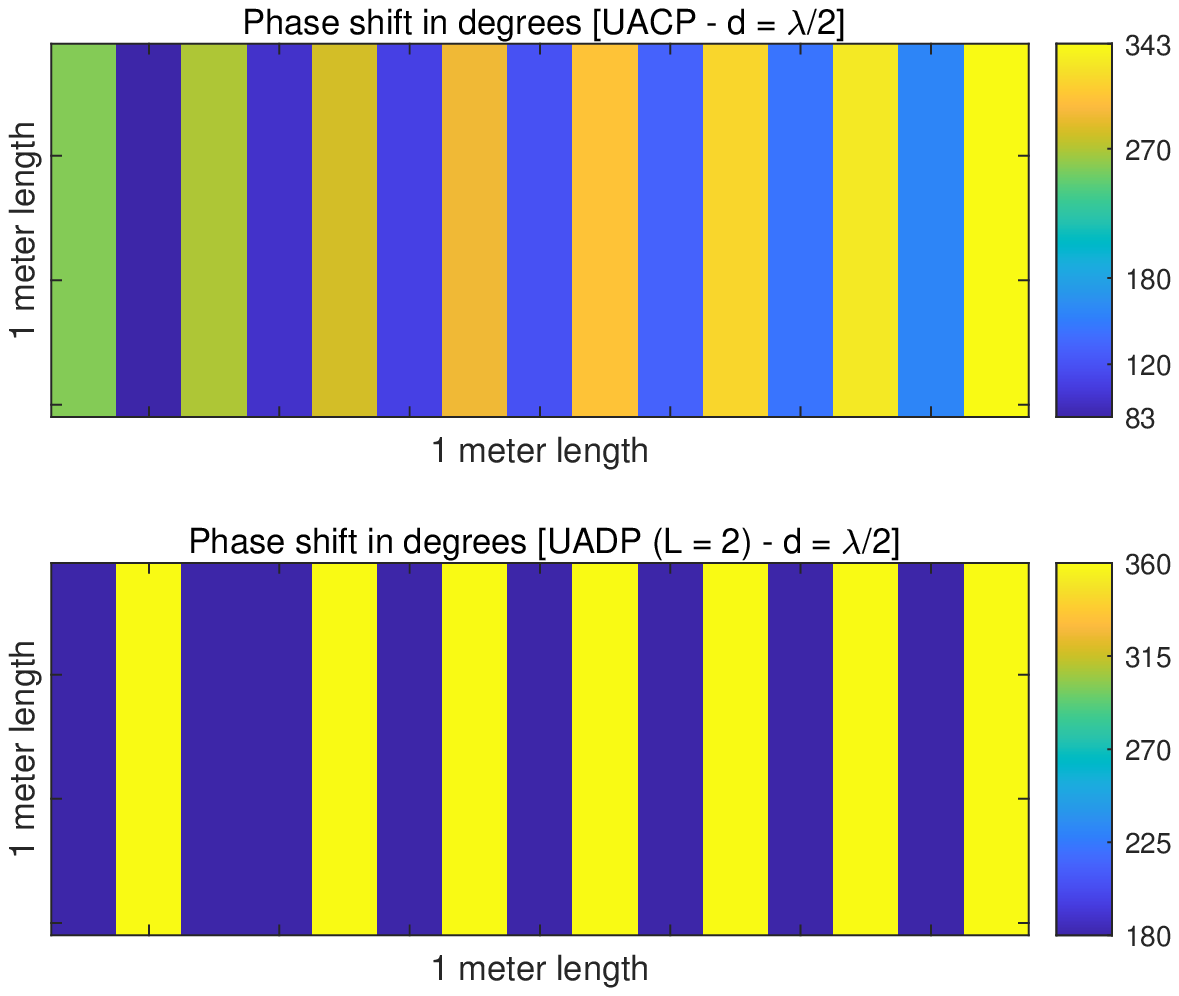}
\vspace{-0.60cm} \caption{Color map representation of $\boldsymbol{\Gamma}$ corresponding to the UACP and UADP (${L}=2$) case studies. The desired angle of reflection is 75 degrees and the inter-distance is $d=\lambda/2$.}\label{fig:Fig1__75deg_05lambda_Q1} \vspace{-0.30cm}
\end{figure}

\begin{figure}[!t]
\includegraphics[width=0.67\columnwidth]{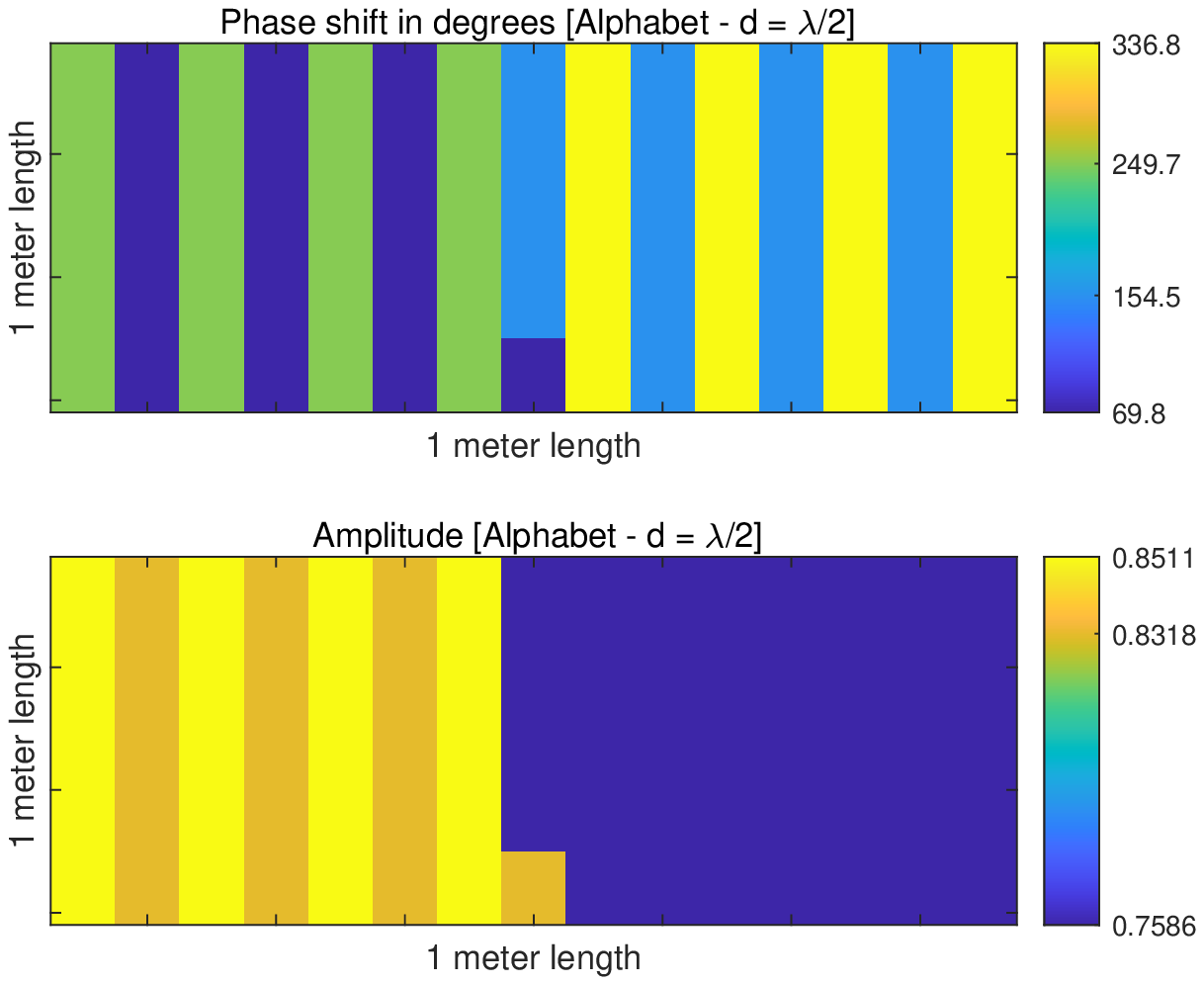}
\vspace{-0.60cm} \caption{Color map representation of $\boldsymbol{\Gamma}$ corresponding to the RIS alphabet in \cite{Linglong_Testbed}. The desired angle of reflection is 75 degrees and the inter-distance is $d=\lambda/2$.}\label{fig:Fig2__75deg_05lambda_Q1} \vspace{-0.30cm}
\end{figure}

\begin{figure}[!t]
\includegraphics[width=0.67\columnwidth]{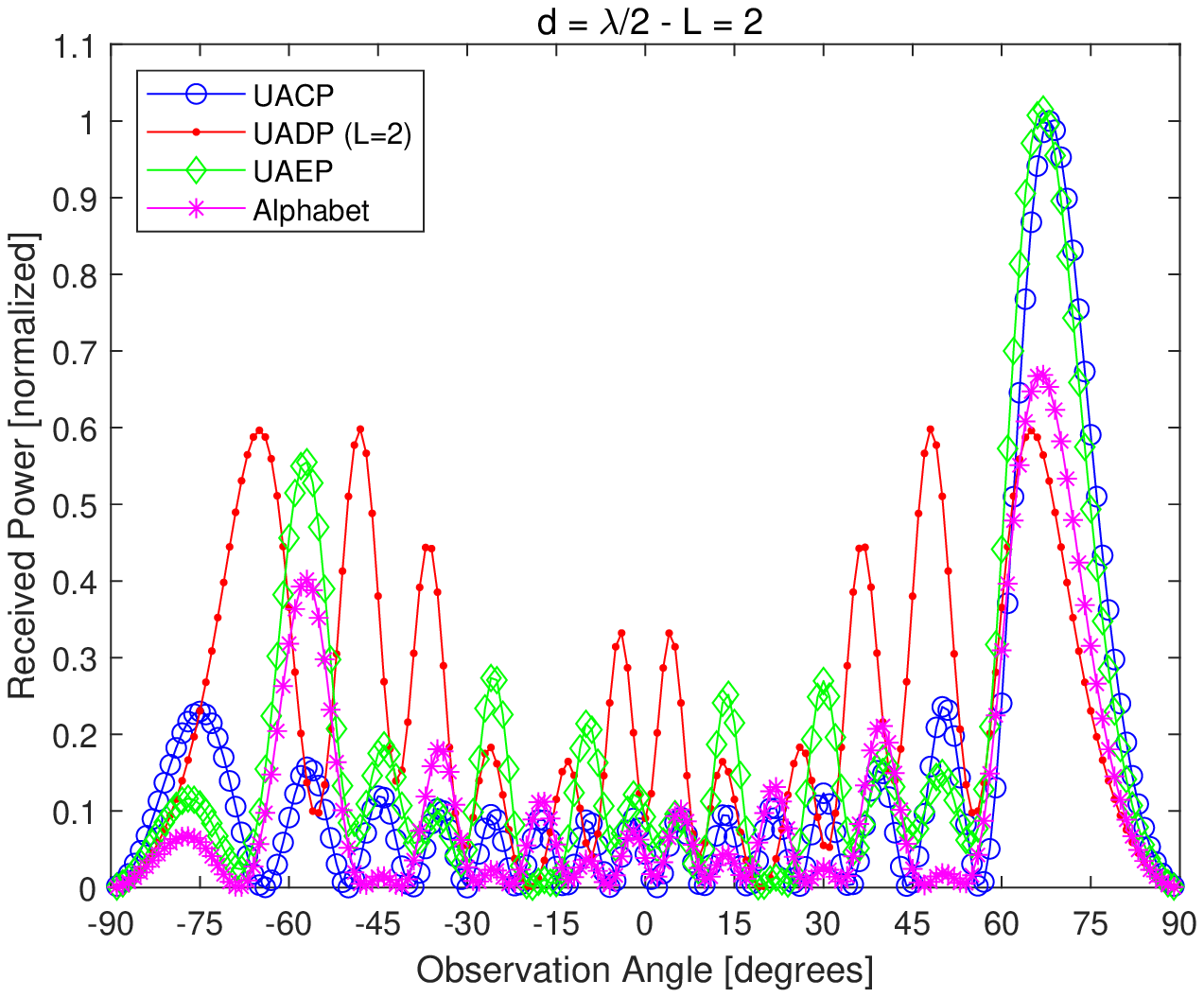}
\vspace{-0.25cm} \caption{Received power as a function of the angle of observation. The RIS alphabet is \cite{Linglong_Testbed}, the desired angle of reflection is 75 degrees, and the inter-distance is $d=\lambda/2$.}\label{fig:Fig3__75deg_05lambda_Q1} \vspace{-0.30cm}
\end{figure}

\begin{figure}[!t]
\includegraphics[width=0.67\columnwidth]{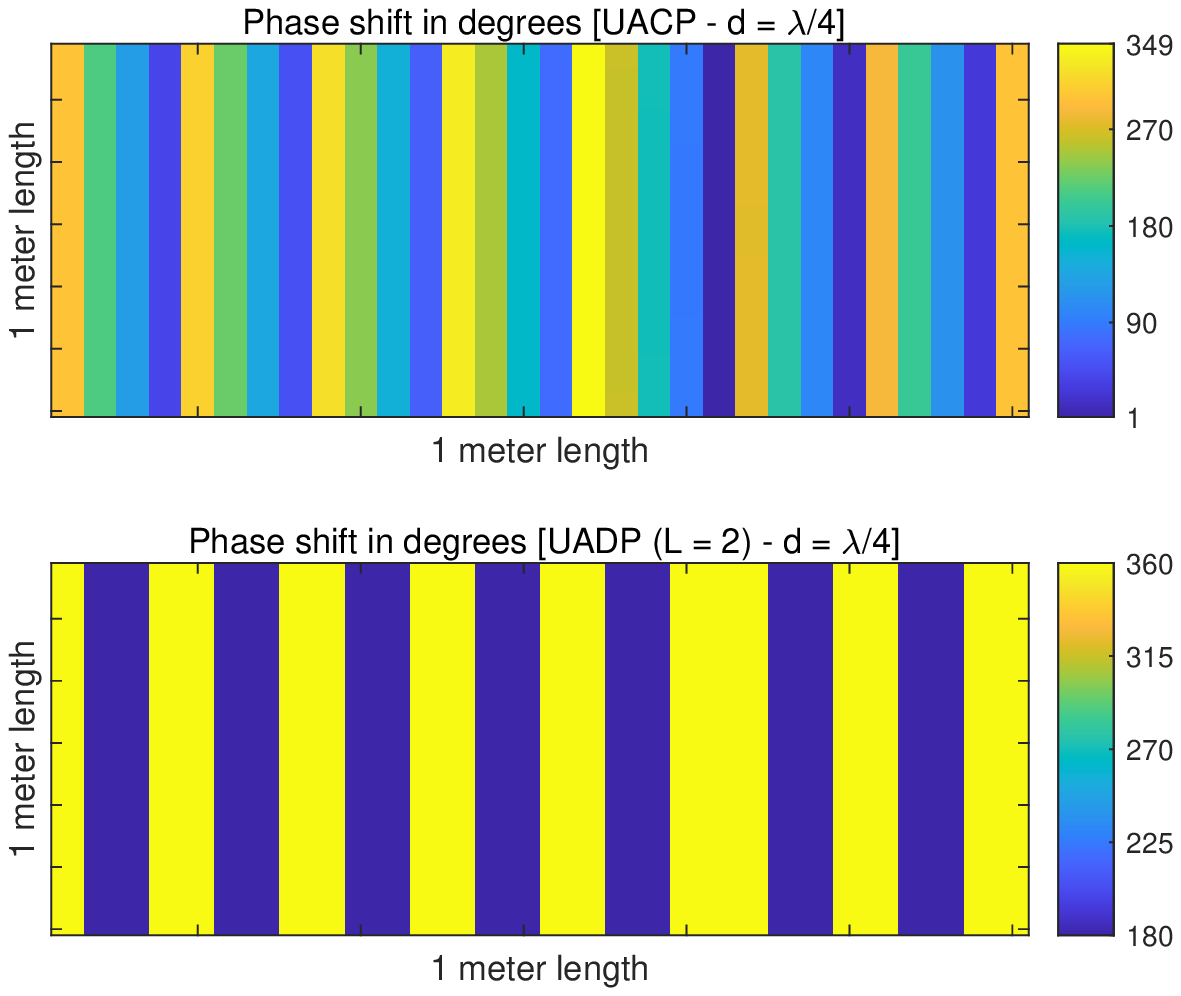}
\vspace{-0.60cm} \caption{Color map representation of $\boldsymbol{\Gamma}$ corresponding to the UACP and UADP (${L}=2$) case studies. The desired angle of reflection is 75 degrees and the inter-distance is $d=\lambda/4$.}\label{fig:Fig1__75deg_025lambda_Q1} \vspace{-0.30cm}
\end{figure}

\begin{figure}[!t]
\includegraphics[width=0.67\columnwidth]{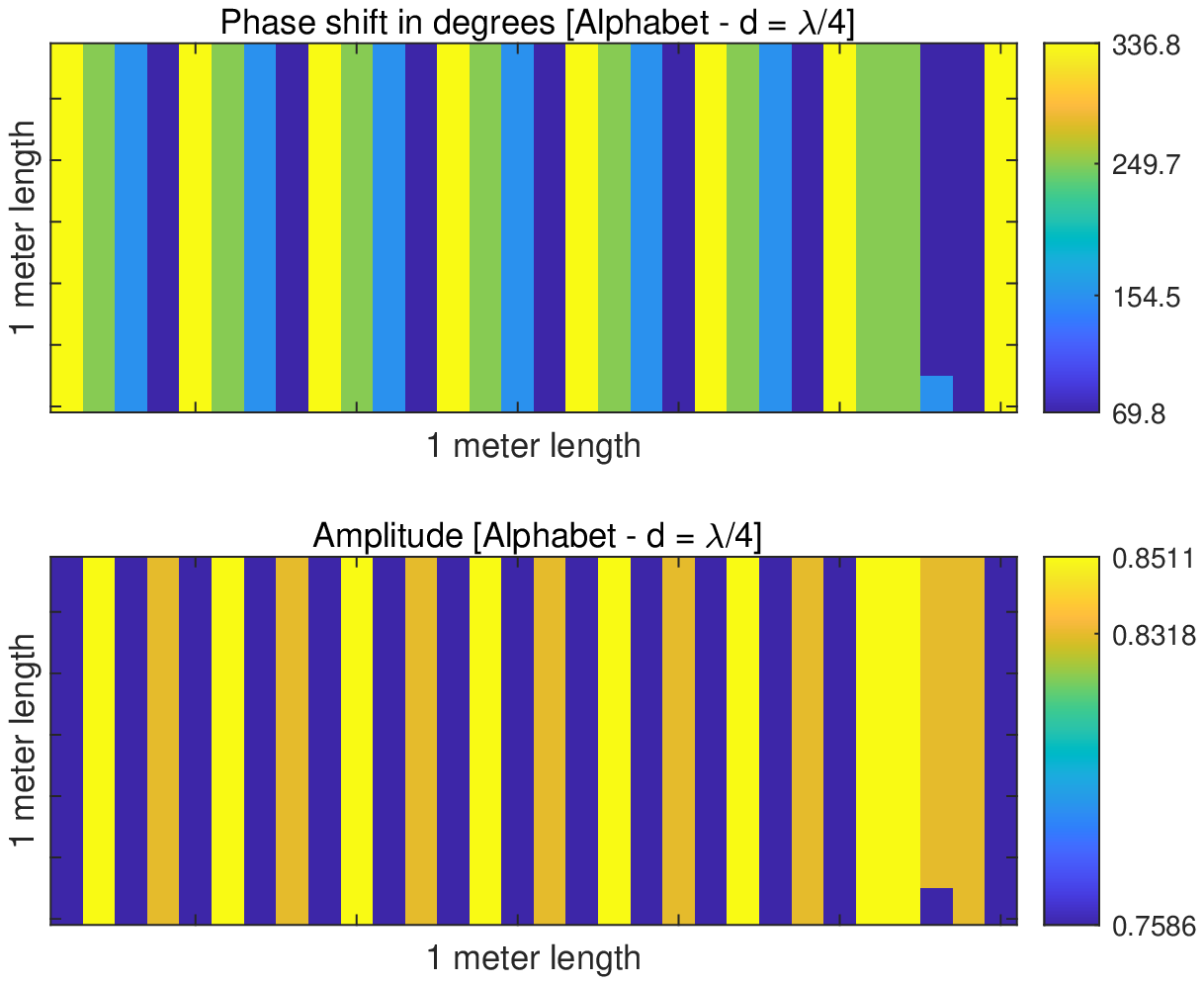}
\vspace{-0.60cm} \caption{Color map representation of $\boldsymbol{\Gamma}$ corresponding to the RIS alphabet in \cite{Linglong_Testbed}. The desired angle of reflection is 75 degrees and the inter-distance is $d=\lambda/4$.}\label{fig:Fig2__75deg_025lambda_Q1} \vspace{-0.30cm}
\end{figure}

\begin{figure}[!t]
\includegraphics[width=0.67\columnwidth]{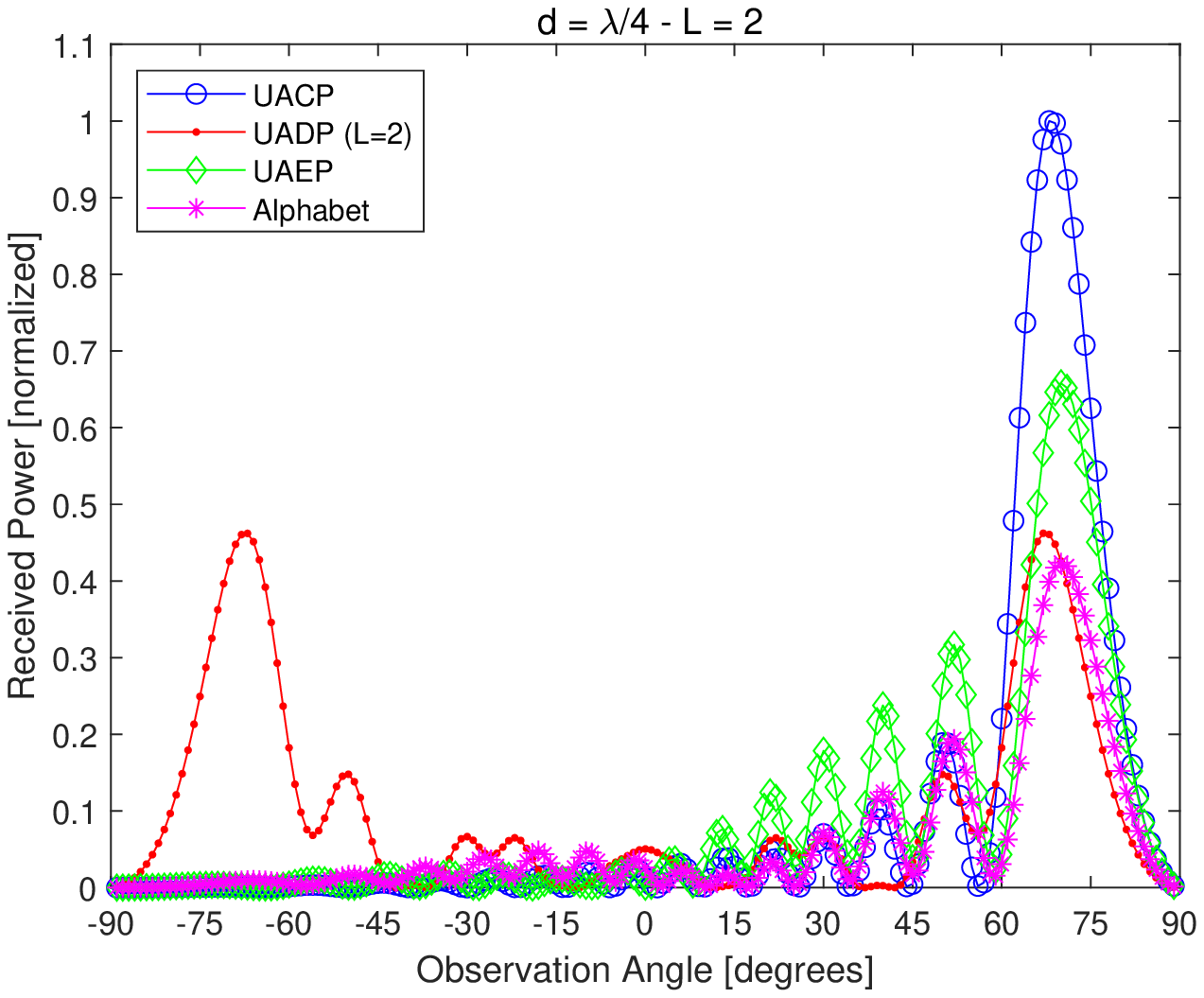}
\vspace{-0.25cm} \caption{Received power as a function of the angle of observation. The RIS alphabet is \cite{Linglong_Testbed}, the desired angle of reflection is 75 degrees, and the inter-distance is $d=\lambda/4$.}\label{fig:Fig3__75deg_025lambda_Q1} \vspace{-0.30cm}
\end{figure}

\begin{figure}[!t]
\includegraphics[width=0.67\columnwidth]{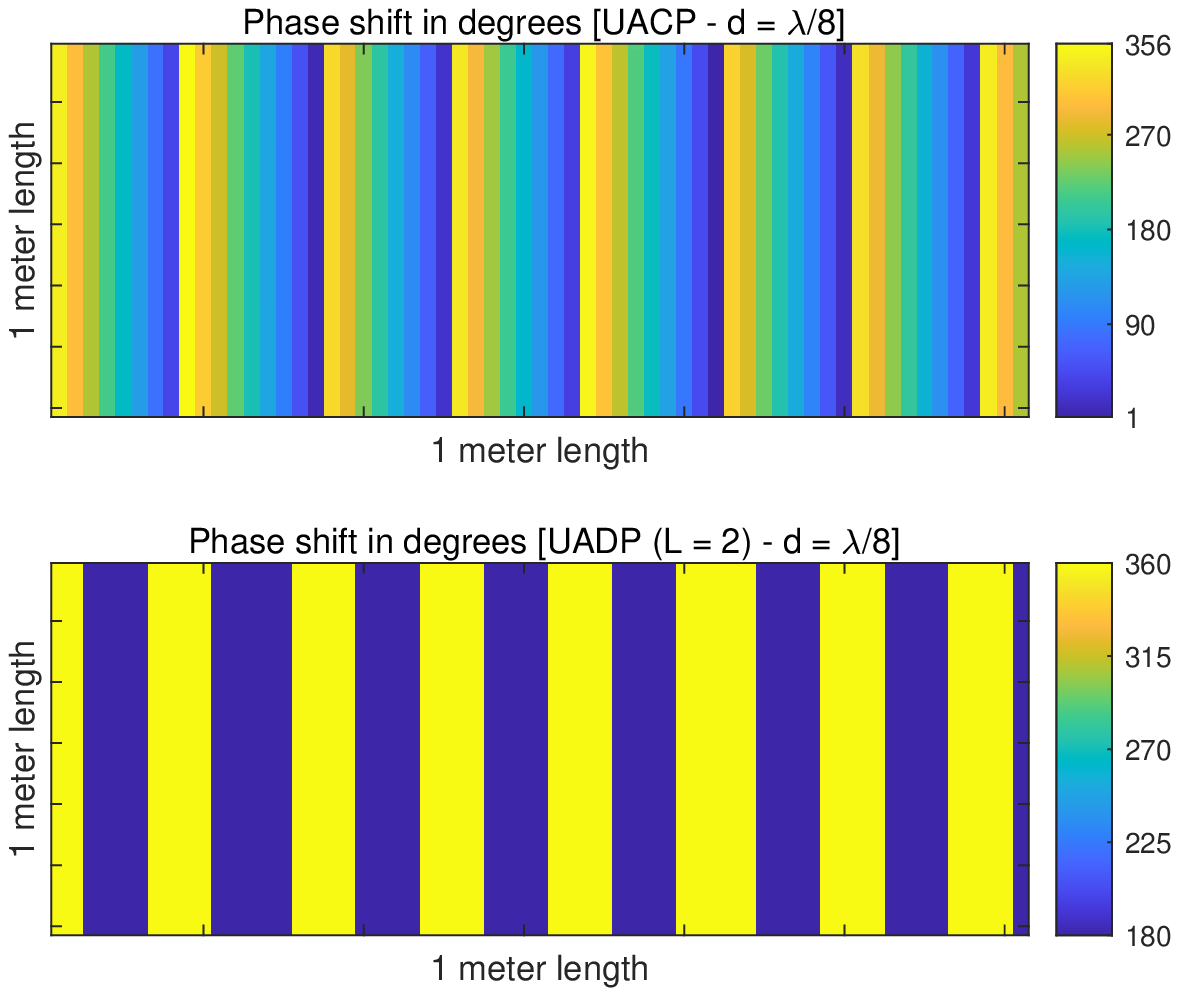}
\vspace{-0.60cm} \caption{Color map representation of $\boldsymbol{\Gamma}$ corresponding to the UACP and UADP (${L}=2$) case studies. The desired angle of reflection is 75 degrees and the inter-distance is $d=\lambda/8$.}\label{fig:Fig1__75deg_0125lambda_Q1} \vspace{-0.30cm}
\end{figure}

\begin{figure}[!t]
\includegraphics[width=0.67\columnwidth]{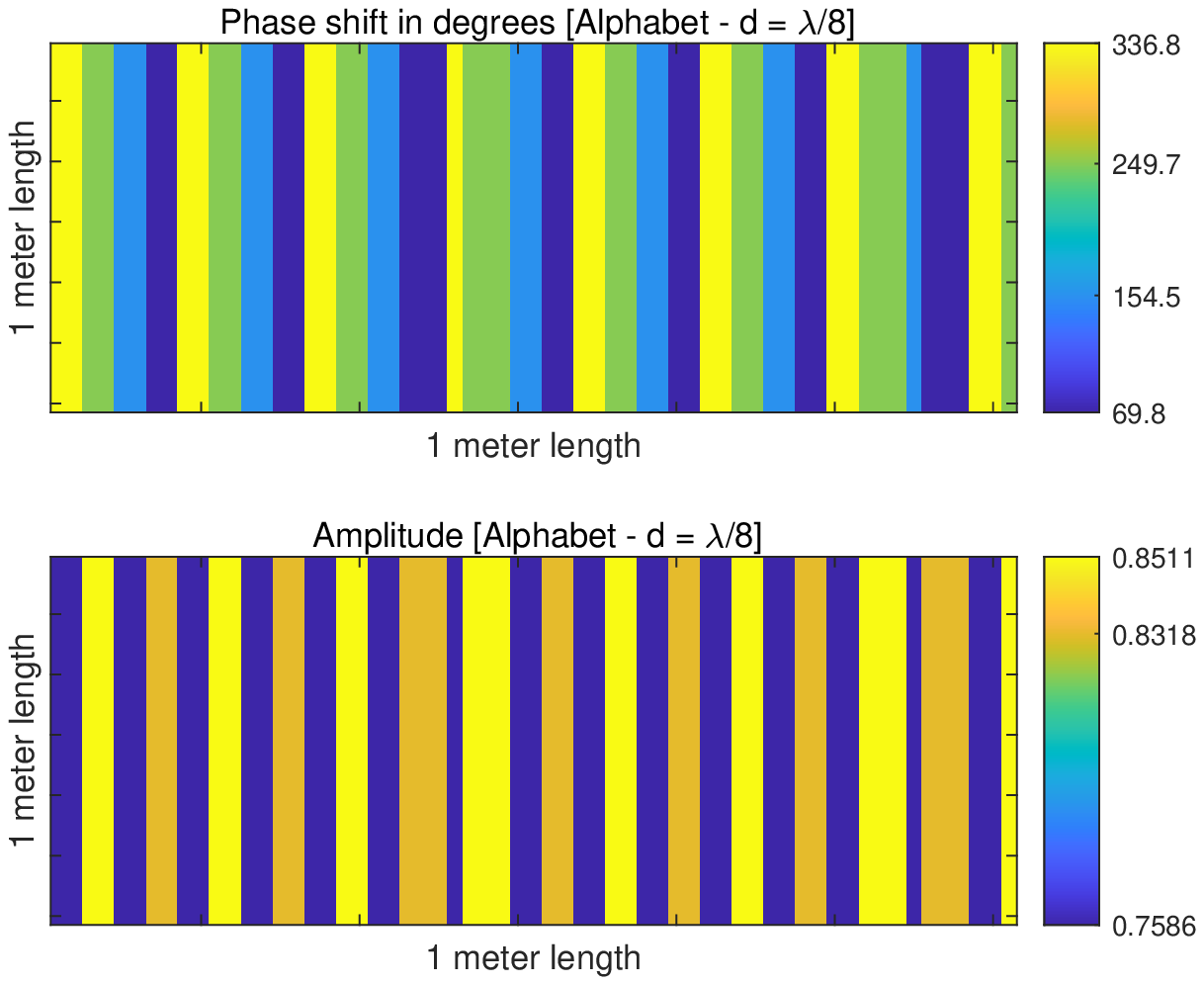}
\vspace{-0.60cm} \caption{Color map representation of $\boldsymbol{\Gamma}$ corresponding to the RIS alphabet in \cite{Linglong_Testbed}. The desired angle of reflection is 75 degrees and the inter-distance is $d=\lambda/8$.}\label{fig:Fig2__75deg_0125lambda_Q1} \vspace{-0.30cm}
\end{figure}

\begin{figure}[!t]
\includegraphics[width=0.67\columnwidth]{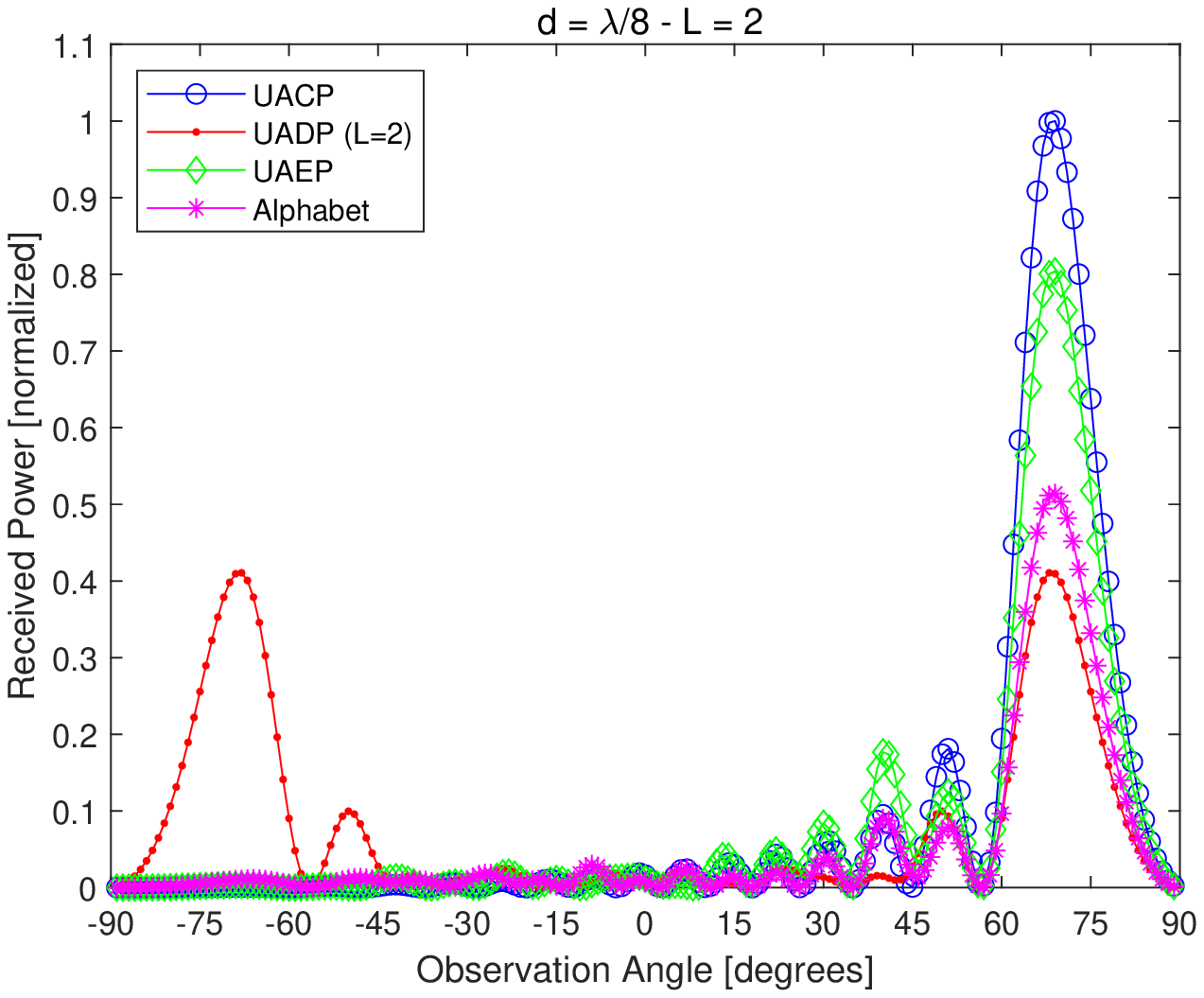}
\vspace{-0.25cm} \caption{Received power as a function of the angle of observation. The RIS alphabet is \cite{Linglong_Testbed}, the desired angle of reflection is 75 degrees, and the inter-distance is $d=\lambda/8$.}\label{fig:Fig3__75deg_0125lambda_Q1} \vspace{-0.30cm}
\end{figure}

\begin{figure}[!t]
\includegraphics[width=0.67\columnwidth]{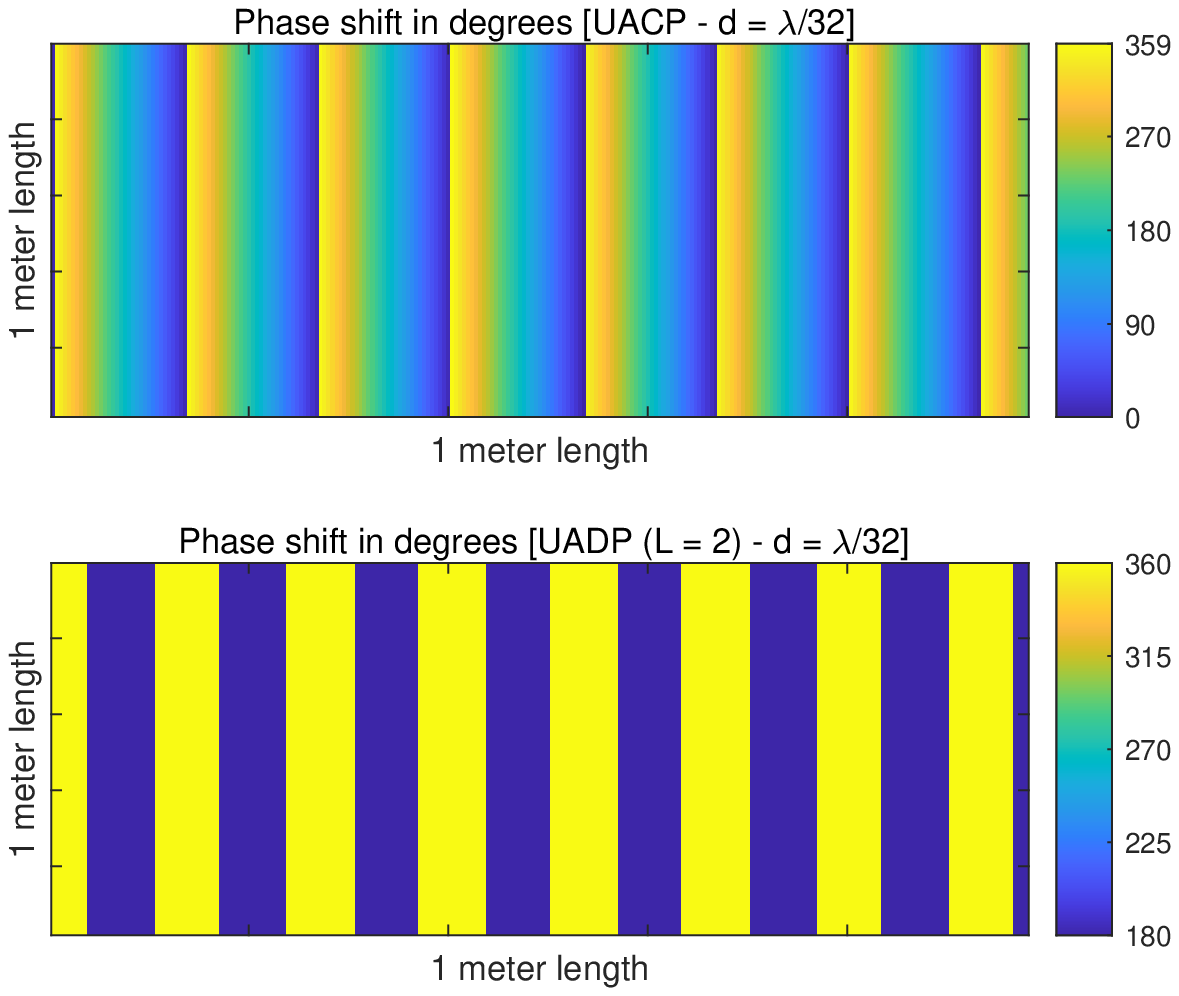}
\vspace{-0.60cm} \caption{Color map representation of $\boldsymbol{\Gamma}$ corresponding to the UACP and UADP (${L}=2$) case studies. The desired angle of reflection is 75 degrees and the inter-distance is $d=\lambda/32$.}\label{fig:Fig1__75deg_lambda32_Q1} \vspace{-0.30cm}
\end{figure}

\begin{figure}[!t]
\includegraphics[width=0.67\columnwidth]{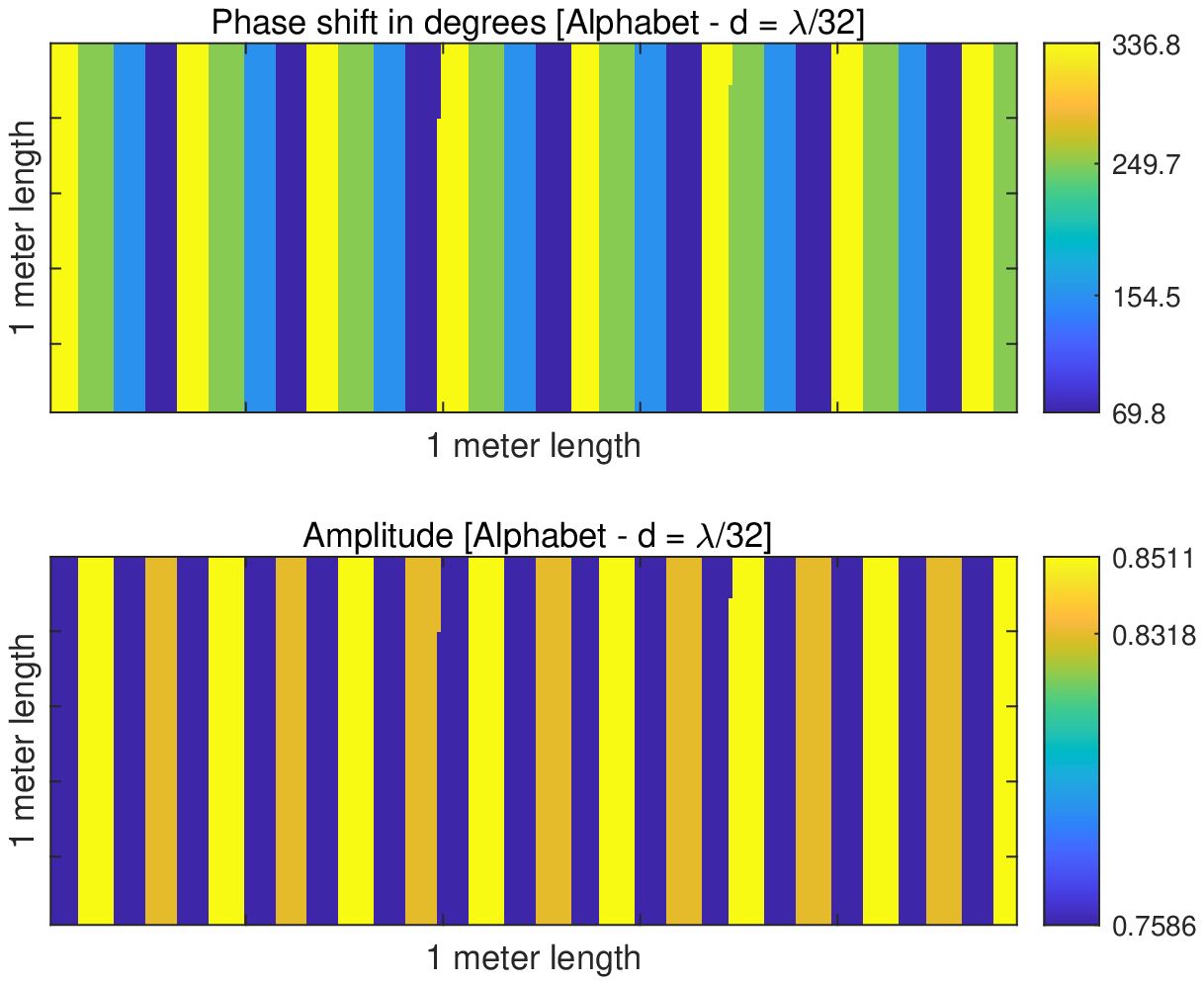}
\vspace{-0.60cm} \caption{Color map representation of $\boldsymbol{\Gamma}$ corresponding to the RIS alphabet in \cite{Linglong_Testbed}. The desired angle of reflection is 75 degrees and the inter-distance is $d=\lambda/32$.}\label{fig:Fig2__75deg_lambda32_Q1} \vspace{-0.30cm}
\end{figure}

\begin{figure}[!t]
\includegraphics[width=0.67\columnwidth]{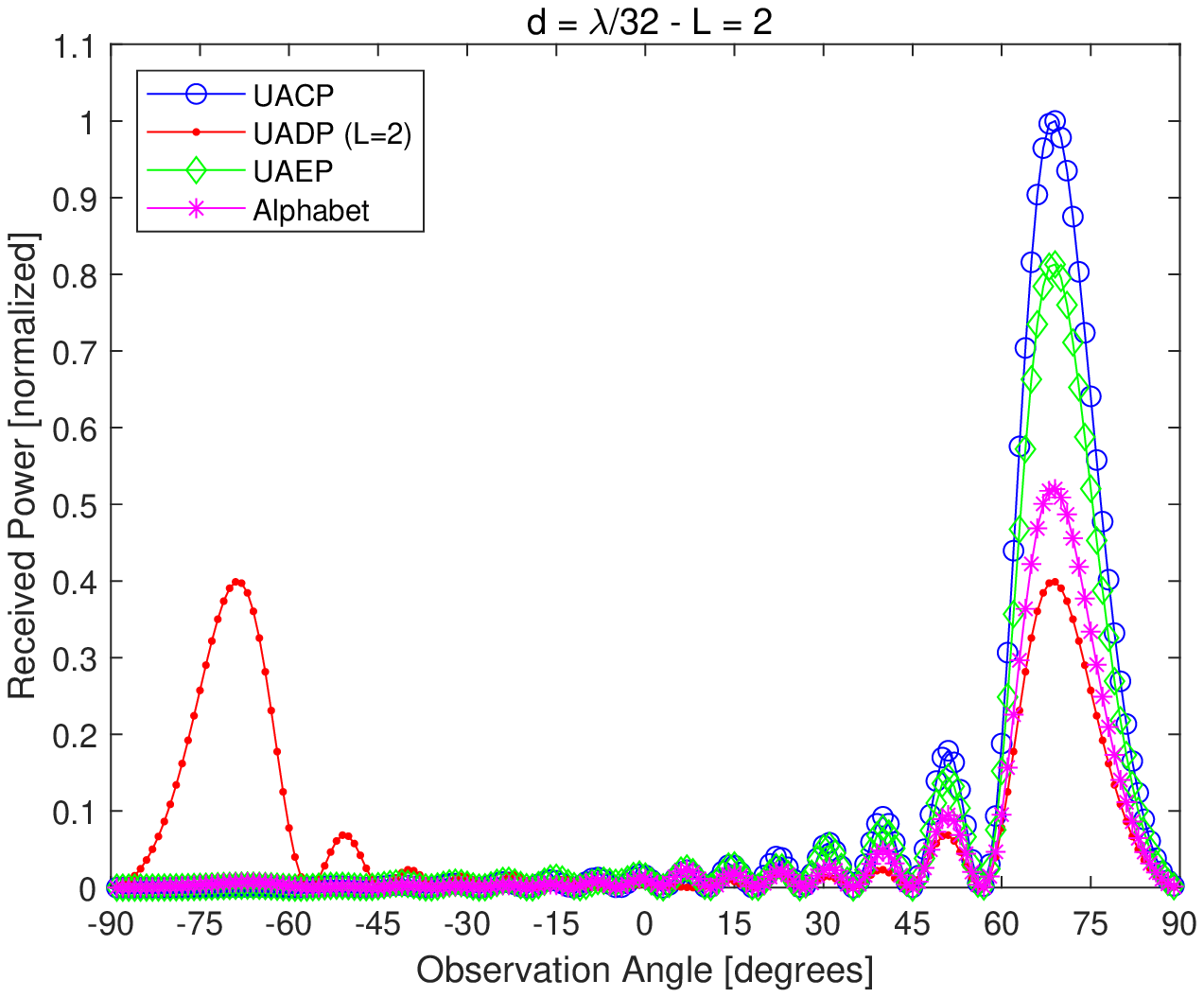}
\vspace{-0.25cm} \caption{Received power as a function of the angle of observation. The RIS alphabet is \cite{Linglong_Testbed}, the desired angle of reflection is 75 degrees, and the inter-distance is $d=\lambda/32$.}\label{fig:Fig3__75deg_lambda32_Q1} \vspace{-0.30cm}
\end{figure}

\begin{figure}[!t]
\includegraphics[width=0.67\columnwidth]{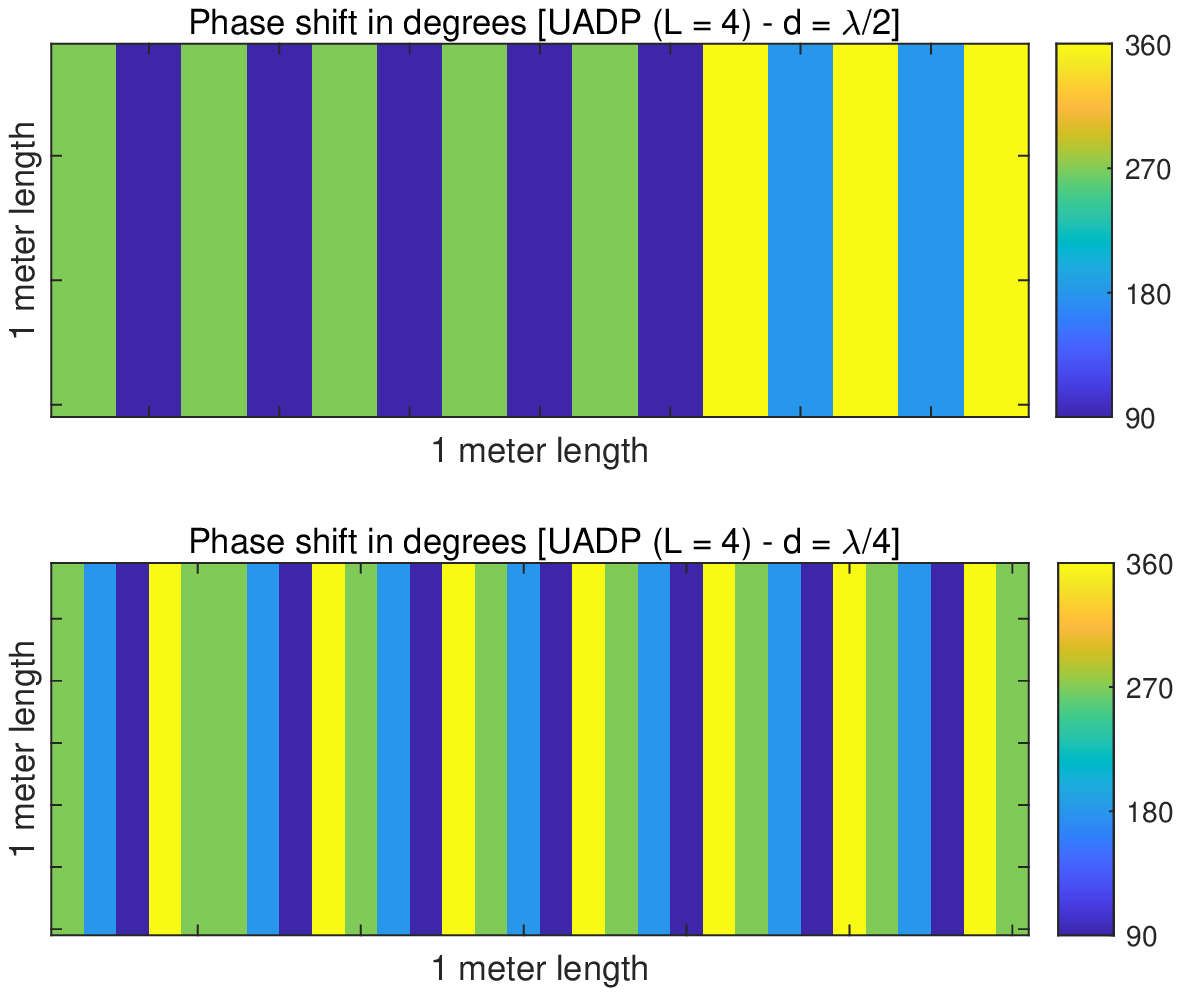}
\vspace{-0.60cm} \caption{Color map representation of $\boldsymbol{\Gamma}$ corresponding to the UADP (${L}=4$) case study. The desired angle of reflection is 75 degrees and the inter-distance is $d=\lambda/2$ and $d=\lambda/4$.}\label{fig:ColorMap__75deg_lambda2and4_Q2} \vspace{-0.30cm}
\end{figure}

\begin{figure}[!t]
\includegraphics[width=0.67\columnwidth]{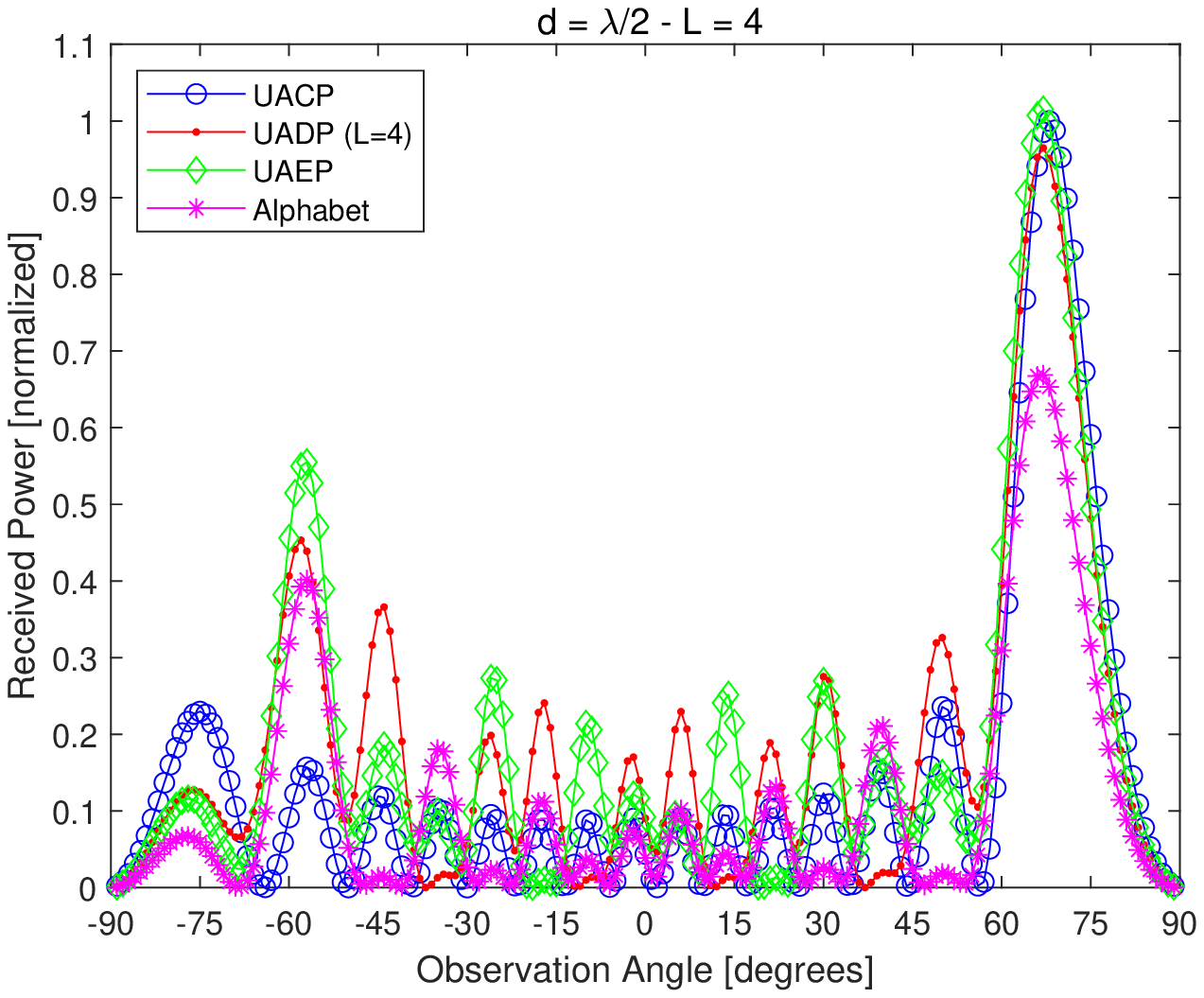}
\vspace{-0.25cm} \caption{Received power as a function of the angle of observation. The RIS alphabet is \cite{Linglong_Testbed}, the desired angle of reflection is 75 degrees, and the inter-distance is $d=\lambda/2$.}\label{fig:Prx__75deg_lambda2_Q2} \vspace{-0.30cm}
\end{figure}

\begin{figure}[!t]
\includegraphics[width=0.67\columnwidth]{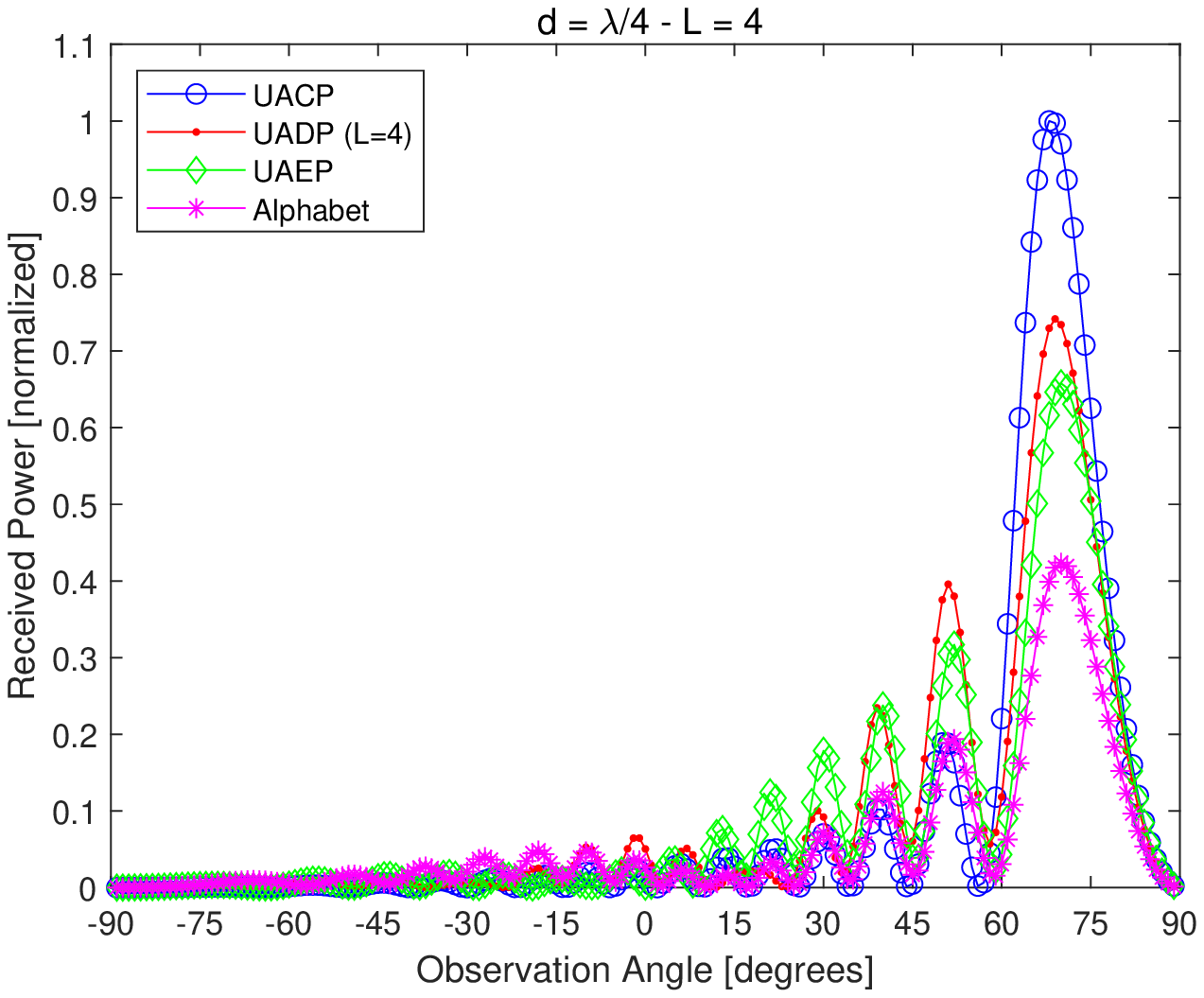}
\vspace{-0.25cm} \caption{Received power as a function of the angle of observation. The RIS alphabet is \cite{Linglong_Testbed}, the desired angle of reflection is 75 degrees, and the inter-distance is $d=\lambda/4$.}\label{fig:Prx__75deg_lambda4_Q2} \vspace{-0.30cm}
\end{figure}

\begin{figure}[!t]
\includegraphics[width=0.67\columnwidth]{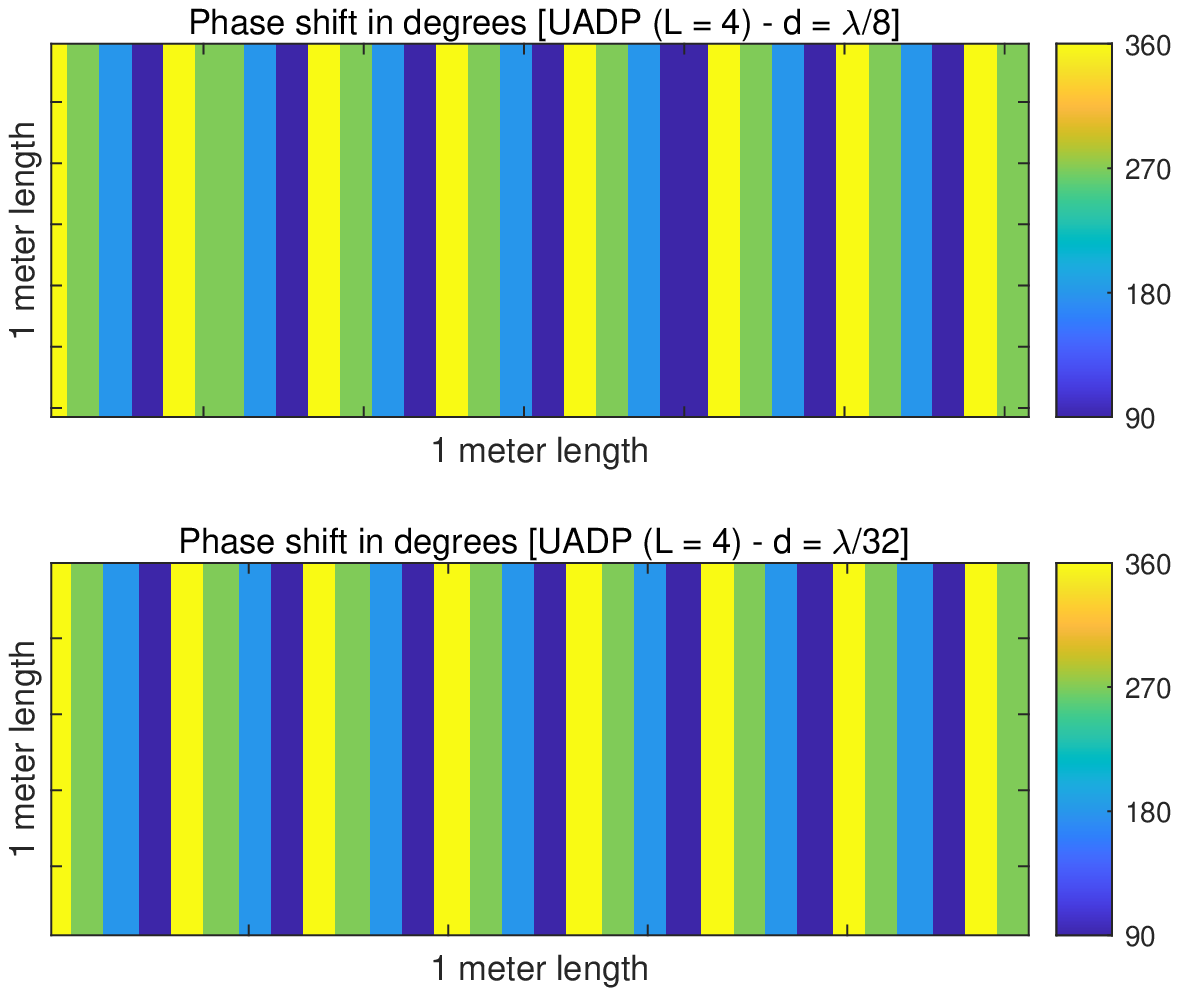}
\vspace{-0.60cm} \caption{Color map representation of $\boldsymbol{\Gamma}$ corresponding to the UADP (${L}=4$) case study. The desired angle of reflection is 75 degrees and the inter-distance is $d=\lambda/8$ and $d=\lambda/32$.}\label{fig:ColorMap__75deg_lambda8and32_Q2} \vspace{-0.30cm}
\end{figure}

\begin{figure}[!t]
\includegraphics[width=0.67\columnwidth]{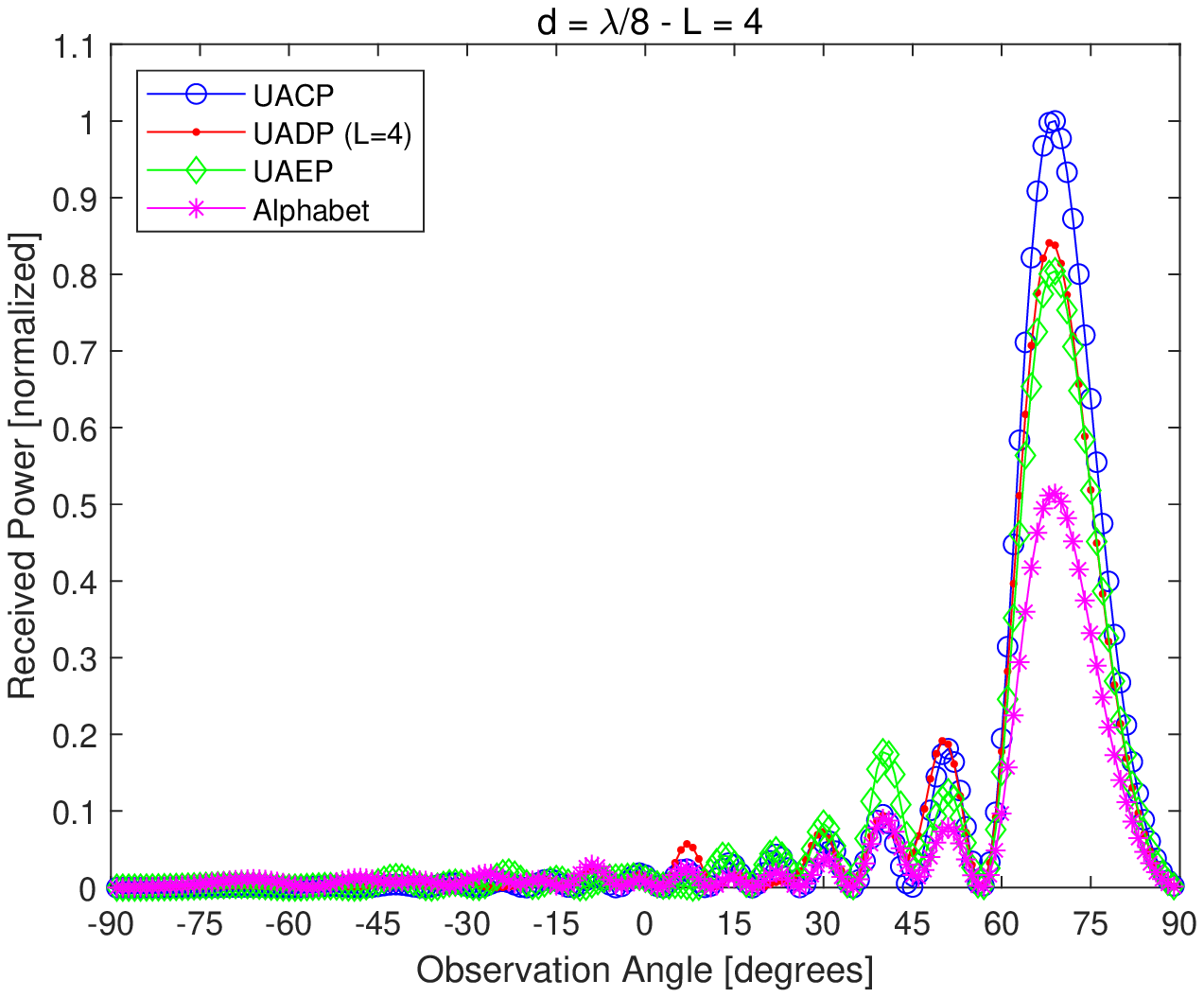}
\vspace{-0.25cm} \caption{Received power as a function of the angle of observation. The RIS alphabet is \cite{Linglong_Testbed}, the desired angle of reflection is 75 degrees, and the inter-distance is $d=\lambda/8$.}\label{fig:Prx__75deg_lambda8_Q2} \vspace{-0.30cm}
\end{figure}

\begin{figure}[!t]
\includegraphics[width=0.67\columnwidth]{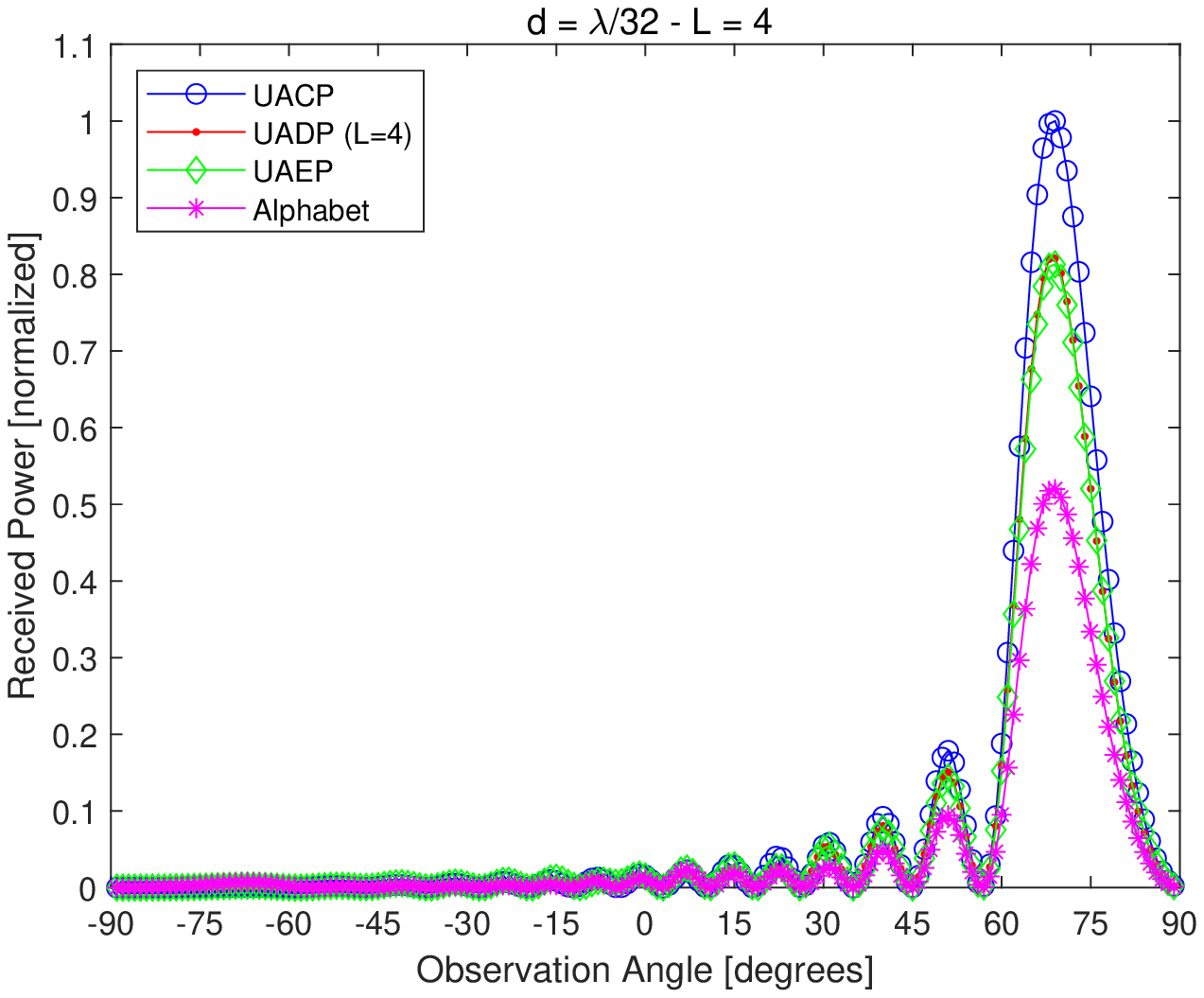}
\vspace{-0.25cm} \caption{Received power as a function of the angle of observation. The RIS alphabet is \cite{Linglong_Testbed}, the desired angle of reflection is 75 degrees, and the inter-distance is $d=\lambda/32$.}\label{fig:Prx__75deg_lambda32_Q2} \vspace{-0.30cm}
\end{figure}

\begin{figure}[!t]
\includegraphics[width=0.67\columnwidth]{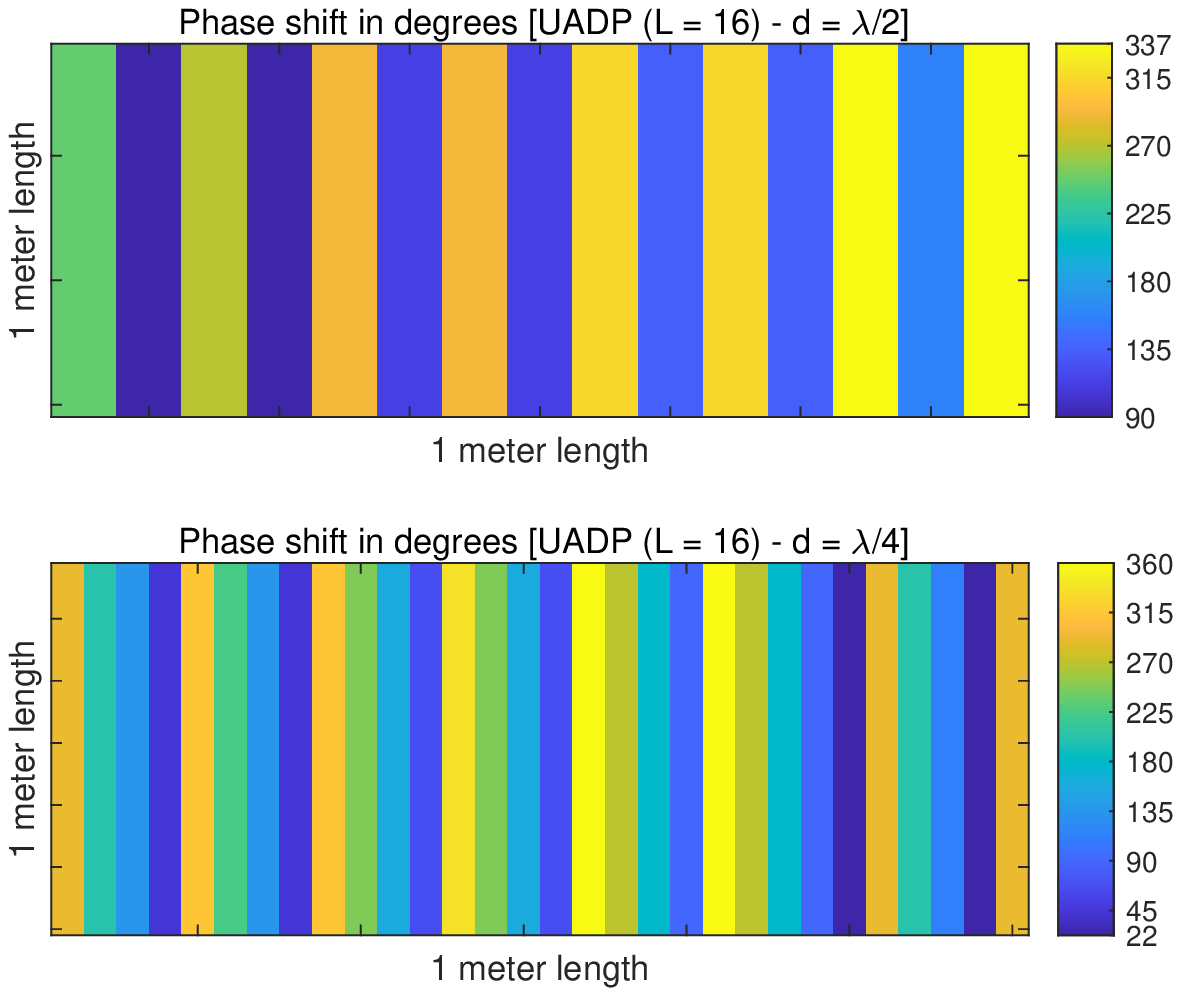}
\vspace{-0.60cm} \caption{Color map representation of $\boldsymbol{\Gamma}$ corresponding to the UADP (${L}=16$) case study. The desired angle of reflection is 75 degrees and the inter-distance is $d=\lambda/2$ and $d=\lambda/4$.}\label{fig:ColorMap__75deg_lambda2and4_Q4} \vspace{-0.30cm}
\end{figure}

\begin{figure}[!t]
\includegraphics[width=0.67\columnwidth]{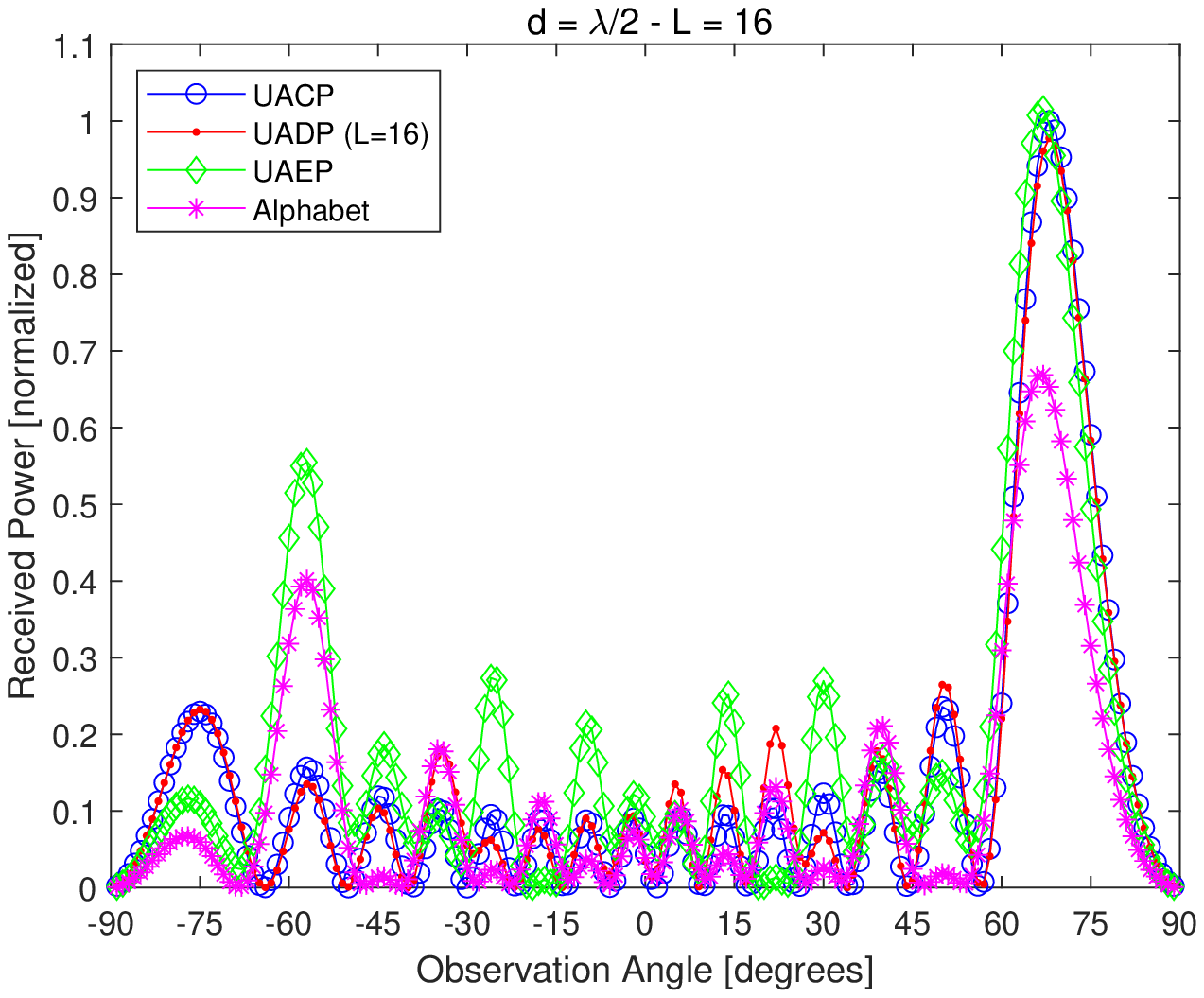}
\vspace{-0.25cm} \caption{Received power as a function of the angle of observation. The RIS alphabet is \cite{Linglong_Testbed}, the desired angle of reflection is 75 degrees, and the inter-distance is $d=\lambda/2$.}\label{fig:Prx__75deg_lambda2_Q4} \vspace{-0.30cm}
\end{figure}

\begin{figure}[!t]
\includegraphics[width=0.67\columnwidth]{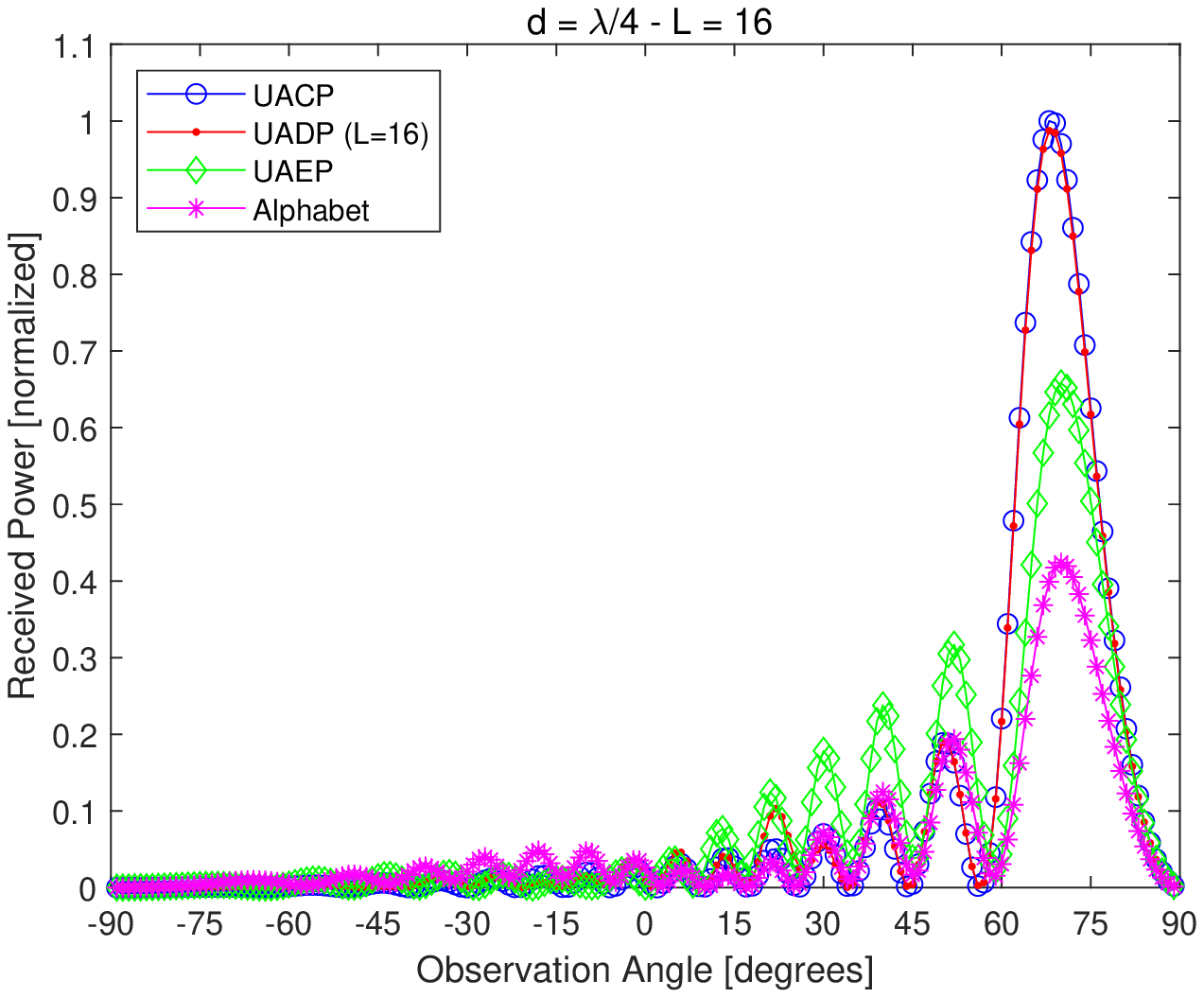}
\vspace{-0.25cm} \caption{Received power as a function of the angle of observation. The RIS alphabet is \cite{Linglong_Testbed}, the desired angle of reflection is 75 degrees, and the inter-distance is $d=\lambda/4$.}\label{fig:Prx__75deg_lambda4_Q4} \vspace{-0.30cm}
\end{figure}

\begin{figure}[!t]
\includegraphics[width=0.67\columnwidth]{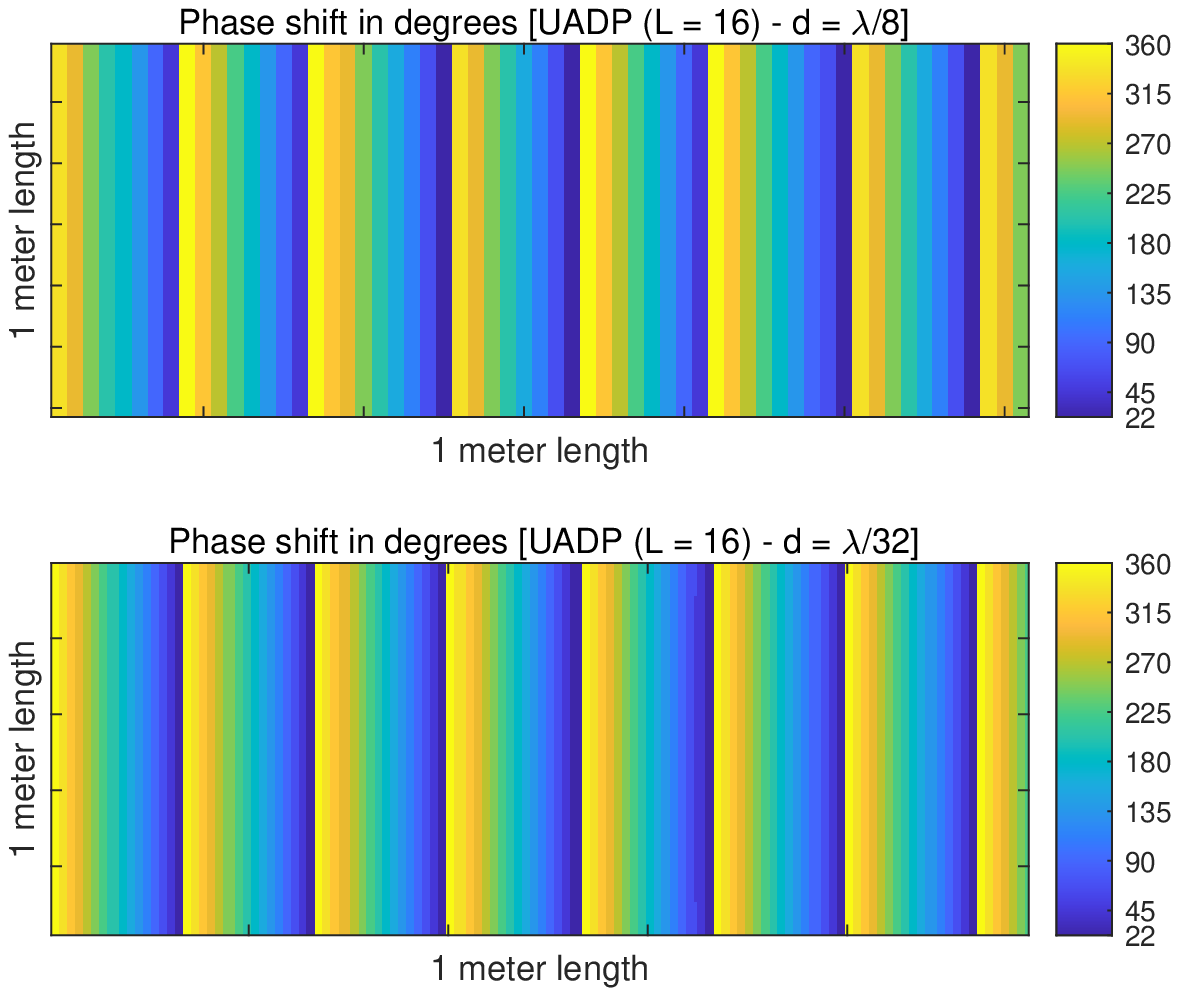}
\vspace{-0.60cm} \caption{Color map representation of $\boldsymbol{\Gamma}$ corresponding to the UADP (${L}=16$) case study. The desired angle of reflection is 75 degrees and the inter-distance is $d=\lambda/8$ and $d=\lambda/32$.}\label{fig:ColorMap__75deg_lambda8and32_Q4} \vspace{-0.30cm}
\end{figure}

\begin{figure}[!t]
\includegraphics[width=0.67\columnwidth]{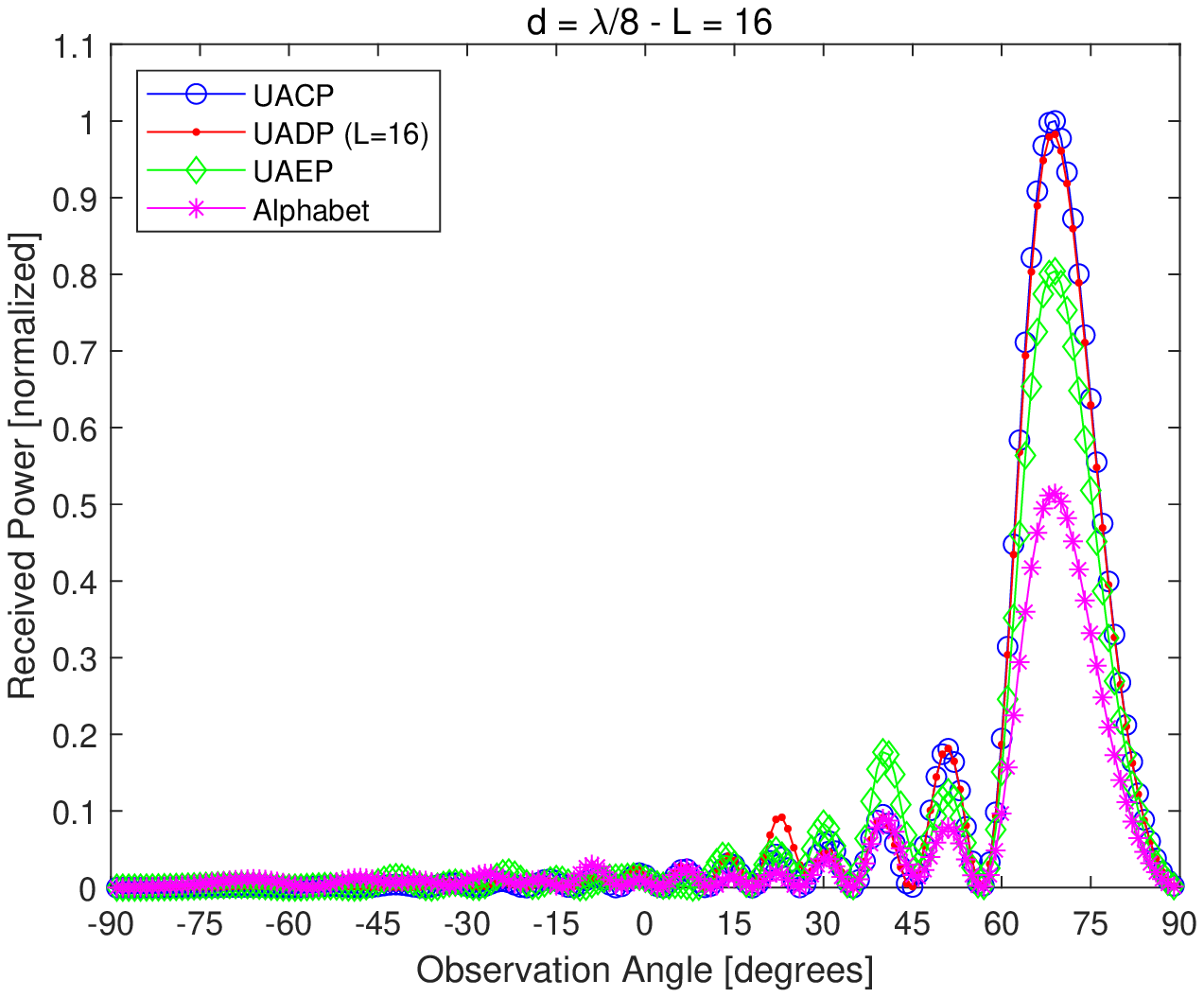}
\vspace{-0.25cm} \caption{Received power as a function of the angle of observation. The RIS alphabet is \cite{Linglong_Testbed}, the desired angle of reflection is 75 degrees, and the inter-distance is $d=\lambda/8$.}\label{fig:Prx__75deg_lambda8_Q4} \vspace{-0.30cm}
\end{figure}

\begin{figure}[!t]
\includegraphics[width=0.67\columnwidth]{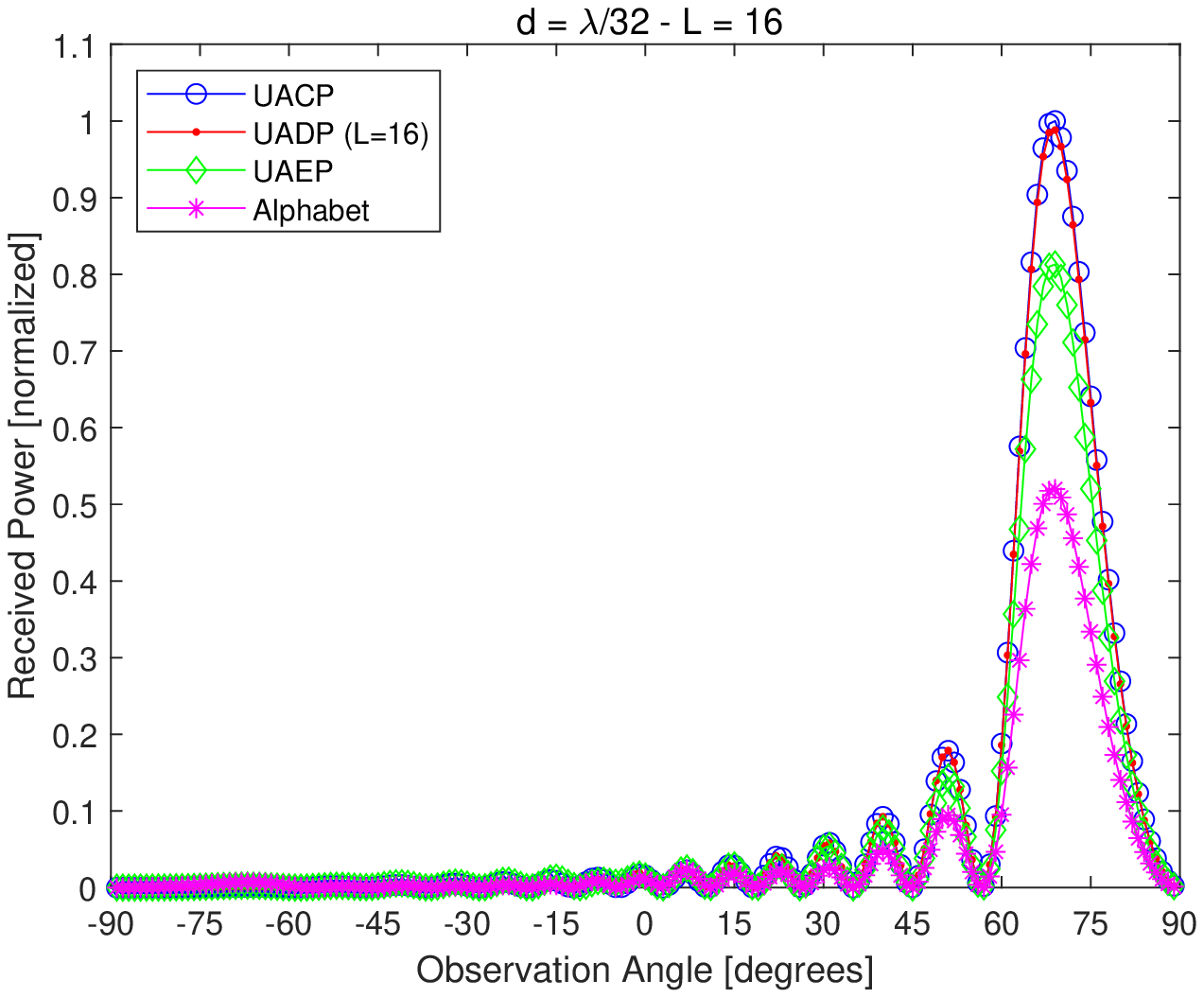}
\vspace{-0.25cm} \caption{Received power as a function of the angle of observation. The RIS alphabet is \cite{Linglong_Testbed}, the desired angle of reflection is 75 degrees, and the inter-distance is $d=\lambda/32$.}\label{fig:Prx__75deg_lambda32_Q4} \vspace{-0.30cm}
\end{figure}

The numerical results that correspond to the RIS alphabet in \cite{Linglong_Testbed}, which is reported in Table \ref{Table_RelectionCoefficientDiscrete}, are illustrated in Figs. \ref{fig:Fig1__45deg_05lambda_Q1}-\ref{fig:Prx__75deg_lambda32_Q4}. For illustrative purposes, we report the color maps that correspond to the optimal configuration of the RIS reconfigurable elements and the corresponding received power as a function of the angle of observation. From the analysis of the color maps, we evince that the design of an RIS with sub-wavelength RIS reconfigurable elements allows us to better approximate the ideal (i.e., continuous-valued) phase profile, which results in an RIS reradiation pattern with fewer and smaller sidelobes. We observe, however, a typical diminishing return law: Reducing the size of the RIS reconfigurable elements from $(1/8)$th of the wavelength to $(1/32)$th of the wavelength does not offer any substantial gains in the considered case studies. Considering the difficulty of realizing RIS reconfigurable elements that are electrically small while ensuring reconfigurability and $360$ degrees of phase modulation, this constitutes a good performance tradeoff. 

By analyzing the curves that correspond to case study ``UADP (${L}=2$)'' with those of the alphabet in \cite{Linglong_Testbed} where ${L}=4$, we note a significantly different performance trend: The curves that correspond to the case study ``UADP (${L}=2$)'' show an undesired beam towards a direction that is symmetrical with respect to the desired direction, i.e., towards $-45$ degrees and $-75$ degrees for the considered case studies. This is related to the fact that an anomalous reflector with ${L}=2$ reduces to a beam splitter. This is apparent from the corresponding color maps while the inter-distance between the RIS reconfigurable elements is kept the same. In other words, regardless of how fine the spatial resolution of an RIS is, an extreme quantization of the phase shift of the RIS reconfigurable elements introduces some fundamental limitations in the reradiation pattern. In addition, this occurs regardless of the practical implementation of an RIS, since the ``UADP (${L}=2$)'' case study is a mathematical abstraction. By considering that many available RISs are designed based on binary RIS reconfigurable elements, i.e., ${L}=2$, this is an important takeaway message, especially because RISs with four-state reconfigurable elements almost completely solve this issue at a reasonable implementation complexity. It needs to be emphasized, in particular, that the undesired beam towards the direction that is symmetrical with respect to the desired direction and the desired beam have the same amplitude, which results in low reradiation performance towards the direction of interest (i.e., the total reradiated power is fixed). 

By analyzing the different figures, we observe that there is a major gain by reducing the size and the inter-distance of the RIS reconfigurable elements from half ($1/2$) of the wavelength to $(1/4)$th (or smaller) of the wavelength: The radiation pattern has fewer and smaller side lobes. This is especially visible when the angle of reradiation (deflection) is very different from the angle of incidence, i.e., if the angle of reflection is $75$ degrees for the considered case studies. If the RIS reconfigurable elements are smaller than $(1/8)$th of the wavelength, the differences are negligible for the considered case studies. As far as the case study for the desired angle of reflection equal to $75$ degrees is concerned, we see that the peak of the desired beam is not exactly steered towards $75$ degrees. This is due to the fact that the channels $g_{nm}$ and $h_{nm}$ in \eqref{Local_Optimization_Problem} are angle-dependent, and the main beam is not perfectly symmetric for large angles of deflection. This issue may be alleviated by designing an RIS based on a global design criterion \cite{arxiv.2110.00833}.

\begin{figure}[!t]
\includegraphics[width=0.66\columnwidth]{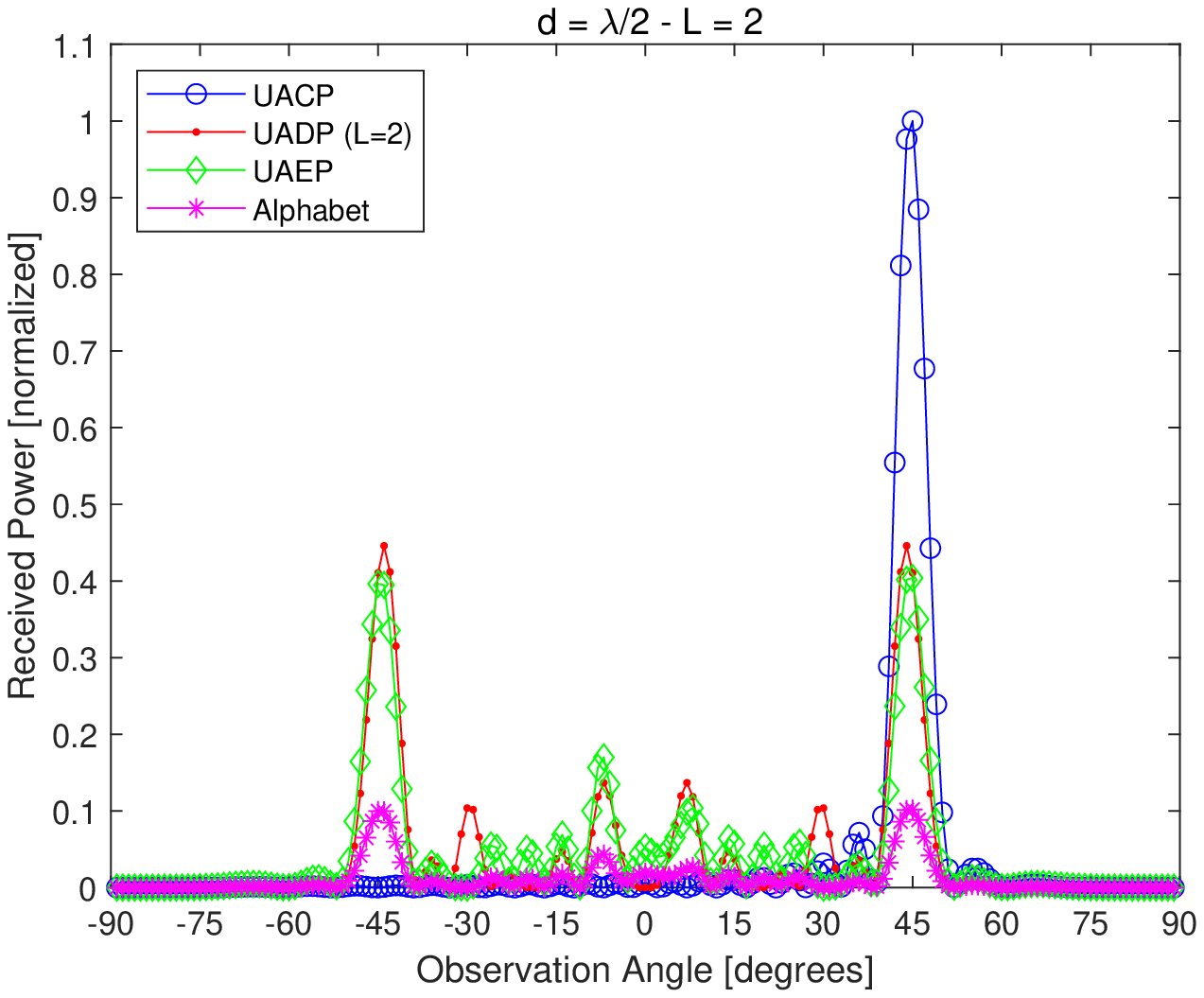}
\vspace{-0.25cm} \caption{Received power as a function of the angle of observation. The RIS alphabet is \cite{Hongliang_OmniSurface}, the desired angle of reflection is 45 degrees, and the inter-distance is $d=\lambda/2$.}\label{fig:Prx__45deg_lambda2_Q1__3p6} \vspace{-0.37cm}
\end{figure}

\begin{figure}[!t]
\includegraphics[width=0.66\columnwidth]{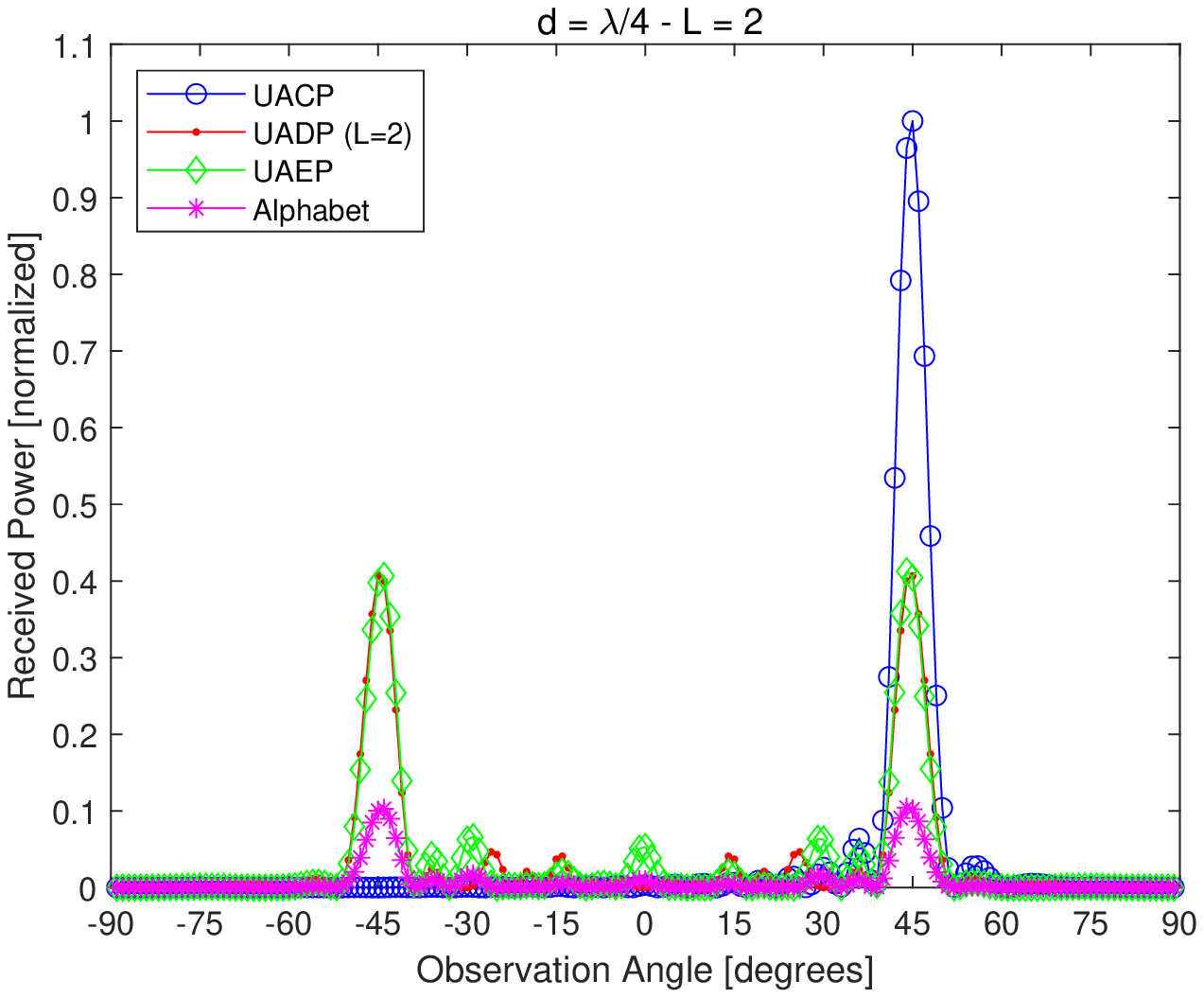}
\vspace{-0.25cm} \caption{Received power as a function of the angle of observation. The RIS alphabet is \cite{Hongliang_OmniSurface}, the desired angle of reflection is 45 degrees, and the inter-distance is $d=\lambda/4$.}\label{fig:Prx__45deg_lambda4_Q1__3p6} \vspace{-0.37cm}
\end{figure}

\begin{figure}[!t]
\includegraphics[width=0.66\columnwidth]{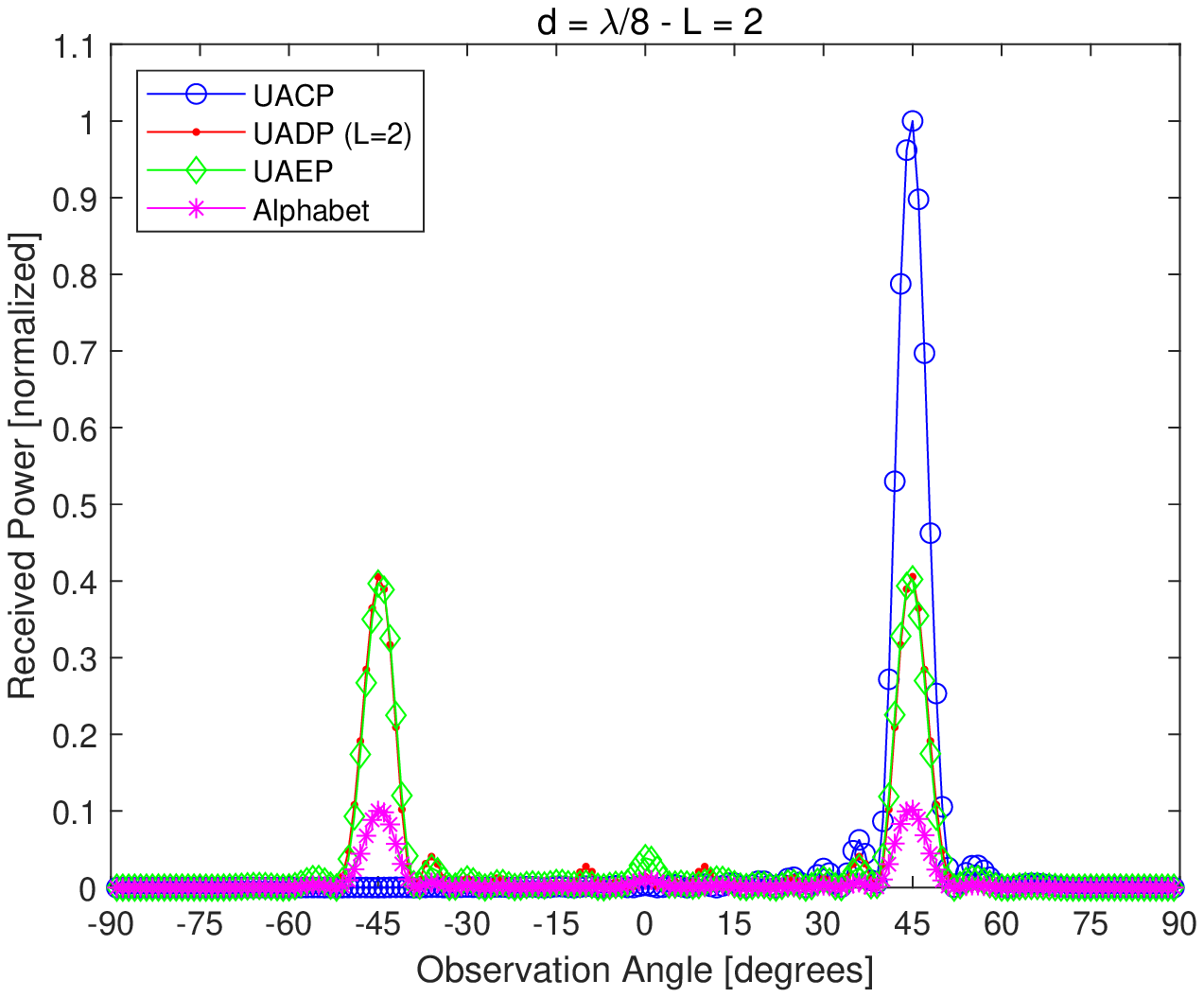}
\vspace{-0.25cm} \caption{Received power as a function of the angle of observation. The RIS alphabet is \cite{Hongliang_OmniSurface}, the desired angle of reflection is 45 degrees, and the inter-distance is $d=\lambda/8$.}\label{fig:Prx__45deg_lambda8_Q1__3p6} \vspace{-0.37cm}
\end{figure}

\begin{figure}[!t]
\includegraphics[width=0.80\columnwidth]{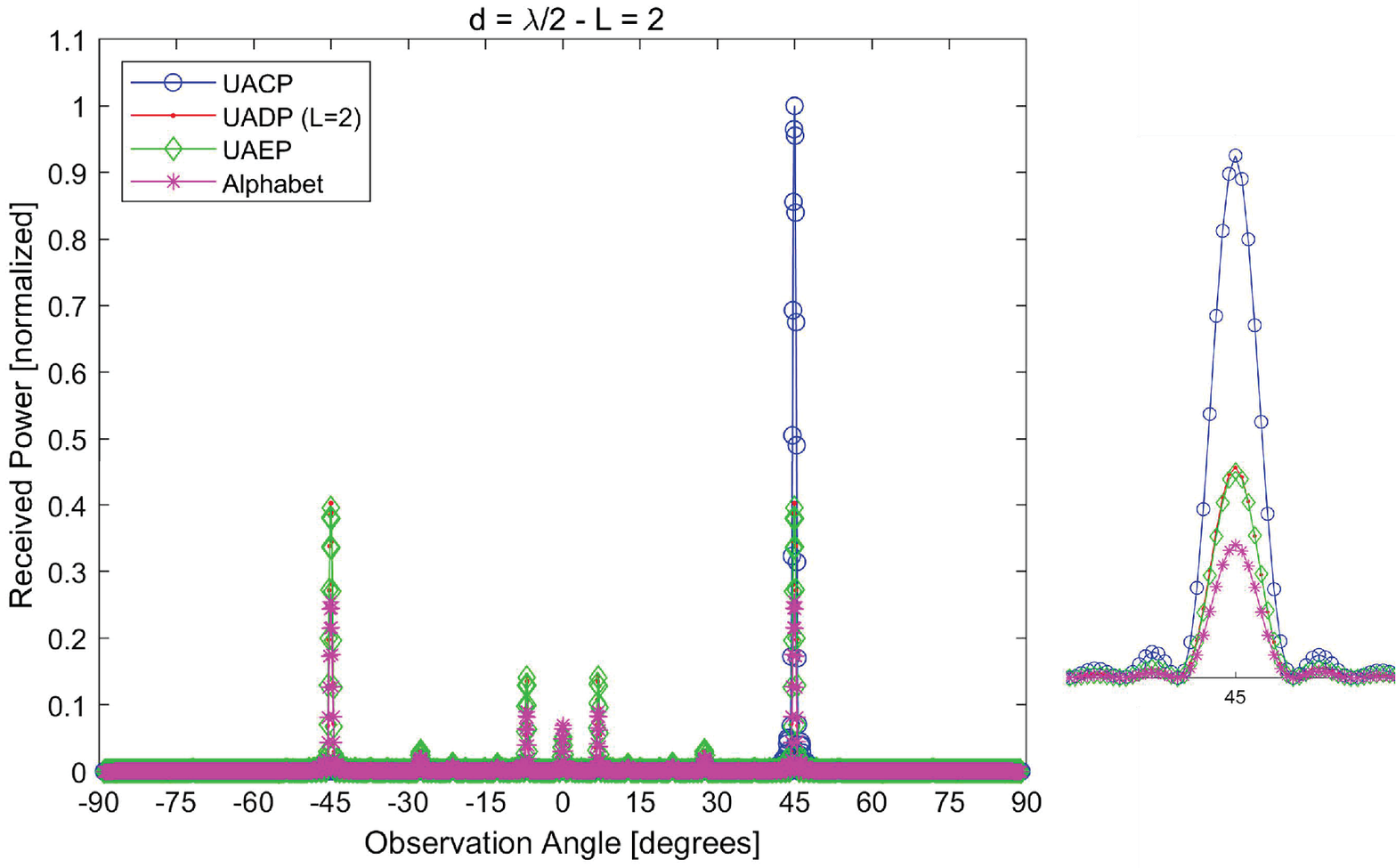}
\vspace{-0.25cm} \caption{Received power as a function of the angle of observation. The RIS alphabet is \cite{Wankai_PathLoss-mmWave} ($f= 27$ GHz), the desired angle of reflection is 45 degrees, and the inter-distance is $d=\lambda/2$.}\label{fig:Prx__45deg_lambda2_Q1__27} \vspace{-0.38cm}
\end{figure}

\begin{figure}[!t]
\includegraphics[width=0.80\columnwidth]{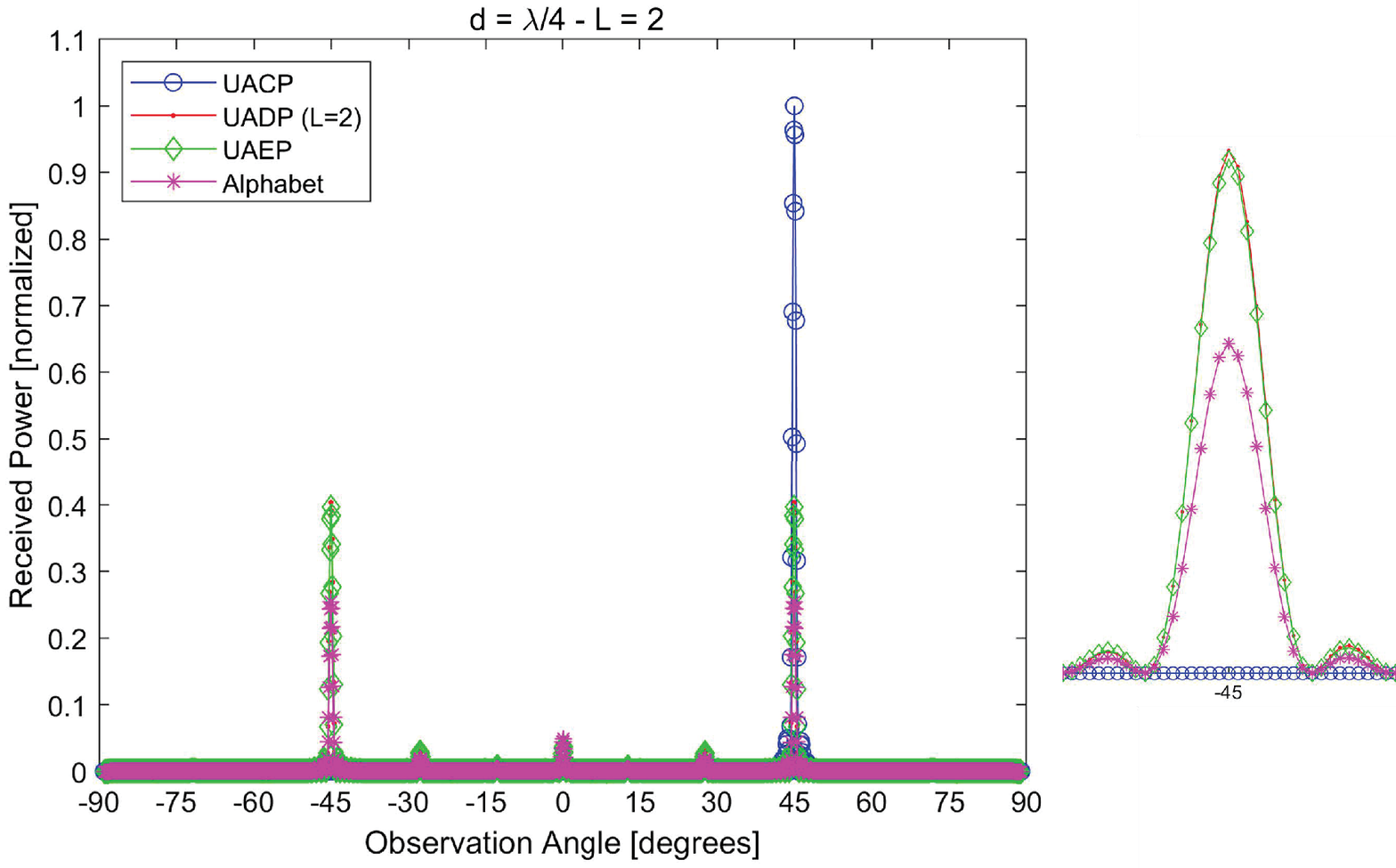}
\vspace{-0.25cm} \caption{Received power as a function of the angle of observation. The RIS alphabet is \cite{Wankai_PathLoss-mmWave} ($f= 27$ GHz), the desired angle of reflection is 45 degrees, and the inter-distance is $d=\lambda/4$.}\label{fig:Prx__45deg_lambda4_Q1__27} \vspace{-0.38cm}
\end{figure}

\begin{figure}[!t]
\includegraphics[width=0.80\columnwidth]{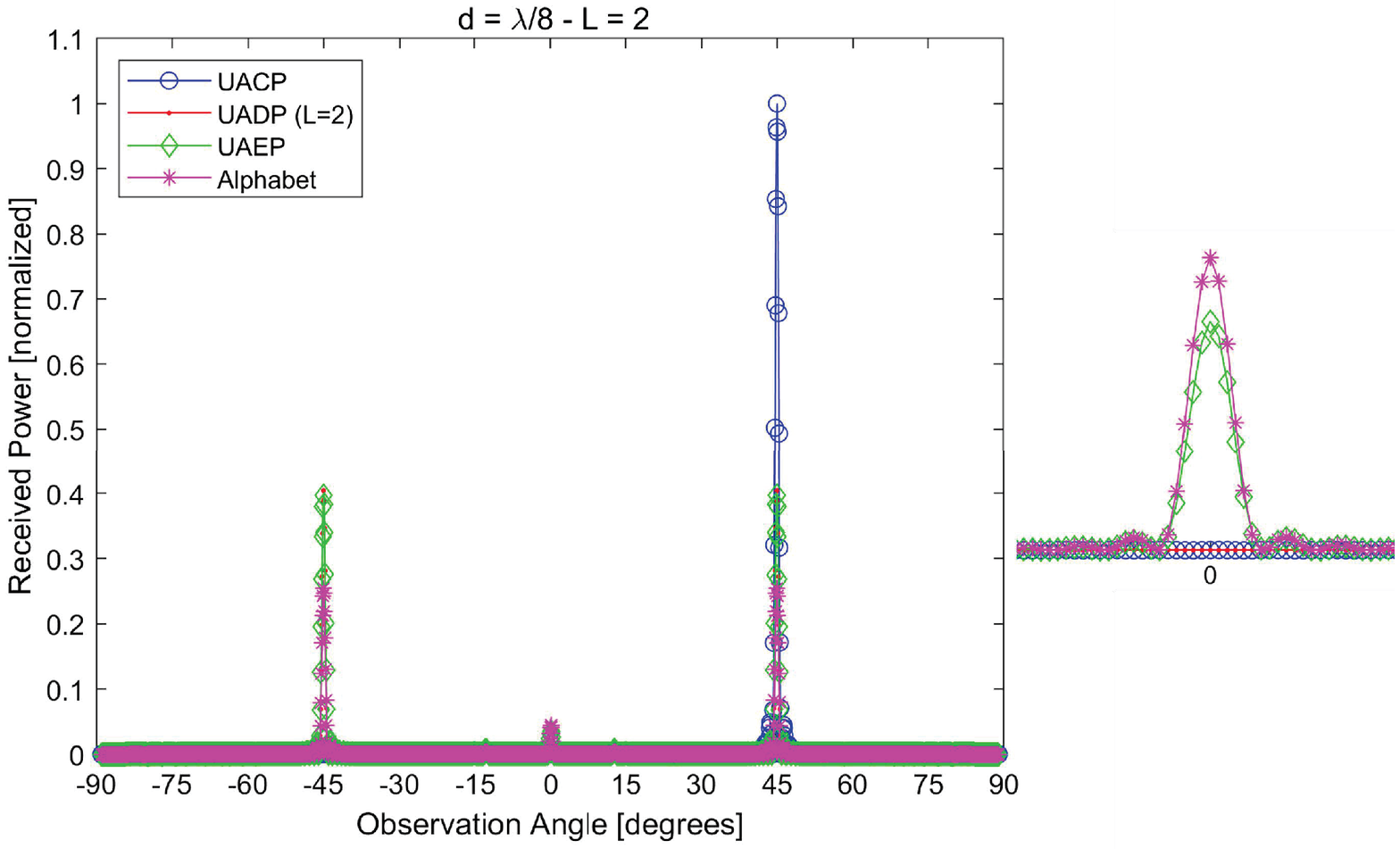}
\vspace{-0.25cm} \caption{Received power as a function of the angle of observation. The RIS alphabet is \cite{Wankai_PathLoss-mmWave} ($f= 27$ GHz), the desired angle of reflection is 45 degrees, and the inter-distance is $d=\lambda/8$.}\label{fig:Prx__45deg_lambda8_Q1__27} \vspace{-0.38cm}
\end{figure}

\begin{figure}[!t]
\includegraphics[width=0.80\columnwidth]{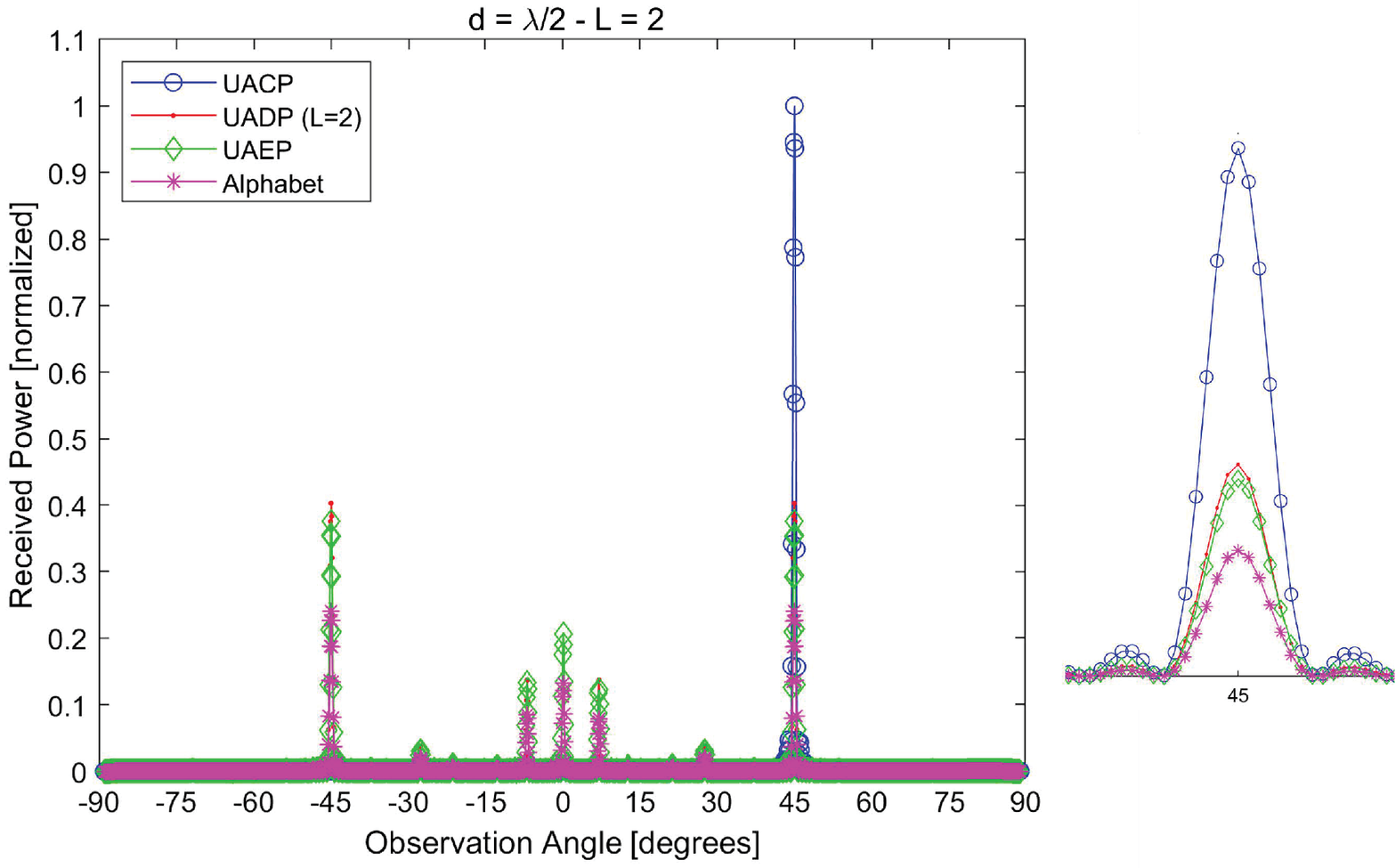}
\vspace{-0.25cm} \caption{Received power as a function of the angle of observation. The RIS alphabet is \cite{Wankai_PathLoss-mmWave} ($f= 33$ GHz), the desired angle of reflection is 45 degrees, and the inter-distance is $d=\lambda/2$.}\label{fig:Prx__45deg_lambda2_Q1__33} \vspace{-0.38cm}
\end{figure}

\begin{figure}[!t]
\includegraphics[width=0.80\columnwidth]{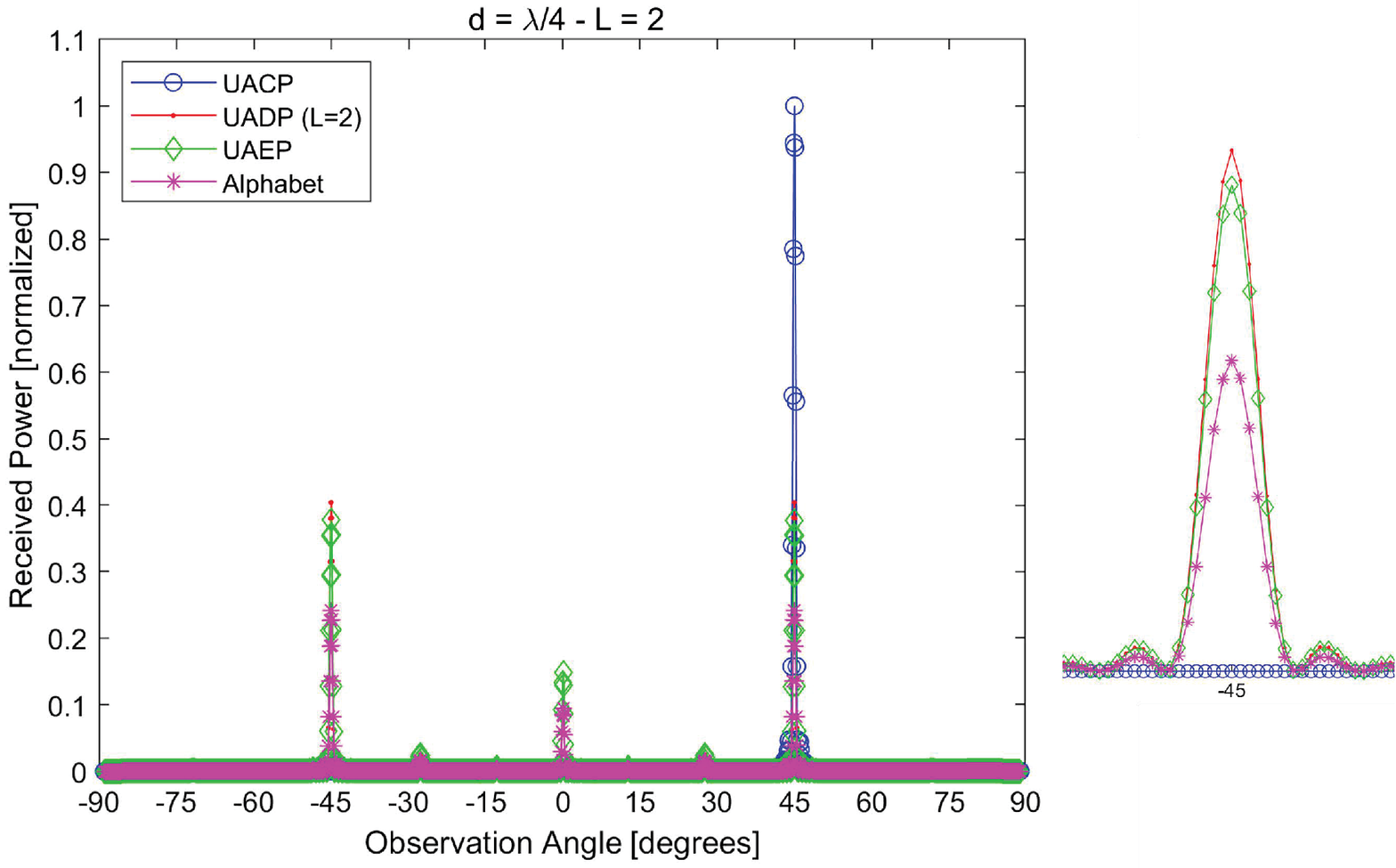}
\vspace{-0.25cm} \caption{Received power as a function of the angle of observation. The RIS alphabet is \cite{Wankai_PathLoss-mmWave} ($f= 33$ GHz), the desired angle of reflection is 45 degrees, and the inter-distance is $d=\lambda/4$.}\label{fig:Prx__45deg_lambda4_Q1__33} \vspace{-0.38cm}
\end{figure}

\begin{figure}[!t]
\includegraphics[width=0.80\columnwidth]{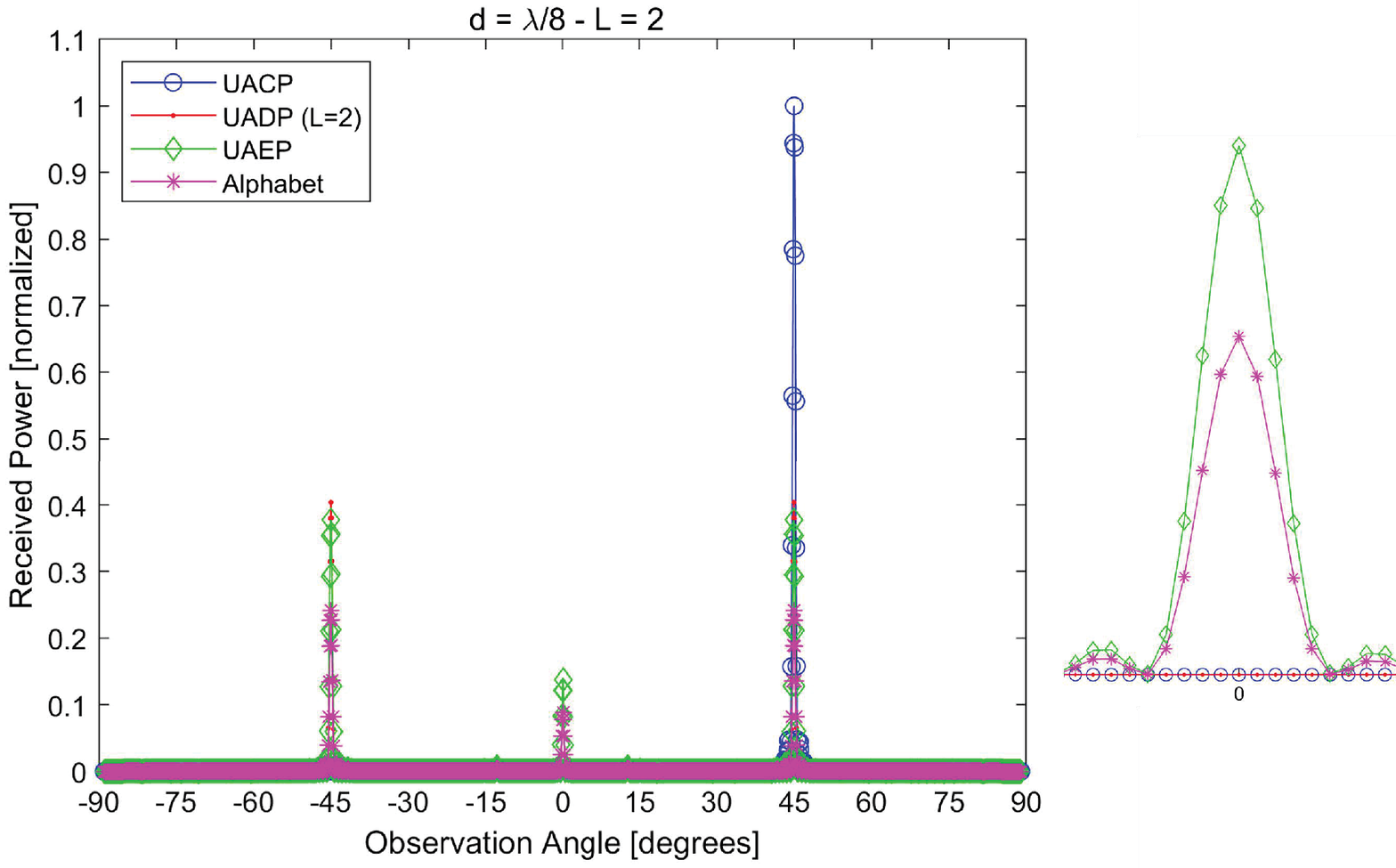}
\vspace{-0.25cm} \caption{Received power as a function of the angle of observation. The RIS alphabet is \cite{Wankai_PathLoss-mmWave} ($f= 33$ GHz), the desired angle of reflection is 45 degrees, and the inter-distance is $d=\lambda/8$.}\label{fig:Prx__45deg_lambda8_Q1__33} \vspace{-0.38cm}
\end{figure}

\begin{figure}[!t]
\includegraphics[width=0.75\columnwidth]{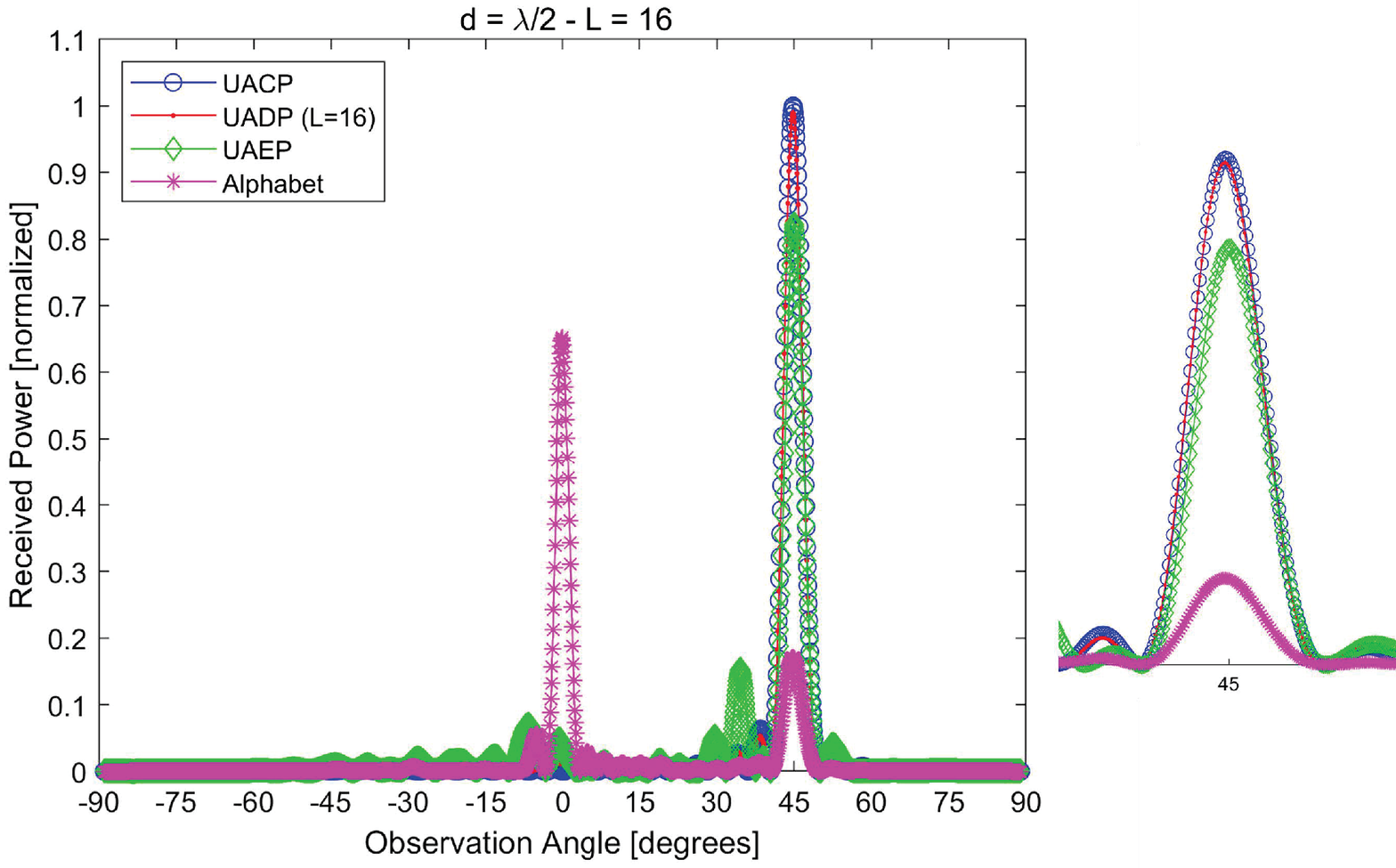}
\vspace{-0.25cm} \caption{Received power as a function of the angle of observation. The RIS alphabet is \cite{Romain_RIS-Prototype}, the desired angle of reflection is 45 degrees, and the inter-distance is $d=\lambda/2$.}\label{fig:Prx__45deg_lambda2_Q4__Orange}\vspace{-0.38cm}
\end{figure}

\begin{figure}[!t]
\includegraphics[width=0.75\columnwidth]{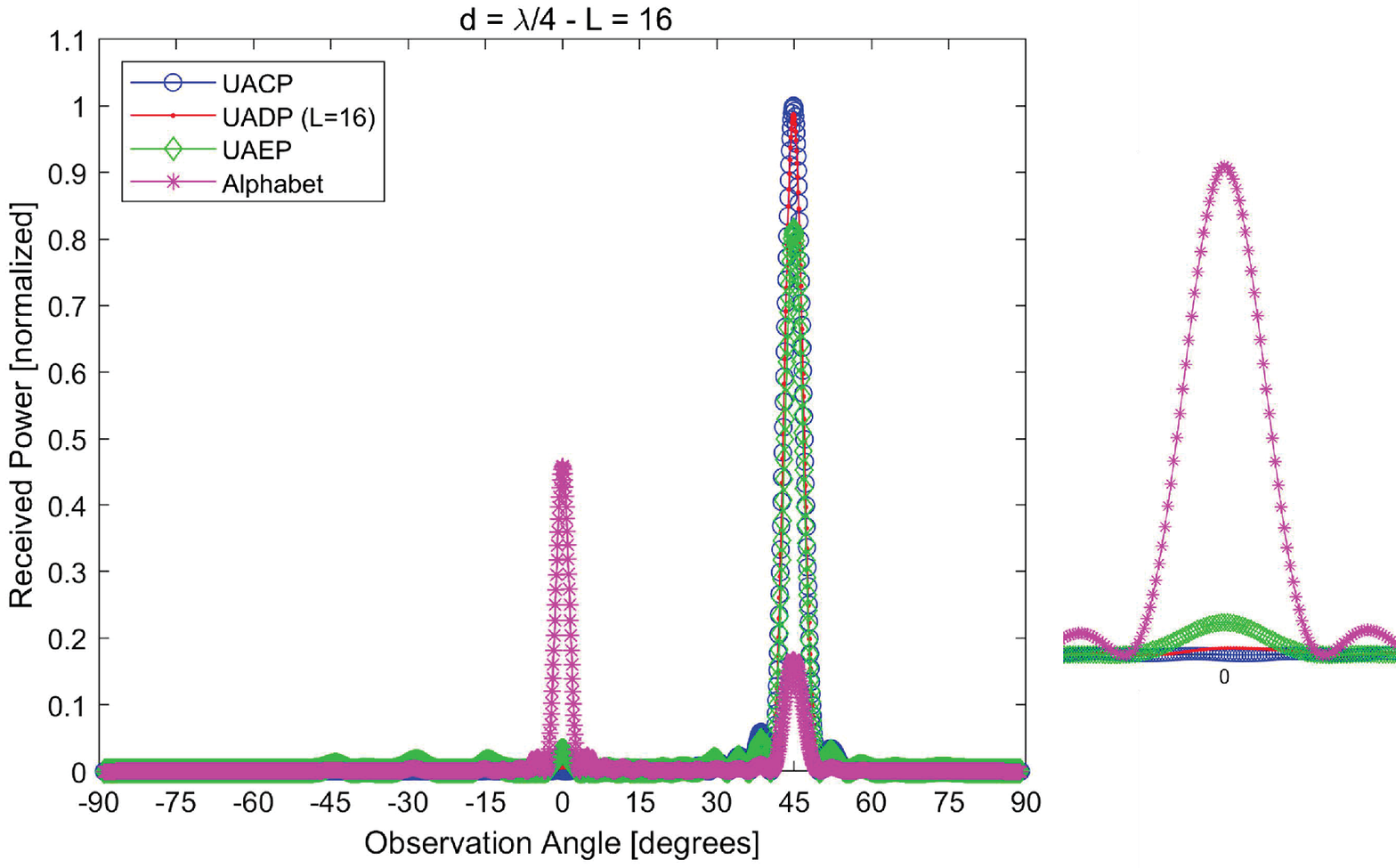}
\vspace{-0.25cm} \caption{Received power as a function of the angle of observation. The RIS alphabet is \cite{Romain_RIS-Prototype}, the desired angle of reflection is 45 degrees, and the inter-distance is $d=\lambda/4$.}\label{fig:Prx__45deg_lambda4_Q4__Orange} \vspace{-0.38cm}
\end{figure}

\begin{figure}[!t]
\includegraphics[width=1.0\columnwidth]{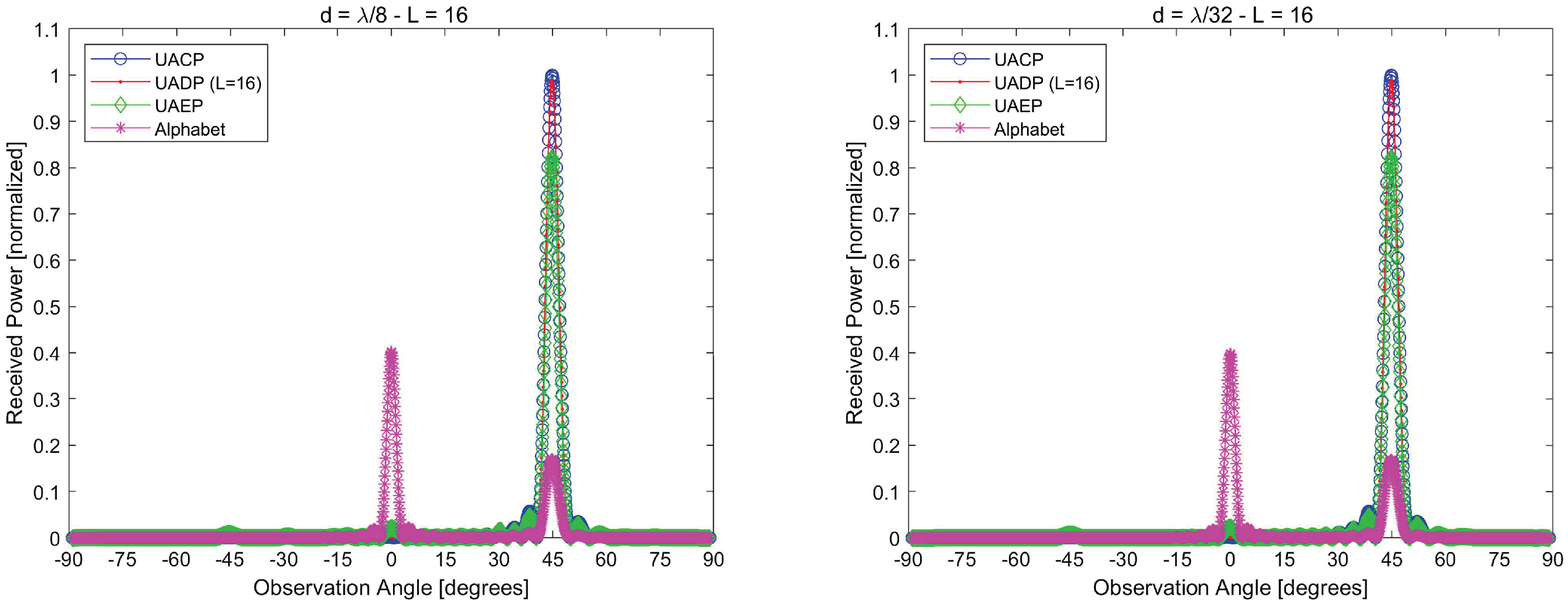}
\vspace{-0.25cm} \caption{Received power as a function of the angle of observation. The RIS alphabet is \cite{Romain_RIS-Prototype}, the desired angle of reflection is 45 degrees, and the inter-distance is $d=\lambda/8$ and $d=\lambda/32$.}\label{fig:Prx__45deg_lambda8_Q4__Orange} \vspace{-0.38cm}
\end{figure}

\begin{figure}[!t]
\includegraphics[width=0.66\columnwidth]{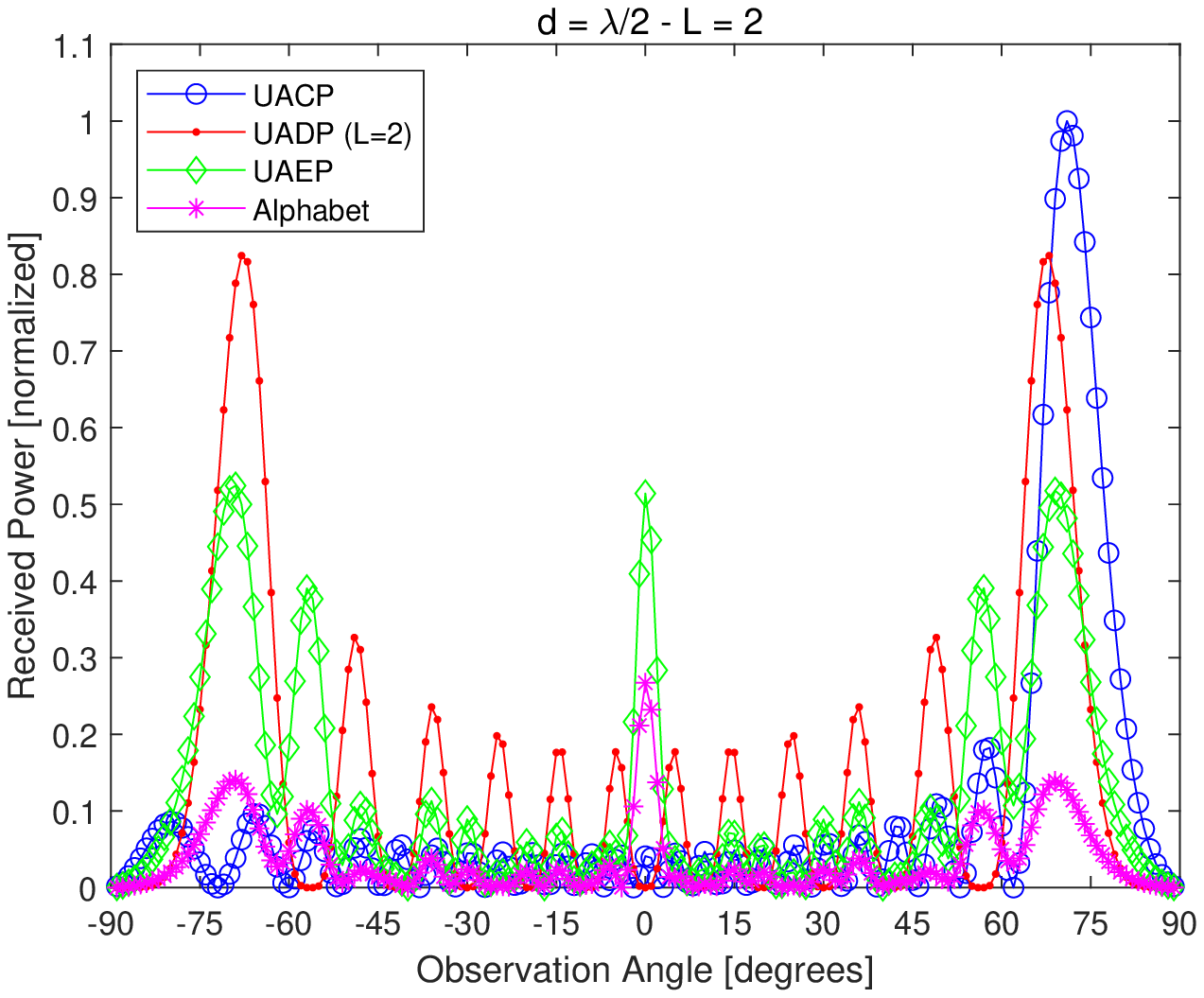}
\vspace{-0.25cm} \caption{Received power as a function of the angle of observation. The RIS alphabet is \cite{Hongliang_OmniSurface}, the desired angle of reflection is 75 degrees, and the inter-distance is $d=\lambda/2$.}\label{fig:Prx__75deg_lambda2_Q1__3p6} \vspace{-0.37cm}
\end{figure}

\begin{figure}[!t]
\includegraphics[width=0.66\columnwidth]{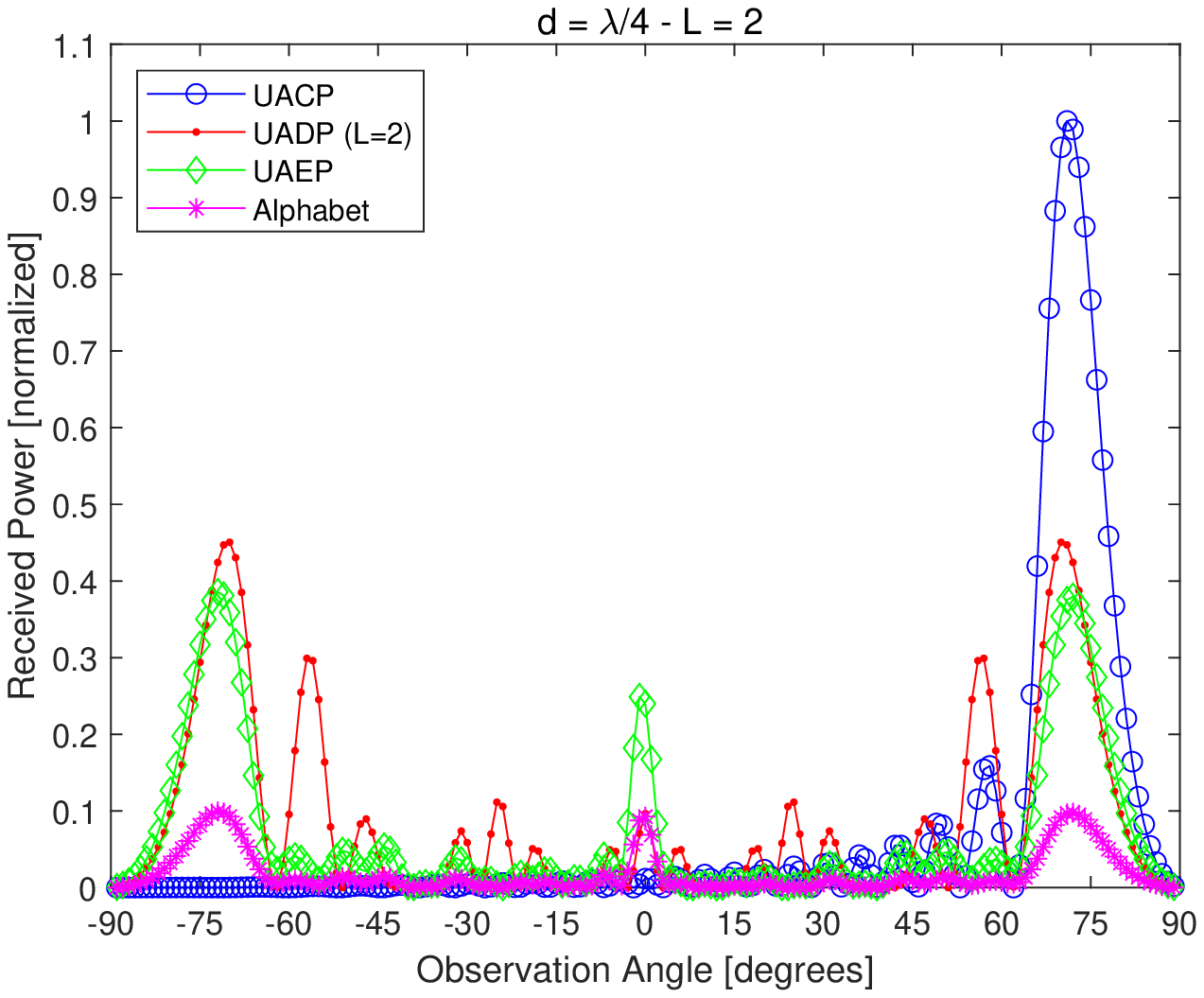}
\vspace{-0.25cm} \caption{Received power as a function of the angle of observation. The RIS alphabet is \cite{Hongliang_OmniSurface}, the desired angle of reflection is 75 degrees, and the inter-distance is $d=\lambda/4$.}\label{fig:Prx__75deg_lambda4_Q1__3p6} \vspace{-0.37cm}
\end{figure}

\begin{figure}[!t]
\includegraphics[width=0.66\columnwidth]{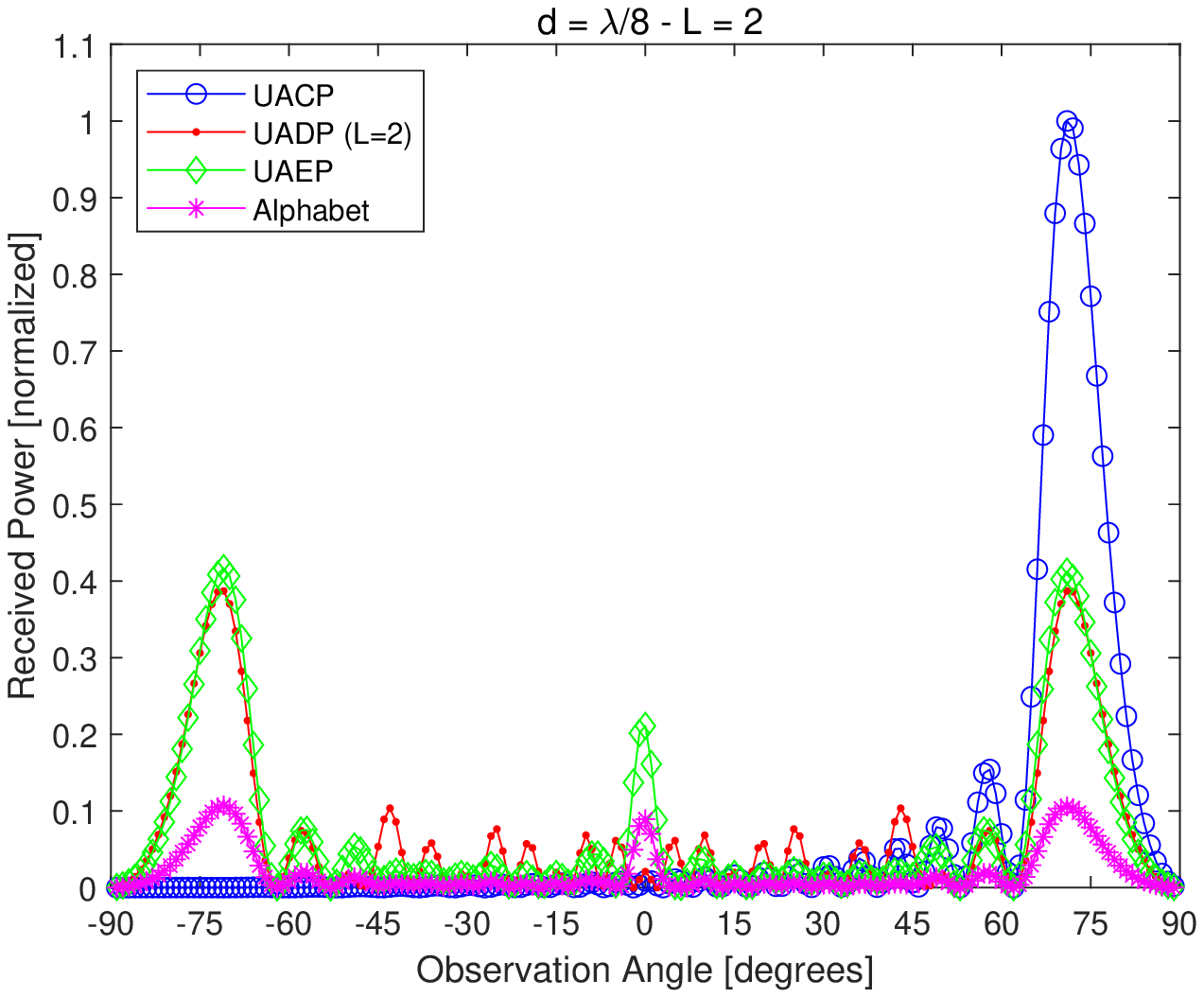}
\vspace{-0.25cm} \caption{Received power as a function of the angle of observation. The RIS alphabet is \cite{Hongliang_OmniSurface}, the desired angle of reflection is 75 degrees, and the inter-distance is $d=\lambda/8$.}\label{fig:Prx__75deg_lambda8_Q1__3p6} \vspace{-0.37cm}
\end{figure}

\begin{figure}[!t]
\includegraphics[width=0.80\columnwidth]{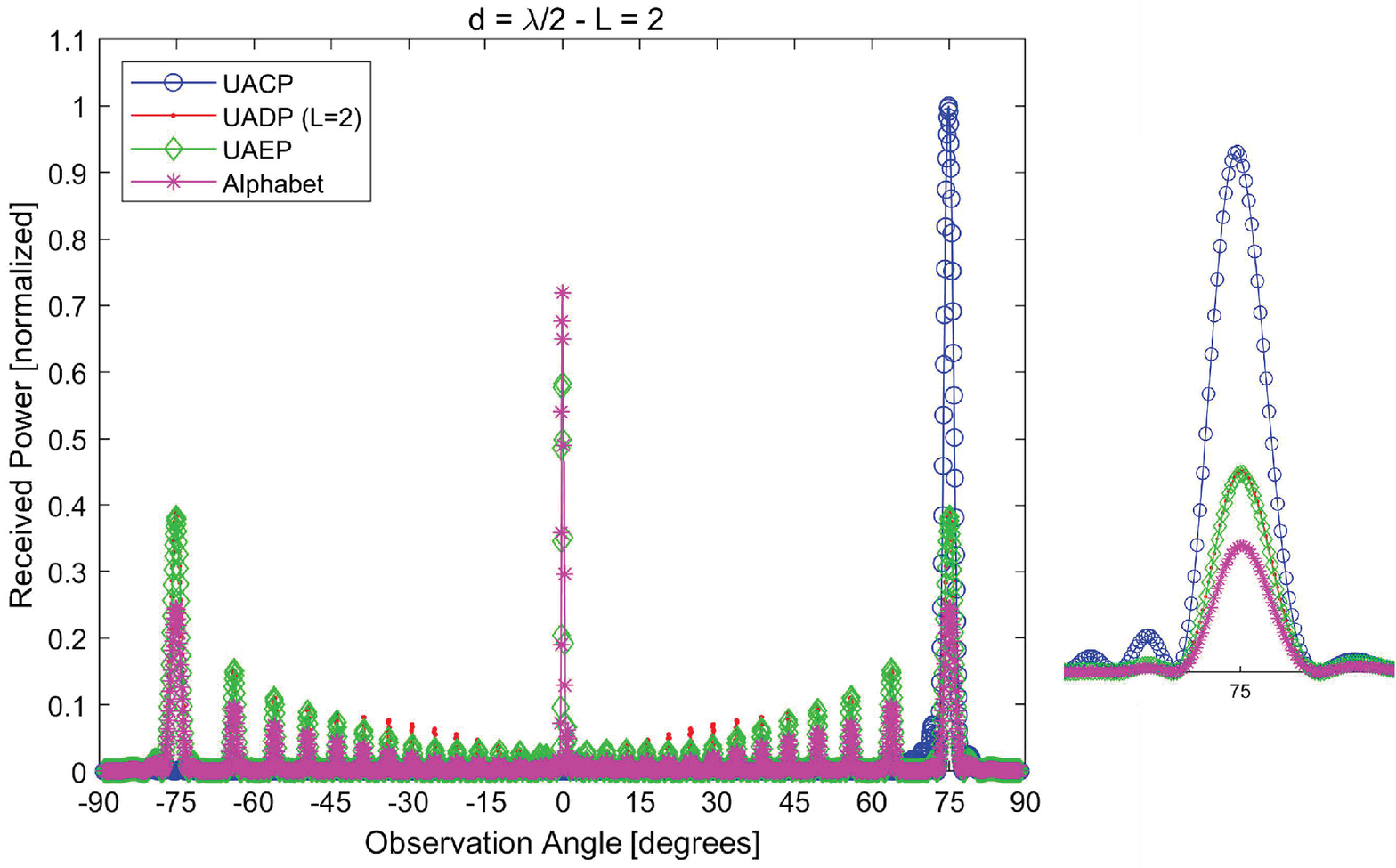}
\vspace{-0.25cm} \caption{Received power as a function of the angle of observation. The RIS alphabet is \cite{Wankai_PathLoss-mmWave} ($f= 27$ GHz), the desired angle of reflection is 75 degrees, and the inter-distance is $d=\lambda/2$.}\label{fig:Prx__75deg_lambda2_Q1__27} \vspace{-0.38cm}
\end{figure}

\begin{figure}[!t]
\includegraphics[width=0.80\columnwidth]{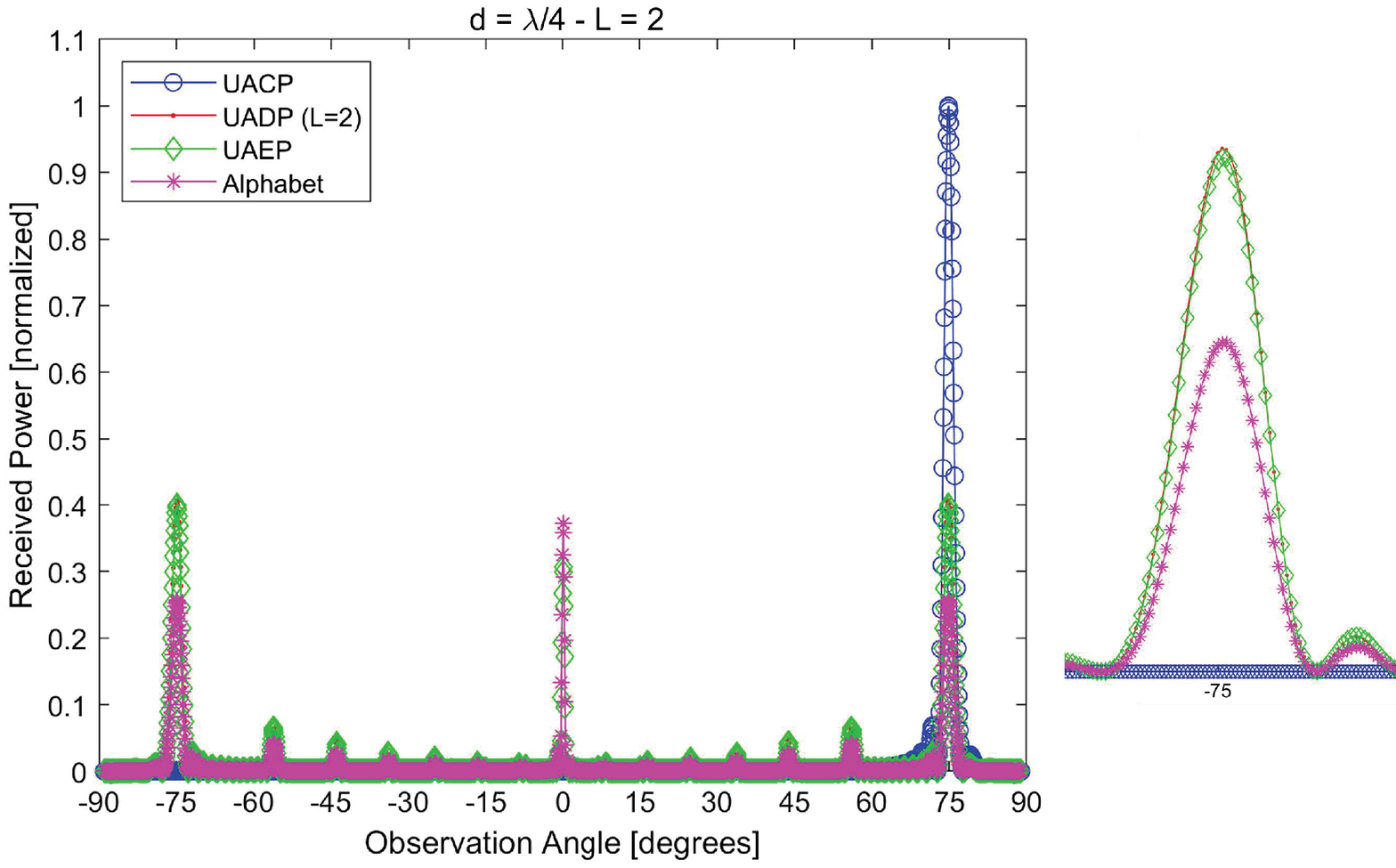}
\vspace{-0.25cm} \caption{Received power as a function of the angle of observation. The RIS alphabet is \cite{Wankai_PathLoss-mmWave} ($f= 27$ GHz), the desired angle of reflection is 75 degrees, and the inter-distance is $d=\lambda/4$.}\label{fig:Prx__75deg_lambda4_Q1__27} \vspace{-0.38cm}
\end{figure}

\begin{figure}[!t]
\includegraphics[width=0.80\columnwidth]{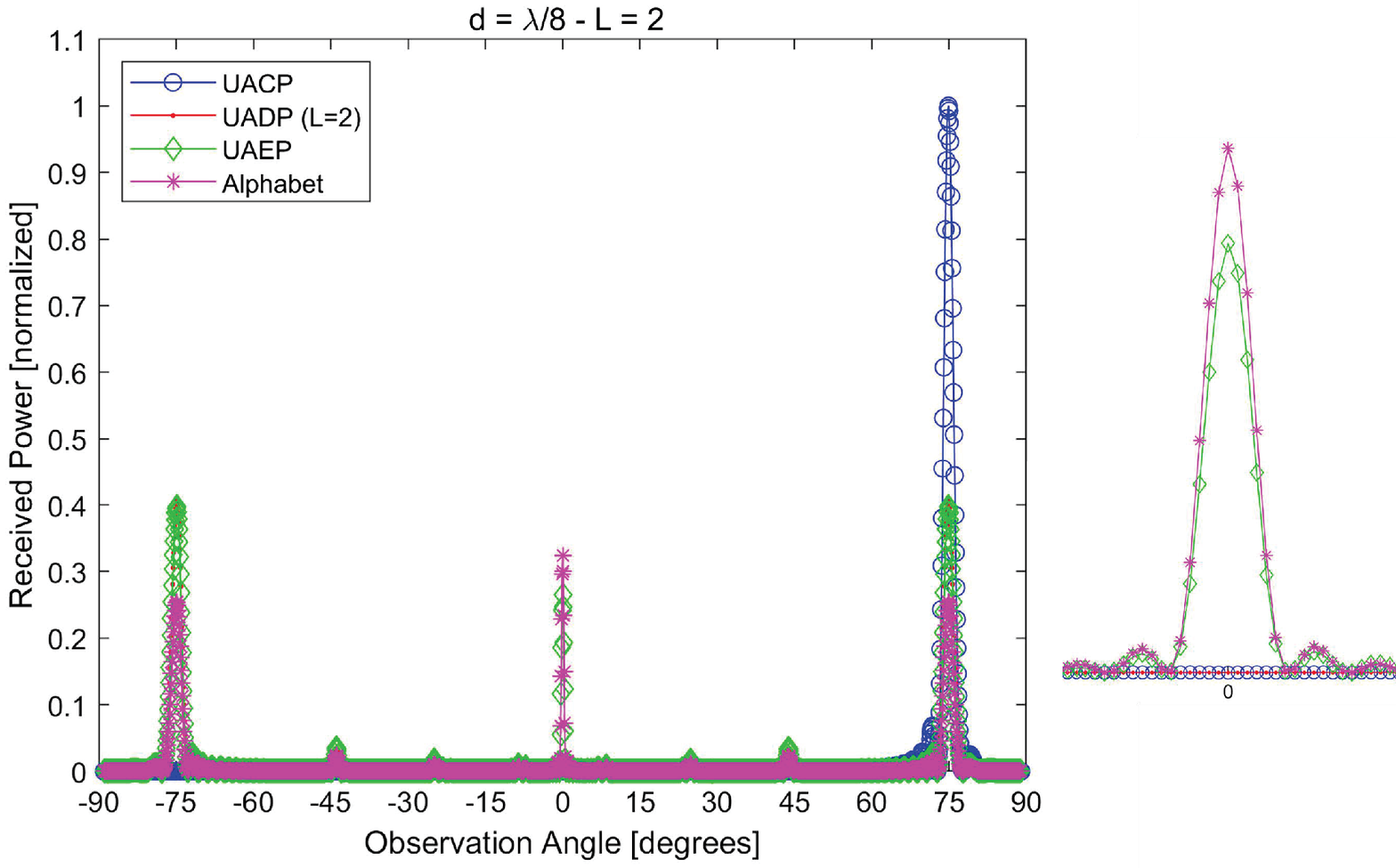}
\vspace{-0.25cm} \caption{Received power as a function of the angle of observation. The RIS alphabet is \cite{Wankai_PathLoss-mmWave} ($f= 27$ GHz), the desired angle of reflection is 75 degrees, and the inter-distance is $d=\lambda/8$.}\label{fig:Prx__75deg_lambda8_Q1__27} \vspace{-0.38cm}
\end{figure}

\begin{figure}[!t]
\includegraphics[width=0.80\columnwidth]{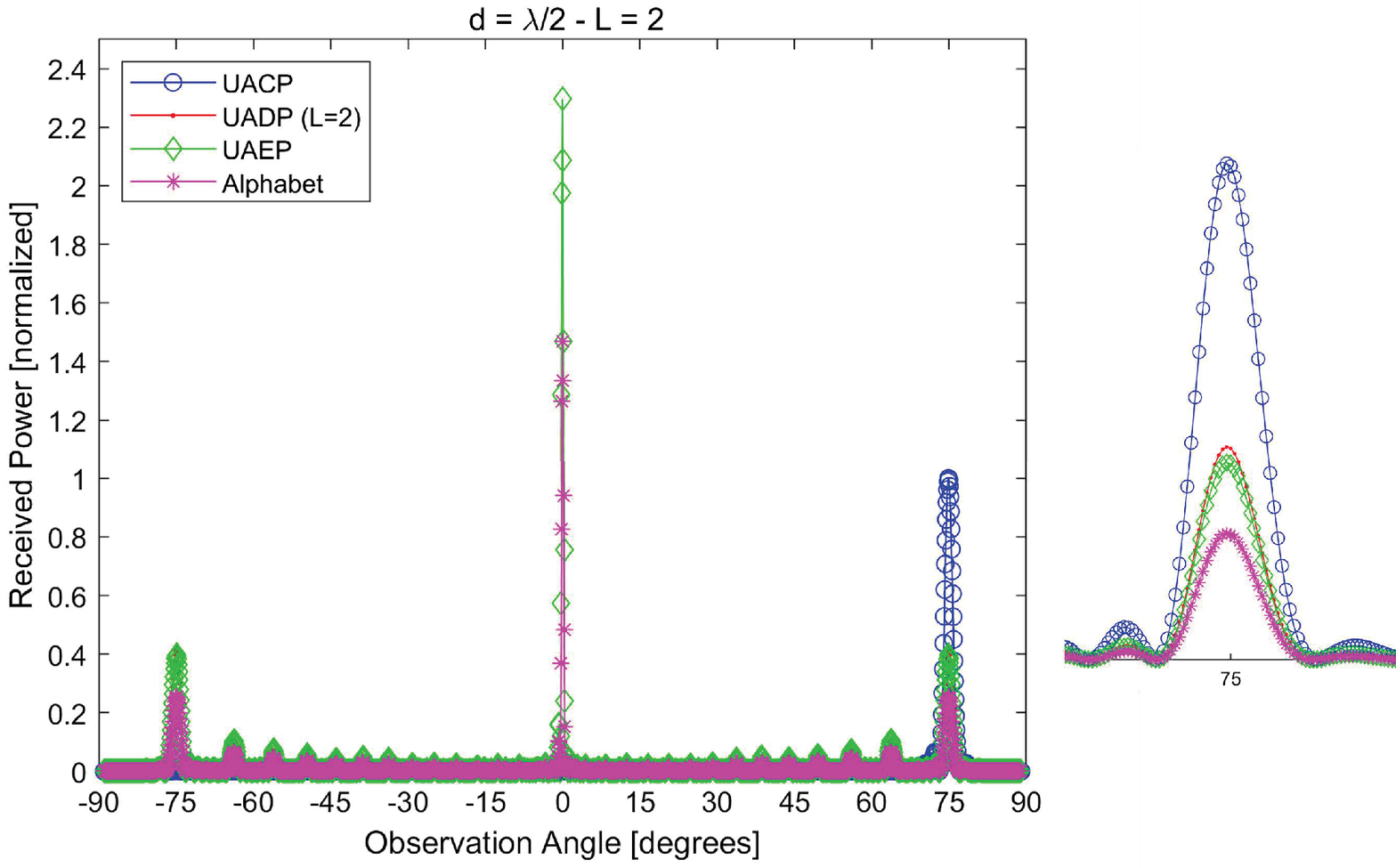}
\vspace{-0.25cm} \caption{Received power as a function of the angle of observation. The RIS alphabet is \cite{Wankai_PathLoss-mmWave} ($f= 33$ GHz), the desired angle of reflection is 75 degrees, and the inter-distance is $d=\lambda/2$.}\label{fig:Prx__75deg_lambda2_Q1__33} \vspace{-0.38cm}
\end{figure}

\begin{figure}[!t]
\includegraphics[width=0.80\columnwidth]{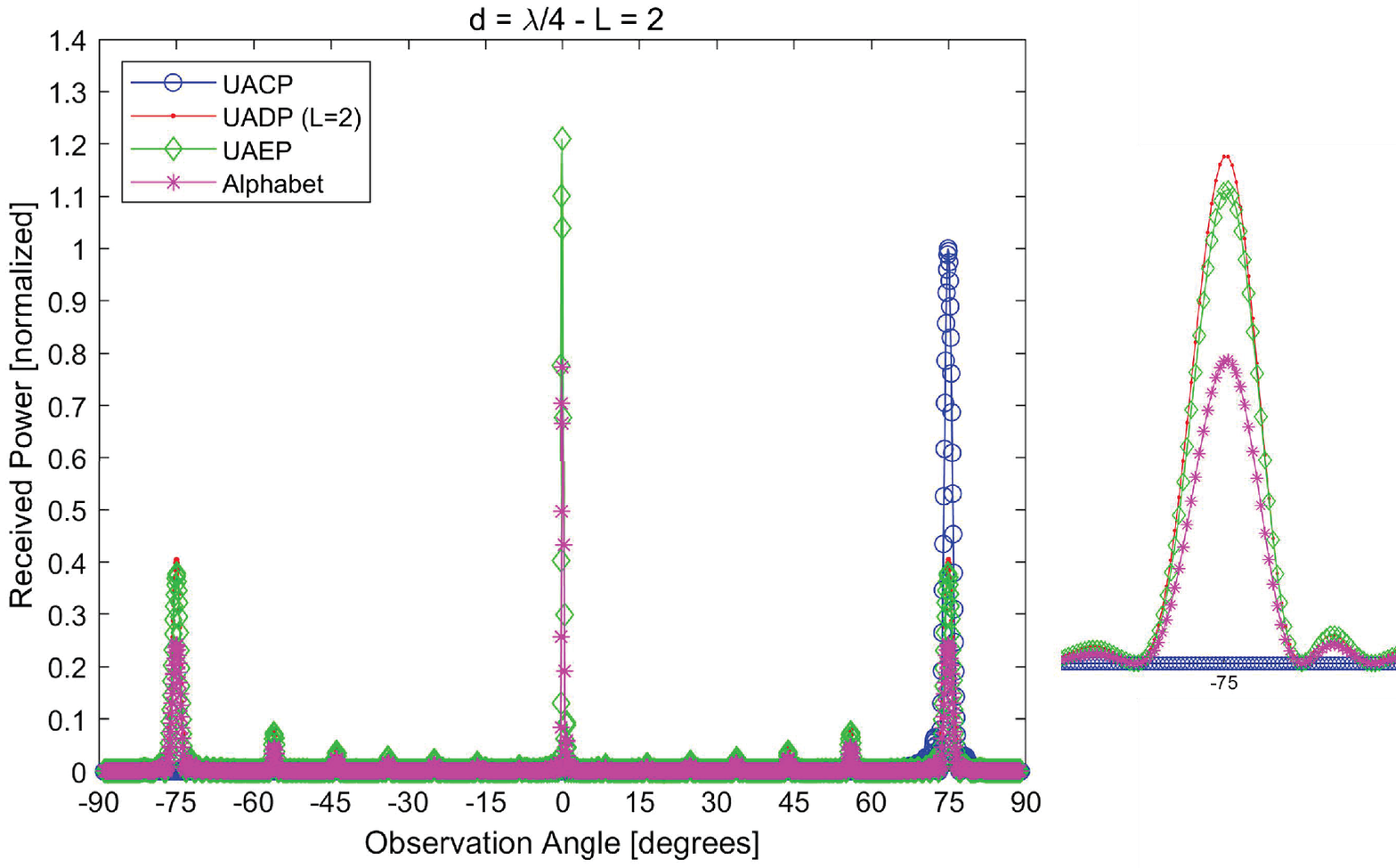}
\vspace{-0.25cm} \caption{Received power as a function of the angle of observation. The RIS alphabet is \cite{Wankai_PathLoss-mmWave} ($f= 33$ GHz), the desired angle of reflection is 75 degrees, and the inter-distance is $d=\lambda/4$.}\label{fig:Prx__75deg_lambda4_Q1__33} \vspace{-0.38cm}
\end{figure}

\begin{figure}[!t]
\includegraphics[width=0.80\columnwidth]{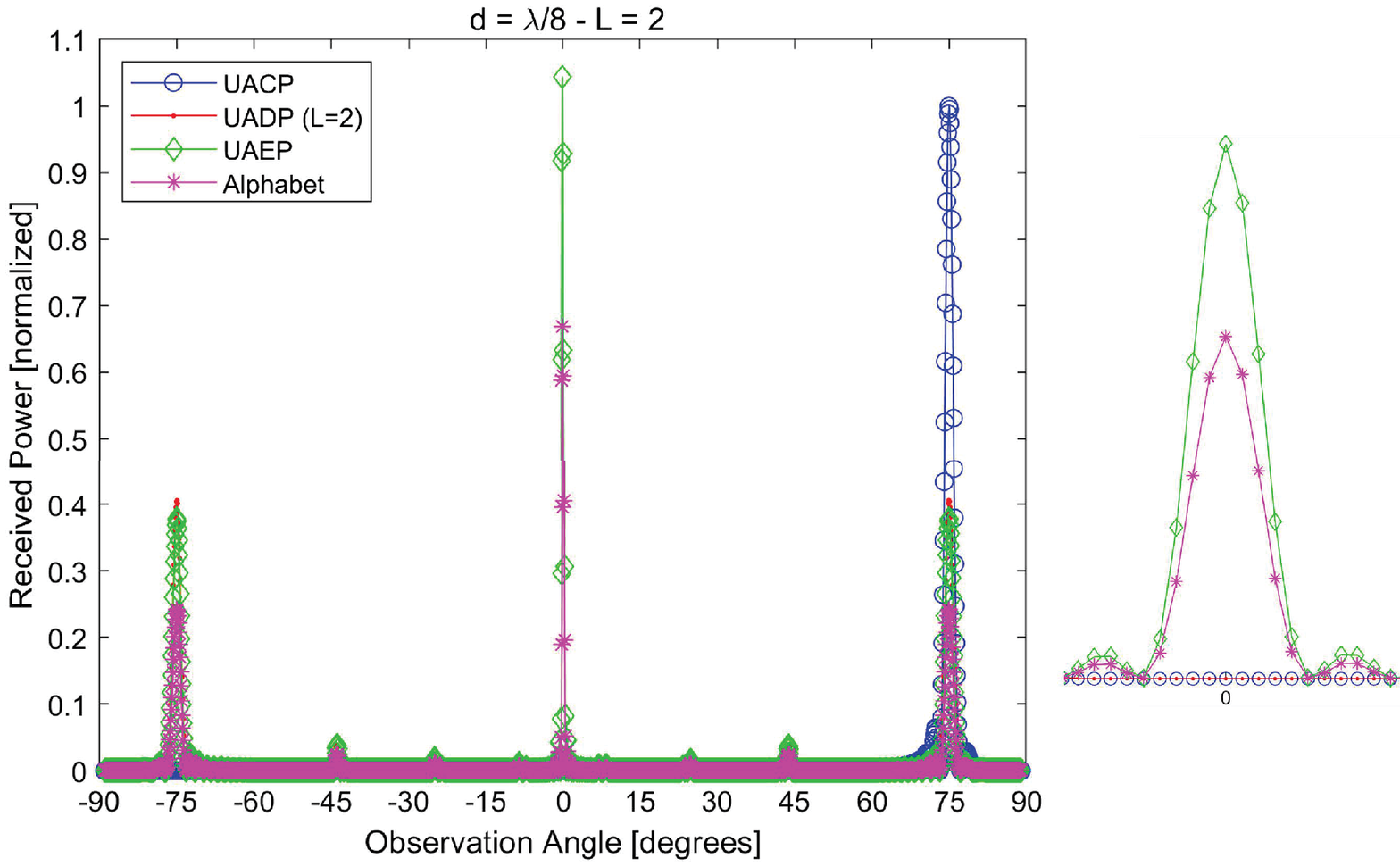}
\vspace{-0.25cm} \caption{Received power as a function of the angle of observation. The RIS alphabet is \cite{Wankai_PathLoss-mmWave} ($f= 33$ GHz), the desired angle of reflection is 75 degrees, and the inter-distance is $d=\lambda/8$.}\label{fig:Prx__75deg_lambda8_Q1__33} \vspace{-0.38cm}
\end{figure}

\begin{figure}[!t]
\includegraphics[width=0.75\columnwidth]{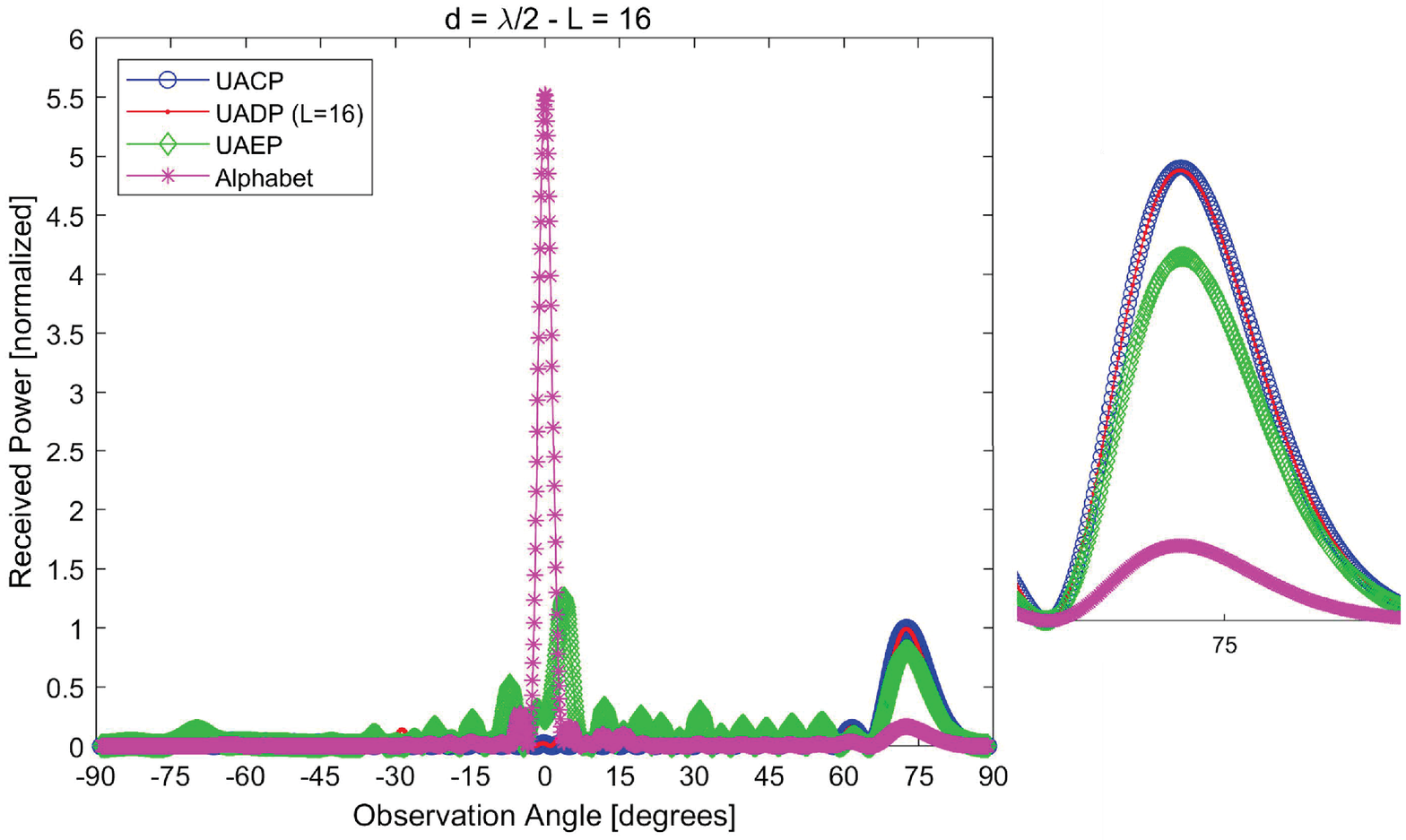}
\vspace{-0.25cm} \caption{Received power as a function of the angle of observation. The RIS alphabet is \cite{Romain_RIS-Prototype}, the desired angle of reflection is 75 degrees, and the inter-distance is $d=\lambda/2$.}\label{fig:Prx__75deg_lambda2_Q4__Orange} \vspace{-0.38cm}
\end{figure}

\begin{figure}[!t]
\includegraphics[width=0.75\columnwidth]{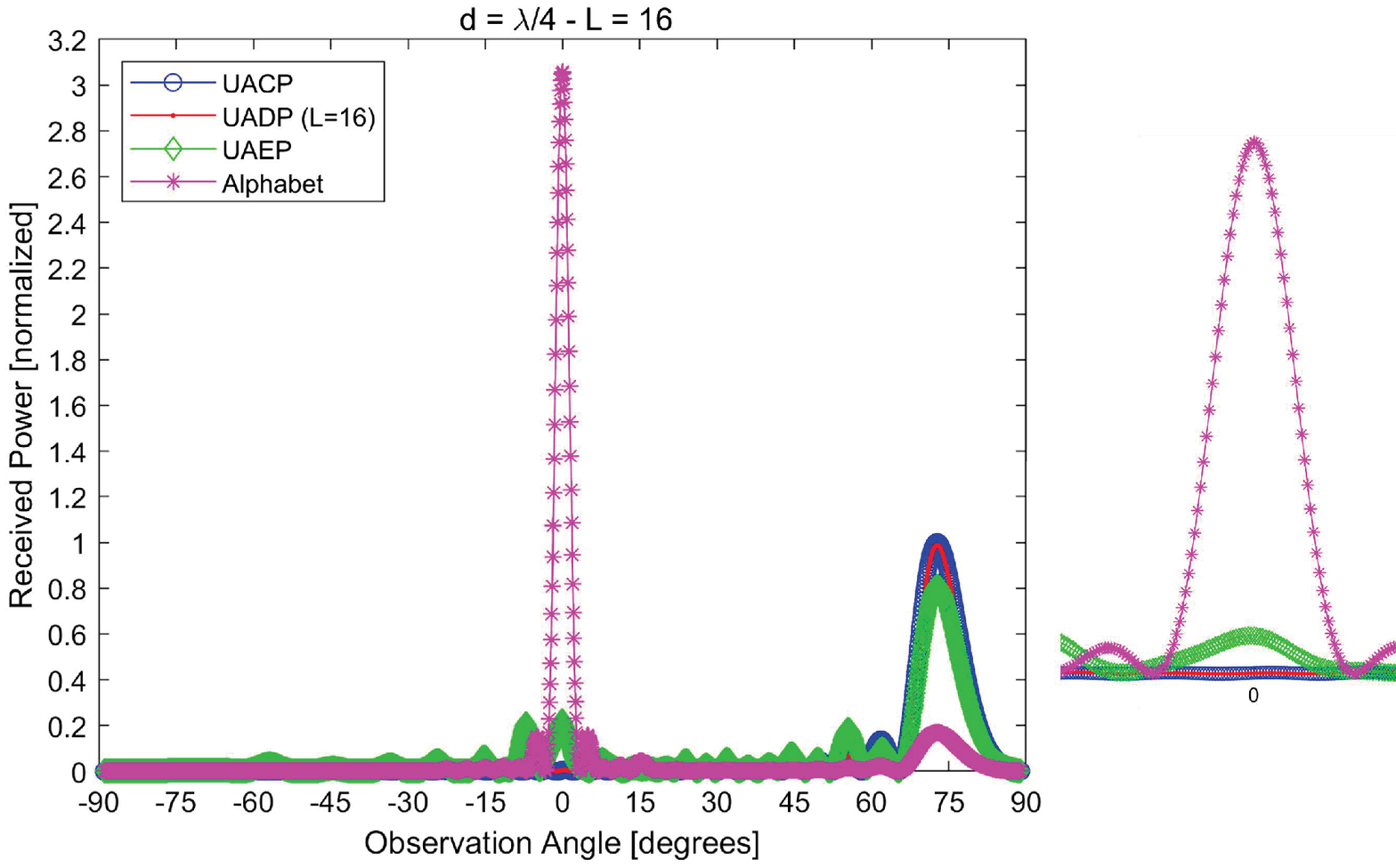}
\vspace{-0.25cm} \caption{Received power as a function of the angle of observation. The RIS alphabet is \cite{Romain_RIS-Prototype}, the desired angle of reflection is 75 degrees, and the inter-distance is $d=\lambda/4$.}\label{fig:Prx__75deg_lambda4_Q4__Orange} \vspace{-0.38cm}
\end{figure}

\begin{figure}[!t]
\includegraphics[width=1.0\columnwidth]{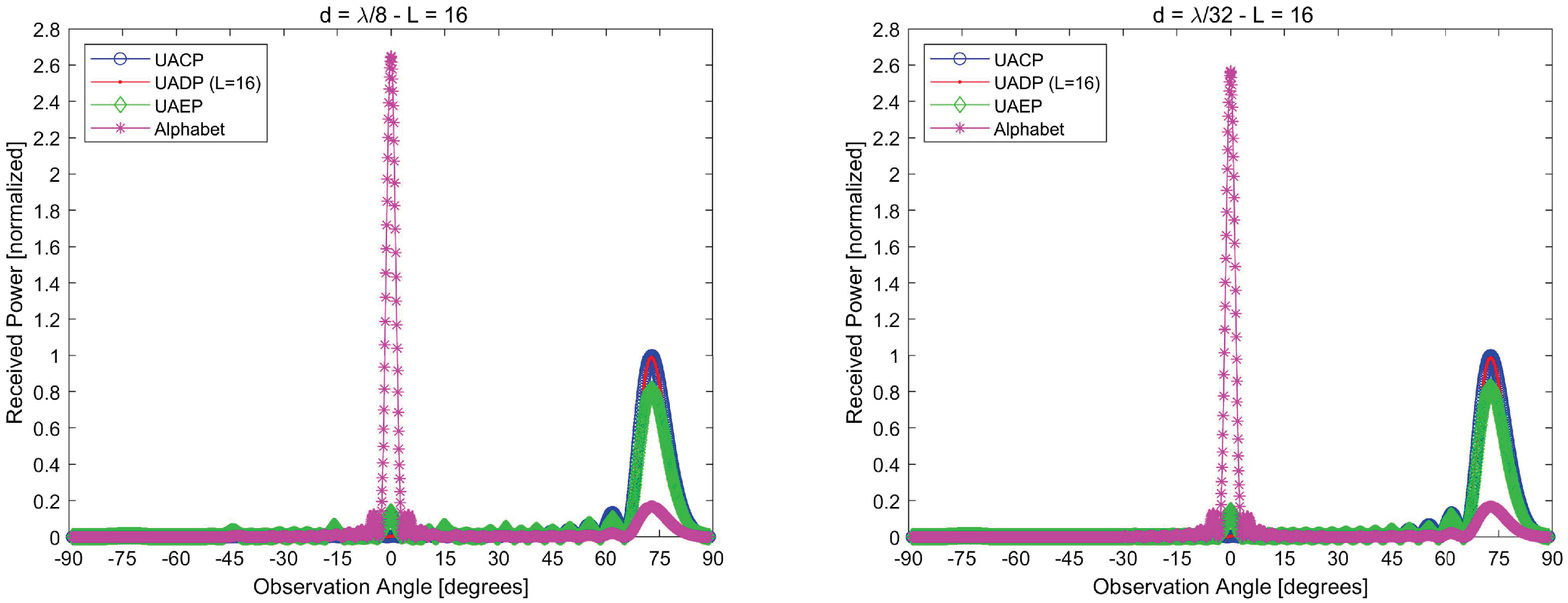}
\vspace{-0.25cm} \caption{Received power as a function of the angle of observation. The RIS alphabet is \cite{Romain_RIS-Prototype}, the desired angle of reflection is 75 degrees, and the inter-distance is $d=\lambda/8$ and  $d=\lambda/32$.}\label{fig:Prx__75deg_lambda8_Q4__Orange} \vspace{-0.38cm}
\end{figure}

The reradiated power as a function of the angle of observation for other RIS alphabets is reported in Figs.
\ref{fig:Prx__45deg_lambda2_Q1__3p6}-\ref{fig:Prx__75deg_lambda8_Q4__Orange}. The general trends observed for the RIS alphabet in \cite{Linglong_Testbed} still hold, but some additional comments are in order. We see that the ``UADP'' case study provides a radiation pattern that does not usually result in undesired beams towards the specular direction, i.e., zero degrees. On the other hand, the ``UAEP'' and ``Alphabet'' case studies do. This is attributed to the perfect symmetry of the ``UADP'' alphabet both in amplitude and phase. A special comment is needed for the RIS prototype in \cite{Romain_RIS-Prototype}. In this case, we note a strong reradiated beam towards the specular direction (zero degrees). This is attributed to the intertwinement between the amplitude and the phase of the reflection coefficient in Table \ref{Table_RelectionCoefficientContinuous}, and, more specifically, to the large attenuation for some values of the phase shift. This remark is confirmed by noting that the reradiated pattern of the ``UAEP''  case study does not have any beam towards the specular direction, since in this case the amplitude of the reflection coefficient is assumed to be equal to one for any phase.

\begin{figure}[!t]
\includegraphics[width=0.62\columnwidth]{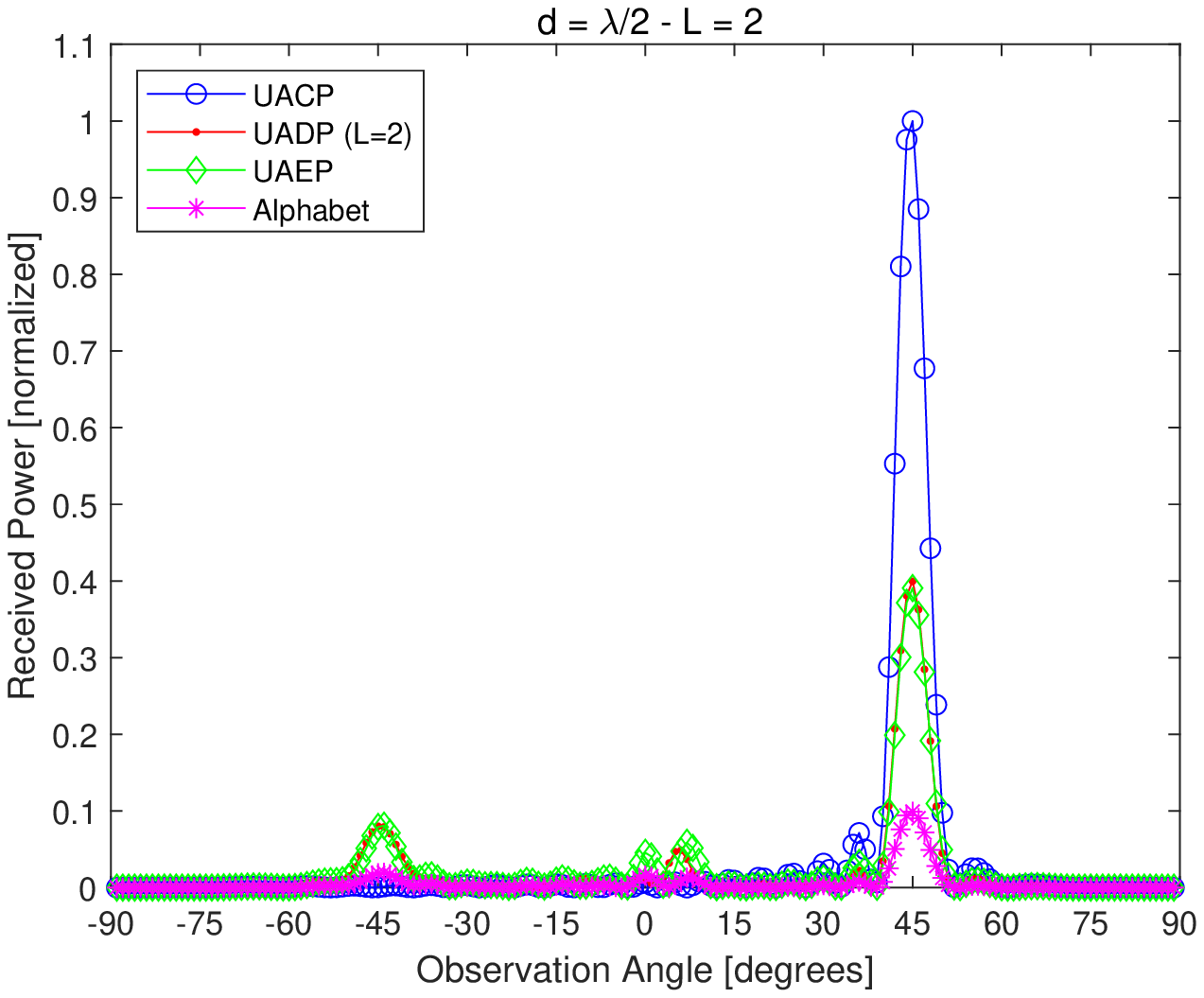}
\vspace{-0.25cm} \caption{Received power as a function of the angle of observation (near-field case with the receiver located 5 meters far from the RIS). The RIS alphabet is \cite{Hongliang_OmniSurface}, the desired angle of reflection is 45 degrees, and the inter-distance is $d=\lambda/2$.}\label{fig:Prx__45deg_lambda2_Q1__3p6__NearField} \vspace{-0.37cm}
\end{figure}

\begin{figure}[!t]
\includegraphics[width=0.62\columnwidth]{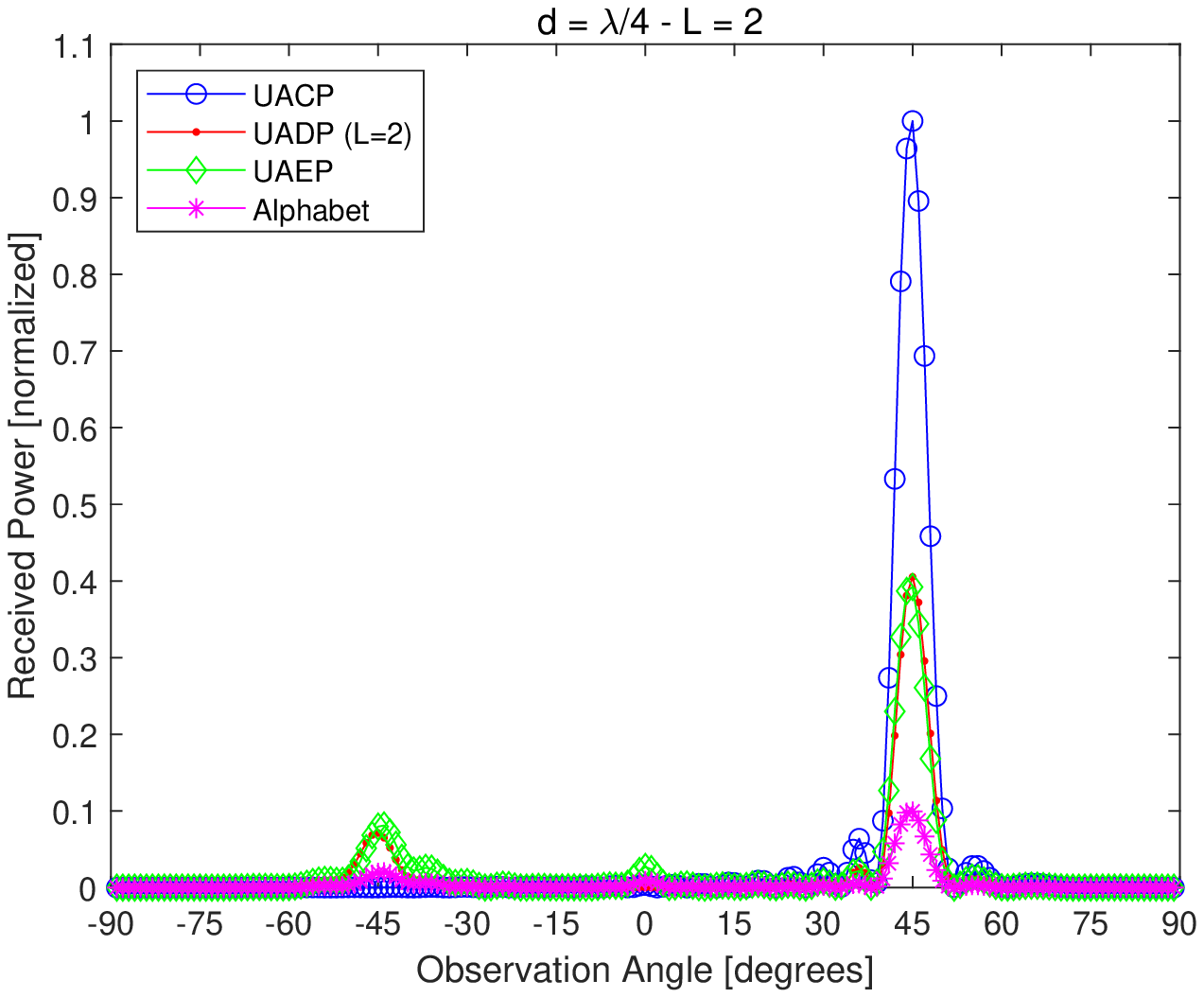}
\vspace{-0.25cm} \caption{Received power as a function of the angle of observation (near-field case with the receiver located 5 meters far from the RIS). The RIS alphabet is \cite{Hongliang_OmniSurface}, the desired angle of reflection is 45 degrees, and the inter-distance is $d=\lambda/4$.}\label{fig:Prx__45deg_lambda4_Q1__3p6__NearField} \vspace{-0.37cm}
\end{figure}

\begin{figure}[!t]
\includegraphics[width=0.62\columnwidth]{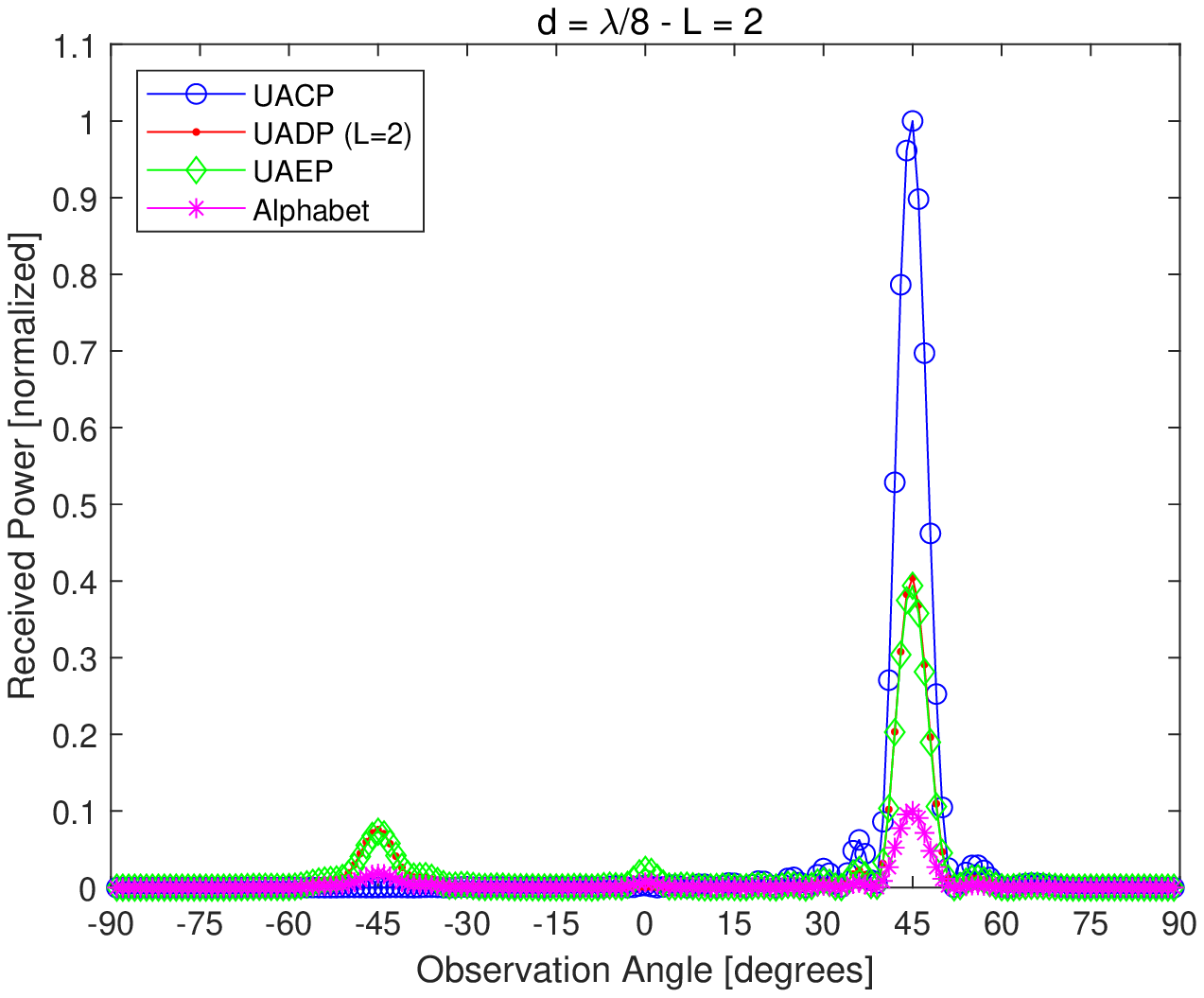}
\vspace{-0.25cm} \caption{Received power as a function of the angle of observation (near-field case with the receiver located 5 meters far from the RIS). The RIS alphabet is \cite{Hongliang_OmniSurface}, the desired angle of reflection is 45 degrees, and the inter-distance is $d=\lambda/8$.}\label{fig:Prx__45deg_lambda8_Q1__3p6__NearField} \vspace{-0.37cm}
\end{figure}

\begin{figure}[!t]
\includegraphics[width=0.62\columnwidth]{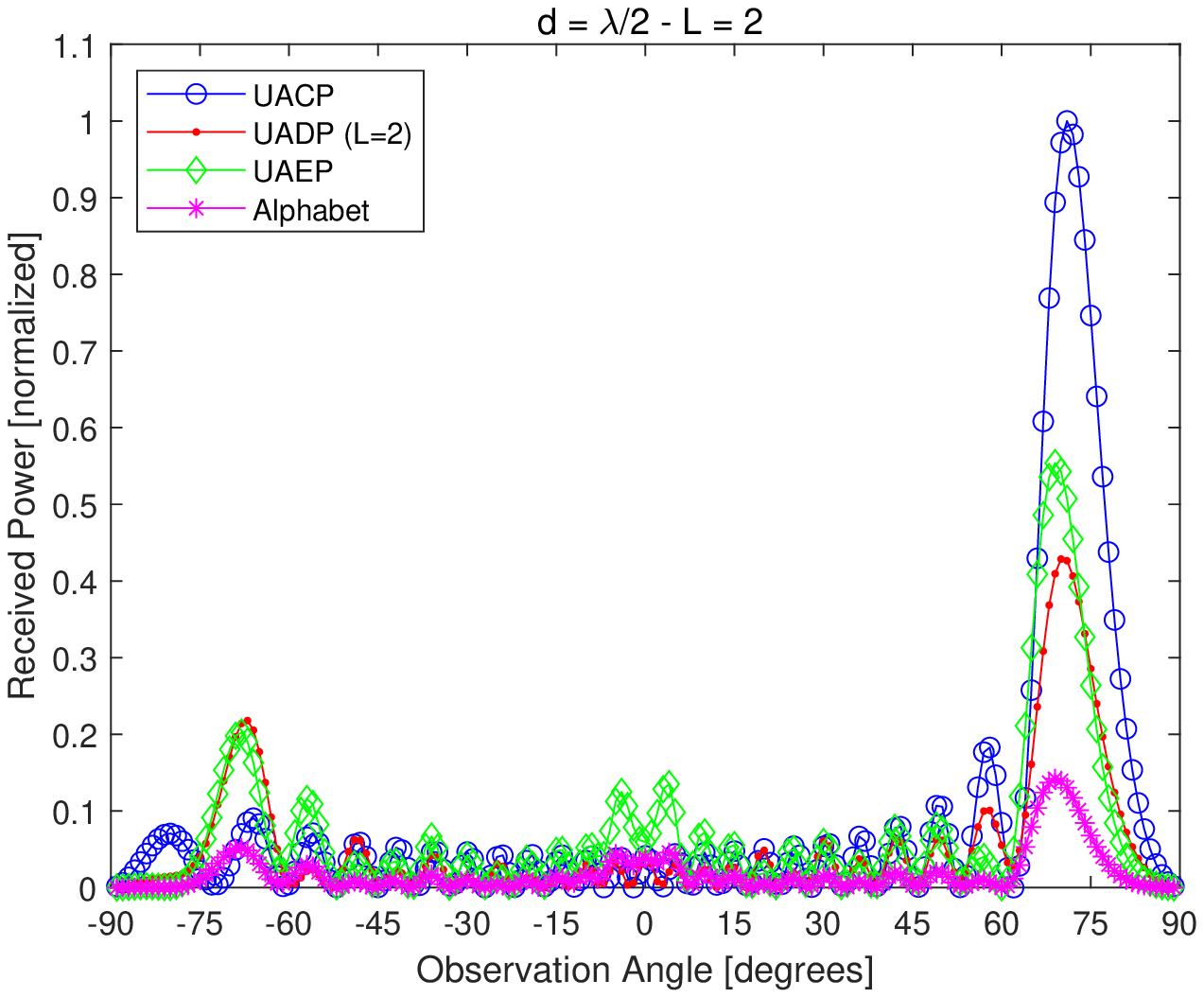}
\vspace{-0.25cm} \caption{Received power as a function of the angle of observation (near-field case with the receiver located 5 meters far from the RIS). The RIS alphabet is \cite{Hongliang_OmniSurface}, the desired angle of reflection is 75 degrees, and the inter-distance is $d=\lambda/2$.}\label{fig:Prx__75deg_lambda2_Q1__3p6__NearField} \vspace{-0.37cm}
\end{figure}

\begin{figure}[!t]
\includegraphics[width=0.62\columnwidth]{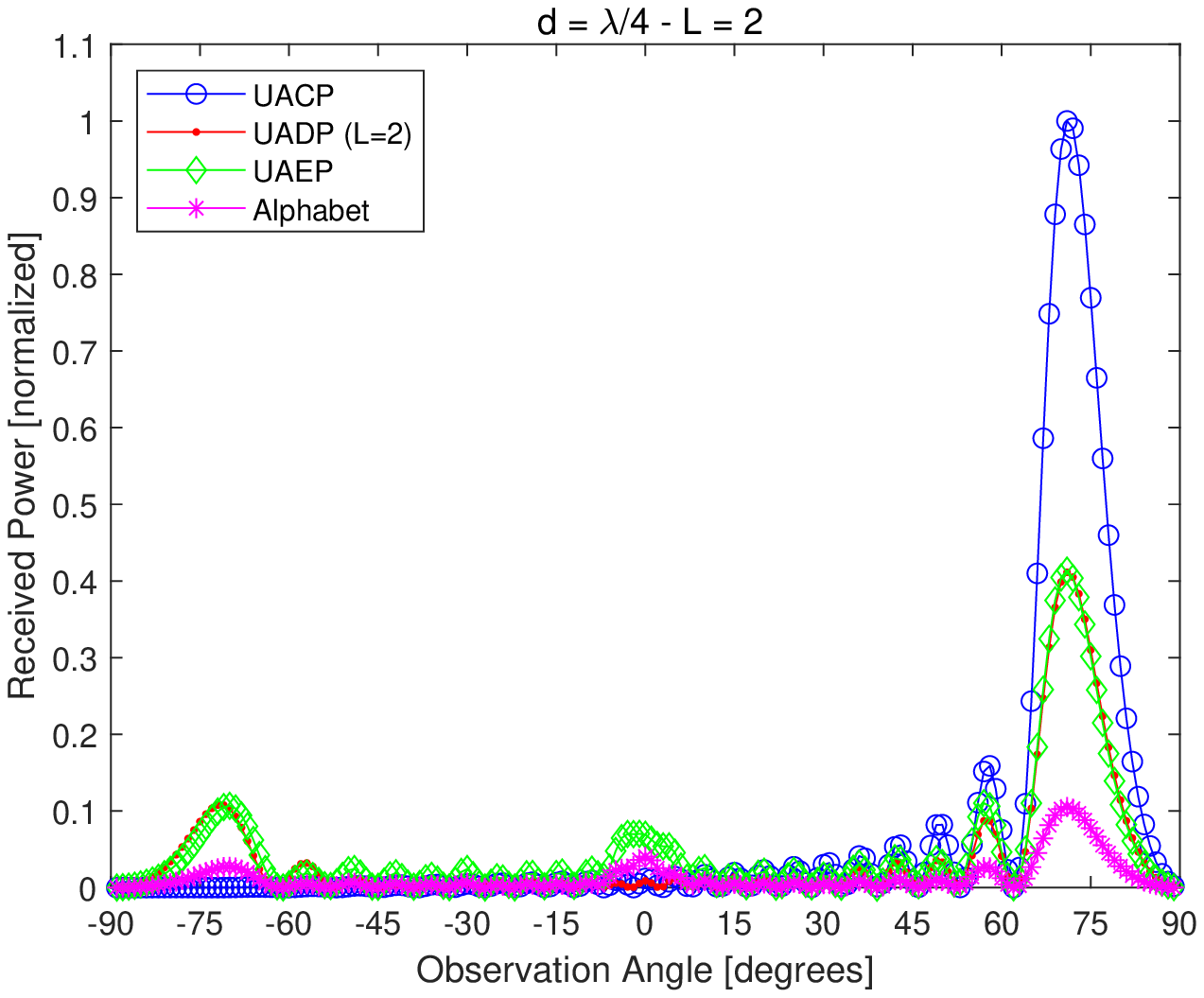}
\vspace{-0.25cm} \caption{Received power as a function of the angle of observation (near-field case with the receiver located 5 meters far from the RIS). The RIS alphabet is \cite{Hongliang_OmniSurface}, the desired angle of reflection is 75 degrees, and the inter-distance is $d=\lambda/4$.}\label{fig:Prx__75deg_lambda4_Q1__3p6__NearField} \vspace{-0.37cm}
\end{figure}

\begin{figure}[!t]
\includegraphics[width=0.62\columnwidth]{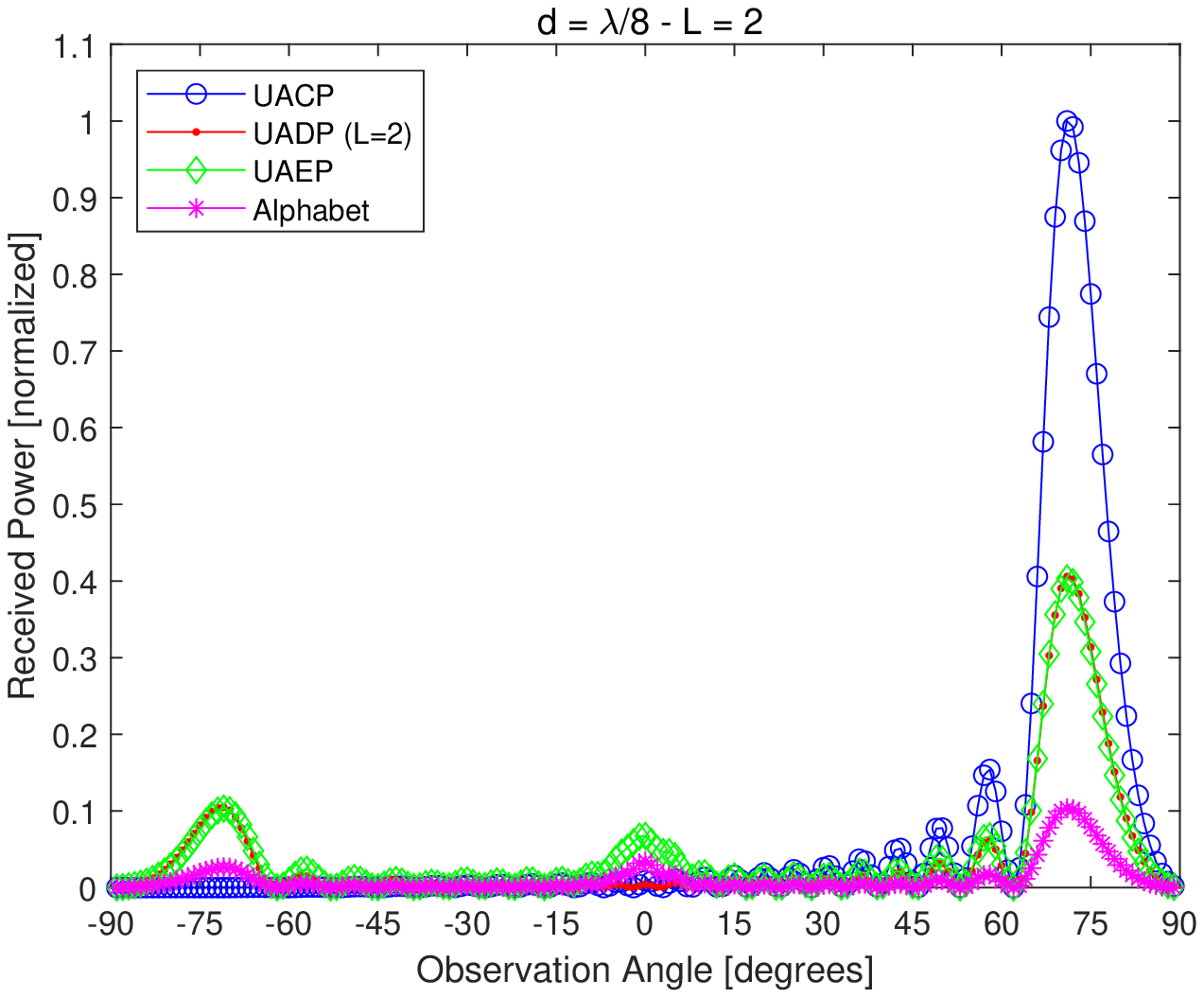}
\vspace{-0.25cm} \caption{Received power as a function of the angle of observation (near-field case with the receiver located 5 meters far from the RIS). The RIS alphabet is \cite{Hongliang_OmniSurface}, the desired angle of reflection is 75 degrees, and the inter-distance is $d=\lambda/8$.}\label{fig:Prx__75deg_lambda8_Q1__3p6__NearField} \vspace{-0.37cm}
\end{figure}

\begin{figure}[!t]
\includegraphics[width=0.62\columnwidth]{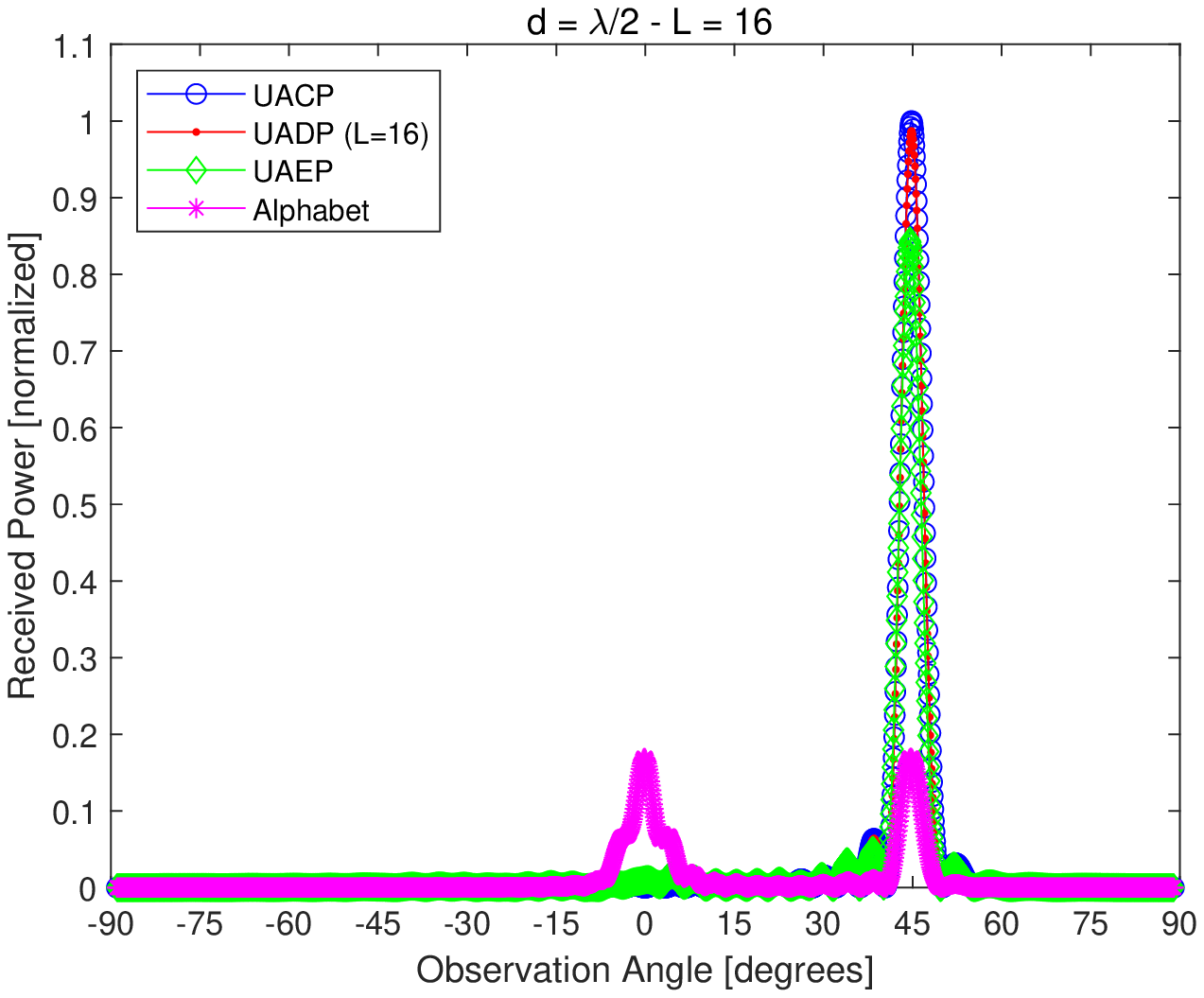}
\vspace{-0.25cm} \caption{Received power as a function of the angle of observation (near-field case with the receiver located 5 meters far from the RIS). The RIS alphabet is \cite{Romain_RIS-Prototype}, the desired angle of reflection is 45 degrees, and the inter-distance is $d=\lambda/2$.}\label{fig:Prx__45deg_lambda2_Q1__5p2Orange__NearField} \vspace{-0.37cm}
\end{figure}

\begin{figure}[!t]
\includegraphics[width=0.62\columnwidth]{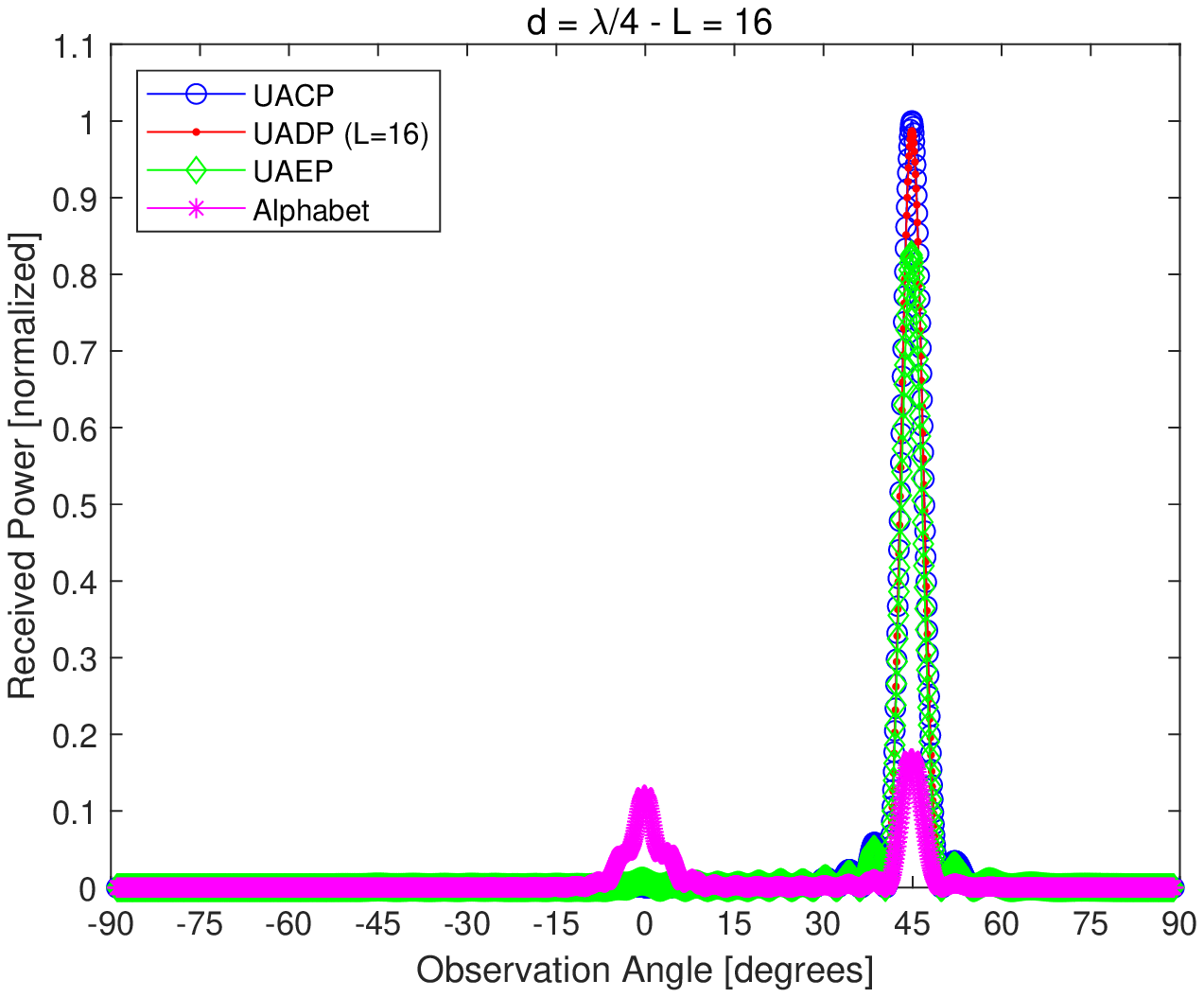}
\vspace{-0.25cm} \caption{Received power as a function of the angle of observation (near-field case with the receiver located 5 meters far from the RIS). The RIS alphabet is \cite{Romain_RIS-Prototype}, the desired angle of reflection is 45 degrees, and the inter-distance is $d=\lambda/4$.}\label{fig:Prx__45deg_lambda4_Q1__5p2Orange__NearField} \vspace{-0.37cm}
\end{figure}

\begin{figure}[!t]
\includegraphics[width=0.62\columnwidth]{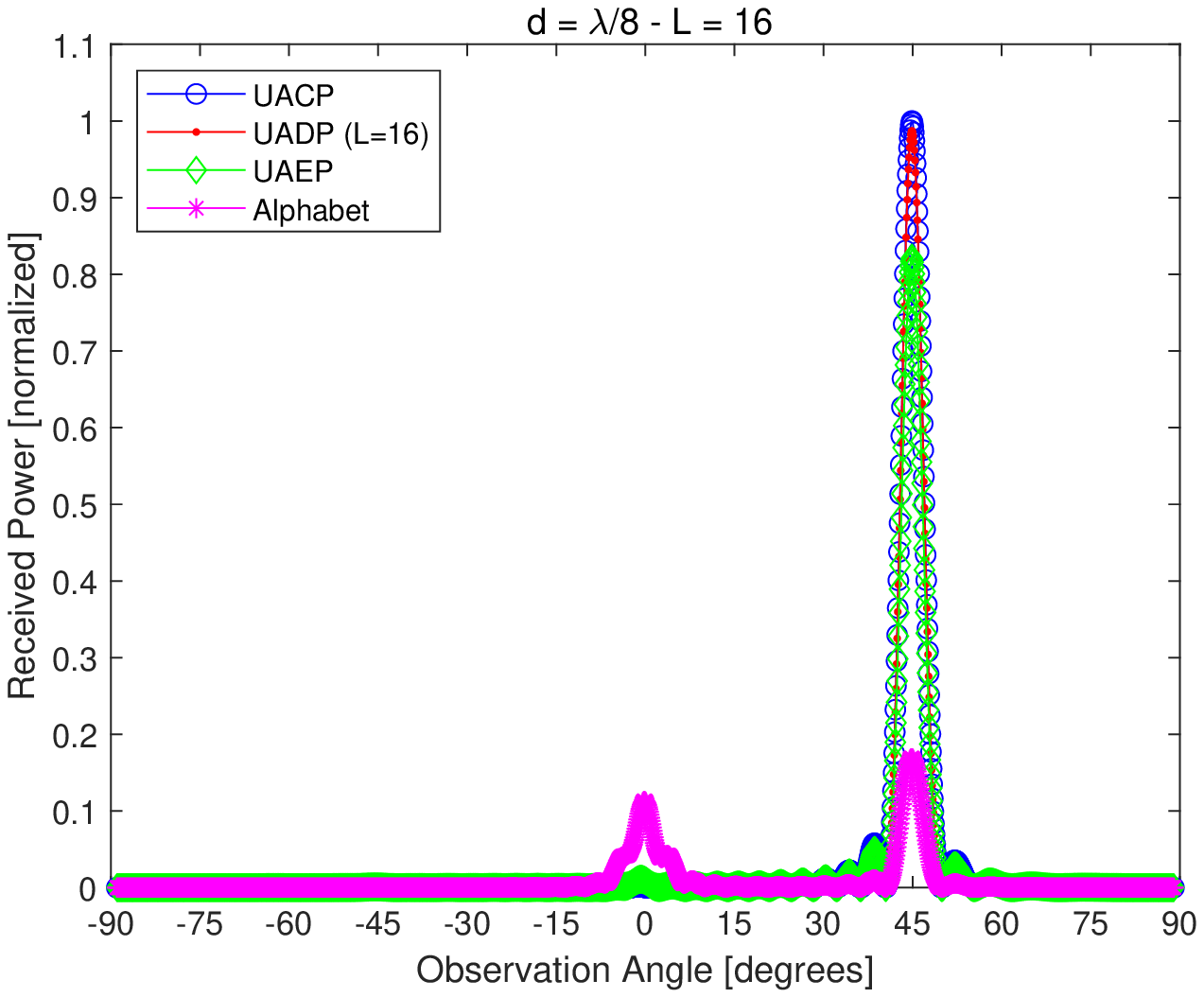}
\vspace{-0.25cm} \caption{Received power as a function of the angle of observation (near-field case with the receiver located 5 meters far from the RIS). The RIS alphabet is \cite{Romain_RIS-Prototype}, the desired angle of reflection is 45 degrees, and the inter-distance is $d=\lambda/8$.}\label{fig:Prx__45deg_lambda8_Q1__5p2Orange__NearField} \vspace{-0.37cm}
\end{figure}

\begin{figure}[!t]
\includegraphics[width=0.62\columnwidth]{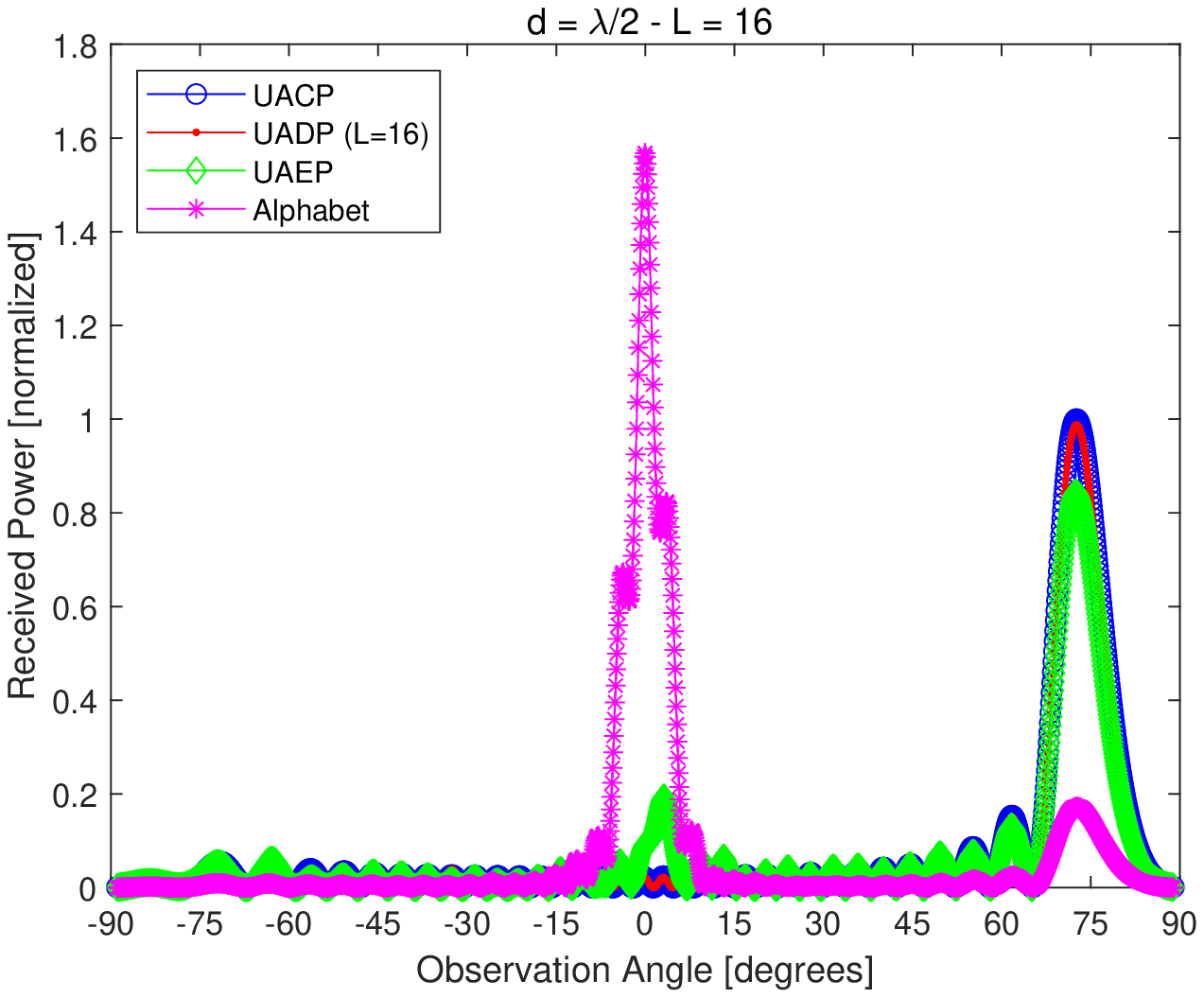}
\vspace{-0.25cm} \caption{Received power as a function of the angle of observation (near-field case with the receiver located 5 meters far from the RIS). The RIS alphabet is \cite{Romain_RIS-Prototype}, the desired angle of reflection is 75 degrees, and the inter-distance is $d=\lambda/2$.}\label{fig:Prx__75deg_lambda2_Q1__5p2Orange__NearField} \vspace{-0.37cm}
\end{figure}

\begin{figure}[!t]
\includegraphics[width=0.62\columnwidth]{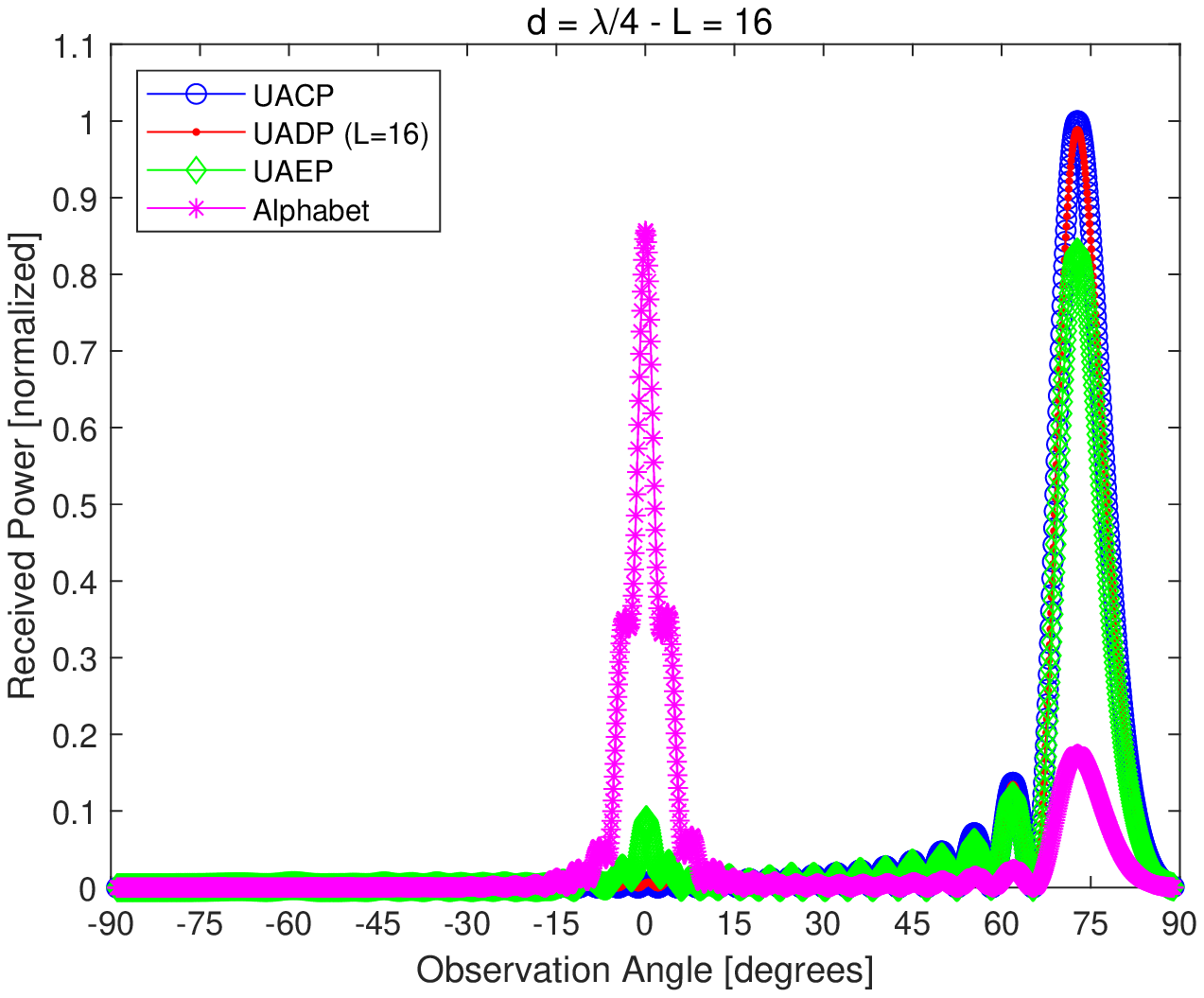}
\vspace{-0.25cm} \caption{Received power as a function of the angle of observation (near-field case with the receiver located 5 meters far from the RIS). The RIS alphabet is \cite{Romain_RIS-Prototype}, the desired angle of reflection is 75 degrees, and the inter-distance is $d=\lambda/4$.}\label{fig:Prx__75deg_lambda4_Q1__5p2Orange__NearField} \vspace{-0.37cm}
\end{figure}

\begin{figure}[!t]
\includegraphics[width=0.62\columnwidth]{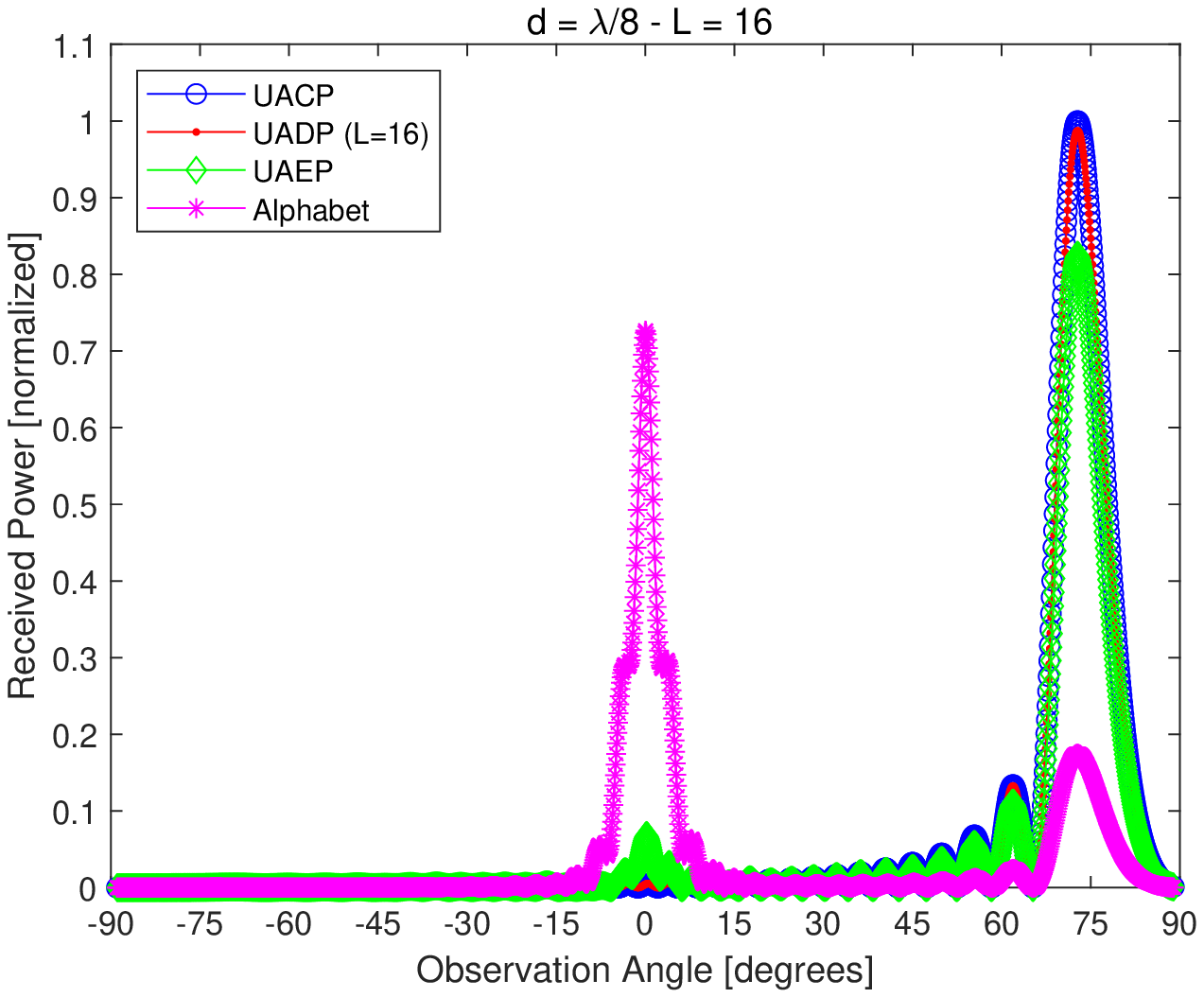}
\vspace{-0.25cm} \caption{Received power as a function of the angle of observation (near-field case with the receiver located 5 meters far from the RIS). The RIS alphabet is \cite{Romain_RIS-Prototype}, the desired angle of reflection is 75 degrees, and the inter-distance is $d=\lambda/8$.}\label{fig:Prx__75deg_lambda8_Q1__5p2Orange__NearField} \vspace{-0.37cm}
\end{figure}

\begin{figure}[!t]
\includegraphics[width=0.7\columnwidth]{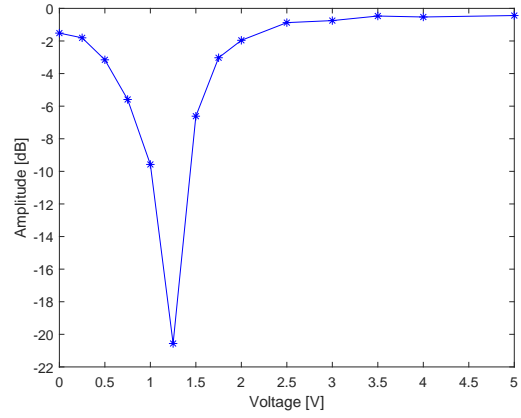}
\vspace{-0.25cm} \caption{Amplitude of the RIS alphabet in Table \ref{Table_RelectionCoefficientContinuous}.}\label{fig:CodebookOrange_Amplitude} \vspace{-0.37cm}
\end{figure}

\begin{figure}[!t]
\includegraphics[width=0.7\columnwidth]{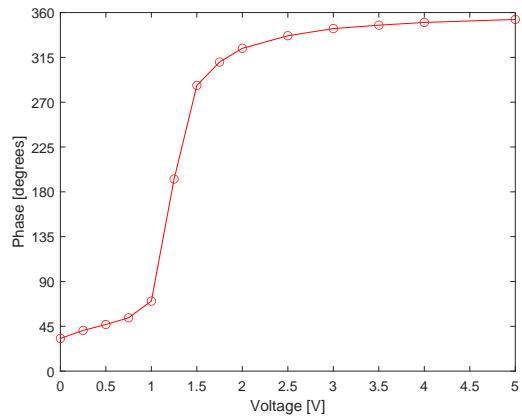}
\vspace{-0.25cm} \caption{Phase of the RIS alphabet in Table \ref{Table_RelectionCoefficientContinuous}.}\label{fig:CodebookOrange_Phase} \vspace{-0.37cm}
\end{figure}

\begin{figure}[!t]
\includegraphics[width=0.7\columnwidth]{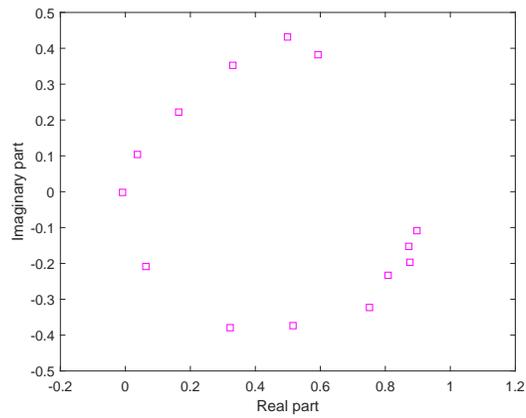}
\vspace{-0.25cm} \caption{Representation in the complex plane of the RIS alphabet in Table \ref{Table_RelectionCoefficientContinuous}.}\label{fig:CodebookOrange_Polar} \vspace{-0.37cm}
\end{figure}

In the considered case studies, we have noted that an RIS with ${L}=2$ usually provides a radiation pattern with an unwanted reradiated beam towards the direction that is symmetrical with respect to the desired direction of reradiation. These conclusions were drawn under the assumption that the receiver and the transmitter are located in the far-field region of the RIS. Therefore, it is interesting to analyze the reradiated electromagnetic field when the receiver is in the near-field region of the RIS, while the transmitter is still kept in the far-field region of the RIS. This corresponds to the case in which the RIS is located close to the receiver. This case study is illustrated in Figs. \ref{fig:Prx__45deg_lambda2_Q1__3p6__NearField}-\ref{fig:Prx__75deg_lambda8_Q1__3p6__NearField} for the RIS alphabet in \cite{Hongliang_OmniSurface}. We note that  the unwanted beams still exist but they are less strong towards the undesired direction of reradiation. Therefore, there exist differences between the reradiated power in the near-field and far-field regions of an RIS, and the use of binary RISs may be acceptable in the near-field region. We observe, however, that the peak of the reradiated power is still not exactly centered towards the desired direction of reradiation, if the angle of deflection is large. The most important observation from the considered case study is that the unwanted reradiated beam towards the direction that is symmetrical with respect to the desired direction of reradiation has a much smaller intensity as compared with the desired beam. Their difference is more than $5$ dB if the desired angle of reradiation is $45$ degrees and approximately $4$ dB if the desired angle of reradiation is $75$ degrees. In the far-field scenario, on the other hand, the two beam have almost the same intensity. A similar trend can be observed in Figs. \ref{fig:Prx__45deg_lambda2_Q1__5p2Orange__NearField}-\ref{fig:Prx__75deg_lambda8_Q1__5p2Orange__NearField} where the RIS alphabet in \cite{Romain_RIS-Prototype} is considered. In this case, we note that the unwanted beam towards the specular direction is significantly attenuated  compared with the corresponding setup in the far-field region. However, we note that the beam towards the specular direction has a smaller amplitude than the desired beam if the inter-distance between the RIS reconfiguable elements is smaller than $d=\lambda/2$ and if the angle of reflection is not too large compared with the angle of incidence. In fact, we observe that a non-negligible specular beam still exists if the desired angle of reradiation is $75$ degrees. This was somehow expected, since the RIS in \cite{Romain_RIS-Prototype} was designed to operate for angles that lie in the range $[-45, 45]$ degrees. The main reason for the presence of a strong beam towards the specular direction can be understood by direct inspection of Table \ref{Table_RelectionCoefficientContinuous}. For ease of discussion, the reflection coefficient in Table \ref{Table_RelectionCoefficientContinuous} is illustrated in Figs. \ref{fig:CodebookOrange_Amplitude}, \ref{fig:CodebookOrange_Phase}, \ref{fig:CodebookOrange_Polar}. We note that the phase shifts that can be realized lie in the range $[30, 350]$ degrees, and, more important, the attenuation for the phases that lie in the range $[90, 270]$ degrees is much higher than the attenuation for the phases that lie in the range $[-90, 90]$ degrees. This results in a non-negligible unbalance of the amplitudes between the values of the reflection coefficient with a positive and negative real part, as illustrated in Fig. \ref{fig:CodebookOrange_Polar}. As a result, the reflection coefficient is not centered around the origin, and this leads to a non-negligible specular reflection for any angle of incidence.  A possible solution to alleviate this issue is to reduce the attenuation for the phases in the range $[90, 270]$. Based on the full-wave simulations reported in \cite{Romain_RIS-Prototype}, the attenuation for this range of phases was expected to be much smaller than that obtained from the experimental measurements. In the near-field region, nevertheless, we note that the beam towards the desired direction of radiation is  $1.44$ dB and $1.74$ dB higher than the beam towards the specular direction for $d=\lambda/4$ and $d=\lambda/8$, respectively, and the desired angle of reradiation is $45$ degrees.

\begin{figure}[!t]
\includegraphics[width=1.0\columnwidth]{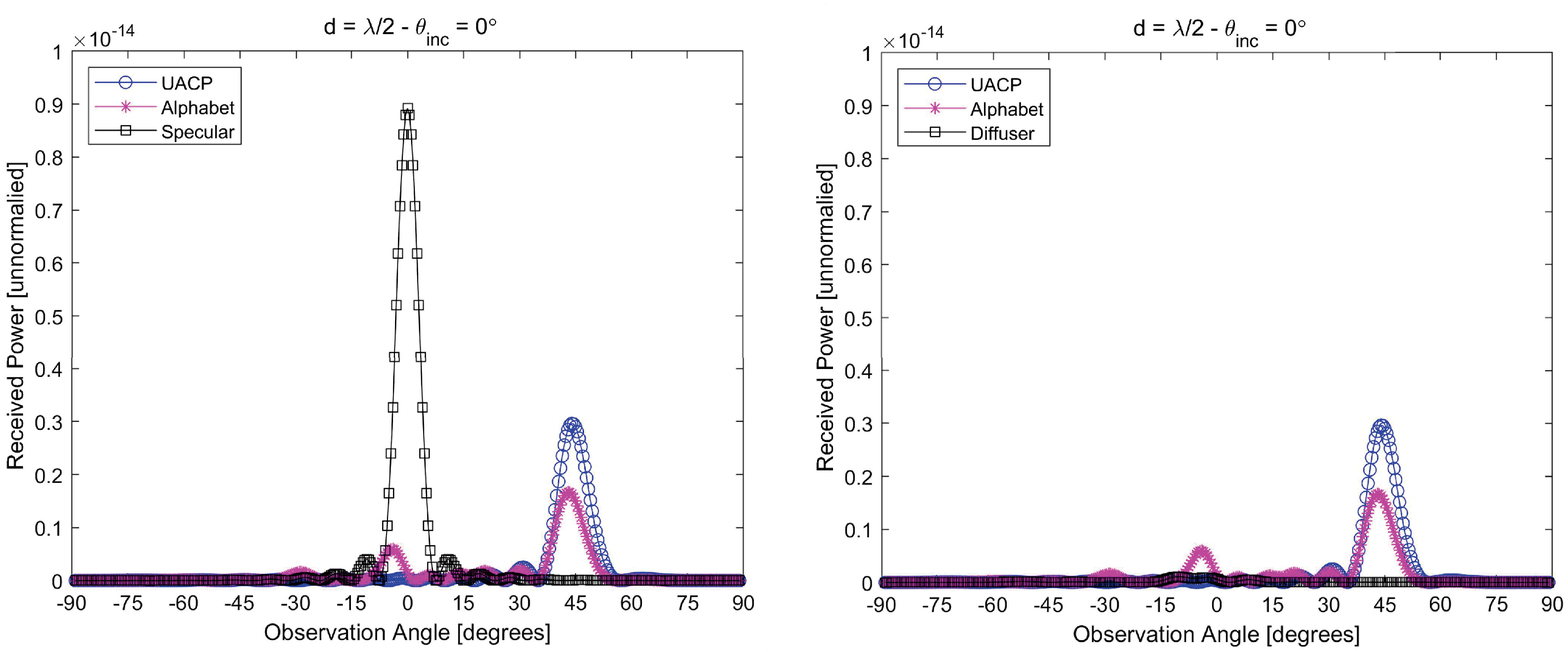}
\vspace{-0.25cm} \caption{Received power as a function of the angle of observation (intended user). The RIS alphabet is \cite{Linglong_Testbed}, the desired angle of reflection is 45 degrees, and the inter-distance is $d=\lambda/2$.}\label{fig:Prx__45deg_lambda2_Interference0} \vspace{-0.37cm}
\end{figure}

\begin{figure}[!t]
\includegraphics[width=1.0\columnwidth]{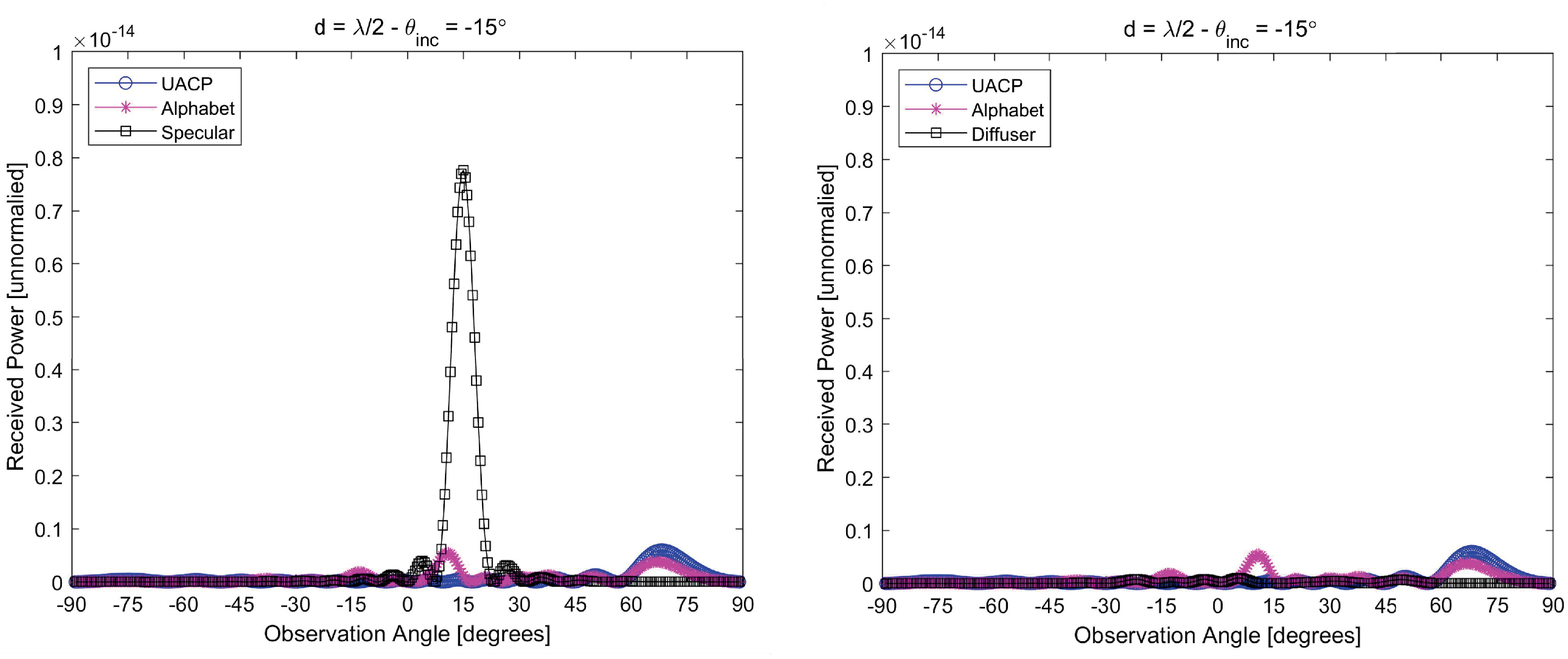}
\vspace{-0.25cm} \caption{Received power as a function of the angle of observation (interferer at $\theta_{\rm{inc}} = - 15$ degrees). The RIS alphabet is \cite{Linglong_Testbed}, the desired angle of reflection is 45 degrees, and the inter-distance is $d=\lambda/2$.}\label{fig:Prx__45deg_lambda2_Interference15} \vspace{-0.37cm}
\end{figure}

\begin{figure}[!t]
\includegraphics[width=1.0\columnwidth]{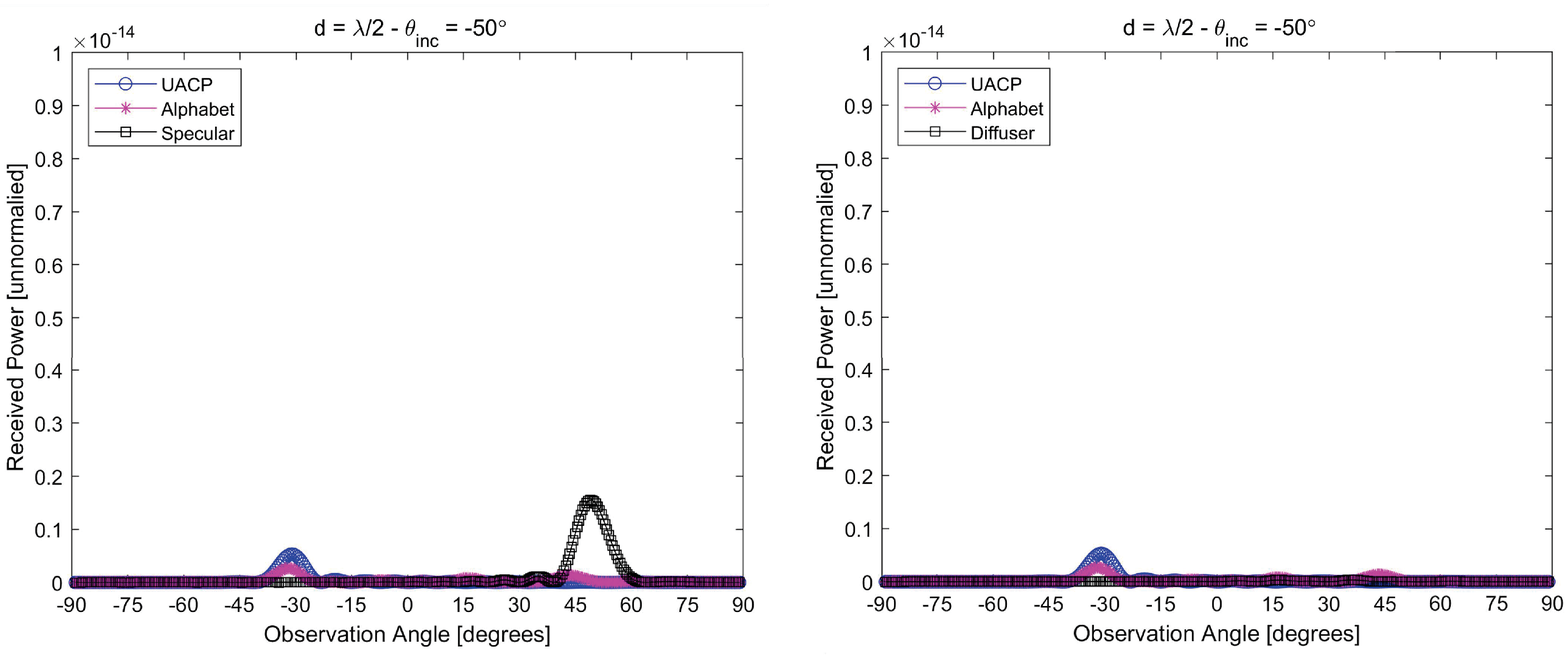}
\vspace{-0.25cm} \caption{Received power as a function of the angle of observation (interferer at $\theta_{\rm{inc}} = -50$ degrees). The RIS alphabet is \cite{Linglong_Testbed}, the desired angle of reflection is 45 degrees, and the inter-distance is $d=\lambda/2$.}\label{fig:Prx__45deg_lambda2_Interference50} \vspace{-0.37cm}
\end{figure}

\begin{figure}[!t]
\includegraphics[width=1.0\columnwidth]{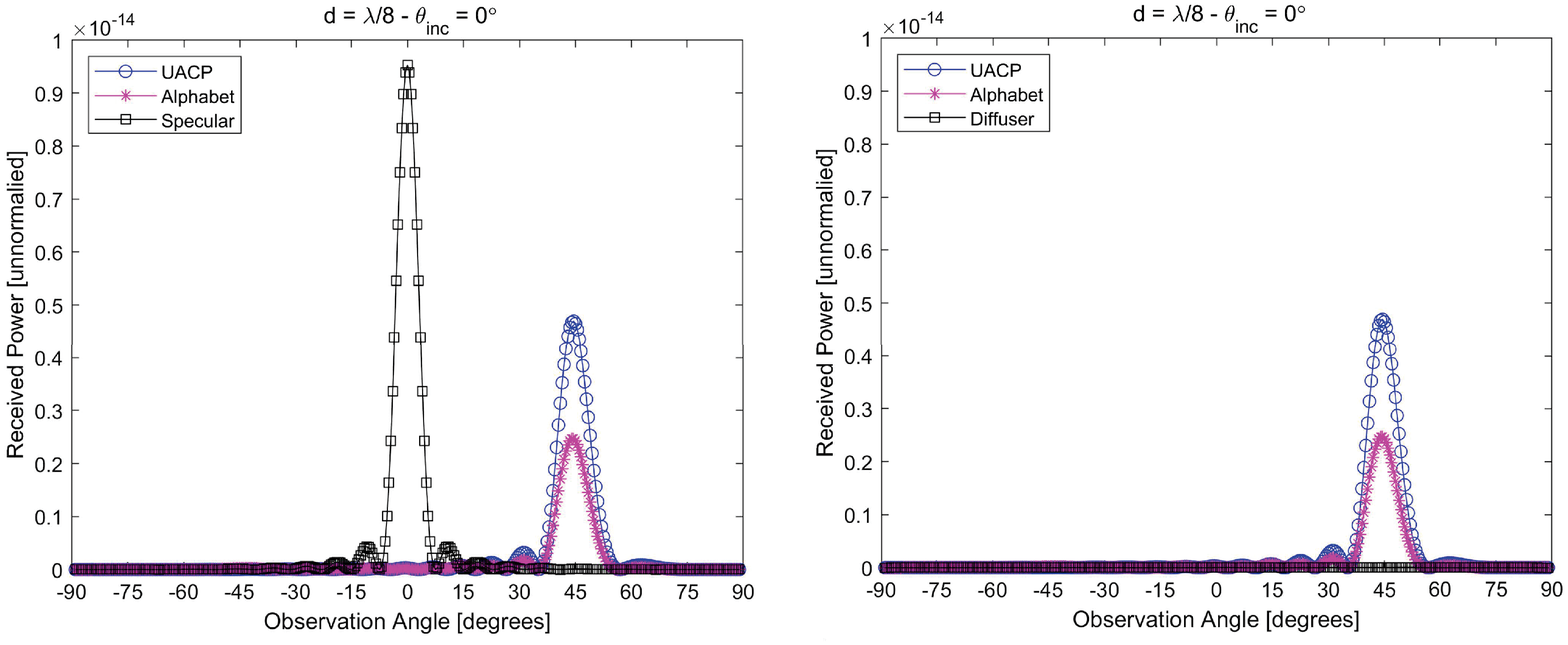}
\vspace{-0.25cm} \caption{Received power as a function of the angle of observation (intended user). The RIS alphabet is \cite{Linglong_Testbed}, the desired angle of reflection is 45 degrees, and the inter-distance is $d=\lambda/8$.}\label{fig:Prx__45deg_lambda8_Interference0} \vspace{-0.37cm}
\end{figure}

\begin{figure}[!t]
\includegraphics[width=1.0\columnwidth]{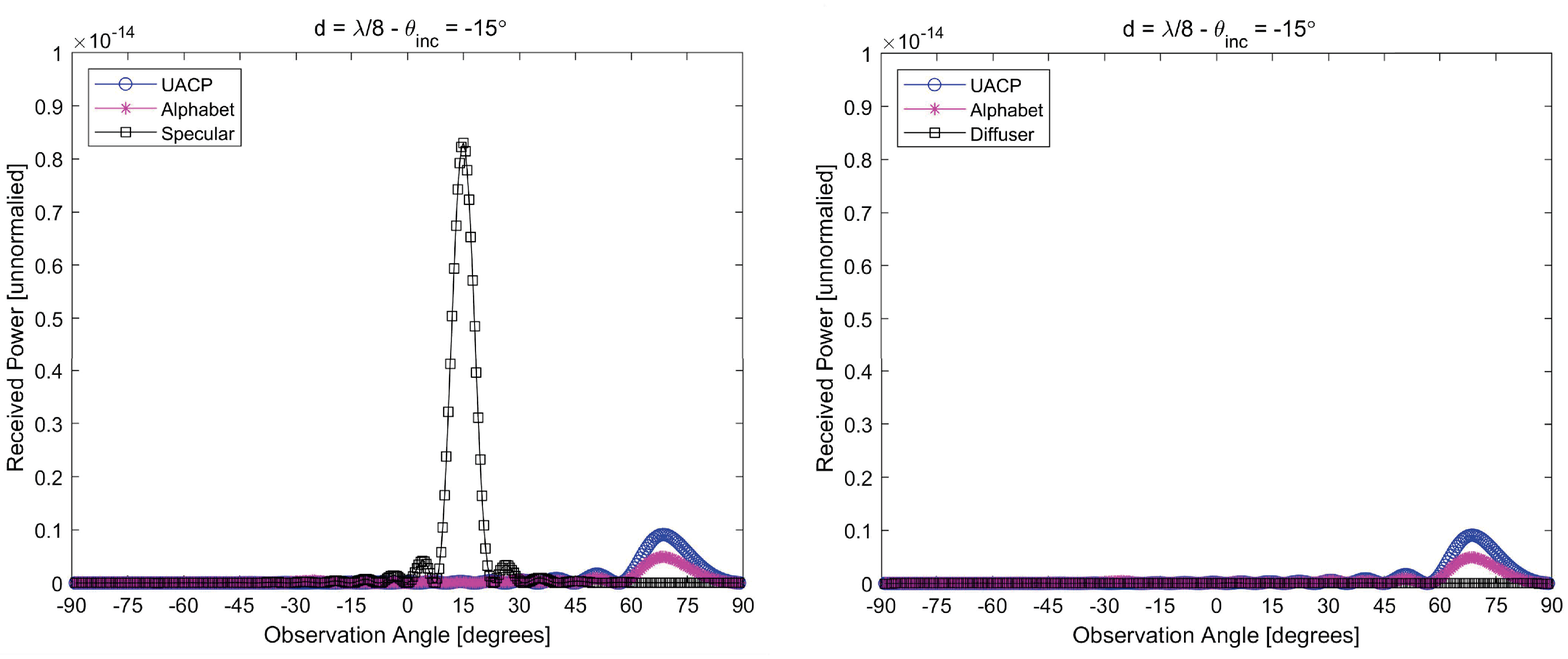}
\vspace{-0.25cm} \caption{Received power as a function of the angle of observation (interferer at $\theta_{\rm{inc}} = - 15$ degrees). The RIS alphabet is \cite{Linglong_Testbed}, the desired angle of reflection is 45 degrees, and the inter-distance is $d=\lambda/8$.}\label{fig:Prx__45deg_lambda8_Interference15} \vspace{-0.37cm}
\end{figure}

\begin{figure}[!t]
\includegraphics[width=1.0\columnwidth]{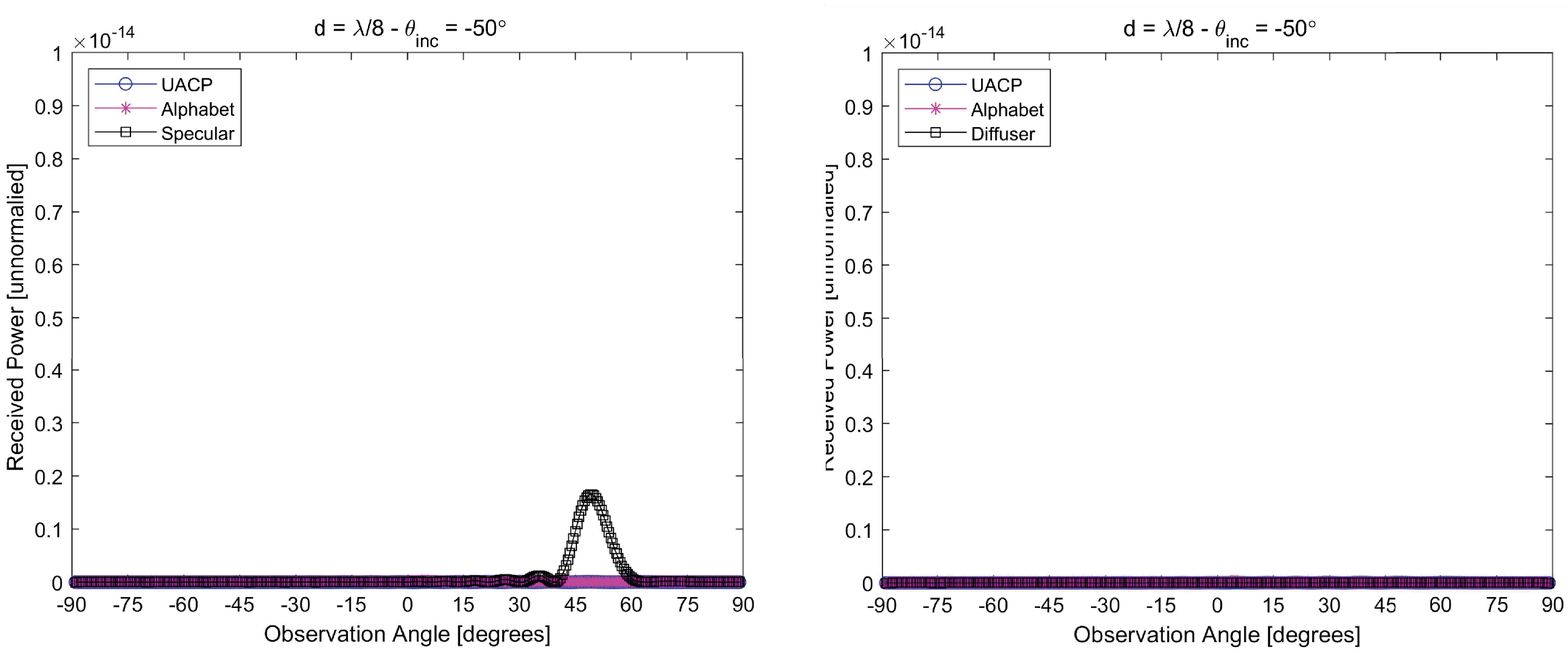}
\vspace{-0.25cm} \caption{Received power as a function of the angle of observation (interferer at $\theta_{\rm{inc}} = -50$ degrees). The RIS alphabet is \cite{Linglong_Testbed}, the desired angle of reflection is 45 degrees, and the inter-distance is $d=\lambda/8$.}\label{fig:Prx__45deg_lambda8_Interference50} \vspace{-0.37cm}
\end{figure}

\begin{figure}[!t]
\includegraphics[width=0.75\columnwidth]{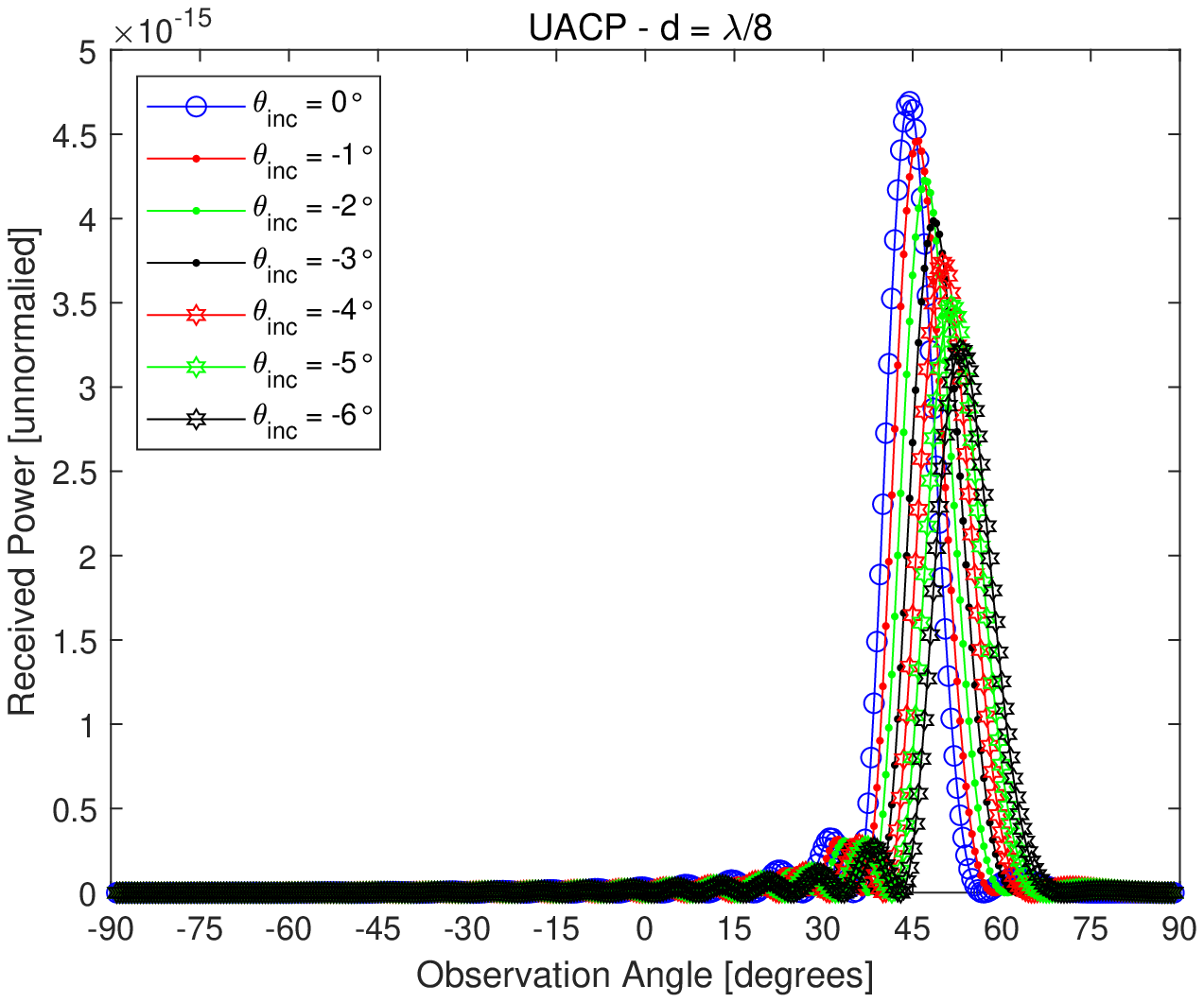}
\vspace{-0.25cm} \caption{Received power as a function of the angle of observation (intended user and interferer from $\theta_{\rm{inc}} = -6$ to $\theta_{\rm{inc}} = - 1$ degrees). The RIS alphabet is UACP, the desired angle of reflection is 45 degrees, and the inter-distance is $d=\lambda/8$.}\label{fig:Prx__45deg_lambda8_InterferenceRangeOPT} \vspace{-0.37cm}
\end{figure}

\begin{figure}[!t]
\includegraphics[width=0.75\columnwidth]{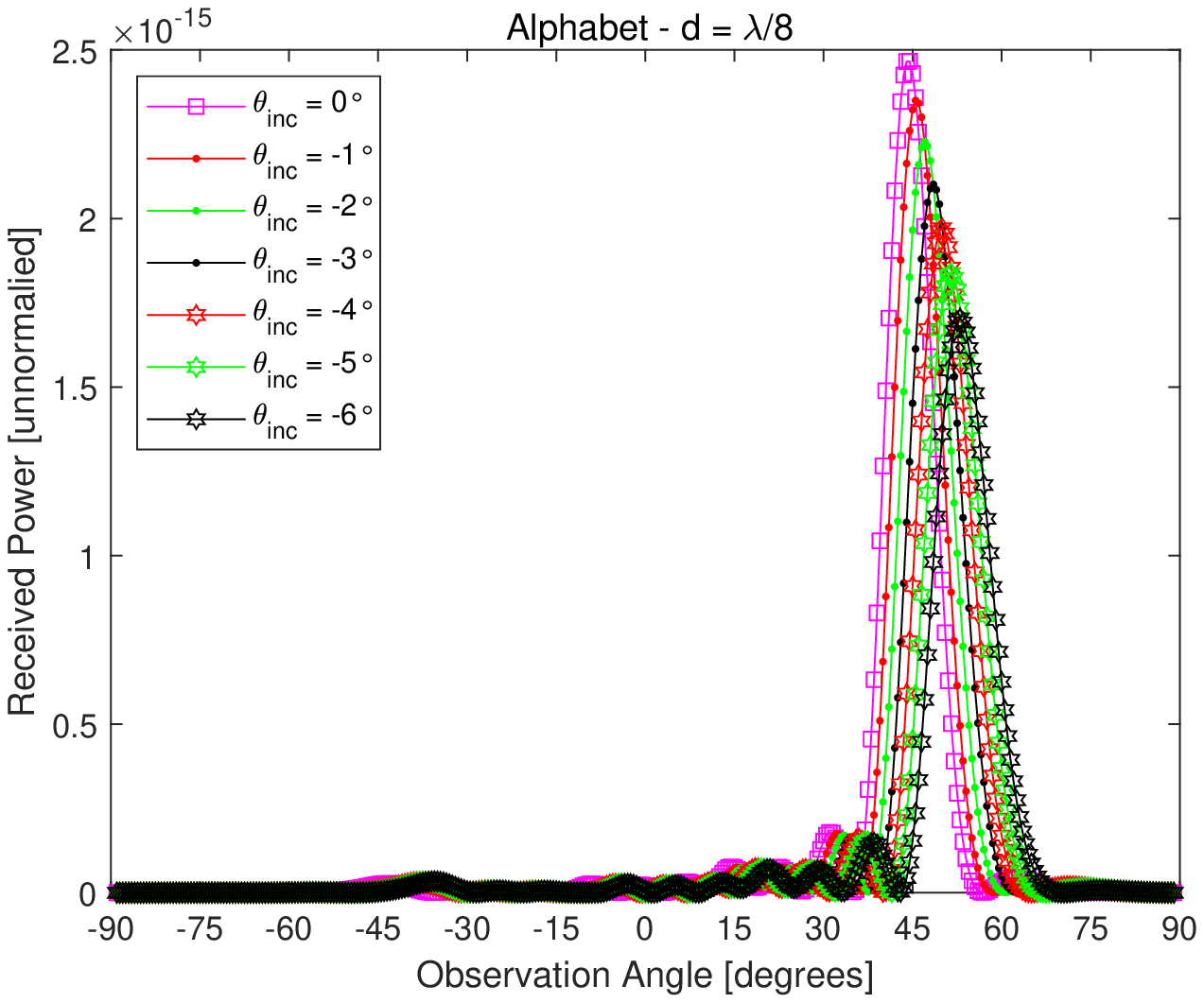}
\vspace{-0.25cm} \caption{Received power as a function of the angle of observation (intended user and interferer from $\theta_{\rm{inc}} = -6$ to $\theta_{\rm{inc}} = - 1$ degrees). The RIS alphabet is \cite{Linglong_Testbed}, the desired angle of reflection is 45 degrees, and the inter-distance is $d=\lambda/8$.}\label{fig:Prx__45deg_lambda8_InterferenceRangeAlphabet} \vspace{-0.37cm}
\end{figure}

The results illustrated so far assume that an RIS is designed and optimized for reradiating an electromagnetic wave that impinges upon the surface from a known direction towards another direction that is known as well. It is important to analyze, however, the reradiation properties of an optimized RIS when it is illuminated by unknown electromagnetic waves that originate from directions that are different from that of design. In wireless communications, this is the typical case study in which an optimized RIS operates in the presence of electromagnetic interference. This case study is illustrated in Figs. \ref{fig:Prx__45deg_lambda2_Interference0}-\ref{fig:Prx__45deg_lambda8_InterferenceRangeAlphabet} by considering the RIS alphabet in \cite{Linglong_Testbed}. More specifically, it is assumed that the RIS is first optimized in order to reradiate an electromagnetic wave from the normal direction towards a given direction of reradiation. Then, the optimized RIS is illuminated by an interfering electromagnetic wave from the direction $\theta_{\rm{inc}} \ne 0$. It needs to be mentioned that the RIS alphabets in Table \ref{Table_RelectionCoefficientDiscrete} and Table \ref{Table_RelectionCoefficientContinuous} are obtained by assuming normal incidence. Usually, however, an RIS alphabet is angle-dependent. Therefore, it changes with the angle of incidence. In our qualitative study this dependency is not considered, since the angle-dependency of the RIS alphabets in Table \ref{Table_RelectionCoefficientDiscrete} and Table \ref{Table_RelectionCoefficientContinuous} is not available. For comparison, in addition, we report the reradiation patterns that correspond to two surfaces that operate as a specular reflector and as a diffuser. In the first case, the reflection coefficients $\gamma_{nm}$ are set equal to one, and in the second case the reflection coefficients $\gamma_{nm}$ have unit amplitude and a random phase. The figures show that the RIS reradiates the interfering signal towards a direction that is different from the location of the intended receiver. Therefore, an interfering electromagnetic wave may not be harmful for the intended receiver. However, the RIS may produce reradiated beams  towards other directions that depend on the angle of incidence of the interfering electromagnetic wave and the phase modulation applied by the RIS. More specifically, the results illustrated in Figs. \ref{fig:Prx__45deg_lambda8_InterferenceRangeOPT} and \ref{fig:Prx__45deg_lambda8_InterferenceRangeAlphabet} show that an interfering electromagnetic wave results in a small interference towards the desired direction of reradiation if the angle of incidence is sufficiently different from the nominal angle of incidence. The example illustrated in Figs. \ref{fig:Prx__45deg_lambda8_InterferenceRangeOPT} and \ref{fig:Prx__45deg_lambda8_InterferenceRangeAlphabet} shows that, given the width of the obtained reradiated beam, an interfering electromagnetic wave that illuminates the RIS from a direction that is at least five degrees different from the nominal direction of reradiation does not interfere with the intended receiver. However, it generates interference towards nearby locations, which need to be accounted for in the context of network-level analysis and optimization. This behavior is inherently due to the phase modulation applied by the RIS.

\section{Conclusions}\label{sec:Conclusions}
In this chapter, we have reported a comprehensive numerical study of the reradiation properties of RISs that are modeled as digital metasurfaces. The considered RISs are characterized by realistic alphabets that are obtained from full-wave simulations and from existing experimental testbeds at different operating frequencies. We have shown that the size and the inter-distance between the RIS reconfigurable elements, and the quantization of the amplitudes and phases of the reflection coefficients both determine the quality of the reradiated pattern. More specifically, it is usually convenient to design an RIS with an inter-distance and a size of the RIS reconfigurable elements smaller than half of the wavelength if the reflection coefficients cannot be continuously adjusted. If the reflection coefficients can be continuously adjusted, on the other hand, the advantage of using sub-wavelength RISs may be negligible in the considered case studies of RISs optimized according to a locally optimum design criterion. The use of binary RIS usually results in the presence of strong unwanted beams towards the direction that is symmetric with respect to the nominal direction of reradiation. This effect is, however, less evident in the near-field region of the RIS. Similarly, RISs that are characterized by alphabets in which the amplitude and the phase are not independent of each other and in which there are strong variations of the amplitude with the phase may result in strong reradiated beams towards other directions. The considered example has revealed the presence of a strong specular reflection. Furthermore, we have shown that the main peak of the reradiated pattern may not be perfectly steered towards the nominal direction of reradiation for large angles of deflection. Finally, we have analyzed the reradiation characteristics of RISs in the presence of electromagnetic interference, and have shown that an RIS inherently filters out the interference towards the desired direction of reradiation, provided that the interfering electromagnetic waves originate from directions that are sufficiently different from the direction of incidence of the intended user. However, the interfering electromagnetic waves may be steered towards other directions, and, therefore, their impact needs to be accounted for in the context of network-level studies. 

In summary, the study conducted in this chapter allows us to conclude that RISs have so far been designed to maximize some objective functions for specific user locations. However, the presence of reradiated beams towards directions different from the nominal ones and the impact of interfering signals have usually been ignored at the design stage. Implementation constraints, such as the need of using quantized reflection coefficients or the inherent interplay between the phase and the amplitude of the reflection coefficients, may result in unwanted reradiated beams and interfering signals that need to be kept under control at the design stage. It is therefore important to optimize an RIS by taking into account the entire reradiation pattern. If large angles of deflection need to be realized, non-local design criteria may need to be applied in order to ensure that the reradiated beam is directed towards the desired direction of reradiation.

\section*{Acknowledgment}
This research work was supported by the European Commission through the H2020 ARIADNE project under grant agreement number 871464, the H2020 RISE-6G project under grant agreement number 101017011, the H2020 MetaWireless project under grant agreement number 956256, the H2020 PAINLESS project under grant agreement number 812991, and the Fulbright Foundation under the ``Programme National Chercheurs 2021'' funding scheme.

\clearpage

%


%
%


\clearpage

%
%
%
%
%
%
%
%
%
%


\clearpage

%


%
%


%
%


\clearpage


\clearpage

\bibliographystyle{elsarticle-num}
\bibliography{Ref.bib}
\end{document}